\documentclass[longauth]{aa}  

\usepackage{txfonts}
\usepackage{mathtools}

\usepackage{graphicx}
\usepackage[dvipsnames]{xcolor}
\usepackage{url}
\usepackage{subcaption}

\usepackage{hyperref}
\hypersetup{
    colorlinks=true,   
    allcolors=blue
}

\begin{document}

   \title{The ALMA-CRISTAL survey: Resolved kinematic studies of main sequence star-forming galaxies at $4$\,$<$\,$z$\,$<$\,$6$}

   \subtitle{}

   \titlerunning{  }

    \author{
    Lilian~L.~Lee\inst{1}
    \and Natascha~M.~Förster~Schreiber\inst{1}
    \and Rodrigo~Herrera-Camus\inst{2,3}
    \and Daizhong~Liu\inst{4}
    \and Sedona~H.~Price\inst{5,6}
    \and Reinhard~Genzel\inst{1,7}
    \and Linda~J.~Tacconi\inst{1}
    \and Dieter~Lutz\inst{1}
    \and Ric~Davies\inst{1}
    \and Thorsten~Naab\inst{8}
    \and Hannah~Übler\inst{1}
    \and Manuel~Aravena\inst{3,9}
    \and Roberto~J.~Assef\inst{9}
    \and Loreto~Barcos-Muñoz\inst{10,11}
    \and Rebecca~A.~A.~Bowler\inst{12}
    \and Andreas~Burkert\inst{13}
    \and Jianhang~Chen\inst{1}
    \and Rebecca~L.~Davies\inst{14}
    \and Ilse~De~Looze\inst{15}
    \and Tanio~Diaz-Santos\inst{16,17}
    \and Jorge~González-López\inst{18}
    \and Ryota~Ikeda\inst{19,20}
    \and Ikki~Mitsuhashi\inst{21}
    \and Ana~Posses\inst{9,22}
    \and Mónica~Relaño~Pastor\inst{23,24}
    \and Alvio~Renzini\inst{25}
    \and Manuel~Solimano\inst{9}
    \and Justin~S.~Spilker\inst{22}
    \and Amiel~Sternberg\inst{26,27}
    \and Kenichi~Tadaki\inst{28}
    \and Kseniia~Telikova\inst{9}
    \and Sylvain~Veilleux\inst{29}
    \and Vicente~Villanueva\inst{2}
    }

    \institute{
    Max-Planck-Institut für Extraterrestrische Physik (MPE), Gießenbachstra. 1, D-85748 Garching, Germany 
    \and Departamento de Astronom\'{\i}a, Universidad de Concepción, Barrio Universitario, Concepción, Chile 
    \and Millenium Nucleus for Galaxies (MINGAL) 
    \and Purple Mountain Observatory, Chinese Academy of Sciences, 10 Yuanhua Road, Nanjing 210023, China 
    \and Space Telescope Science Institute, 3700 San Martin D, MD 21218, USA 
    \and Department of Physics and Astronomy and PITT PACC, University of Pittsburgh, Pittsburgh, PA 15260, USA 
    \and Departments of Physics and Astronomy, University of California, Berkeley, CA 94720, USA 
    \and Max-Planck-Institut für Astrophysik (MPA), Karl-Schwarzschild-Str. 1, D-85748 Garching, Germany 
    \and Instituto de Estudios Astrof\'isicos, Facultad de Ingenier\'ia y Ciencias, Universidad Diego Portales, Av.  Ej\'ercito Libertador 441, Santiago, Chile [C\'odigo Postal 8370191] 
    \and Department of Astronomy, University of Virginia, 530 McCormick Road, Charlottesville, VA 22903, USA 
    \and National Radio Astronomy Observatory, 520 Edgemont Road, Charlottesville, VA 22903, USA 
    \and Jodrell Bank Centre for Astrophysics, Department of Physics and Astronomy, School of Natural Sciences, The University of Manchester, Manchester, M13 9PL, UK 
    \and Universitäts-Sternwarte Ludwig-Maximilians-Universität (USM), Scheinerstr. 1, München, D-81679, Germany 
    \and Centre for Astrophysics and Supercomputing, Swinburne University of Technology, Hawthorn 3122, Australia 
    \and Sterrenkundig Observatorium, Ghent University, Krijgslaan 281 S9, B-9000 Ghent, Belgium 
    \and Institute of Astrophysics, Foundation for Research and Technology - Hellas (FORTH), Heraklion 70013, Greece 
    \and School of Sciences, European University Cyprus, Diogenes Street, Engomi 1516, Nicosia, Cyprus 
    \and Las Campanas Observatory, Carnegie Institution of Washington, Casilla 601, La Serena, Chile 
    \and Department of Astronomy, School of Science, SOKENDAI (The Graduate University for Advanced Studies), 2-21-1 Osawa, Mitaka, Tokyo 181-8588, Japan 
    \and National Astronomical Observatory of Japan, 2-21-1 Osawa, Mitaka, Tokyo 181-8588, Japan 
    \and Department for Astrophysical \& Planetary Science, University of Colorado, Boulder, CO 80309, USA 
    \and Department of Physics and Astronomy and George P. and Cynthia Woods Mitchell Institute for Fundamental Physics and Astronomy, Texas A\&M University 
    \and Dept. Fisica Teorica y del Cosmos, E-18071 Granada, Spain 
    \and Instituto Universitario Carlos I de Fisica Teorica y Computacional, Universidad de Granada, E-18071 Granada, Spain 
    \and Osservatorio Astronomico di Padova, Vicolo dell’Osservatorio 5, Padova, I-35122, Italy 
    \and School of Physics and Astronomy, Tel Aviv University, Tel Aviv 69978, Israel 
    \and Centre for Computational Astrophysics, Flatiron Institute, 162 5th Avenue, New York, NY 10010, USA 
    \and Faculty of Engineering, Hokkai-Gakuen University, Toyohira-ku, Sapporo 062-8605, Japan 
    \and Department of Astronomy and Joint Space-Science Institute, University of Maryland, College Park, Maryland USA 20742 
    }

   \date{Received XXX; accepted YYY}

  \abstract
   {We present a detailed kinematic study of a sample of 32 massive ($9.5$\,$\leqslant$\,$\log(M_*/{\rm M_{\odot}})$\,$\leqslant$\,$10.9$) main-sequence star-forming galaxies (MS SFGs) at $4$\,$<$\,$z$\,$<$\,$6$ from the ALMA-CRISTAL program. 
The data consist of deep (up to 15\,hr observing time per target), high-resolution ($\sim$\,$1$kpc) 
ALMA observations
of the [\ion{C}{II}]158$\mu$m\ line emission. 
This data set enables the first systematic kpc-scale characterisation of the kinematics nature of typical massive SFGs at these epochs.
   We find that $\sim$\,$50\%$ of the sample are disk-like, 
   with a number of galaxies located in systems of multiple components. 
   Kinematic modelling reveals these main sequence disks exhibit high-velocity dispersions ($\sigma_0$), with a median disk velocity dispersion of $\sim$\,$70\,$${\rm km\,s^{-1}}$\ and $V_{\rm rot}/\sigma_0$\,$\sim$\,$2$, 
   and consistent with dominant gravity driving.
   The elevated disk dispersions are in line with the predicted evolution based on Toomre theory 
   and the extrapolated trends 
   from $z\sim0$--$2.5$ MS star-forming disks.
   The inferred dark matter (DM) mass fraction within the effective radius $f_{\rm DM}(<R_{\rm e})$ for the disk systems decreases with the central baryonic mass surface density, 
   and is consistent with the trend reported by kinematic studies at $z\lesssim3$; 
   roughly half the disks have $f_{\rm DM}(<R_{\rm e})\lesssim$\,${30\%}$.
   The CRISTAL sample of massive MS SFGs provides a reference of the kinematics of a representative population and extends the view onto typical galaxies beyond previous kpc-scale studies at $z\lesssim3$.}

\maketitle
%-------------------------------------------------------------------

\section{Introduction}\label{sec:intro}
Studying the kinematics of high-redshift galaxies offers a direct tracer of the distribution of stars, gas, and dark matter on galactic scales.
Observations at multiple epochs provide constraints on the evolution of 
the contribution of rotation and turbulence to the dynamical support of galaxies.
Galaxy kinematics are thus a powerful probe of the dominant mechanisms governing the growth and structural formation of galaxies, with processes including gas accretion, non-circular motions, galaxy interactions, and feedback.

Thanks to the advent of near-IR IFU and slit spectroscopy on 8--10m telescopes
\citep[e.g.][]{Eisenhauer2003,Sharples2013}, 
ionised gas kinematics from rest-frame optical emission lines have become routinely accessible at cosmic noon $z$\,$\sim$\,$1$--$3$.
This has enabled the census of 
resolved kinematics of massive star-forming galaxies (SFGs) at $z$\,$\sim$\,$1$--$3$\ on the `main sequence' (MS) of SFGs \citep[e.g.][]{nmfs2009,Kassin2012,Wisnioski2015}, 
which dominate the population and cosmic star formation rate density \citep{MD2014}. 
Cold gas kinematics of the same population of galaxies
traced primarily via CO lines are increasingly available through
the Atacama Large Millimeter/submillimeter Array (ALMA) 
and Northern Extended Millimetre Array (NOEMA) 
\citep{Uebler2018,Nestor2023,Rizzo2023,Liu2023}.

The general findings at
$z$\,$\lesssim$\,$3$
suggests a majority of disks among massive 
(stellar masses $M_{\star}$\,$\ga$\,$10^{10}$$M_{\rm \odot}$) SFGs
have increasing gas velocity dispersion $\sigma_{0}$\
and decreasing rotational-to-dispersion support $V_{\rm rot}/\sigma_{0}$ towards higher redshift.
These findings are in line with the increase of galactic gas mass fractions, which plays an important role in the emergence of the `equilibrium growth' model of galaxy evolution
\citep[see reviews by][and references therein]{Tacconi2020} 
in the framework of marginally stable gas-rich disks \citep[e.g.][]{Genzel2008,Dekel2009,DekelBurkert2014,Zolotov2015,Ginzburg2022}. 
These works have highlighted the important role of internal processes, alongside accretion and merging, 
in galaxy stellar mass and structural buildup during the peak epoch of cosmic star formation activity around $z$\,$\sim$\,2 \citep[e.g. reviews by][]{Glazebrook2013,fs2020}.

The exploration of kinematics beyond cosmic noon at $z$\,$>$\,$3$ has been opened by facilities such as ALMA and NOEMA,
primarily via the bright [\ion{C}{II}]$^2P_{3/2}-^2P_{1/2}$ ([\ion{C}{II}]158$\mu$m, hereafter {[\ion{C}{II}]}) line emission. 
The [\ion{C}{II}] line serves as a key coolant for the interstellar medium (ISM) \citep[e.g.][]{Wolfire2003}.
Its relatively low ionisation potential of $11.3\,$eV (compared to $13.6$\,eV for hydrogen) 
allows [\ion{C}{II}] to arise from various ISM phases, 
primarily from the photo-dissociation regions (PDRs) \citep[e.g.][]{Vallini2015,Clark2019}.
Consequently, the [\ion{C}{II}] line is an ideal tracer for 
kinematics, enabling 
measurements reaching
well beyond the effective radii of these early galaxies \citep[e.g.][]{Lelli2021,Jones2021,Tsukui2021,Umehata2025}.

\begin{figure*}
\centering
    \includegraphics[width=1.0\textwidth]{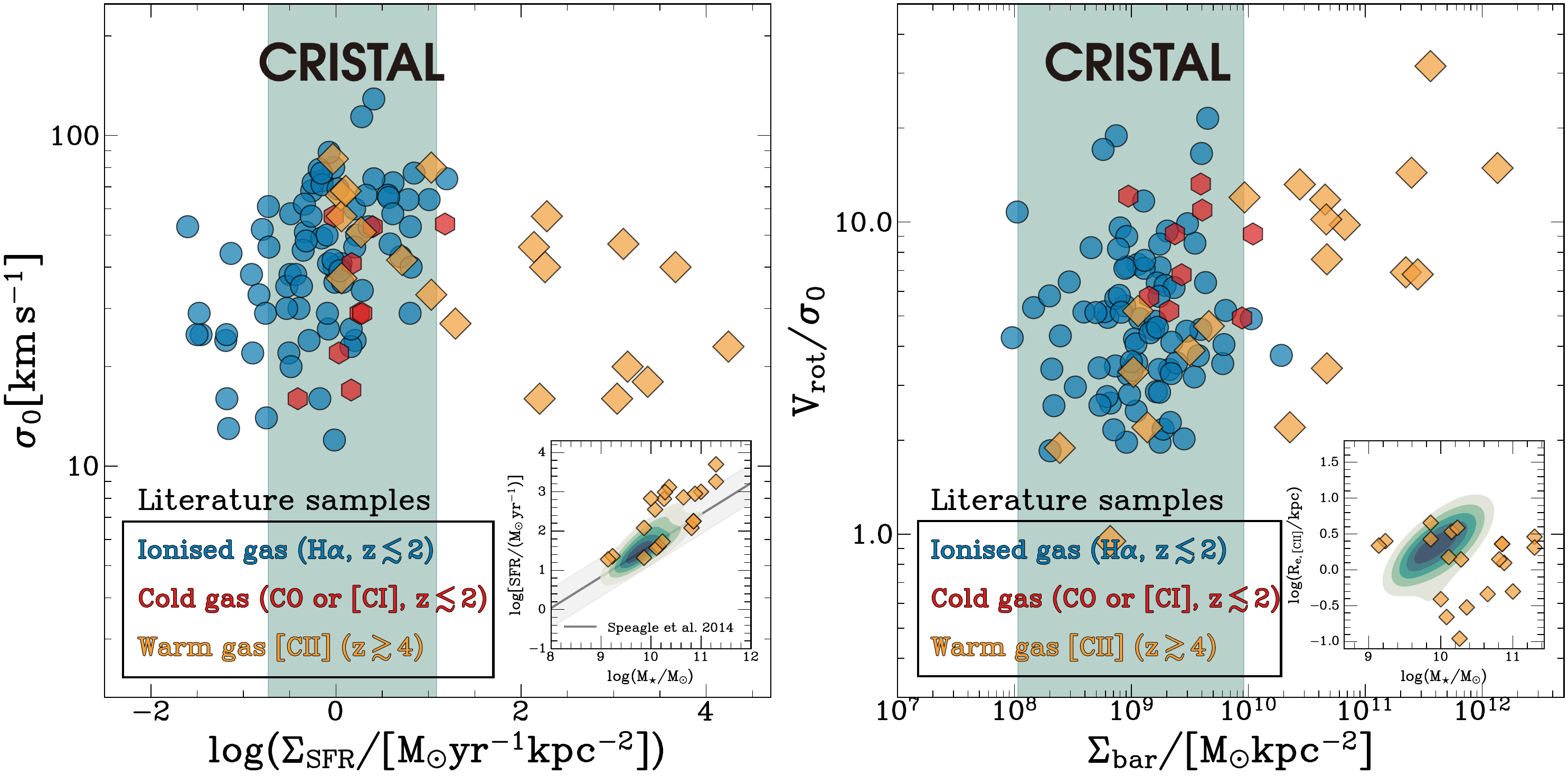}
      \caption{Parameter space of the CRISTAL sample in terms of surface densities of star formation rate 
      ($\Sigma_{\rm SFR}$, left) and baryons 
      ($\Sigma_{\rm bar}$, right). 
      For comparison, the $z\sim2$ main-sequence galaxies observed in H${\alpha}$ (blue circles) and CO (red hexagons) by \citet{Nestor2023} are also shown. 
        The inset of the left and right panels show the SFR-$M_{\star}$ and [\ion{C}{II}]-sizes-$M_{\star}$ distributions of CRISTAL (green contour) and other studies at $z\gtrsim4$ (yellow diamonds) \citep{Neeleman2020,Rizzo2020,Rizzo2021,Fraternali2021,Herrera-Camus2022,
        Parlanti2023,RomanOliveira2023}
        , respectively.
        The solid line in the left inset is the main-sequence (MS) relation from \citet{Speagle2014} at $z=5$ with the shading representing $\pm0.5\,$dex offset from the relation.
      For the purpose of this figure, the values of CRISTAL shown on the plots are adopted from ALPINE \citep{Bethermin2020,Faisst2020a,LeFevre2020}.
      CRISTAL provides a higher-resolution sample of MS SFGs that fills up the parameter space currently poorly explored by existing $z$\,$\gtrsim$\,$4$ samples observed with [\ion{C}{II}], which predominantly comprises starburst and compact galaxies.
      }
      \label{fig:sample_diff}
\end{figure*}

So far, studies suggest that disks appear to be sub-dominant among the bulk of SFGs at 
$z$\,$\sim$\,4$-$7 
(from rest-UV-selected samples, e.g. \citealt{Smit2018,LeFevre2020,Jones2021,Herrera-Camus2022,Posses2023}). 
While some of them appear to be quite turbulent,
high-resolution observations of small samples of individual galaxies, 
primarily infrared-luminous dusty SFGs, 
reveal the existence of dynamically cold disks at those early epochs \citep[e.g.][]{Sharda2019,Fraternali2021,Lelli2021,Rizzo2021,Tsukui2021,RomanOliveira2023,Rowland2024}.
Larger, more systematic samples covering typical galaxies, such as those provided by 
ALPINE ($4$\,$<$\,$z$\,$<$\,$6$; \citealt{LeFevre2020,Bethermin2020,Faisst2020a}) and 
REBELS ($7$\,$<$\,$z$\,$<$\,$8$; \citealt{Bouwens2022}), 
are available, albeit at lower spatial resolution and shallower depths. 
Nevertheless, the typical SFGs targeted by ALPINE formed an ideal sample for high-resolution follow-up studies, and provided the motivation for the CRISTAL 
({\textbf{[\ion{C}{II}]}} \textbf{R}esolved \textbf{I}SM in \textbf{ST}ar forming galaxies with \textbf{AL}MA) survey.

CRISTAL is designed to address four primary science goals: (i) kinematics, (ii) outflows, (iii) spatial distribution of the ISM and star formation, and (iv) the ISM conditions. 
\citet{Herrera-Camus2022} presented a pilot study of the kinematics of one of the CRISTAL galaxies, HZ4 (CRISTAL-20). 
\citet{Posses2025} and \citet{Telikova2025} presented detailed case studies of the kinematics of 
CRISTAL-05 and CRISTAL-22, respectively, and highlighted the improvement in kinematics modelling of complex systems, 
thanks to higher resolution.
The outflow properties of one of the CRISTAL galaxies are presented in \citet{Davies2025}
and the overall outflow demographics through stacking in \citet{Birkin2025}. 
Works addressing goals (iii) and (iv) are presented in \citet{Ikeda2025}, \citet{Li2024}, \citet{Lines2025}, \citet{Mitsuhashi2024b}, \citet{Solimano2025}, and \citet{Villanueva2024}.
We refer the readers to the overview survey paper by \citet{HerreraCamus2025} for more details.
In this paper, we present the systematic study of the kinematics of the CRISTAL sample.

The structure of this paper is as follows. 
In \S~\ref{sec:data_sample}, we describe the sample selection and data reduction process. 
\S~\ref{sec:kins_class} presents the classification of the kinematic types and \S~\ref{sec:diskfrac} discusses the disk fraction of CRISTAL galaxies. 
\S~\ref{sec:diskkins} discusses the kinematic properties of the disks from forward modelling, and compares them with literature results in the context of the dynamical evolution of MS SFGs. 
\S~\ref{sec:dispersion_props} investigates the intrinsic velocity dispersion of CRISTAL disks in the broader context of the dynamical evolution of galaxies with redshift, and explores possible drivers of turbulence, particularly gravitational instabilities and stellar feedback.
In \S~\ref{sec:dm}, we discuss the dark matter fraction within the effective radius of the CRISTAL disks, and its dependence on the circular velocity and baryonic surface density.
\S~\ref{sec:nondisk} presents a brief discussion on the properties of non-disk galaxies. 
Finally, \S~\ref{sec:summary} summarises the key findings and outlines ways forward to improve constraints on the nature and kinematics of $z$\,$\sim$\,$4$--$6$ galaxies.

Throughout, we adopt a flat $\Lambda$CDM cosmology with $H_0$\,$=$\,$70{\rm\,km\,s^{-1}\,Mpc^{-1}}$, 
and $\Omega_m$\,$=$\,$0.3$. 
Physical size is always reported in physical kiloparsecs (pkpc), but we use kpc henceforth for brevity.
Where relevant, we quote the rest-frame air wavelengths if not specified otherwise.

\section{Data and sample selection}\label{sec:data_sample}
\subsection{Galaxy sample}\label{Sec:samples}
The full CRISTAL sample is presented by \citet{HerreraCamus2025}.
In brief, the survey builds on an ALMA Cycle-8 Large Program (2021.1.00280.L; PI: R. Herrera-Camus), targeting [\ion{C}{II}] emission of 19 MS SFGs drawn from ALPINE \citep{LeFevre2020}, 
selected at $\log(M_\star/{\rm M_\odot})$\,$\geqslant$\,$9.5$ and within a factor of 3 of the main sequence of SFGs in their specific SFRs (sSFR/sSFR(MS)\,$\leqslant$\,3).
At the higher resolution of the CRISTAL ALMA observations, some of these sources split into separate components. 
This sample was expanded with six systems satisfying the same criteria taken from pilot programs and the literature \citep[see details in][]{HerreraCamus2025}.
There are 39 galaxies in the final CRISTAL sample in total.

The sample studied here consists of the 32 galaxies near the MS that have sufficient S/N and are sufficiently well-resolved to perform quantitative measurements and model their kinematics properties.
Table~\ref{tab:main_table} lists the galaxies included in this kinematics sample, along with their redshifts, coordinates, stellar masses ($M_\star$) and SFRs, as well as the beam size and sensitivity of the ALMA data sets.
Fig.~\ref{fig:sample_diff} shows the range of $M_\star$, SFRs, and surface densities in SFR and baryonic masses ($\Sigma_{\rm SFR}$ and $\Sigma_{\rm bar}$) covered by the CRISTAL MS SFGs, in comparison to those from selected detailed kinematics studies of disks at lower redshifts in ionised and cold molecular gas, and literature work at $z\gtrsim4$ from [\ion{C}{II}] data \citep{Neeleman2020,Rizzo2020,Rizzo2021,Fraternali2021,Herrera-Camus2022,Parlanti2023,RomanOliveira2023}.
CRISTAL MS SFGs are distinct from most existing samples that predominantly comprise starburst or compact galaxies,
allowing for more direct comparison to $z$\,$\sim$\,$1$--$3$\ kinematics studies of MS star-forming disks at lower redshifts \citep[e.g.][]{Wisnioski2015,Wisnioski2019,Genzel2020,Nestor2023,Puglisi2023}.
For the purpose of the figures, we adopt the values from the ALPINE catalogues \citep{Bethermin2020,Faisst2020a,LeFevre2020} to demonstrate the original sample selection. We adopt the values in Table~\ref{tab:main_table} for our subsequent analysis in this work.

\begin{sidewaystable*}
\caption{Properties of the 32 galaxies in kinematics sample of ALMA-CRISTAL}
\label{tab:main_table}
\centering
\begin{tabular}{l l l c c c c c c}
\hline\hline  
  CRISTAL ID & Full name & $z_{\rm [\ion{C}{II}]}$ & R.A. & Decl. & $\log{(M_\star/\rm{M_\odot})}$ & $\log{[{\rm SFR}/(\rm{M_\odot}\,{\rm yr}^{-1}})]$\tablefootmark{a} & Beam size & Cube Noise\tablefootmark{e} \\
  &&&($^{\circ}$)&($^{\circ}$)&(dex)&(dex)&& (mJy\,beam$^{-1}$)

\\
\hline  
CRISTAL-01a & DEIMOS\_COSMOS\_842313 & 4.554 & 150.2271 & 2.5762 & 10.65 & 2.31 & 0\farcs42$\times$0\farcs47 & 0.13\\
CRISTAL-01b & \ldots & 4.530 & 150.2282 & 2.5745 & 9.81 & 1.71 & 0\farcs42$\times$0\farcs47 & 0.13 \\
CRISTAL-02 & DEIMOS\_COSMOS\_848185, HZ6, LBG-1 & 5.294 & 150.0896 & 2.5864 & 10.30 & 2.25 & 0\farcs45$\times$0\farcs55 & 0.13 \\
CRISTAL-03 & DEIMOS\_COSMOS\_536534, HZ1 & 5.689 & 149.9719 & 2.1182 & 10.40 & 1.79 & 0\farcs68$\times$0\farcs82 & 0.15 \\
CRISTAL-04a & vuds\_cosmos\_5100822662 & 4.520 & 149.7413 & 2.0809 & 10.15 & 1.89 & 0\farcs57$\times$0\farcs75 & 0.18 \\
CRISTAL-04b & vuds\_cosmos\_5100822662 & 4.520 & 149.7414 & 2.0813 & 8.91 & 0.63 & 0\farcs57$\times$0\farcs75 & 0.16 \\
CRISTAL-05 & DEIMOS\_COSMOS\_683613, HZ3 & 5.541 & 150.0393 & 2.3372 & 10.16 & 1.83 & 0\farcs31$\times$0\farcs38 & 0.16 \\
CRISTAL-06a & vuds\_cosmos\_5100541407 & 4.562 & 150.2538 & 1.8094 & 10.09 & 1.62 & 0\farcs45$\times$0\farcs54 & 0.13 \\
CRISTAL-06b & \ldots & 4.562 & 150.2542 & 1.8097 & 9.19 & 1.07 & 0\farcs45$\times$0\farcs54 & 0.13 \\
CRISTAL-07a & DEIMOS\_COSMOS\_873321, HZ8 & 5.154 & 150.0169 & 2.6266 & 10.00 & 1.89 & 0\farcs49$\times$0\farcs74 & 0.18 \\
CRISTAL-07b & \ldots & 5.154 & 150.0166 & 2.6268 & \ldots & \ldots & 0\farcs49$\times$0\farcs74 & 0.18 \\
CRISTAL-07c & \ldots & 5.155 & 150.0134 & 2.6271 & 10.21 & 1.92 & 0\farcs49$\times$0\farcs74 & 0.18 \\
CRISTAL-08 & vuds\_efdcs\_530029038 & 4.430 & 53.0793 & -27.8771 & 9.85 & 1.88 & 0\farcs54$\times$0\farcs80 & 0.17 \\
CRISTAL-09a & DEIMOS\_COSMOS\_519281 & 5.575 & 149.7537 & 2.091 & 9.84 & 1.51 & 0\farcs33$\times$0\farcs36 & 0.16 \\
CRISTAL-10a & DEIMOS\_COSMOS\_417567, HZ2 & 5.671 & 150.5172 & 1.929 & 9.99 & 1.86 & 0\farcs44$\times$0\farcs48 & 0.09 \\
CRISTAL-10a-E & \ldots & 5.671 & 150.5186 & 1.9304 & \ldots & \ldots & 0\farcs44$\times$0\farcs48 & 0.09 \\
CRISTAL-11 & DEIMOS\_COSMOS\_630594 & 4.439 & 150.1358 & 2.2579 & 9.68 & 1.57 & 0\farcs38$\times$0\farcs47 & 0.22 \\
CRISTAL-12 & CANDELS\_GOODSS\_21 & 5.572 & 53.0498 & -27.6993 & 9.30 & 0.98 & 0\farcs42$\times$0\farcs58 & 0.12 \\
CRISTAL-13 & vuds\_cosmos\_5100994794 & 4.579 & 150.1715 & 2.2873 & 9.65 & 1.51 & 0\farcs44$\times$0\farcs52 & 0.18 \\
CRISTAL-14 & DEIMOS\_COSMOS\_709575 & 4.411 & 149.9461 & 2.3758 & 9.53 & 1.45 & 0\farcs11$\times$0\farcs12 & 0.15 \\
CRISTAL-15 & vuds\_cosmos\_5101244930 & 4.580 & 150.1986 & 2.3006 & 9.69 & 1.44 & 0\farcs36$\times$0\farcs42 & 0.17 \\
CRISTAL-16a & CANDELS\_GOODSS\_38 & 5.571 & 53.0662 & -27.6901 & 9.60 & 1.30 & 0\farcs42$\times$0\farcs58 & 0.14 \\
CRISTAL-19 & DEIMOS\_COSMOS\_494763 & 5.233 & 150.0213 & 2.0534 & 9.51 & 1.45 & 0\farcs31$\times$0\farcs40\tablefootmark{b} & 0.15 \\
CRISTAL-20\tablefootmark{c} & DEIMOS\_COSMOS\_494057, HZ4 & 5.545 & 149.6188 & 2.0518 & 10.11 & 1.82 & 0\farcs41$\times$0\farcs45 & 0.06 \\
CRISTAL-21\tablefootmark{c} & HZ7 & 5.255 & 149.8769 & 2.1341 & 10.11 & 1.80 & 0\farcs32$\times$0\farcs35 & 0.22 \\
CRISTAL-22a\tablefootmark{c} & HZ10 & 5.653 & 150.2471 & 1.5554 & 10.35 & 2.13 & 0\farcs27$\times$0\farcs34 & 0.22 \\
CRISTAL-22b\tablefootmark{c} & HZ10 & 5.653 & 150.2469 & 1.5554 & \ldots & & 0\farcs27$\times$0\farcs34 & 0.22  \\
CRISTAL-23a\tablefootmark{d} & DEIMOS\_COSMOS\_818760 & 4.560 & 150.4786 & 2.5421 & 10.55 & 2.50 & 0\farcs25$\times$0\farcs30 & 0.35 \\
CRISTAL-23b\tablefootmark{d} & DEIMOS\_COSMOS\_818760 & 4.562 & 150.4790 & 2.5421 & 10.46 & 1.72 & 0\farcs25$\times$0\farcs30 & 0.35 \\
CRISTAL-23c\tablefootmark{d} & \ldots & 4.565 & 150.4778 & 2.5421 & \ldots & \ldots & 0\farcs25$\times$0\farcs30 & 0.35 \\
CRISTAL-24\tablefootmark{d} & DEIMOS\_COSMOS\_873756 & 4.546 & 150.0113 & 2.6278 & 10.53 & 2.06 & 0\farcs26$\times$0\farcs31 & 0.46 \\
CRISTAL-25\tablefootmark{d} & vuds\_cosmos\_5101218326 & 4.573 & 150.3021 & 2.3146 & 10.90 & 2.75 & 0\farcs25$\times$0\farcs30 & 0.38 \\
  \hline
\end{tabular}
\tablefoot{
\tablefoottext{a}{Adopted from \citet{Mitsuhashi2024b} and \citet{Li2024}. 
They are in broad agreement with the values inferred from existing $L_{[\ion{C}{II}]}$-SFR relations as demonstrated by \citet{Li2024} and references therein.}
\tablefoottext{b}{Briggs-weighted data is used.}
\tablefoottext{c}{CRISTAL pilot targets.}
\tablefoottext{d}{From ALMA archive.}
\tablefoottext{e}{Cube noise are measured in cubes with 20\,${\rm km\,s^{-1}}$\ channel width.}
CRISTAL-01c, 07d, 09b, 10b, 16b, 17, 18 are not included in this study because they are either too faint for kinematics analysis or not detected. SMG~J1000+0234 and CRLE are excluded because they are starburst galaxies. A full table with all CRISTAL source is available in Table 1 of \citet{HerreraCamus2025}.
}
% \end{table*}
\end{sidewaystable*}

\subsection{ALMA observations and data}\label{Sec:obs}
We refer the readers to \citet{HerreraCamus2025} for the observation set-up.
In brief, each observation was reduced using the standard \texttt{CASA} (Common Astronomy Software Applications; \citealt{CASA2022}) pipeline. 
Each cube was continuum-subtracted in the $uv$-plane 
resulting in a line-only, continuum-free data cube. 
For the kinematics modelling, 
we use the cubes with channel with of $20$\,${\rm km\,s^{-1}}$ and natural weighting.\footnote{Except for CRISTAL-19, in which the higher resolution Briggs-weighted (robust $=0.5$) cubes have sufficient S/N. There is, however, no significant difference in the kinematics modelling results between the natural-weighted and Briggs-weighted cubes of the two galaxies.} 
We do not apply the `JvM' correction \citep{JvM1995} in our analysis, which mostly affects the measurement of integrated properties.

\subsection{Space-based ancillary data}\label{Sec:anc_data}
All of the galaxies in the kinematics sample benefit from high-resolution broad-band optical to mid-IR imaging obtained with the \textit{Hubble Space Telescope (HST)} and in most cases also the \textit{James Webb Space Telescope (JWST)}.  Covering stellar continuum emission at rest-frame UV to near-IR wavelengths, 
this imaging data provides important complementary information for a full morpho-kinematic classification and important priors in the dynamical modelling of {[\ion{C}{II}]} kinematics.  These data sets were reduced in a homogeneous fashion for the CRISTAL sample, as described by \citet{Li2024} and \citet{HerreraCamus2025}.

\section{(Morpho)-kinematic classification}\label{sec:kins_class}

% \begin{widetext}
\begin{table*}
\caption{Kinematic Classification and Properties of CRISTAL galaxies}
\label{tab:kins_props}
\centering
\begin{tabular}{l c c c c c c c}
\hline\hline
ID &
    Classification~\tablefootmark{\scriptsize a}&
    PA$_{\rm k}$~\tablefootmark{\scriptsize b}&
    $V_{\rm obs}/2$~\tablefootmark{\scriptsize c}&
    $V_{\rm obs}/ 2\sigma_{\rm int}$~\tablefootmark{\scriptsize d}&
    $f_{\rm molgas}$~\tablefootmark{\scriptsize e} &
    $K_{\rm asym}$~\tablefootmark{\scriptsize f} &
    Disk Score~\tablefootmark{\scriptsize g} \\
   &
   &
  \scriptsize($^{\circ}$) &
  \scriptsize(km\,s$^{-1}$) &
   &
   &
   & \scriptsize(Total: 8)\\
\hline
01a & Non-Disk & 295.6 & 116.54$\pm$26.56 & 0.41 $\pm$0.09  & 0.10$_{-0.06}^{+0.21}$ & 0.29$\pm$0.05 & 3 \\ 
01b & Non-Disk & 325.8 & 54.54$\pm$5.64 & 0.46 $\pm$0.12  & \ldots& 0.63$\pm$0.78 & 3 \\ 
02 & Disk & 126.9 & 65.19$\pm$6.24 & 0.45 $\pm$0.04  & 0.43$_{-0.11}^{+0.21}$ & 0.23$\pm$0.03 & 5 \\ 
03 & Best Disk & 126.6 & 50.0$\pm$11.99 & 0.35 $\pm$0.09  & 0.13$_{-0.06}^{+0.14}$ & 0.15$\pm$0.03 & 7 \\ 
04a & Non-Disk & 171.5 & 47.16$\pm$13.87 & 0.58 $\pm$0.17  & 0.39$_{-0.12}^{+0.25}$ & 0.71$\pm$0.12 & 1 \\ 
04b & Non-Disk & 207.9 & 27.6$\pm$20.78 & 0.68 $\pm$0.51  & \ldots& 0.54$\pm$0.15 & 1 \\ 
05a & Disk & \ldots & \ldots & \ldots & $0.39^{+0.29}_{-0.13}$& \ldots& \ldots\tablefootmark{\scriptsize h} \\ 
05b & Non-Disk & \ldots & \ldots & \ldots &\ldots &\ldots & \ldots\tablefootmark{\scriptsize h} \\ 
06a & Non-Disk & 268.0 & 44.18$\pm$2.09 & 0.57 $\pm$0.03  & 0.51$_{-0.13}^{+0.26}$ & 2.10$\pm$0.28 & 3 \\ 
06b & Disk & 79.0 & 48.11$\pm$7.29 & 0.56 $\pm$0.09  & \ldots& 0.35$\pm$0.17 & 5 \\ 
07a & Disk & 359.0 & 83.3$\pm$6.03 & 0.88 $\pm$0.08  & 0.43$_{-0.13}^{+0.28}$ & 0.20$\pm$0.03 & 5\\ 
07b & Non-Disk & 66.8 & 63.37$\pm$15.79 & 0.62 $\pm$0.16  & \ldots& 0.84$\pm$0.15 & 3 \\ 
07c & Non-Disk & 32.0 & 100.56$\pm$19.01 & 0.91 $\pm$0.18  & 0.21$_{-0.09}^{+0.20}$& 0.81$\pm$0.42 & 4 \\ 
08 & Best Disk & 123.0 & 110.43$\pm$28.70 & 0.94 $\pm$0.25  & 0.63$_{-0.13}^{+0.29}$ & 0.18$\pm$0.06 & 7 \\ 
09a & Disk & 170.0 & 84.27$\pm$59.79 & 0.63 $\pm$0.45  & 0.74$_{-0.12}^{+0.32}$ & 0.25$\pm$0.27 & 6 \\ 
10a & Non-Disk & 226.0 & 127.79$\pm$24.82 & 0.78 $\pm$0.16  & 0.73$_{-0.11}^{+0.23}$ & 0.36$\pm$0.06 & 3 \\ 
10a-E & Disk & 123.0 & 30.54$\pm$15.88 & 0.32 $\pm$0.16  & \ldots& 0.35$\pm$0.03 & 6\\ 
11 & Best Disk & 335.0 & 63.19$\pm$45.60 & 0.45 $\pm$0.32  & 0.73$_{-0.11}^{+0.23}$ & 0.13$\pm$0.05 & 8 \\ 
12 & Disk & 34.3 & 25.72$\pm$18.66 & 0.43 $\pm$0.31  & 0.67$_{-0.15}^{+0.43}$ & 0.45$\pm$0.08 & 7 \\ 
13 & Non-Disk & 17.0 & 145.26$\pm$13.16 & 0.96 $\pm$0.11  & 0.63$_{-0.13}^{+0.27}$ & 0.37$\pm$0.03 & 4 \\ 
14 & Non-Disk & 61.9 & 73.79$\pm$27.03 & 0.55 $\pm$0.20  & 0.78$_{-0.10}^{+0.23}$ & 0.49$\pm$0.33 & 3 \\ 
15 & Best Disk & 308.2 & 37.27$\pm$3.89 & 0.28 $\pm$0.03  & 0.41$_{-0.13}^{+0.27}$ & 0.51$\pm$0.22 & 7 \\ 
16 & Non-Disk & 14.0 & 52.58$\pm$9.06 & 0.44 $\pm$0.08  & 0.69$_{-0.14}^{+0.35}$ & 2.78$\pm$0.70 & 1 \\ 
19 & Best Disk & 313.0 & 110.58$\pm$37.79 & 0.89 $\pm$0.31  & 0.70$_{-0.12}^{+0.27}$ & 0.17$\pm$0.32 & 8 \\ 
20 & Best Disk & 17.0 & 68.31$\pm$13.26 & 0.73 $\pm$0.14  & 0.47$_{-0.15}^{+0.35}$ & 0.14$\pm$0.08 & 8 \\ 
21 & Non-Disk & 41.0 & 27.82$\pm$18.31 & 0.19 $\pm$0.12  & 0.59$_{-0.13}^{+0.28}$ & 2.17$\pm$1.43 & 4 \\ 
22a & Disk & \ldots & \ldots & \ldots & $0.71^{+0.28}_{-0.12}$ & \ldots& \ldots\tablefootmark{\scriptsize i} \\ 
22b & Non-Disk & \ldots & \ldots & \ldots &\ldots &\ldots & \ldots\tablefootmark{\scriptsize i} \\ 
23a & Non-Disk & 354.0 & 45.79$\pm$1.27 & 0.40 $\pm$0.02  & 0.40$_{-0.12}^{+0.23}$ & 1.12$\pm$0.29 & 3 \\ 
23b & Disk & 260.0 & 51.31$\pm$2.32 & 0.34 $\pm$0.02  & 0.25$_{-0.09}^{+0.18}$ & 0.21$\pm$0.02 & 6 \\ 
23c & Best Disk & 45.0 & 92.78$\pm$4.82 & 0.83 $\pm$0.06  & 0.17$_{-0.07}^{+0.13}$ & 0.17$\pm$0.04 & 8 \\ 
24 & Non-Disk & 60.0 & 140.69$\pm$22.69 & 0.43 $\pm$0.07  & 0.51$_{-0.05}^{+0.06}$ & 0.38$\pm$0.11 & 4 \\ 
25a & Non-Disk & 309.0 & 60.03$\pm$30.54 & 0.61 $\pm$0.31  & \ldots& 0.45$\pm$0.10 & 4 \\ 
25b & Non-Disk & 335.0 & 98.95$\pm$26.06 & 0.86 $\pm$0.23  & \ldots& 0.32$\pm$0.08 & 4 \\ 

\hline

\end{tabular}
\tablefoot{
\tablefoottext{a}{\footnotesize Classification of the sample based on the Disk Score.}
\tablefoottext{b}{\footnotesize Position angle along the kinematic major axis pointing to the blue side.}
\tablefoottext{c}{\footnotesize Half difference between the observed maximum and minimum velocities:  $V_{\rm obs}/2$\,$=$\,$(V_{\rm max} - V_{\rm min})/2$.}
\tablefoottext{d}{\footnotesize Ratios between (c) and the [\ion{C}{II}] integrated line widths $\sigma_{\rm int}$.} 
\tablefoottext{e}{Molecular gas fraction inferred in Appendix~\ref{app:dust2gas} for systems 
with dust continuum detection } 
\tablefoottext{f}{Asymmetry measure of the velocity and velocity dispersion fields following \footnotesize \citet{Shapiro2008} demarcation.} 
\tablefoottext{g}{\footnotesize 
The sum of points from the metrics outlined in \S\S~\ref{subsec:pv}--\ref{subsec:morphs}, with the following scheme:
Best Disk if Disk Score $\geqslant7$, Disk if $5$--$6$, and Non-Disk if $\leqslant4$.
}
\tablefoottext{h}{\footnotesize See \citet{Posses2025}.}
\tablefoottext{i}{\footnotesize See \citet{Telikova2025}.}
}

\end{table*}

In this section, we describe the classification of the galaxies from the CRISTAL kinematics sample.  We use several methods that have been devised for applications to IFU and interferometric observations of high redshift galaxies, with comparable S/N and resolution as our ALMA data.  
We determine the final classification by combining the results from each method detailed in \S\S~\ref{subsec:pv}--\ref{subsec:morphs}.
% We assign a double weighting to three metrics involving kinematic information, 
% except for $V_{\rm obs}/2\sigma$, compared to morphological information alone.
% We assign a score to each galaxy ranging from 0 to 8.
{We assign 2 points to three metrics (PV, kinemetry and spectro-astrometry) and 1 point to $V_{\rm obs}/2\sigma$ and morphological information, totalling 8 points.}

Table~\ref{tab:kins_props} compiles the relevant measurements and resulting classification, and Appendix~\ref{app:globalprops} provides details of the individual galaxies. In summary, the systems in the CRISTAL sample can be broadly classified into three general groups:

\begin{enumerate}
\item  Best Disk (22\%): Score $\geqslant7$. 
They are clear disks with no clear sign of nearby interacting companion within 
a projected distance of $\sim$\,$20\,{\rm kpc}$ in the \textit{HST}, \textit{JWST} and ALMA data. 
The velocity gradients are monotonic with a well-defined kinematic position angle (PA$_{\rm k}$), 
and the location of steepest slope coinciding with a central peak in observed velocity dispersion, defining the kinematic centre.

\item  Disk (28\%): Score $=5$--$6$. 
They show features of a rotating disk but also irregularities.
These systems meet most criteria but exhibit some deviations from one or the other pure disk rotation features.
Except for two cases, they belong to
systems with visible companions in both [\ion{C}{II}] and \textit{HST} or \textit{JWST} imaging data.

\item Non-Disk (50\%): The rest of the systems with a score $\leqslant4$.
They do not have an apparent velocity gradient across $\geqslant$\,$2$ beams and no centralised dispersion peak. Some of them have a visible companion. 
\end{enumerate}

\subsection{Position-velocity diagrams}\label{subsec:pv}
We extract position-velocity (p-v) diagrams from the original reduced data cubes along the kinematic major axis (${\rm PA_{\rm k}}$) defined as the direction of the largest observed velocity difference across the source. 
The width of the synthetic slit is equal to the FWHM of each beam, which is taken to be the geometric average of the major and minor axes values listed in Table~\ref{tab:main_table}. 
The slit is positioned to pass through the dynamical centre of the galaxies. 
We then integrate the light along the spatial direction perpendicular to the slit orientation.
These p-v diagrams are presented in Figs.~\ref{fig:kins_map_bestdisk} to \ref{fig:kins_map_nondisk3} in Appendix~\ref{app:globalprops}. 
For visualisation purposes, we median-filtered the p-v diagrams with a kernel size of 3 pixels.

We classify a system as disk-like if there are no detached and distinct velocity structures in the p-v diagrams. 
This metric contributes to two points in the disk score in Table~\ref{tab:kins_props}.
We do not consider deviations (e.g. CRISTAL-02 and 08) from a standard S-shape as indicative of a merger, as they may instead reflect the possible origin of non-circular gas flow, which could be common for gas-rich systems at higher redshifts.

\subsection{Kinematics profiles and $V_{\rm obs}/2\sigma_{\rm int}$}\label{subsec:v2sig}

The ratio between the full observed velocity difference across a source ($V_{\rm obs}$) 
and the source-integrated line width ($\sigma_{\rm int}$) 
has been used as proxy to distinguish systems with dominant support from rotational/orbital motions versus random motions 
\citep[e.g.][]{nmfs2009,Wisnioski2015}. 
The $V_{\rm obs}$ and $\sigma_{\rm int}$ are measured from the data, without beam-smearing and inclination corrections.
The boundary at $V_{\rm obs}/2\sigma_{\rm int}$\,$=$\,$0.4$ adopted in previous work, based on mock disk models, is also applicable for the typical range of galaxy sizes relative to beam sizes for our sample.

\begin{figure}
    \centering
    \includegraphics[width=0.45\textwidth]{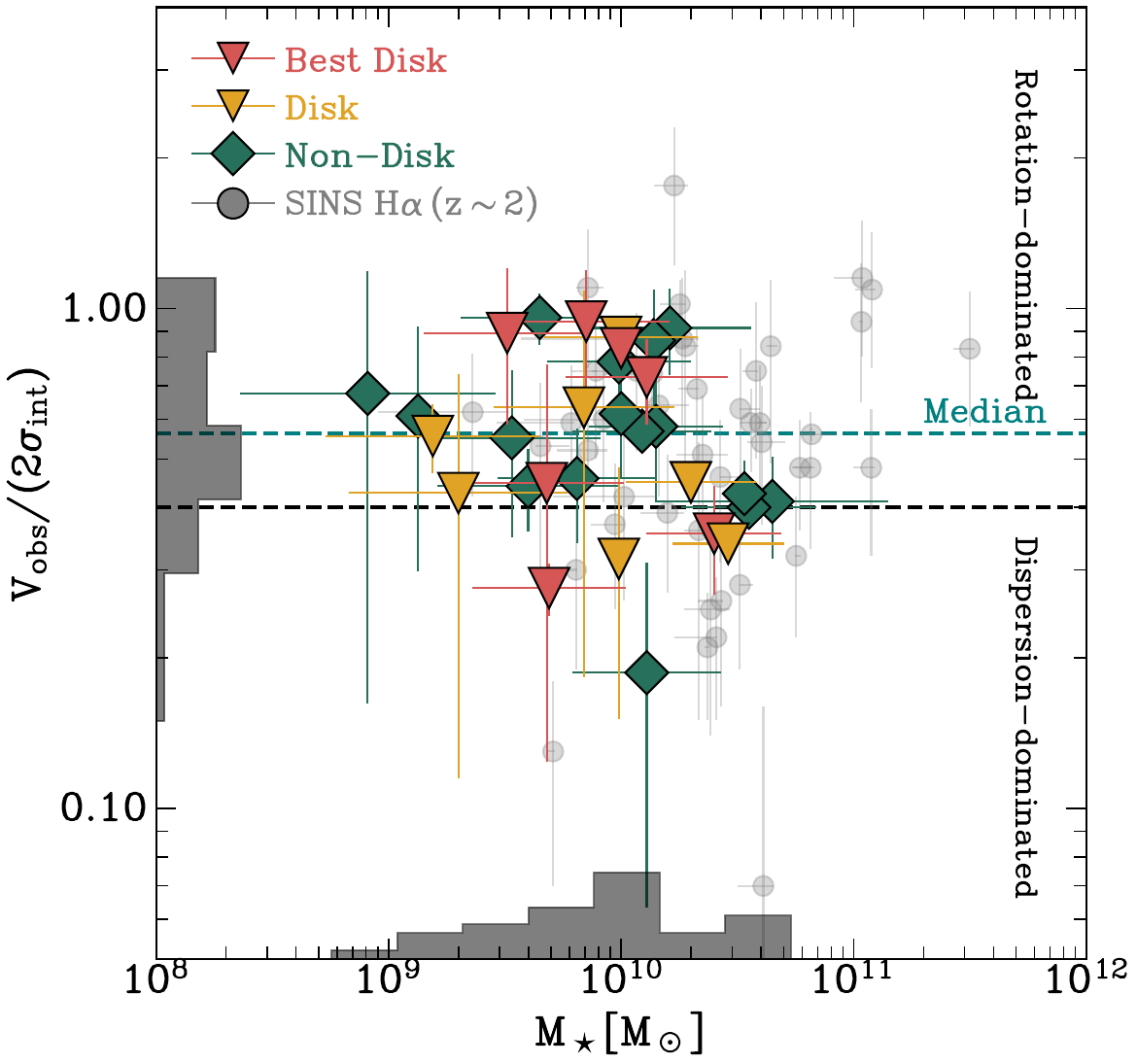}
    \caption{Ratios of observed half velocity gradient $V_{\rm obs}/2$ and integrated line width ($\sigma_{\rm int}$) of [\ion{C}{II}] as a function of
    stellar mass $M_*$. 
    The black horizontal dashed line represents the distinction value ($V_{\rm obs}/2\sigma_{\rm int}=0.4$), which demarcates the boundary between rotation- and dispersion-dominated systems, as commonly used in the literature.
    }
    \label{fig:vd2s_mdyn_msta}
\end{figure}

\begin{figure*}
    \centering
    \includegraphics[width=0.92\textwidth]{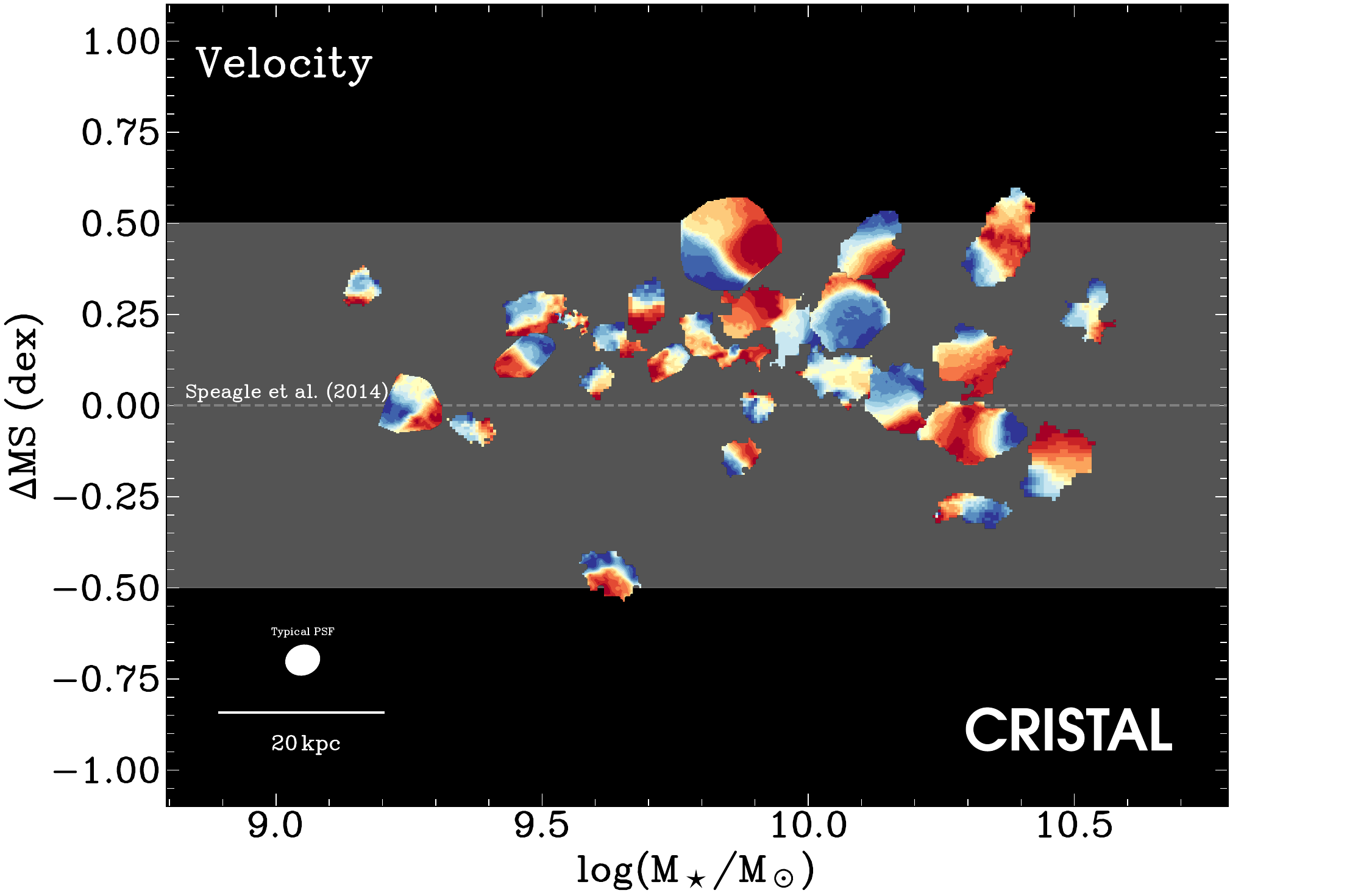}
    \includegraphics[width=0.92\textwidth]{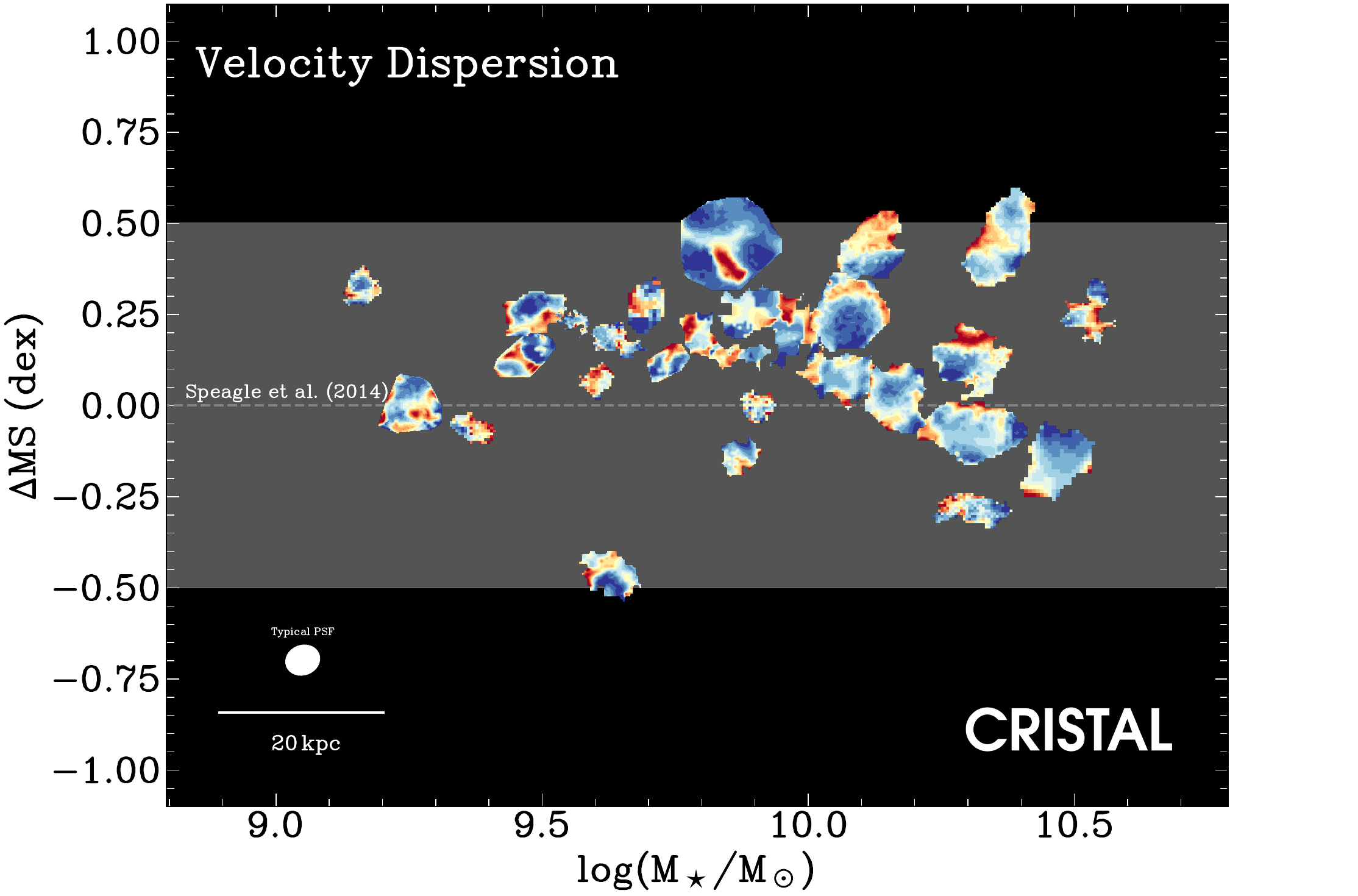}
    \caption{
    Velocity (\textit{Top}) and velocity dispersion (\textit{bottom}) fields of CRISTAL galaxies 
    placed on the $\Delta$MS offset relative to the \citet{Speagle2014} main-sequence (MS) relation. 
    The velocity and dispersion fields correspond to that derived from the [\ion{C}{II}] emission described in \S~\ref{subsec:vmaps}. 
    For the velocity fields, the colour coding represents the relative velocity of the 
    line emission with respect to the systemic velocity. 
    For the dispersion fields, the colour coding indicates the widths (in standard deviation) of the 1D Gaussian fitted to the spectra of individual spaxels.
    All sources are shown in the same field of view of $3\arcsec$ ($\sim$20\,kpc at $z$\,$\sim$\,$5$). The median beam size (0\farcs43$\times$0\farcs36) with position angle (104$^\circ$ counter-clockwise from north) is shown at the bottom left; the individual values are listed in Table~\ref{tab:main_table}.
    }
    \label{fig:mom1_ms}
\end{figure*}

We measure the integrated line width $\sigma_{\rm int}$ from
spatially integrated spectra extracted from the reduced data cubes.
The cubes are at the original spatial and spectral resolution and with the channel size of $\Delta V$\,$=$\,$20\,$${\rm km\,s^{-1}}$. 
The circular apertures for the extraction are positioned at the 
in order to capture the contributions from both velocity gradients and local velocity dispersion to $\sigma_{\rm int}$.
The apertures' sizes roughly followed those determined by \citet{Ikeda2025}.
We then sum the spectra of individual pixels within the apertures to obtain the integrated spectrum. 
The extracted spectra are shown in the last column of Figs.~\ref{fig:kins_map_bestdisk} to \ref{fig:kins_map_nondisk3} in Appendix~\ref{app:globalprops}. 

We fit the spectrum with a single Gaussian with \texttt{emcee} \citep{ForemanMackey2013} to extract the 
line widths, except for CRISTAL-02, where we fit a double Gaussian profile as there is a broad component possibly associated with an outflow \citep{Davies2025}. 
The emission from the narrow component is always the one used in this analysis.

The fitted values of integrated line widths ($\sigma_{\rm int}(\text{[\ion{C}{II}]})$) are annotated in Figs.~\ref{fig:kins_map_bestdisk} to \ref{fig:kins_map_nondisk3} along with the best-fit model overlaid on the extracted spectra. 
Uncertainties of $\sigma_{\rm int}(\text{[\ion{C}{II}]})$ are taken as the $[16,84]$-th percentile ($1\sigma$) bounds of the marginalised posterior distributions. 
We stress that the $\sigma_{\rm int}$ quantity determined here does not represent the local intrinsic disk velocity dispersion, but rather a global measure of the dynamical support combining rotation/orbital motions and random motions.

The $V_{\rm obs}$ is defined as the maximum observed velocity difference $V_{\rm obs}$\,$=$\,$V_{\rm max} - V_{\rm min}$.
We extract velocity profiles from the p-v diagrams obtained in \S~\ref{subsec:pv} by fitting a single Gaussian profile column-by-column (i.e., collapsed emission of the velocity channels at the same position) 
using again \texttt{emcee} \citep{ForemanMackey2013}. 
A one-pixel-wide vertical pseudo-slit is moved along the position axis. 
The centroids and widths of the fitted Gaussian models will then be the velocity and velocity dispersion at the locations of the slits. 
Finally, the extracted profiles are 
down-sampled by averaging to a resolution of one-half to one-fourth of the beam FWHM.
The velocity and velocity dispersion profiles 
will then serve as an input for the dynamical modelling in \S~
\ref{sec:diskkins}.

We list the $V_{\rm obs}/2\sigma_{\rm int}$ 
of our sample in Table~\ref{tab:kins_props}. 
We plot in Fig.~\ref{fig:vd2s_mdyn_msta} the distribution of $V_{\rm obs}/2\sigma_{\rm int}$ 
as a function of the 
stellar mass.
For comparison, we show the values for the SINS {H${\alpha}$} IFU sample at $z$\,$\sim$\,$2$ \citep{nmfs2009}.
The median $V_{\rm obs}/2\sigma_{\rm int}$ of all samples is $0.56$. 
For the Best Disk, Disk and Non-Disk samples, the median values are 
$0.73$, $0.45$ and $0.57$, 
respectively.

We observe that many Non-Disk systems have 
$V_{\rm obs}/2\sigma_{\rm int}$\,$>$\,$0.4$, 
which can be attributed to the fact that mergers may exhibit 
a substantial projected velocity gradient from orbital motions depending on the orientation of the merging system.
While the $V_{\rm obs}/2\sigma_{\rm int}$ ratio is useful, especially in cases where the sources are less well resolved, it is not sufficient to unambiguously distinguish disks from mergers.
Therefore, this metric contributes only one point to the disk score in Table~\ref{tab:kins_props}.

\subsection{Velocity and velocity dispersion maps and their asymmetry}\label{subsec:vmaps}
To derive the flux, velocity, and velocity dispersion maps, we fit
a single Gaussian profile to the [\ion{C}{II}] emission line of each spaxel in the continuum-subtracted line cube in velocity units, with the amplitude, mean and standard deviation of the profile as free parameters. 
In the resulting [\ion{C}{II}] kinematic maps, we mask pixels with S/N\,$<3$ and pixels resulting in unphysical outlier values.
For the velocity map, we determine the systemic velocity of the galaxy by symmetrising the red-shifted and blue-shifted peak velocities.
Fig.~\ref{fig:mom1_ms} displays the derived velocity and velocity dispersion maps, plotted 
in the $M_{\star}$ versus offset from the MS in SFR.
The line flux, velocity, and velocity dispersion maps of individual galaxies are shown in the fourth and fifth columns of Figs.~\ref{fig:kins_map_bestdisk} to \ref{fig:kins_map_nondisk3} in Appendix~\ref{app:globalprops}.

Under the assumption of a single Gaussian profile, our fits primarily capture the narrower line emission component dominated by star formation. 
Such single-component fits of individual pixel spectra will be little sensitive to possible emission from broader lines originating, 
e.g. from outflowing gas as long as the amplitude of the broad component is sufficiently low \citep[e.g.][]{nmfs2018}.  
Examination of our CRISTAL data shows this is the case 
for all galaxies considered here except for CRISTAL-02, where more prominent outflow components are detected \citep{Davies2025}. 
In the kinematic maps of these galaxies, the regions (largely outside of the main body of the sources) are masked out for quantitative analysis.

We then use kinemetry \citep{Krajnovic2006} to quantify asymmetries in the velocity and velocity dispersion maps of our galaxies, 
following the method described 
by \citet{Shapiro2008} for applications in studies of galaxy kinematics at $z$\,$\sim$\,$2$ \citep[see also, e.g.][]{Swinbank2012,Genzel2023}.

Kinemetry performs Fourier analysis on the velocity and dispersion maps, decomposed into concentric ellipses, with the centre, position angle (PA), and inclination determined a priori through methods detailed in \S~\ref{subsec:pv}. 
Given the limited S/N and angular resolution of our data, we fix the centre, PA$_{\rm k}$, 
and inclination to the adopted values in \S\S~\ref{subsec:pv} and \ref{sec:diskkins}, respectively.
and require at least 75\% of valid pixels in an annulus. We follow the Fourier expansion up to the fifth order term, similar to \citet{Shapiro2008}.

We use the demarcation set by \citet{Shapiro2008} at $K_{\rm asym}$\,$=$\,$\sqrt{(v_{\rm asym}^2 + \sigma_{\rm asym}^2)}$\,$=$\,$0.5$, above which the system is classified as a merger and below which is a disk. 
The $v_{\rm asym}$ and $\sigma_{\rm asym}$ are dimensionless measures of the average higher-order kinematic coefficients in the Fourier expansion relative to the coefficients corresponding to the regular rotation.

\begin{figure}
    \centering
    \includegraphics[width=0.5\textwidth]{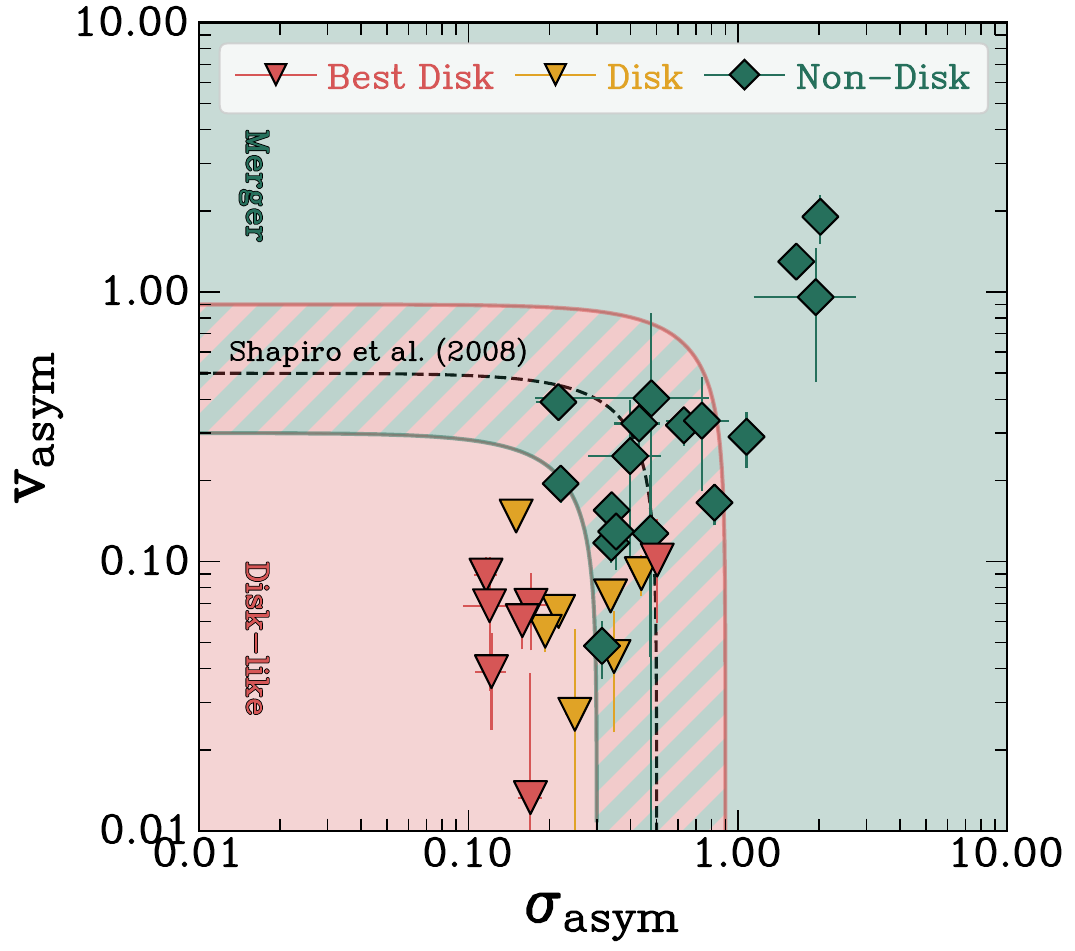}
    \caption{Asymmetry measure of the velocity and velocity dispersion fields for the CRISTAL galaxies from \texttt{Kinemetry} \citep{Krajnovic2006} is shown in Fig.~\ref{fig:km_summary}. 
   The black dashed line represents the demarcation in \citet{Shapiro2008} at $K_{\rm asym}$\,$=$\,$0.5$. 
   With this metric alone, we classify a system as disk-like if it falls below the dashed line and as a merger if it falls above. Over half of the samples fall below the dashed line.
   The pink ($K_{\rm asym}$\,$=$\,$0.9$) and green ($K_{\rm asym}$\,$=$\,$0.3$) lines indicate the modified demarcations.
    }
    \label{fig:km_summary}
\end{figure}

We plot the $v_{\rm asym}$ and $\sigma_{\rm asym}$ of the CRISTAL galaxies in Fig.~\ref{fig:km_summary}. 
There are 20 and 10 galaxies (out of 30 with measurements) fall into the `disk' and `merger' regime, respectively, according to the fiducial $K_{\rm asym}$\,$=$\,$0.5$.
Some Disks and Non-disks (according to the overall classification) 
overlap in the region around the boundary, 
which can reflect intrinsic deviations from pure circular motions caused by minor merging, non-axisymmetric structures such as bars/spirals even in the absence of interactions, 
noise in kinematic maps or incomplete coverage of the objects due to regions with lower S/N and fainter surface brightness, 
beam smearing, 
or a combination of these factors.  
For illustrative purposes, we indicate in Fig.~\ref{fig:km_summary} the band corresponding to 
$K_{\rm asym}$\,$=$\,$0.3$ to $0.9$, 
which for the simulated galaxies used by \citet{Shapiro2008} to calibrate the threshold would result in 
6\% higher merger fraction or 
3\% higher disk fraction, 
respectively.  
Adopting the fiducial threshold for CRISTAL, we classify galaxies with $K_{\rm asym}$\,$\leqslant$\,$0.5$ as disk, counting for 2 points in the disk score from this metric.

\subsection{Spectro-astrometry}\label{subsec:sa}
We also applied spectro-astrometry (SA) to classify our sample. This technique, commonly used to study compact, marginally resolved stellar binary systems 
\citep[e.g.][]{Christy1981,Beckers1982}
was successfully applied to IFU observations of high-$z$ galaxies \citep[e.g.][]{Gnerucci2010,Perna2025}.
SA operates on the principle that closely spaced sources with projected separation below the angular resolution element can be spectrally separated if their relative velocities differ.
This technique leverages the 3D information in data cubes and can be especially useful in retrieving velocity gradients when beam smearing is important.

For each velocity channel map, we derive the spatial offsets along the $X$ and $Y$ directions by fitting a 2D Gaussian without any a priori assumptions of the kinematics major axis.
In cases where multiple emission peaks are observed and are separated by multiple beams in FWHM, 
we fit multiple 2D Gaussians to the emission blob. 
We derive the positional uncertainty following Eq.~(1) in \citet{Condon1998}, 
which depends on the amplitude of the emission and the beam size, in addition to the formal fitting errors. 
We show the locus traced by the SA measurements in the sixth column in Figs.~\ref{fig:jwst_hst_gallery_best_disk} 
to ~\ref{fig:jwst_hst_gallery_nondisk2} in Appendix~\ref{app:globalprops}, 
overlaid on either \textit{HST} or \textit{JWST} colour images. 

For a galaxy to be considered Best Disk, its SA locus should be uni-directional, i.e., moving along monotonically in one direction, as in the case of CRISTAL-15 and 20 (Fig.~\ref{fig:jwst_hst_gallery_best_disk}).
However, for systems with non-circular motion, the loci would deviate from a straight line near the centre, as in the case of CRISTAL-02 and 08, but would overall follow a single direction.
In all cases, the overall direction of the velocity gradients agrees with the velocity map with consistent ${\rm PA_{kin}}$.
For Non-disks, the SA loci would exhibit a more zig-zag shape, characterised by sudden changes in opposite direction; when companions are present, the loci show discontinuities with abrupt jumps from one location to another, often separated by one to two resolution elements.

The inherent nature of SA results in uneven spatial sampling, while spectral sampling remains constant. 
The spectral resolution of our SA measurement is naturally determined by the channel width, 
which is set at $20$\,${\rm km\,s^{-1}}$. We do not consider a wider channel width, such as e.g. $50$\,${\rm km\,s^{-1}}$\
because it would have poorly compromised spectral sampling for several sources with observed velocities $v_{\rm obs}$\,$<$\,$200$\,${\rm km\,s^{-1}}$\ (Table~\ref{tab:kins_props}). 
In the case of Non-Disk sources, which could potentially be mergers, there could be a few channels with emission peaks that are spatially close to each other. 
Although this may appear to be a spatial sampling that is too high on the SA curve, we have chosen to retain these data, as they could indicate unresolved line-of-sight mergers.

The fainter outer regions of the sources, often associated with the most blue-/redshifted emission, tend to have too low S/N for a robust centroid measurement; consequently, the full velocity gradient may not be probed for some of our targets.

This metric adds two points to the disk score in Table~\ref{tab:kins_props}.

\subsection{Morphology of rest-frame UV-optical and {[\ion{C}{II}]} line emission}\label{subsec:morphs}

We complement the kinematic classification methods described in the previous sections with morphological information.  
We consider the \textit{HST} along with \textit{JWST}/NIRCam data 
and [\ion{C}{II}] data.
The longest wavelength NIRCam filter F444W corresponds to the rest-frame optical emission ($\lambda$\,$=$\,$0.7\,\mu$m) at $z$\,$=$\,$5$, redwards of the Balmer break, 
in contrast to the rest-frame $0.3~\mu$m provided by \textit{HST}/F160W. The longest 
The depth of the \textit{JWST} data varies across the sample, with CRISTAL-08, 11, 13, and 15 being the deepest.

We consider source multiplicity in our inspection of the imaging data.  
In most cases, our ALMA data already indicate the single or multiple nature of the galaxies, 
with 14 of the multiple systems associated with Non-Disks according to the kinematic criteria applied in \S\S~\ref{subsec:pv}--\ref{subsec:sa}.  
The higher angular resolution of \textit{JWST} can provide a more detailed view of the morphology and deblending unresolved companions within the ALMA beam that could explain the observed perturbations in the [\ion{C}{II}] kinematics.
We emphasise that the companions we defined in 
Table~\ref{tab:main_table} are unlikely to be multiple clumps within a single galaxy, 
as their closest separations is on average $\sim8$\,kpc, ranging from 5 to 10\,kpc, which is much larger than the typical size of SFGs at $z$\,$\sim$\,$5$ \citep{Miller2024,Varadaraj2024}.
We further stress that multiplicity on a few kpc-scales could be ambiguous, 
as clumpy disks may mimic multiple systems especially if the sensitivity in insufficient to detect a fainter host galaxy underlying bright clumps.

The first two columns of Figs.~\ref{fig:jwst_hst_gallery_best_disk} to \ref{fig:jwst_hst_gallery_nondisk2} present a comparison between \textit{HST} and \textit{JWST} colour-composite images. 
We observe a marked difference between the rest-frame optical and UV images of galaxies in 
CRISTAL-01a, 07c, 08, 11, 12, 13, 16, 24, and 25.
For systems CRISTAL-02 and 04a, the rest-frame optical resembles that in the UV. 
However, all galaxies retain their substructures or clumpy appearance. 
The single-pair classification using the [\ion{C}{II}] and \textit{HST}-based morphologies of \citet{Ikeda2025} are unchanged with the additional information from \textit{JWST} data and the multiplicity remain the same.

To highlight the clumpiness and substructures, we subtract the F444W image by a smooth S\'ersic \citep{Sersic1968} model in Appendix~\ref{app:f444w_imfit}. 
The clumpy appearance of CRISTAL-02, 08, and 15 is apparent in the residual images shown in Fig.~\ref{fig:f444w_imfit}. 

Since many of the galaxies are not well-fitted by 
a S\'ersic model, 
we do not consider the difference of \textit{JWST} (or \textit{HST}) 
morphological PA and ${\rm PA}_{\rm kin}$ (defined in \S~\ref{subsec:pv}) as a disk criterion because the former is not well-constrained.

There are systems with visible [\ion{C}{II}] companions or extended emission, such as CRISTAL-01b, 12, and 13, 
but no associated counterparts in either \textit{HST} or \textit{JWST} images. 
This suggests that we are still missing the more evolved stellar population due to extinction
or simply fainter emission with the shallow NIRCam data. 
Therefore, we consider the imaging data as complementary but not decisive evidence for the kinematics nature.
With the visual inspection of \textit{JWST} 
and \textit{HST} images, alongside [\ion{C}{II}] line maps, 
we classify a system as a disk if it is a single-component system with smooth underlying emission, 
possibly featuring bright clumpy substructures that are closely spaced (typically within $\lesssim2\,$kpc, generally consisting of more than two clumps). Additionally, there should be no detectable companion within $\lesssim20\,$kpc across the observed wavelengths. This contributes one point towards the disk score.

\begin{figure}
    \centering
    \includegraphics[width=0.48\textwidth]{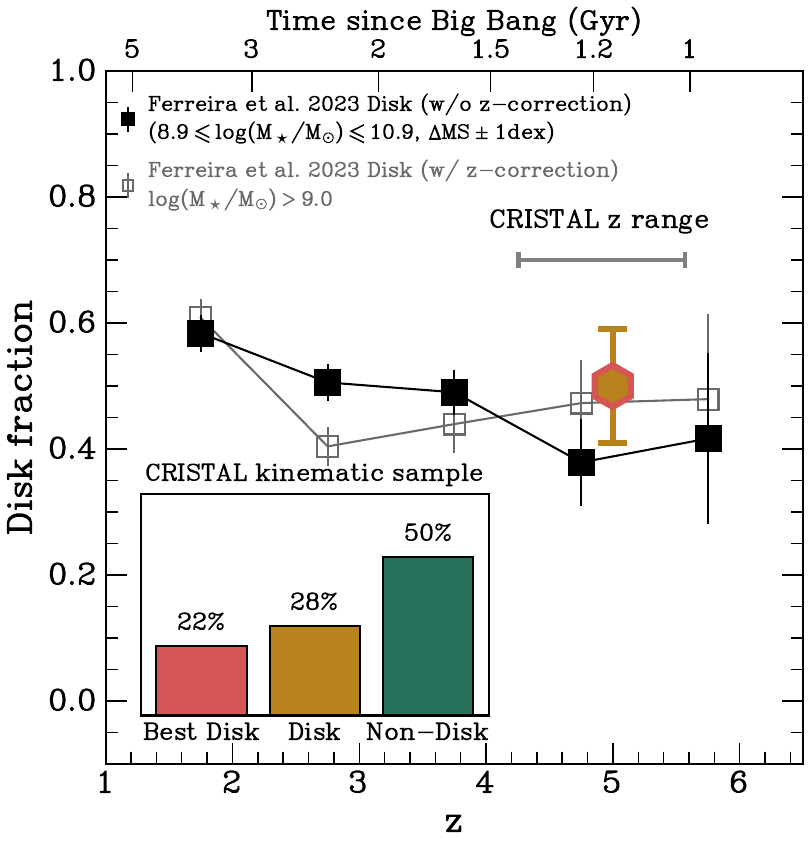}
    \caption{Comparison of the CRISTAL {total} disk fraction {(yellow hexagon with red outline)} to the mass-selected sample from \citet{Ferreira2023} based on CRISTAL mass range. 
    {
    The grey curve with square marker shows the `redshift-corrected' trend of their high-mass bin. To more fairly compare with our sample, we }
    further selected so that their star formation rates are within 1\,dex from the \citet{Speagle2014} main sequence relation.
    The grey horizontal line indicates the redshift range of 
    CRISTAL ($4.4$\,$<$\,$z$\,$<$\,$5.7$).
    The black solid curve with square markers represents the redshift evolution of disk fraction as reported by \citet{Ferreira2023} before their application of `redshift corrections', which primarily account for surface brightness effects in higher redshift objects. It would be a more equal comparison since we do not apply any such correction to the CRISTAL disk fraction. The inset shows the distribution of Best Disk, Disk and Non-Disk among the CRISTAL sample.
    }
    \label{fig:ferreira+23}
\end{figure}

\section{Disk fraction}\label{sec:diskfrac}
Considering the systems with disk score $\geqslant5$,
the total disk fraction among the 32 CRISTAL kinematics samples is $50\pm9\%$ ($N$\,$=$\,$16$),
encompassing both Best Disk and Disk (Table~\ref{tab:kins_props}).
Best Disk and Disk make up of 
$22\pm7\%$ ($N$\,$=$\,$7$) and 
$28\pm8\%$ ($N$\,$=$\,$9$) of the sample, respectively. 
The errors are binomial errors.
The distribution of these types is shown in the inset in Fig.~\ref{fig:ferreira+23}.
The disk fraction in our study is higher than that reported in the previous ALPINE work \citep{LeFevre2020}, which found that for the overlapping CRISTAL sample, $\lesssim$\,$40\%$ 
of systems are classified as `Rotator' or `Extended Dispersion Dominated', 
with only $<$\,$20\%$
being the former,
and the remaining systems being `Pair-Merger'.

The 50\% disk fraction of CRISTAL 
($4.4$\,$<$\,$z$\,$<$\,$5.7$) 
is consistent with morphological studies based on NIRCam/\textit{JWST} data, 
which reveal a high fraction of disks of $\sim$\,$35\%$ on average across studies \citep[][]{Ferreira2023,Jacobs2023,Kartaltepe2023,Huertas-Company2024,JHLee2024,Pandya2024,Tohill2024}.
A smaller sample using \textit{JWST}/MIRI also supports this finding \citep{Costantin2025}.
These results suggest an early establishment of the Hubble sequence \citep{Ferreira2023,XuD2024,HuertasCompany2025}. 

In particular, we compare our CRISTAL disk fraction with \citet{Ferreira2023}, a morphological study of $\sim$\,4,000 galaxies from the CEERS survey, classified in rest-frame optical observed with \textit{JWST}.
Fig.~\ref{fig:ferreira+23} plots our disk fractions against their evolutionary trend.
For consistency we select galaxies from \citet{Ferreira2023} in the same mass and $\Delta$MS ranges as our kinematic sample, using the galaxy parameters from the CANDELS-EGS catalogue of \citet{Stefanon2017}.  
We do not apply surface brightness corrections to the fractions as we compare with the direct fraction from our kinematic classification.  
Fig.~\ref{fig:ferreira+23} shows these derived morphology-based disk fractions over $z$\,$\sim$\,$1$--$6$;
our kinematics-based disk fraction for the CRISTAL MS SFGs are in very good agreement.

The presence of Disk (disks in an interacting system) is perhaps not surprising, as hinted from simulation (see also the classic example of M51); gravitational interactions between galaxies and the presence of rotating disks are not inherently contradictory \citep{SpringelHernquist2005,Robertson2006}. 
The rotation of a disk is relatively resilient to minor mergers. 
For gas-rich systems,
the stellar disk can rapidly reform and sustain itself through the formation of new stars from the remaining gas, 
even if the pre-existing stellar disk is destroyed in the process \citep{Ubler2014,Peschken2020,SotilloRamos2022}.

\section{Kinematics modelling and properties of the disk samples}\label{sec:diskkins}
\subsection{Forward modelling with \texttt{DysmalPy}}\label{subsec:dysmalpy}
To extract the intrinsic kinematics and mass distribution of CRISTAL disks, 
we use the public forward-modelling code \texttt{DysmalPy} \footnote{\url{https://www.mpe.mpg.de/resources/IR/DYSMALPY}} 
\citep{Davies2004a,Davies2004b,Cresci2009,Davies2011,Wuyts2016,Lang2017,Price2021,Lee2025}. 
Table~\ref{tab:dysmalpy_results} reports the best-fit results.
We refer the readers to the earlier cited
works for a detailed description of \texttt{DysmalPy}. 
In short, it is a forward modelling tool that starts from a parametrised input mass distribution to establish the best-fit models for the data. 
The models consist of a baryonic disk, bulge (optional), and DM halo. The disk component is parametrised as a S\'ersic profile of index $n_{\rm d}$\,$=$\,$1$ (exponential disk) adopted for all fits, 
with flattening $q_{\rm d}$ and effective radius $R_{\rm e,disk}$. 

The baryonic disk component is assumed to be a thick oblate disk,
treated as a flattened spheroid of 
intrinsic axis ratio $q$\,$=$\,$c/a$, 
and the rotation curve (RC) is derived 
accordingly following the \citet{Noordermeer2008} parametrisation. 
We assume the velocity dispersion is locally isotropic and radially uniform, representing a dominant turbulence term, $\sigma_{0}$.

\begin{figure*}
    \centering
    \includegraphics[width=\textwidth]{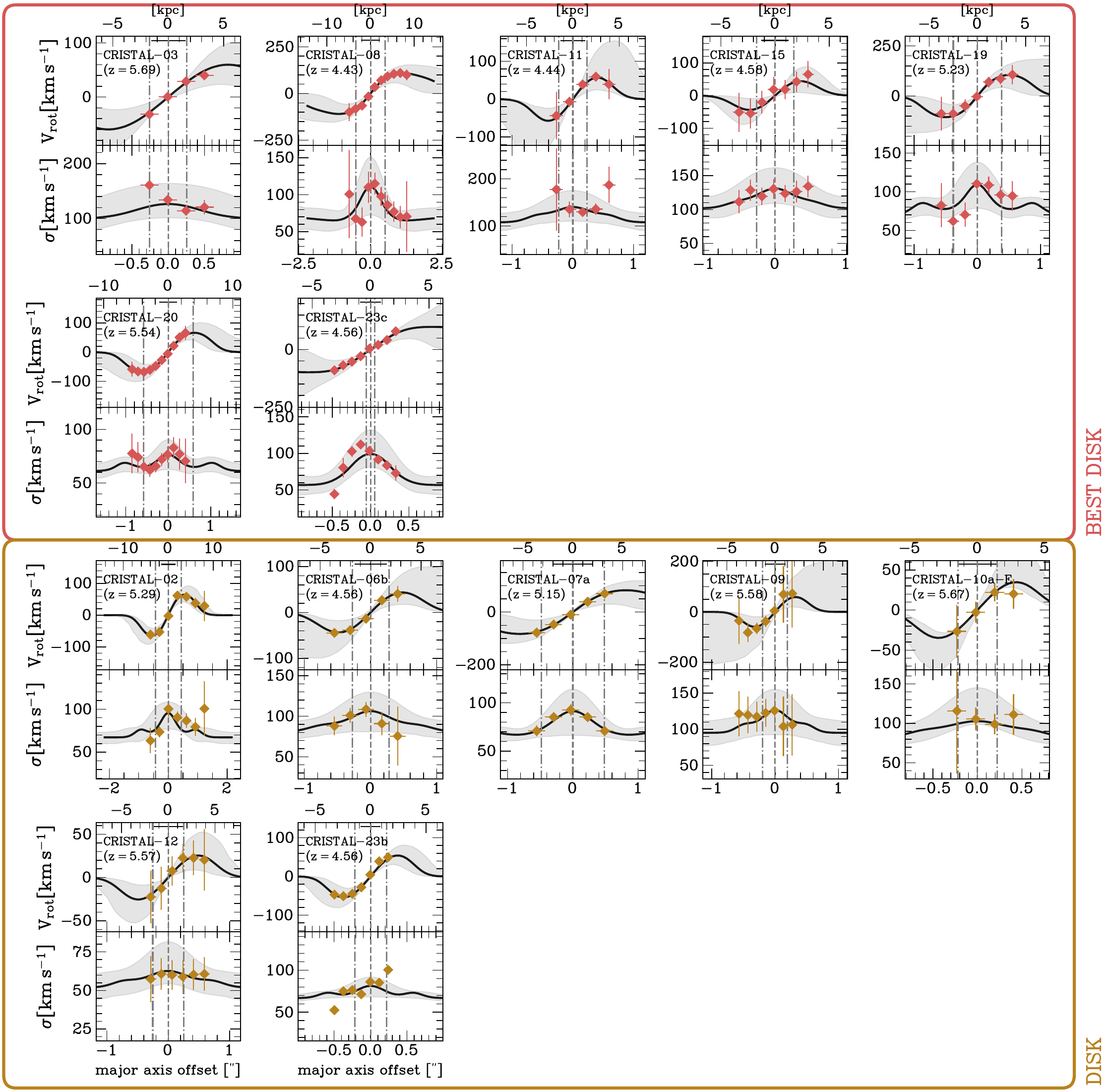}
    \caption{Observed rotation curves (RCs) of the CRISTAL disk sample. 
    The RCs are the fitted line velocity centroids 
    of the position-velocity diagram extracted along the kinematic major axis.
    The RCs are grouped according to their kinematics types. 
    {The black curves are the extracted 1D model (in the same way as data) from the best-fit 3D model cube from {\texttt{DysmalPy}}. The 3D model cubes are projected and convolved with the beam, which give rise to the apparent central peak in the velocity dispersion profile and the shallower velocity profiles.}
    The two symmetric grey vertical lines about the dynamical centre indicate the effective radius.
    The synthesised beam size is shown as the horizontal black line at the top.
    CRISTAL-05 and 22a belong to Disk, their velocity profiles are shown in \citet{Posses2025} and \citet{Telikova2025}, respectively.
    }
    
    \label{fig:allRC}
\end{figure*}

\begin{table*}
\caption{Best-fit properties from our dynamical models and molecular gas fractions}
\label{tab:dysmalpy_results}
\centering
\begin{tabular}{lccccccccc}
\hline\hline  
  ID &
  $\log_{10}(M_{\rm tot}/{\rm M_{\odot}})$\tablefootmark{\scriptsize a} &
  $R_{\rm e, disk}$\tablefootmark{\scriptsize b} &
  B/T\tablefootmark{\scriptsize c} &
  $V_{\rm rot}(R_{\rm e})$\tablefootmark{\scriptsize d} &
    $\sigma_0$\tablefootmark{\scriptsize e} &
   $f_{\rm DM}$($R_{\rm e}$)\tablefootmark{\scriptsize f} &
  $i$\tablefootmark{\scriptsize g}  & 
  $R_{\rm out}/R_{\rm e, disk}$\tablefootmark{\scriptsize h} &
  $R_{\rm out}$/beam\tablefootmark{\scriptsize i}\\
    &
  \scriptsize(dex) &
  \scriptsize(kpc) &
   & \scriptsize(${\rm km\,s^{-1}}$)
   & \scriptsize(${\rm km\,s^{-1}}$)
   &  
  &
  \scriptsize$(^{\circ})$
  &
  % &
  & \\ \hline
02 & $10.5^{+0.1}_{-0.2}$ & $2.7^{+0.5}_{-0.5}$ & 0.1$^{+0.1}_{-0.1}$ & $129.0^{+36.6}_{-37.2}$ & $67.1^{+7.8}_{-2.4}$  & $0.07_{-0.06}^{+0.31}$ &  [47.0]  &  3.0 & 5.5\\ 
03 & $10.4^{+0.1}_{-0.2}$ & $1.5^{+0.9}_{-0.2}$ & [0.0] & $165.3^{+20.6}_{-65.5}$ & $77.8^{+28.5}_{-0.2}$  & $0.09_{-0.08}^{+0.28}$ &  [67.0]  &  2.3 & 1.6\\ 
06b & $10.5^{+0.1}_{-0.3}$ & $1.8^{+0.7}_{-0.7}$ & [0.3] & $136.3^{+44.6}_{-136.3}$ & $80.8^{+12.5}_{-6.4}$  & $0.08_{-0.07}^{+0.46}$ &  [43.0]  &  2.3 & 2.6\\ 
07a & $10.2^{+0.4}_{-0.2}$ & $3.0^{+0.9}_{-1.0}$ & 0.1$^{+0.1}_{-0.1}$ & $132.6^{+41.7}_{-73.5}$ & $63.1^{+5.0}_{-7.2}$  & $0.62_{-0.29}^{+0.28}$ &  [54.7]  &  1.4 & 2.2\\ 
08 & $10.5^{+0.3}_{-0.2}$ & $3.4^{+0.9}_{-0.9}$ & 0.1$^{+0.1}_{-0.1}$ & $183.6^{+50.2}_{-61.0}$ & $62.4^{+10.9}_{-12.8}$  & $0.52_{-0.29}^{+0.28}$ &  [46.0]  &  2.6 & 4.1\\ 
09 & $10.4^{+0.1}_{-0.6}$ & $1.2^{+0.2}_{-0.2}$ & 0.1$^{+0.1}_{-0.0}$ & $153.1^{+70.0}_{-153.1}$ & $93.2^{+26.2}_{-13.2}$  & $0.08_{-0.07}^{+0.52}$ &  [55.0]  &  3.1 & 3.7\\ 
10a-E & $10.3^{+0.0}_{-0.2}$ & $[1.3]$ & [0.0] & $111.8^{+40.3}_{-20.6}$ & $79.2^{+13.1}_{-9.0}$  & $0.05_{-0.04}^{+0.27}$ &  [80.0]  &  2.2 & 2.2\\ 
11 & $10.6^{+0.1}_{-0.3}$ & $1.5^{+0.6}_{-0.9}$ & [0.0] & $160.3^{+47.9}_{-98.9}$ & $107.8^{+16.7}_{-12.2}$  & $0.10_{-0.09}^{+0.27}$ &  [82.0]  &  2.9 & 3.3\\ 
12 & $9.7^{+0.1}_{-0.1}$ & $[1.5]$ & [0.0] & $69.6^{+32.9}_{-69.6}$ & $48.5^{+14.8}_{-14.8}$  & $0.43_{-0.12}^{+0.12}$ &  [58.0]  &  2.6 & 2.7\\ 
15 & $10.5^{+0.1}_{-0.5}$ & $[1.7]$ & [0.1] & $117.8^{+45.1}_{-117.8}$ & $100.5^{+27.0}_{-1.0}$  & $0.17_{-0.16}^{+0.35}$ &  [65.0]  &  2.4 & 3.3\\ 
19 & $10.5^{+0.1}_{-0.4}$ & $2.3^{+0.4}_{-0.2}$ & 0.1$^{+0.1}_{-0.1}$ & $156.0^{+50.3}_{-81.5}$ & $70.3^{+12.0}_{-7.4}$  & $0.18_{-0.13}^{+0.45}$ &  [65.0]  &  1.7 & 3.7\\ 
20 & $10.3^{+0.2}_{-0.1}$ & $3.5^{+0.9}_{-0.6}$ & [0.0] & $98.7^{+37.2}_{-33.9}$ & $61.2^{+3.8}_{-6.7}$  & $0.43_{-0.37}^{+0.10}$ &  [64.4]  &  1.5 & 4.2\\ 
23b & $9.8^{+0.1}_{-0.1}$ & $1.4^{+0.2}_{-0.2}$ & [0.0] & $93.3^{+24.2}_{-27.1}$ & $66.8^{+2.0}_{-2.5}$  & $0.57_{-0.11}^{+0.09}$ &  [60.9]  &  2.5 & 3.9\\ 
23c & $9.2^{+0.1}_{-0.3}$ & $0.4^{+0.1}_{-0.1}$ & 0.1$^{+0.2}_{-0.1}$ & $114.4^{+17.3}_{-34.8}$ & $56.5^{+11.0}_{-15.7}$  & $0.53_{-0.05}^{+0.23}$ &  [73.5]  &  9.2 & 3.7\\ 
\hline
\end{tabular}
\tablefoot{
From (a) to (h) are the best-fit (maximum a posteriori) parameters from \texttt{DysmalPy}\ modelling. 
Values in square brackets are fixed. 
The rotation velocity in (d) is the intrinsic total (baryons and dark matter) velocity defined in Equation~\ref{eqn:dy_pressure_support}.
Inclinations in (g) are inferred from the \textit{JWST}/F444W image, except for CRISTAL-10a, 20 and 23, which we infer from [\ion{C}{II}] line flux map due to lack of NIRCam data.
The gas fraction $f_{\rm molgas}$\ listed in (h) is estimated using the method outlined in Appendix~\ref{app:dust2gas} for galaxies with available dust continuum measurements. 
The 1$\sigma$ uncertainties from (a) to (f) are the distance to the shortest interval containing $68\%$
of the marginalised posterior for each parameter (See Appendix~A.2 in \citealt{Price2021}).
The ratios in (h) and (i) represent the outermost measurable radius $R_{\rm out}$ (maximum radius on either the approaching or receding side) relative to the disk effective radius $R_{\rm e,disk}$ and the beam size in half-width-half-maximum, respectively.
{Best-fit values of CRISTAL-05 and 22a are presented in \citet{Posses2025} and \citet{Telikova2025}, respectively.}
} 
\end{table*}

We adopt the \citet{Burkert2010} pressure support (asymmetric drift) correction to circular velocity $V_{\rm circ}$.
We use the option for a self-gravitating exponential disk with
constant velocity dispersion $\sigma(R)=\sigma_0$, such that
\begin{equation}
    V_{\rm rot}^2(R) = V_{\rm circ}^2(R) - 3.36\sigma_{0}^2(R/R_{\rm e}).
    \label{eqn:dy_pressure_support}
\end{equation}

{We note that the pressure support corrections 
derived from local galaxies \citep{DalcantonStilp2010}
and from simulations of high-$z$ galaxies \citep{Kretschmer2021} predict more moderate corrections \citep[see][]{Bouche2022,Price2022}.
As discussed below, due to the lack of empirical evidence of strong radially varying velocity dispersions in SFGs at cosmic noon, here we choose to adopt the \citet{Burkert2010} prescription derived for isotropic dispersion and self-gravitating disks.
% We treat the non-spherical potential following the methods laid out in \citet{Price2022} in which $V_{\rm circ}$ is derived directly from deprojected S\'ersic profiles.
}
% correction that is based on a deprojected S\'ersic profile 

We choose the two-parameter NFW \citep*{Navarro1996} profile for the DM halo. 
The virial mass $M_{\rm vir}$ is tied to 
the variable $f_{\rm DM}(<R_{\rm e})$. 
The initial guess of $M_{\rm vir}$ is set by the expected value from the stellar-mass-halo-mass (SMHM) 
scaling relation 
from abundance matching \citep{Moster2018}, and $\log(M_{\rm vir}/{\rm M_{\odot}})\in[11.7,12.1]$.
We then allow $M_{\rm vir}$ to 
vary by tying it to $f_{\rm DM}$.
The concentration parameter $c_{\rm vir}$ is fixed at a value following the $M_{\rm vir}$--$c_{\rm vir}$ relation from \citet{Dutton2014}, such that $c_{\rm vir}\in[3.4,3.6]$
. We do not apply adiabatic contraction in our fits \citep{Burkert2010}.

% Because of obs of 
% constant dispersion, we assume Burkert10 
% (which assumes self-grav disks and fully isotropic 
% dispersions). 

\texttt{DysmalPy}\ assumes an isotropic velocity dispersion profile $\sigma(R) = \sigma_0$.
It is motivated by empirical results from MS SFGs at $z$\,$\sim$\,$1$--$3$\ \citep[e.g.][]{Genzel2011,Wuyts2016,Uebler2019,Liu2023}, in which $\sigma(R)$ do not show strong trends with inclination and radius in high-resolution and high-S/N IFU observations, after accounting for beam-smearing effects.
They also do not exhibit significant residuals after subtracting a constant profile, 
that would otherwise justify using a more complicated model for the dispersion profile.
The $\sigma_0$ is sensitive to the masking of spectral channels, especially for the S/N of the CRISTAL data \citep{Davies2011,deBlok2024,Lee2025}. 
Overly aggressive masking, which removes the fainter wings of the line emission, 
can result in a bias towards lower $\sigma_0$ values. 
To avoid this bias entirely, 
we do not apply masking along the spectral axis. 
Instead, we evaluate the integrated S/N for each spaxel, and if the S/N falls below a threshold of $\sim$\,$3$, we mask the entire spaxel.

The inclination ($i$) of the galaxies is inferred from the intrinsic axis ratio of the \textit{JWST}/NIRCam F444W image when available, or from the {[\ion{C}{II}]} line emission map if not.
The F444W-inferred inclinations are, on average, 
$10^\circ$ more face-on than those inferred from the {[\ion{C}{II}]} line emission map suggesting a 
possible overestimation of inclination when using the {[\ion{C}{II}]} line emission map alone, 
even after accounting for beam convolution due to the elongated beam sizes and shapes.

The inclination is then derived from the axis ratio ($\epsilon$) using the equation
\begin{equation}
\cos^2(i) = (\epsilon^2-\epsilon_0^2) / (1-\epsilon_0^2),
\label{eqn:i2e}
\end{equation}
where $\epsilon_0$ is the intrinsic axis ratio which we assume to be $0.25$ \citep[e.g.][]{Wisnioski2019}. The median inclination is $54^\circ$, which is essentially the same as the average over a population of randomly oriented disks \citep{Law2009}.

We simultaneously fit the 1D velocity and dispersion profiles extracted in \S~\ref{subsec:v2sig} along the kinematic major axis.
This approach is preferred for our data over 2D and 3D methods, 
the latter are more demanding in terms of per-spaxel S/N and are more sensitive to non-circular motions.
{Since this work primarily focuses on the first-order kinematics of disks, the 1D approach is sufficient since the motion along the major axis best captures these properties \citep[e.g.][]{vanderKruit1978,Genzel2017,Genzel2020,Price2021}. }
The extended radial coverage provided by the 1D method allows us to constrain $\sigma_0$ at larger distances from the central region, thereby mitigating the effects of beam smearing and helping to resolve degeneracies in the model parameters, particularly those related to the relative contributions of baryons and dark matter to the observed RCs.

As demonstrated by \citet{Price2021} for $z$\,$\sim$\,$1$--$2.5$ MS galaxies,
such a 1D approach is in broad agreement with 2D modelling.
We additionally verify that the 3D and 1D methods agree within $\sim$\,$10\%$ if the per-pixel S/N within the effective radius is on average $\gtrsim$\,$20$ within $R_{\rm e}$, 
and in the worst case $\sim$\,$20\%$ if such S/N $\lesssim$\,$3$.
We note that while the terms 1D and 2D refer to the 
method of profile extraction from the data, {
\texttt{DysmalPy}\ always construct the model cube in full hypercube space when accounting for beam-smearing, projection, and spectral-broadening effects, as described above, irrespective of the extraction approach, and the full 3D information is used to identify the kinematic major axis.}
The model profile is then extracted from a 3D model cube in the same fashion as the data profiles are extracted from the observed data cube (see Fig.~6 in \citealt{Price2021}).

% \texttt{DysmalPy}\ always operates in full 3D space when accounting for beam-smearing, projection, and spectral-broadening effects, as described above. 

Since for all galaxies, the resolution and S/N of our data cannot provide constraints on many parameters, 
we leave four parameters free: 
(i) the baryonic mass $\log_{10}(M_{\rm bary}/{\rm M_{\odot}})$, 
(ii) the disk effective radius $R_{\rm e,disk}$ (kpc),  
(iii) the enclosed dark matter fraction $f_{\rm DM}(<R_{\rm e, disk})$, and
(iv) the velocity dispersion $\sigma_0$ (${\rm km\,s^{-1}}$). 
We employ Gaussian priors for $\log_{10}(M_{\rm bary}/{\rm M_{\odot}})$, 
with a standard deviation of 1\,dex centred on the sum of the stellar mass reported in \citet{Mitsuhashi2024b} and \citet{Li2024}
and the molecular gas mass derived in Appendix~\ref{app:dust2gas}.
For $R_{\rm e,disk}$ (henceforth $R_{\rm e}$), 
we adopt Gaussian priors of standard deviation 1\,kpc, centred on the fitted value of $R_{\rm e}$\ of the [\ion{C}{II}] emission measured in \citet{Ikeda2025}. For CRISTAL-10a-E, 12 and 15, the quality of the data is not sufficient to constrain the $R_{\rm e}$, we fix $R_{\rm e}$ to the [\ion{C}{II}]-based radius.
The prior range is tailored for each galaxy but generally spans $[0,10]$\,kpc.
We assume flat bounded priors for the intrinsic dispersion $\sigma_0$\,$\in$\,$[20,200]\,$${\rm km\,s^{-1}}$ and dark matter fraction $f_{\rm DM}(<R_{\rm e, disk})$\,$\in$\,$[0,1]$. 
Finally, we fix the geometrical parameters $i$ and PA inferred from Eq.~(\ref{eqn:i2e}) and \S~\ref{subsec:pv}, respectively. 
Other parameters are either tied, such as the disk scale height (through $R_{\rm e}$) and halo virial mass (through $f_{\rm DM}$), or fixed.
We run \texttt{DysmalPy}\ with the \texttt{emcee} sampler, employing 512 walkers and a minimum of 200 burn-in steps followed by 1000 iterations. 
{For all our fit, the final acceptance fraction is between $0.2$ and $0.5$ (mean $=0.32$) and the chain is run for $>10\times$ (mean $=23\times$) the maximum estimated parameter autocorrelation time \citep{ForemanMackey2013}.}

We begin the first modelling without the bulge component 
 given that the F444W/NIRCam data show no strong indication of a bulge based on the relatively low S\'ersic indices (Appendix~\ref{app:f444w_imfit}).
For some galaxies, we observe a fair level of residuals between the data and the best-fit velocity and dispersion models 
in the central region, which could be evidence of a concentrated mass distribution deviating from the pure exponential disk profile. 
 We therefore introduce a small, low-mass de Vaucouleurs bulge component with a fixed S\'ersic index ($n_{\rm bulge}=4$) and an effective radius ($R_{\rm e,b}\leqslant1$\,kpc). 

 We iteratively increment the bulge-to-total ratio $B/T$ in steps of $0.1$. For all but 2 cases, a $B/T=0.1$ leads to the best fit in terms of reduced-$\chi^2$.  
 For two galaxies the preferred $B/T$\,$\geqslant$\,$0.3$ (CRISTAL-06b, as well as CRISTAL-05 modelled by \citealt{Posses2025}).

\begin{figure}
    \centering
    \includegraphics[width=0.5\textwidth]{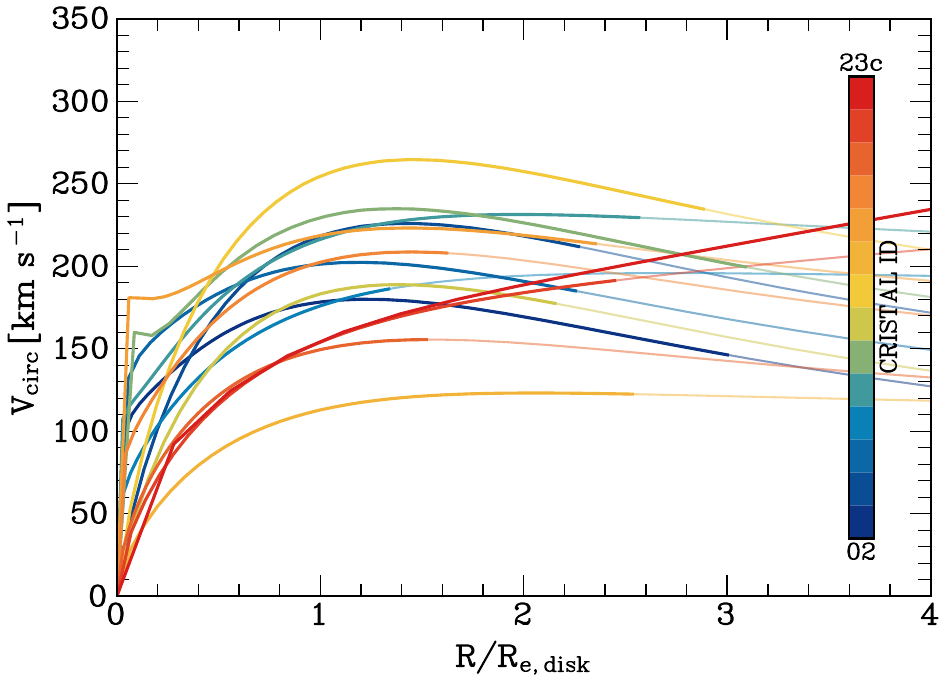}
    \caption{Intrinsic total circular velocity profiles ${V_{\rm circ}(R/R_{\rm e,disk})}$ of all CRISTAL disks, corrected for beam-smearing and projection effects. 
    Colours in ascending order represent the CRISTAL ID (first column in Table~\ref{tab:dysmalpy_results}).
    The individual $V_{\rm circ}(R)$ of baryons and dark matter are presented in Figure~\ref{fig:mass_profile} in Appendix~\ref{app:mdl_prof}.
    Thin, lighter-coloured lines indicate the radial range beyond which the data is not covered. 
    }
    \label{fig:allRC_mdl}
\end{figure}

\begin{figure}
    \centering
    \includegraphics[width=0.48\textwidth]{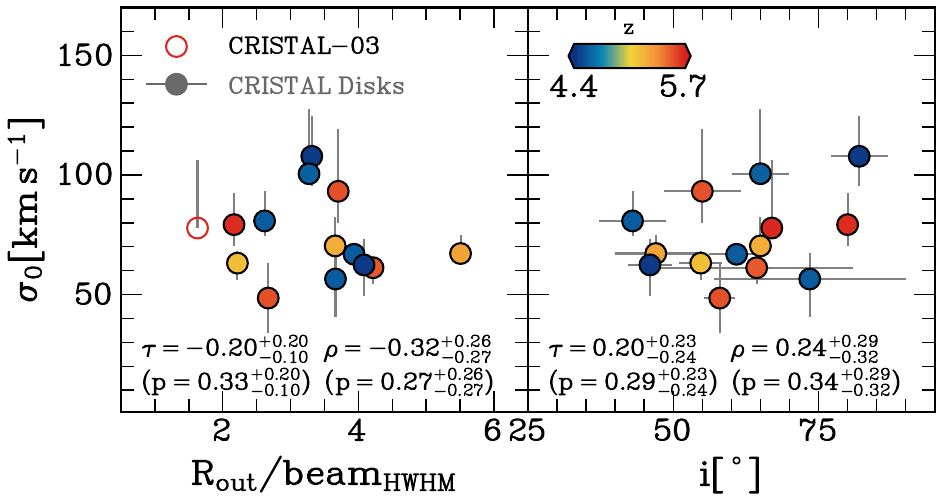}
    \caption{Intrinsic velocity dispersion $\sigma_0$ as a function of the number of resolution elements within the outermost measurable radius $R_{\rm out}$ 
    and inclination ($i$). The Spearman and Kendall rank correlation coefficients ($\rho$ and $\tau$, respectively) 
    do not indicate a significant correlation of $\sigma_0$ with the inclination and resolution effects, 
    though the uncertainties in both the coefficients and $p$-value are large.
    CRISTAL-03 which has $R_{\rm out}/{\rm beam_{HWHM}}$\,$=$\,$1.6$ is excluded in the correlation analysis in \S~\ref{sec:trends}.
    }
    \label{fig:sigma_inc}
\end{figure}

We compare in Fig.~\ref{fig:allRC} the observed and best-fit (projected and beam-smeared) model RCs for the CRISTAL disks.
The intrinsic circular velocity profiles $V_{\rm circ}(R)$ of the models are shown in Fig.~\ref{fig:allRC_mdl}, with the maximum radial coverage of the data indicated.
Fig.~\ref{fig:mass_profile} in Appendix~\ref{app:mdl_prof} shows the intrinsic $\sigma_0$, circular velocity profiles of the DM and the baryonic components.
For all systems, except CRISTAL-01b, 07a, 08, 19, 23b and 23c, we observe a fall-off of circular velocities, 
indicative of masses dominated by the baryonic components. 
We will discuss the DM fractions of the samples later in \S~\ref{sec:dm}.

We examine the potential dependence of $\sigma_0$ on the angular resolution relative to the galaxies' sizes and $i$. 
Table~\ref{tab:dysmalpy_results} lists the ratios of 
$R_{\rm out}$ to the beam size (geometric average of the values in Table~\ref{tab:main_table}) 
in terms of half-width-half maximum (${\rm beam_{HWHM}}$). 
$R_{\rm out}$ represents the outermost radius at which we can reliably extract velocity and dispersion profiles using the method described in \S~\ref{subsec:v2sig}.
Overall, the kinematics profiles are traced out to $\sim$1.5--$3R_{\rm e}$ ($\sim9\,R_{\rm e}$ for CRISTAL-23c), with $R_{\rm out}$/beam$_{\rm HWHM}\sim2$--$4$ (5.5 for CRISTAL-02).  
\citet{Lee2025} has tested these requirements are sufficient to recover $V_{\rm rot}$ and $\sigma_0$ with a large suite of mock galaxies having comparable angular resolution and S/N to the CRISTAL data, 
provided that the adopted parametric profiles are close to the intrinsic profiles.

We perform Spearman and Kendall rank correlation tests (Fig.~\ref{fig:sigma_inc}) to investigate the relationships between $\sigma_0$--$R_{\rm out}/{\rm beam_{HWHM}}$ as well as $\sigma_0$--$i$. We do not detect significant correlations in either case (similarly for $\sigma_0$--$R_{\rm e}/{\rm beam_{HWHM}}$). 
The small sample size, however, would only allow to detect stronger correlations if present.

We conservatively exclude the least-resolved CRISTAL-03 in the correlation analysis in \S~\ref{sec:trends}, which has $R_{\rm out}/{\rm beam_{HWHM}}$\,$=$\,$1.6$.
For this galaxy, the 5$\times$ better angular resolution NIRSpec/\textit{JWST} data reveals a consistent rotational pattern in H${\alpha}$\ (W. Ren et al., in prep.).

\begin{figure*}
    \centering
    \includegraphics[width=0.84\textwidth]{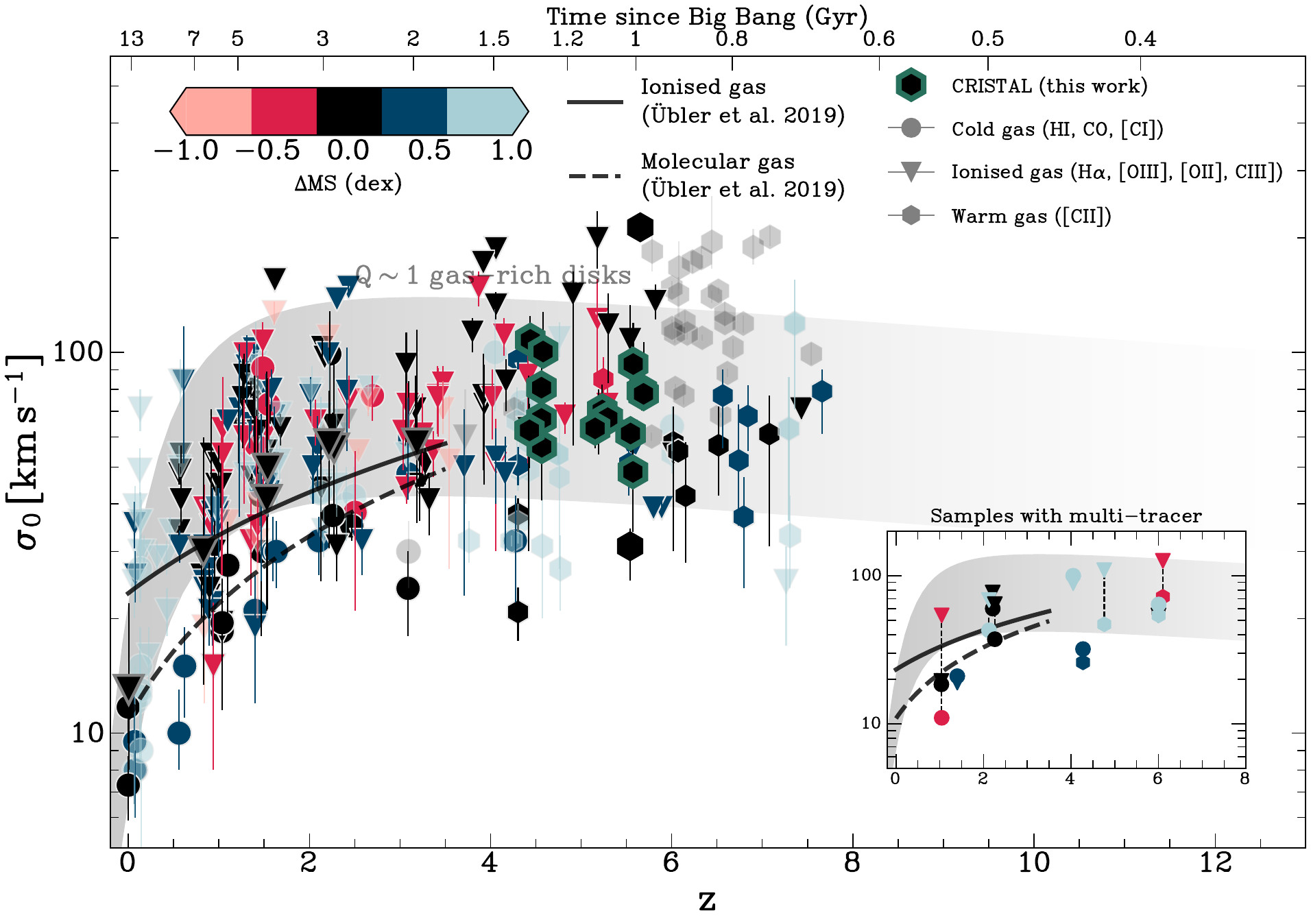}\\
    \includegraphics[width=0.84\textwidth]{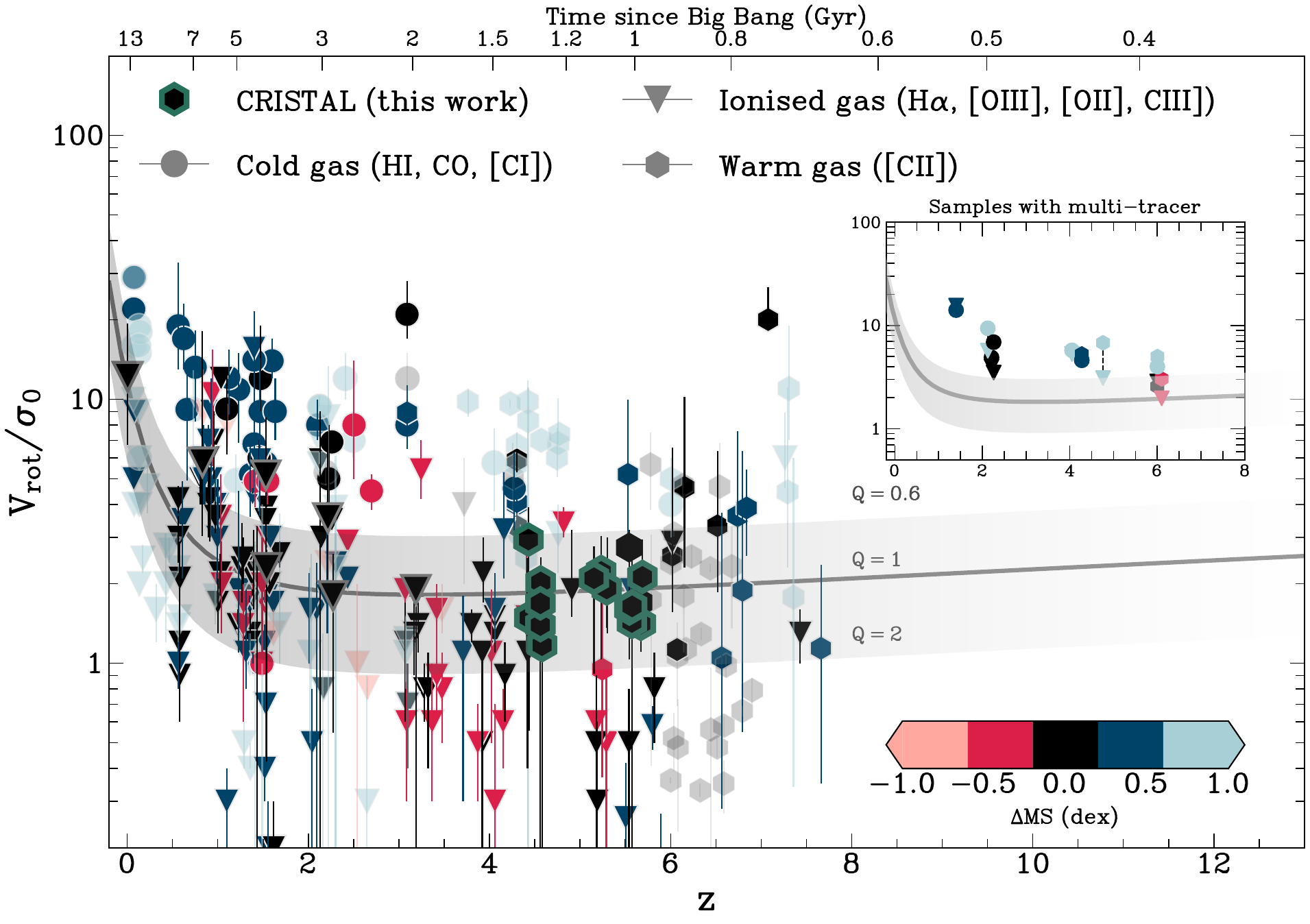}
    \caption{
    Dynamical evolution of high-$z$ galaxies. 
    \textit{Top}: Intrinsic velocity dispersion 
    $\sigma_0$ as a function of redshift for our disk sample (black hexagons with green outline) 
    compared with the literature values (see Appendix~\ref{app:literature_refs} Table~\ref{tab:literature_sample}). 
    \textit{Bottom}: Same but for the ratio between rotational velocity and $\sigma_0$, $V_{\rm rot}/\sigma_0$.
    The solid and dashed black lines are the \citet{Uebler2019}'s best-fit relations for galaxies at $z$\,$<$\,$4$, for ionised and cold molecular gas, respectively. 
    Where available, the points are colour-coded by their main-sequence offset $\Delta$MS relative to the \citet{Speagle2014} relation, otherwise in grey.
    The insets shows the literature sample (see Table~\ref{tab:literature_sample_multiphase}) of 
    the same galaxy with two gas tracers tracing different phases.
    The insets include the galaxies with large uncertainties which are omitted in the main plots.
    The grey shading encloses the corresponding range in $\sigma_0$ and $V_{\rm rot}/\sigma_0$ for $Q\in[0.6,2.0]$.
    The definition of the Toomre $Q$ parameter is presented in \S~\ref{subsec:toomre}.
    }
    \label{fig:s0z}
\end{figure*}
\begin{figure*}
    \centering
    \includegraphics[width=1.01\textwidth]{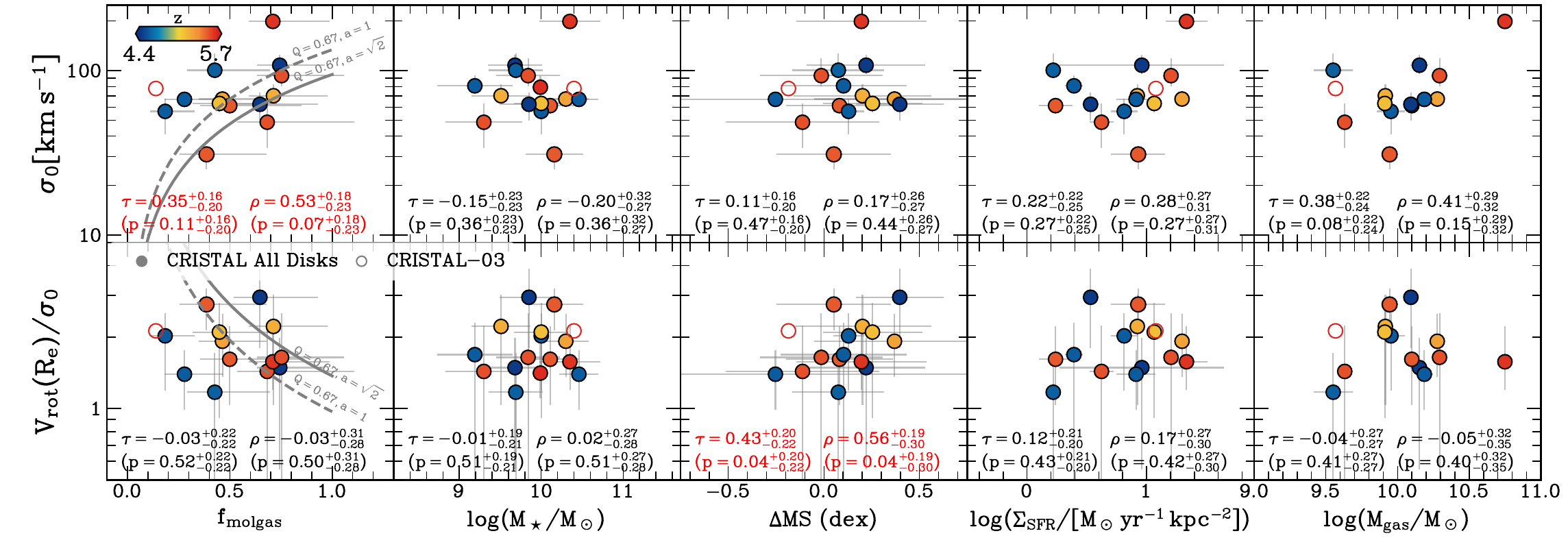}
    \caption{Intrinsic velocity dispersion ($\sigma_0$) and the ratio of rotation velocity at the effective radius ($V_{\rm rot}(R_{\rm e})$) to intrinsic velocity dispersion ($\sigma_0$) as a function of five galaxy properties: 
    molecular gas fraction ($f_{\rm molgas}$), 
    stellar mass ($\log(M_*/M_\odot)$), offset from the main sequence ($\Delta$MS), 
    star formation rate surface density ($\Sigma_{\rm SFR}$), and 
    molecular gas mass ($\log(M_{\rm gas}/M_\odot)$). 
    The dashed and solid grey curves in the first column are the predicted trends based on Equation~\ref{eqn:toomreQ} with $Q$ fixed at $Q_{\rm crit}=0.67$, with $a$\,$=$\,$1$ and $a=$\,$\sqrt{2}$, respectively.
    The Kendall and Spearman's rank correlation coefficients 
    ($\tau$ and $\rho$) and their corresponding $p$-values are shown alongside the 1-$\sigma$ percentile errors. 
     Taken $\sigma_0$ at face values, it shows a tentative correlation (highlighted in red) with $f_{\rm molgas}$, 
    while $V_{\rm rot}(R_{\rm e})$ correlates most significantly with $\Delta$MS. 
    The most poorly resolved galaxy CRISTAL-03 (empty red circle), 
    with an extent covered by $<$\,$2$ beams, is excluded from the correlation analysis. 
    The limited statistics and narrow parameter range of our data lead to large uncertainties in the correlation coefficients and their $p$-values, and the trends should be interpreted with caution.
    }
    \label{fig:svsphy}
\end{figure*}

\section{Disk turbulence and dynamical support}\label{sec:dispersion_props}
\subsection{Comparison to other samples and redshift trends}\label{sec:turbulence_z}
To put CRISTAL disks' kinematics in the context of dynamical evolution over cosmic time, 
we present in Fig.~\ref{fig:s0z} the intrinsic dispersion 
$\sigma_0$ and the dynamical support from $V_{\rm rot}/\sigma_0$, compared with literature values from local to distant galaxies up to $z$\,$\lesssim$\,$8$. 
The data points are coded according to the MS offset $\Delta$MS of the galaxies, and different symbols distinguish measurements based on tracers of atomic, cold molecular, warm, and ionised gas phases.

The literature compilation is listed in Appendix~\ref{app:literature_refs} Table~\ref{tab:literature_sample}.
It includes studies of local galaxies and local analogues observed in \ion{H}{I}, CO, H${\alpha}$\ or [\ion{O}{II}], as well as unlensed and lensed galaxies at 
$0.5$\,$<$\,$z$\,$<$\,$4$ 
traced by CO, [\ion{C}{I}], H${\alpha}$, [\ion{O}{III}]\ or [\ion{C}{II}], and 
at $z$\,$\geqslant$\,$4$ traced by H${\alpha}$, CO, [\ion{C}{II}], [\ion{O}{III}], or [\ion{C}{III}].
We only consider the systems classified as disk-like.
We exclude those measurements with uncertainties in $\sigma$ and $V_{\rm rot}$ greater than $50\%$, but include all disks in CRISTAL without this cut.

The definition and methodology employed for $V_{\rm rot}$ and $\sigma_0$ vary between studies.  
In cases where $V_{\rm rot}$ is not available at $R_{\rm e}$ is not available, $V_{\rm rot}$ is taken as the maximum velocity $V_{\rm max}$ or $V_{\rm rot}(2.2R_{\rm e})$.
For parametric modelling, 
the profiles adopted for $V_{\rm rot}$ include $\arctan$ and the multi-parameter function from \citet{Courteau1997}.
The definition of $\sigma$ also varies across literature; for non-parametric modelling, such as using \texttt{$^{\rm 3D}$Barolo}\ \citep{DiTeodoro2015} or \texttt{KinMS} \citep{Davis2013,Davis2017}, 
it would be either the median or mean of the radial profile; 
for parametric modelling, which assumed either a constant profile $\sigma(R)$\,$=$\,$\sigma_0$, $\sigma(R)$\,$=$\,$\sigma_0 \exp(-R/R_\sigma)$ 
or other functions, we adopt the $\sigma_0$ or the median in the latter cases, following the choice of the original authors. 
The literature values are also a mix of data obtained from various observational methods, including IFU, interferometry, and slit spectroscopy. The slit-based method tends to give higher $\sigma$ values than the other two \citep{Uebler2019}. Different CO transitions can also trace gas with various kinematic and spatial properties.

We recalculate $\Delta$MS using the relation of \citet{Speagle2014}, 
extrapolated to the redshift range of the samples. Recent studies of the star-forming main sequence at $z$\,$\gtrsim$\,$5$, utilising \textit{JWST} imaging data, have provided support for this extrapolated relationship \citep{Cole2025,Koprowski2024}. 
When available, the stellar mass $M_\star$ values are taken directly from the literature, 
which was derived from spectral energy distribution (SED)  fitting using various tools and assuming different initial mass functions (IMFs) 
or decomposition of RC. 
In cases where $M_\star$ is not reported, 
we estimate it from the dynamical and gas mass ($M_\star$\,$=$\,$M_{\rm dyn} - M_{\rm gas}$).

In Fig.~\ref{fig:s0z}, the literature values of $\sigma_0$ in ionised and molecular gas tracers are both displaying 
an overall increasing trend with redshift. 
For MS galaxies up to $z$\,$\sim$\,$4$, 
the trends are well-described by the best-fit relations derived by \citet{Uebler2019}.
Qualitatively, extrapolating these relations matches the evolutionary trend at even higher redshifts for the MS galaxies and agrees well with the CRISTAL values. 
On the other hand, some starburst galaxies observed with [\ion{C}{II}] at similar epochs lie below the extrapolated relationships.

Sample selection differences between studies could partly explain the large spread in $\sigma_0$ (and $V_{\rm rot}/\sigma_0$). As extensively discussed by \citet{Wisnioski2025}, the interpretation of the dispersion 
should also consider the different ISM phases probed by the various kinematic tracers and, relatedly, the varying contributions of different gas phases to the [\ion{C}{II}] line emission as a function of $\Sigma_{\rm SFR}$ and other properties \citep[e.g.][]{Cormier2019,Wolfire2022,Ikeda2025}.
To better understand the potential dependence on ISM phases, it is essential to study the same object using multiple tracers; currently, this has only been done for a limited number of samples (Table~\ref{tab:literature_sample_multiphase}), as highlighted in the insets of Fig.~\ref{fig:s0z}.

Compared to unlensed MS SFGs observed with [\ion{C}{II}] at similar epochs, 
CRISTAL disks have comparable values of $\sigma_0$ within uncertainties, 
with a median difference of $11\,$${\rm km\,s^{-1}}$, 
and a lower 
$V_{\rm rot}/\sigma_0$ by $0.8$ in median. 
This is possibly because CRISTAL disks are less massive in $M_\star$ than the literature samples by an average of $0.15\,$dex. 
Given the mass-dependence of $V_{\rm rot}/\sigma$ in simulations that span a wider dynamic range \citep[e.g.][]{Dekel2020,Kohandel2024} than allowed by our data, 
the lower mass of our sample may explain the lower $V_{\rm rot}/\sigma_0$ values.
When compared with the same population observed in ionised gas, now possible thanks to \textit{JWST}, CRISTAL disks are in very good agreement with the `gold' sample in \citet{Danhaive2025}, with median differences of only $\sim$\,2\% in $\sigma_0$ and $\sim$\,10\% in $V_{\rm rot}(R_{\rm e})/\sigma_0$.

On the other hand, compared with the lensed samples,
CRISTAL disks have higher $\sigma_0$, by 
$34\,$${\rm km\,s^{-1}}$ in median, 
and significantly lower $V_{\rm rot}/\sigma_0$ by $-6$.
In particular, the lensed samples tend to be starburst galaxies, 
and have smaller sizes (with $R_{\rm e}$\ typically $\lesssim$\,$1.5\,$kpc, see also Fig.~\ref{fig:sample_diff}) 
that are $\sim$\,$50\%$ and $\sim$\,$70\%$ smaller than the unlensed galaxies and CRISTAL disks, respectively. 
The starburst and compact nature of these galaxies suggest that they have experienced a distinct assembly history \citep{Stach2018,Hayward2021}, differing from that of the more typical galaxy populations in CRISTAL.

\subsection{Trends with galaxy properties}\label{sec:trends} 
We explore here trends with galaxy properties, including molecular gas mass fraction ($f_{\rm molgas}$), stellar mass ($M_{\star}$), MS offset ($\Delta$MS), SFR surface density ($\Sigma_{\rm SFR}$), and molecular gas mass ($M_{\rm gas}$).  
Fig.~\ref{fig:svsphy} plots the derived $\sigma_0$ and $V_{\rm rot}/\sigma_0$ as a function of these properties for the CRISTAL disk sample.  
We quantify the correlations by computing the Spearman's $\rho$ and Kendall's $\tau$, and their $p$-values to assess the significance of any possible correlations.
The resulting coefficients and the $p$-values with confidence intervals are annotated in Fig.~\ref{fig:svsphy}. 

For $\sigma_0$, the strongest correlation observed is with $f_{\rm molgas}$.  The $V_{\rm rot}/\sigma_0$ appears to correlate most importantly with $\Delta$MS. 
No other obvious trend is detected with the other galaxy properties.  
The dependence of $\sigma_0$ with $f_{\rm molgas}$ is in line with expectations for marginally stable, gas rich disks as discussed in \S~\ref{sec:turbulence_z}. 
The trend between $V_{\rm rot}/\sigma_0$ and $\Delta$MS may reflect an underlying dependence on $\Sigma_{\rm bar}$ (see also the right panel of Fig.~\ref{fig:sample_diff}). However, although the correlation coefficient $\rho$ between $V_{\rm rot}/\sigma_0$ and $\Sigma_{\rm bar}$ is $\sim$\,0.5, the large accompanying $p$-value suggests that more precise size measurements are required to confirm this relationship.
Clearly, the CRISTAL disk sample is small and only the strongest correlations can be discerned.  
Future larger samples of near MS SFGs at $z$\,$\sim$\,$4$--$6$ will be important to strengthen the results,
{such as the literature compilation efforts by \citet{Wisnioski2025}.}

% \Lilian{Some afterthoughts from NMFS review: could this correlation arise from the short-timescale ($\sim10^{7-8}$ years) variations in SFR at a given mass, induced by the breathing cycles of star-formation feedback and temporal fluctuations in the rate of gaseous inflows and/or minor mergers, in which case, the $V_{\rm rot}/\sigma$ would be driven up temporarily?}

\subsection{Turbulence in the framework of marginally Toomre-stable disks}\label{subsec:toomre}
The observed evolutionary trend of $\sigma_0$ and $V_{\rm rot}/\sigma_0$ discussed above has been attributed to 
the increasing gas fraction at higher redshifts \citep[e.g.][]{Tacconi2010,Tacconi2020}, 
as predicted by Toomre theory.
The correlation of $\sigma_0$ with $f_{\rm molgas}$ for the CRISTAL disks discussed above also is in line with expectations for marginally stable gas-rich disks.  
In this framework, the stability of the disks against fragmentation and local gravitational collapse is directly linked to the level of turbulence in the ISM. Turbulence is driven by both \textit{ex-situ}, such as accretion from the cosmic web, and \textit{in-situ}, including radial flows and clump migration, which release gravitational potential energy. This creates a self-regulating cycle that maintains the disk in a state of marginal stability.
Following Eq.~(3) in \citet{Genzel2014a} \citep[see also,][]{Uebler2019,Genzel2011,Genzel2023,Liu2023}, the classical \citet{Toomre64} parameter $Q$ can be formulated as:
\begin{equation}
Q_{\rm gas} = \frac{\sigma_0\kappa}{\pi G\Sigma_{\rm gas}} = \Big(\frac{\sigma_0}{v_c}\Big)\Big(\frac{a}{f_{\rm gas}}\Big),    
\label{eqn:toomreQ}
\end{equation}
where the epicyclic frequency is 
$\kappa$\,$=$\,$\sqrt{Rd\Omega^2/dR+4\Omega^2}$ and $\Omega$\,$=$\,$V_{\rm rot}/R$. 
The constant $a$ depends on the rotational structure of the disk: $a$\,$=$\,$\sqrt{1}$ for Keplerian-like rotation and $a$\,$=$\,$\sqrt{2}$ for a disk with constant rotational velocity. 
For a quasi-stable thick gas disk, $Q_{\rm crit}$\,$=$\,$0.67$ \citep[e.g.][]{Behrendt2015}. 
The two panels in the first column of Fig.~\ref{fig:svsphy} plot the predicted trends of $\sigma(f_{\rm gas})$ and $V_{\rm rot}/\sigma(f_{\rm gas})$ based on Eq.~(\ref{eqn:toomreQ}) with $Q$\,$=$\,$Q_{\rm crit}$ and $a$\,$=$\,$\sqrt{1...2}$, 
taking into account that some RCs show a drop-off. Overall, there is a good match between the predicted trend and the CRISTAL values.

Specifically, taking the median values of CRISTAL disks,
$\sigma_0/v_c$\,$=$\,$70/200$\,$=$\,$0.4$ and $f_{\rm molgas}$\,$\sim$\,$0.5$.
The corresponding values of $Q_{\rm gas}$ is then $0.6$, 
and the entire sample has $Q$ in the range $[0.4, 2.0]$, indicating that the CRISTAL disks are, 
on average, marginally gravitationally stable. 
The $Q$ values are broadly similar to the results of \citet{Uebler2019} for their $z$\,$\sim$\,$1$--$3$ samples. 
The similar Toomre $Q$ values inferred for MS SFGs from $z$\,$\sim$\,$5$ to $z$\,$\sim$\,$1$ 
in MS SFGs suggests that this galaxy population has grown in a marginally stable and self-regulating manner for at least 5 billion years of cosmic time.

From a broader evolutionary perspective based on the Toomre theory for gas-rich disks, the observed trends from the literature combining with CRISTAL are in remarkably good agreement.
The grey bands in both panels of Fig.~\ref{fig:s0z} show the prediction 
in the Toomre framework for the evolution of dispersion for $\log(M_\star/{\rm M_\odot})$\,$=$\,$10.0$ galaxies with $Q_{\rm crit}$\,$\in$\,$[0.4,2.0]$ and $v_c$\,$=$\,$200{\,\rm km\,s^{-1}}$. 
These values are appropriate for the CRISTAL disk sample (and differ from more massive samples studied at lower redshifts, e.g., \citealt{Wisnioski2015}).
The gas fraction $f_{\rm molgas}$\ adopted evolves according to the scaling relation in \citet{Tacconi2020} 
which is a function of stellar mass, SFR, and size. 
The SFR and size evolution with redshift is determined from the \citet{Speagle2014} MS relation and \citet{vanderWel2014} mass-size relation. 
CRISTAL galaxies have $f_{\rm molgas}$$\sim$\,$51\%$, consistent with the expected value from \citet{Tacconi2020}'s relation at $z$\,$=$\,$5$ for $M_\star$\,$=$\,$10^{10}\,{\rm M_{\odot}}$, which is $\sim$53\% (see discussions in Appendix~\ref{app:dust2gas}).

\subsection{Drivers of the gas turbulence}\label{subsec:KH18}
To explore the relative contribution of star formation- versus gravitational instability-driven turbulence in CRISTAL disks, we compare our results with the analytic model of \citet{Krumholz2018}. 
This model combines stellar feedback and gravitational processes to drive turbulence, incorporating prescriptions for star formation, stellar feedback, and gravitational instabilities into a unified `transport+feedback' framework.
In the model, gas is in vertical hydrostatic equilibrium and energy equilibrium, with energy losses through turbulence decay balanced by energy input from stellar feedback and the release of gravitational energy via mass transport through the disk. 
Based on the model, there is a critical value of gas velocity dispersion ($\sigma_g$), $\sigma_{\rm sf}$, at which the amount of turbulence can be sustained by star formation alone, without the need for gravitational instability or radial transport. In such a case, $\sigma_g$ is related to $\sigma_{\rm sf}$ by {(Eq.~(39) in \citealt{Krumholz2018})}
\begin{equation}
\begin{split}
\sigma_{\rm g}=\sigma_{\rm sf} \equiv & \frac{4f_{\rm SF}\epsilon_{\rm ff}}{\sqrt{3f_{g,P}}\pi\eta\phi_{\rm mp}\phi_Q\phi_{\rm nt}^{3/2}}\left\langle\frac{p_*}{m_*}\right\rangle \\ &  \cdot {\rm max}\left[1,\,\sqrt{\frac{3f_{g,P}}{8(1+\beta)}}\frac{Q_{\rm min}\phi_{\rm mp}}{4f_{g,Q}\epsilon_{\rm ff}}\frac{t_{\rm orb}}{t_{\rm sf,max}}\right].
\end{split}
\label{eqn:sigma_g}
\end{equation}
Following Table~3 in \citet{Krumholz2018} for high-$z$ galaxies, the fraction of ISM in the star-forming molecular phase, $f_{\rm SF}$, is set to 1.0; $t_{\rm SF, max}$\,$=$\,$2\,$Gyr; 
the fractional contribution of gas to the mid-plane
pressure and $Q$, $f_{g,P}$ and $f_{g,Q}$, respectively, are both assumed to be $0.7$; the slope index of the RC, $\beta$\,$=$\,$d\ln v_{\phi}/d\ln r$, is set to 0.0 (i.e., flat), in which $v_\phi$ is the circular velocity $V_{\rm circ}$; 
the Toomre parameter, $Q$, is fixed at 1, following the fiducial value. 
The orbital period $t_{\rm orb}$\,$=$\,$2\pi r/ V_{\rm circ}$\,$\in$\,$[30,120]\,$Myr, and is adjusted to the values of our sample.
The other values that we adopt are listed in Table~\ref{tab:k18_fid_params} in Appendix~\ref{app:krumholz18_SFR}.
The $\sigma_g$ from Eq.~(\ref{eqn:sigma_g}) is therefore $\sim$\,$10\,$${\rm km\,s^{-1}}$. 
Dispersion much larger than this critical value ($\gtrsim$\,20\,${\rm km\,s^{-1}}$) requires gravitational instability or radial mass transport for moderate SFR.

\begin{figure}
    \centering
    \includegraphics[width=0.45\textwidth]{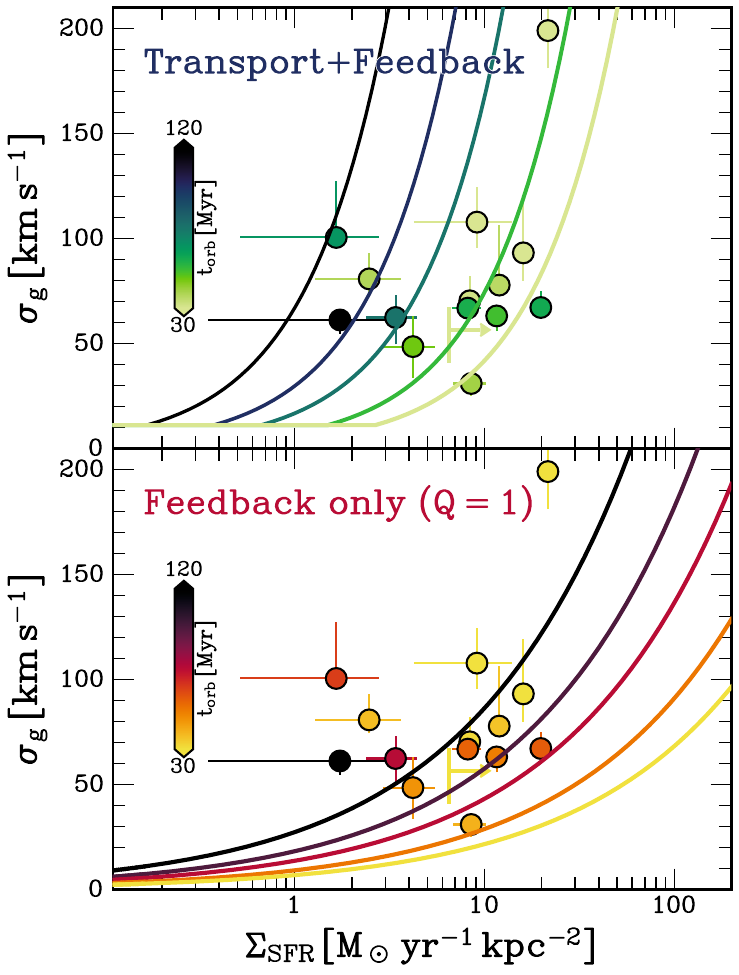}
    \caption{$\sigma_0$ vs $\Sigma_{\rm SFR}$ of CRISTAL disks compared with analytical models from \citet{Krumholz2018}. The solid lines in the upper panel show the $\sigma_0$ values predicted from the Transport+Feedback model with orbital period $t_{\rm orb}=[30,120]\,$Myr. Similarly for the lower panel but for the `Feedback-only' model.
    The results are broadly consistent with the `feedback+transport' model, suggesting that the elevated velocity dispersion of normal star-forming galaxies at this epoch requires additional gravitational energy from mass transport across the disk.
    }

    \label{fig:kh18_comp}
\end{figure}

\begin{figure*}
    \centering
    \includegraphics[width=\textwidth]{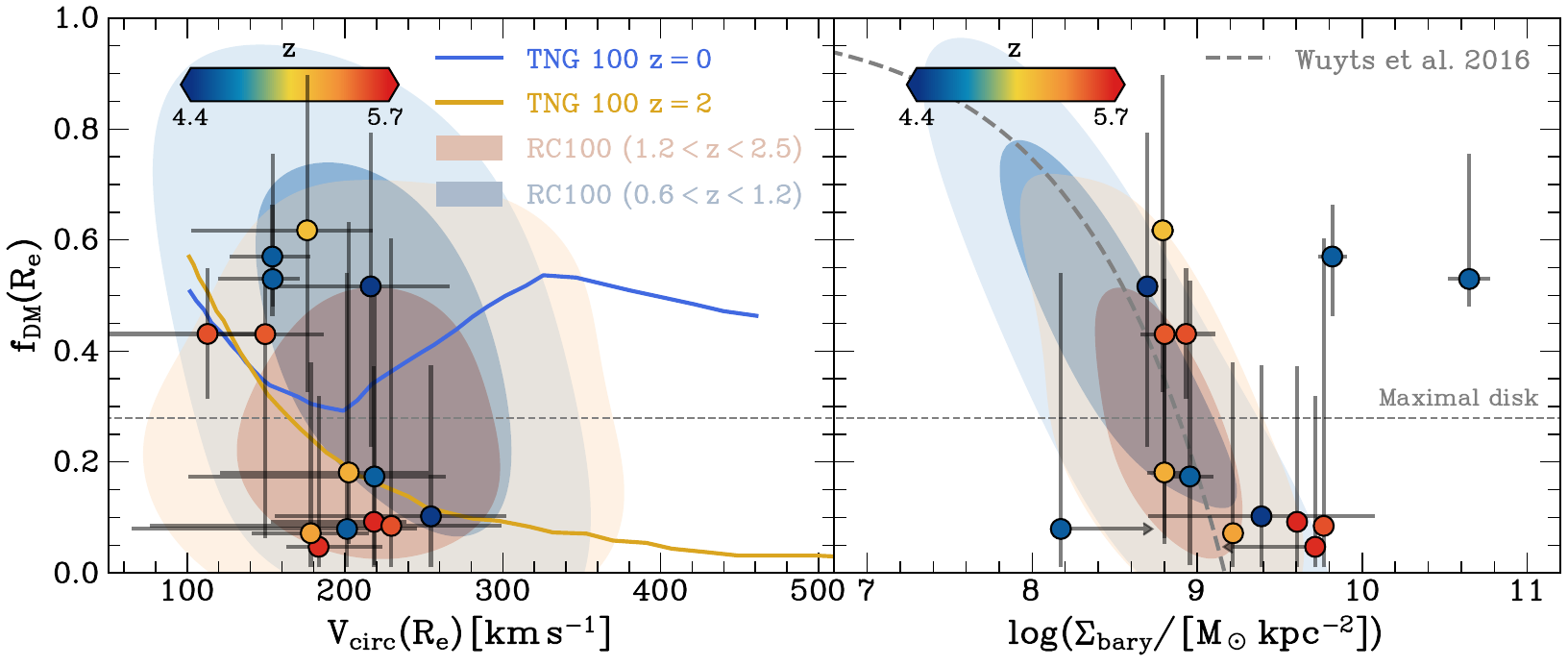}
    \caption{
    Dark matter (DM) fraction $f_{\rm DM}$\ as a function of the circular velocity at the effective radius ($V_{\rm circ}$($R_{\rm e}$), \textit{left}) and baryonic surface density ($\Sigma_{\rm bary}$) (\textit{right}). 
    The yellow and blue curves show the TNG100 (without adiabatic contraction) relations at $z$\,$=$\,$0$ and $z$\,$=$\,$2$ \citep{Lovell2018}, respectively. The pale blue and yellow shading are measurements from RC100 at $z$\,$<$\,$2.5$ \citep{Nestor2023}. The grey dashed curve is the best-fit relation of $z\sim2$ star-forming galaxies from \citet{Wuyts2016}.
    The CRISTAL sample statistics suggest a tentative anti-correlation with $f_{\rm DM}$($R_{\rm e}$) and $\Sigma_{\rm bary}$, although individual galaxy uncertainties are limited by the data depth. 
    The horizontal dashed line denotes $f_{\rm DM}$ 
    for a maximal disk $f_{\rm DM}$\,$\coloneqq$\,$28\%$ \citep{vanAlbada1985}.
    {In the \textit{right} panel, the two symbols plotted with only limit arrows represent CRISTAL-10a-E (upper limit) and CRISTAL-06b (lower limit). The stellar mass is only for CRISTAL-10a as a whole, which includes CRISTAL-10a-E, while no gas mass is available for CRISTAL-06b.
}
    }
    \label{fig:fdm_vc}
\end{figure*}

\begin{figure}
    \centering
    \includegraphics[width=0.48\textwidth]{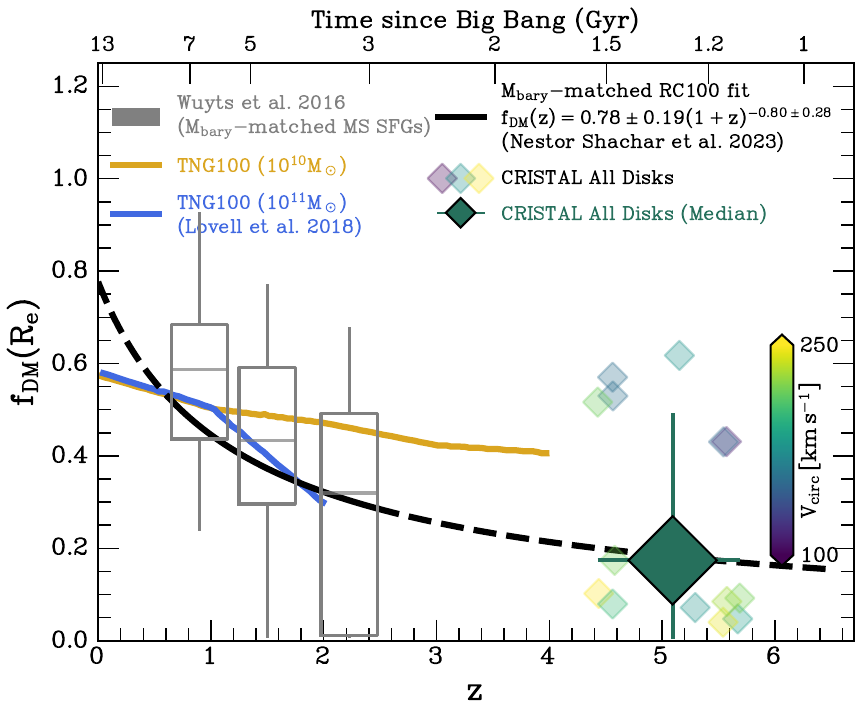}
    \caption{The dark matter (DM) fraction at effective radius ($f_{\rm DM}$($R_{\rm e}$)) as a function of redshift. 
    The $f_{\rm DM}$($R_{\rm e}$) of individual CRISTAL disks ($4$\,$<$\,$z$\,$<$\,$6$) are represented by small diamonds, colour-coded by their $V_{\rm circ}$($R_{\rm e}$).
    For comparison, we include lower redshift studies around $z$\,$=$\,$2$
    from \citet{Wuyts2016} (grey box-and-whisker) and the best-fit relation to the RC100 data set \citep{Nestor2023} (black solid line for the redshift range covered and dashed line for the extrapolated range).
    For both observational studies, we consider only star-forming galaxies on the main sequence (MS SFGs) and the baryonic mass $M_{\rm bary}$ matched within 0.3\,dex of the CRISTAL disks. The new best-fit relation to RC100, using the $M_{\rm bary}$-matched sample, is shown in the legend. 
    The trends predicted from the TNG100 simulation for $M_*$\,$=$\,$10^{10}M_\odot$ and $M_*$\,$=$\,$10^{11}M_\odot$ are shown in yellow and blue, respectively \citep{Lovell2018}.
    Overall, the CRISTAL disks tend to be baryon-dominated on galactic scales, with a median DM fraction of $\sim$\,${18}\%$ (large green diamond) that is tentatively consistent with the extrapolated relation based on RC100.
    However, there is a significant scatter in the values among the sample, partly driven by the scattered distributions of circular velocity $V_{\rm circ}$($R_{\rm e}$) (and $\Sigma_{\rm bary}$) shown in Figure~\ref{fig:fdm_vc}.
    }
    \label{fig:fdm_z}
\end{figure}
In the `transport+feedback' model, the star formation rate surface density $\Sigma_{\rm SFR}$ and the gas velocity dispersion $\sigma_g$ can be related as (Eq.~(59) in \citealt{Krumholz2018}):
\begin{equation}
\begin{split}
    \Sigma_{\rm SFR} =& f_{\rm SF}\frac{\sqrt{8(1+\beta)}f_{g,Q}}{GQ} \frac{\sigma_g}{t^2_{\rm orb}} \\ 
    & \cdot \text{max} \left[\frac{8\epsilon_{\rm ff} f_{g,Q}}{Q} \sqrt{\frac{2(1+\beta)}{3f_{g,P}\phi_{\rm mp}}}, \frac{t_{\rm orb}}{t_{\rm SF, max}} \right],
\end{split}
\label{eqn:sigma_sfr}
\end{equation}
while for `feedback-only' (fixed $Q$) model (Eq.~(61) in \citealt{Krumholz2018}): 
\begin{equation}
\begin{split}
    \Sigma_{\rm SFR} = \frac{8(1+\beta)\pi\eta\sqrt{\phi_{\rm mp}{\phi_{\rm nt}^3}}\phi_{Q}}{GQ^2\langle p_*/m_*\rangle f_{g,P}}\frac{\sigma_{\rm g}^2}{t_{\rm orb}^2}.
\end{split}
\label{eqn:sigma_sfr_no_transport}
\end{equation}

In Fig.~\ref{fig:kh18_comp} we show the $\sigma_0$--$\Sigma_{\rm SFR}$\footnote{For CRISTAL-23c in which the $\Sigma_{\rm SFR}$ is not available, we adopt the $\Sigma_{\rm SFR,IR}$ 
from \citet{Bethermin2023} as a lower limit, although the contribution from unobscured star formation is not significant.}
measurements of CRISTAL disks, compared with the `transport+feedback' and `feedback only' model of \citet{Krumholz2018}. 
For reference, we also compare $\sigma_0$ and SFR in Appendix~\ref{app:krumholz18_SFR}.
Overall, our results are broadly consistent with the `feedback+transport' model of \citet{Krumholz2018}, 
which suggests that the high-velocity dispersion 
of normal SFGs can be predominantly attributed to the release of gravitational energy from mass transport across the disk.

This result differs from some previous studies at similar epochs, which found that star formation feedback alone can sustain the observed dispersion in starburst-like galaxies \citep[e.g.][]{RomanOliveira2023,Rowland2024}.
However, our analysis of CRISTAL MS disks, characterised by modest star formation activity, indicates that a different dominant mechanism drives turbulence in the ISM of MS SFGs.

The result is nevertheless consistent with the weak correlation of $\sigma$ with global or local SFR ($\Sigma_{\rm SFR}$) of our sample as shown in \S~\ref{sec:trends}.
Such a weak correlation is also found in
\citet{Genzel2011,Johnson2018} and \citet{Uebler2019} for cosmic noon galaxies (after redshift normalisation), 
and in the nearby universe \citep[e.g.][]{Elmegreen2022}.  
This is also in agreement with the theoretical works \citep[e.g.][]{Shetty2012,Kim2018}, which have derived a weak dependence of gas velocity dispersion on the supernova explosion rate.

We note, however, that for CRISTAL-05 with its relatively low 
$\sigma_0$\,$\approx$\,$31\,$${\rm km\,s^{-1}}$ 
and $\Sigma_{\rm SFR}$\,$=$\,$8.5\,{\rm M_{\odot}\,yr^{-1}\,kpc^{-2}}$ ($\rm{SFR}$\,$=$\,$68\,{\rm M_{\odot}}\,{\rm yr}^{-1}$), 
the stellar feedback-only model would better match the observed values of $\sigma_0$ and $\Sigma_{\rm SFR}$ (and SFR), suggesting that different mechanisms among the disk samples may contribute to varying degrees of the observed velocity dispersion, {as seen also in simulations \citep[e.g.][]{Jimenez2023}}. 
Additionally, spatial variation of different mechanisms within a single galaxy is also possible, 
but the resolution of our data is currently insufficient to reveal such variation conclusively.
In the future, higher resolution observations of kinematics and SFR maps would enable to test more directly the coupling (or lack thereof) between $\sigma_0$ and stellar feedback.
Other simulation works also show that stellar feedback can sustain higher dispersions compared to the \citealt{Krumholz2018}'s analytical treatment \citep{Gatto2015,Orr2020,Rathjen2023}.
The relative contribution of stellar feedback versus gas transport depends on halo mass and redshift, in which gas transport plays a more dominant role in the high redshift systems \citep{Ginzburg2022}.

\section{Exploration of galactic DM fraction and mass budget}\label{sec:dm}
On the galactic scale, CRISTAL disks tend to be baryonic-dominated, 
with low $f_{\rm DM}$ at $R_{\rm e}$\ (Table~\ref{tab:dysmalpy_results}), 
having a {median value of ${18\%}$ (mean\,${=27\%}$)}, comparable to or less than maximal disks \citep[$f_{\rm DM}$\,$\coloneqq$\,$28\%$][]{vanAlbada1985}, 
albeit with significant scatter among the samples that span a wider range from a few \% to $\sim60\%$.
In comparison, the Galaxy's $f_{\rm DM}(R_{\rm e})$\,$=$\,$0.38\pm0.1$ \citep{Bovy2013,BH2016}.

The radial profiles of $f_{\rm DM}$ of CRISTAL disks are shown in Fig.~\ref{fig:mass_profile}, which is defined in \texttt{DysmalPy} as \footnote{We note that in our analysis, ${f_{\rm DM}}$ at ${R_{e,\mathrm{disk}}}$ is a free parameter --- by instead calculating the total halo mass given the specified baryonic mass distribution and ${f_{\rm DM}(R_{\rm e,\mathrm{disk}})}$, the full ${f_{\rm DM}}$ radial profile for each galaxy is then computed given the best-fit parameters (${M_{\rm bar}}$, ${R_{\rm e,disk}}$, ${f_{\rm DM}}$ enclosed within ${R_{\rm e}}$, and ${\sigma_0}$).}
\begin{equation}
f_{\rm DM}(<R)\coloneqq V_\mathrm{circ,DM}^2(R)/V^2_\mathrm{circ,tot}(R).
\label{eqn:fdm}
\end{equation}

We observe a tentative inverse relationship between $f_{\rm DM}(R_{\rm e})$ 
and circular velocity ($V_{\rm circ}$) in Fig.~\ref{fig:fdm_vc}, which is pressure-support corrected (Eq.~(\ref{eqn:dy_pressure_support})), and similarly for the baryonic surface density $\Sigma_{\rm bary}$ based on the values in Tables~\ref{tab:main_table} and \ref{tab:kins_props}.
This trend is similar to that observed in \citet{Nestor2023} for cosmic noon MS SFGs. 
Such an inverse correlation is well-established for local SFGs, 
where the most DM-dominated disks are those with low 
baryonic mass surface density and circular velocity 
\citep[e.g.][and references therein]{Martinsson2013a,Martinsson2013b,Courteau2015}.
We also compare our results to \citet{Wuyts2016}, who derived $f_{\rm DM}$ at the inner disk by subtracting the sum of stellar and gas masses from the dynamical mass 
obtained from RCs of 240 galaxies, assuming $f_{\rm bary}$\,$=$\,$M_{\rm bary}/M_{\rm dyn}$\,$=$\,$1-f_{\rm DM}$. 
We find that most of the CRISTAL disks follow the \citet{Wuyts2016} relation on the $\Sigma_{\rm bary}$--$f_{\rm DM}$ plane, 
except for CRISTAL-23b, and 23c, which are both disk-like galaxies
in an interacting system.

The low $f_{\rm DM}$($R_{\rm e}$) of our sample is broadly consistent with the general trend of decreasing DM fraction towards higher redshifts. 
Fig.~\ref{fig:fdm_z} shows the median $f_{\rm DM}$($R_{\rm e}$) of our sample aligns with the extrapolated trend toward higher redshift from \citet{Nestor2023} 
and \citet{Wuyts2016}, when considering a sample matched in $M_{\rm bary}$ and $\Delta$MS.
Both $f_{\rm DM}$ are below the expectations from the TNG100 simulation from \citet{Lovell2018}.
This could be due to the insufficient physical resolution of large-scale cosmological simulation to resolve sub-galactic processes \citep{Uebler2021} and the effect of adiabatic contraction \citep{Blumenthal1986}.
The possible drivers of DM deficit in MS SFGs have been discussed thoroughly by, e.g. \citealt{Genzel2020} and \citealt{Nestor2023}. 
It is potentially linked to kinetic heating due to the efficient transport of baryons to central regions in gas-rich systems \citep[e.g.][]{ElZant2001},
and/or strong feedback processes redistributing DM to larger radii \citep[e.g.][]{Freundlich2020}.
One example is CRISTAL-02, which has $f_{\rm DM}$($R_{\rm e}$)$<0.1$ and drives a vigorous outflow detected in [\ion{C}{II}] \citep{Davies2025}.

The widely scattered distribution in Fig.~\ref{fig:fdm_z} with large errors associated with individual galaxies prevents a definite conclusion on the physical origin of this distribution.
We investigated whether the inhomogeneous radial coverage of the RCs (Fig.~\ref{fig:allRC_mdl}) could systematically drive up the $f_{\rm DM}$($R_{\rm e}$) for galaxies with limited radial coverage. 
However, we do not find a straightforward one-to-one correspondence between $f_{\rm DM}$($R_{\rm e}$) and 
the ratio of 
$R_{\rm out}/$$R_{\rm e}$ (Kendall's $\tau=-0.19^{+0.23}_{-0.25}$, $p=0.32^{+0.23}_{-0.25}$).

We emphasise that $f_{\rm DM}$ is measured at the effective radius, and we lack constraints of the DM distribution on the halo scale ($\gg$$R_{\rm e}$) with our current data, except for the very compact CRISTAL-23c. 
Extrapolating the DM mass from the inner disk to the virial scale with an NFW distribution will 
result in an unphysically large baryon fraction larger than the cosmic baryon fraction \citep{Genzel2017}. 
% Stacking techniques \citep[e.g.][]{Lang2017,Tiley2019b} would help to increase the S/N at the outer regions.
{As discussed in \S~\ref{subsec:dysmalpy}}, the intrinsic, circular velocities would also depend on the assumption of the pressure support corrections, 
which would explain the differences found in the literature (cf. {\citet{Sharma2021} with, e.g. \citet{Genzel2020,Price2021} and \citet{Nestor2023}}).
If we were to assume a non-constant $\sigma$, 
such as an exponential decline, for the pressure support correction, 
rather than using Eq.~(\ref{eqn:dy_pressure_support}), 
the correction to $V_{\rm rot}$ would be even larger \citep[Eq.~(12) in][]{Price2022}, 
leading to a more steeply declining $V_{\rm circ}$, 
which would further exacerbate the discrepancies between our observations and simulations. 

Although we have adopted the NFW profile for the DM profile, the $f_{\rm DM}(<R_{\rm e})$ results for the sample change {less than $10\%$ (in terms of absolute difference)} for alternative DM mass profile assumption, {such as the two-power halo (2PH) profile \citep{BinneyTremaine2008} with a variable inner slope.}
Improved constraints on the DM fraction in the future would benefit from deeper observations of individual galaxies and/or kinematic stacking analysis \citep[e.g.][]{Lang2017,Tiley2019b}.

\section{Nature of Non-Disks}\label{sec:nondisk}
The origin of the kinematic perturbations in the Non-Disk subset of the sample could arise from galaxy interactions and mergers.
Most Non-Disks, except CRISTAL-21, 24 and 25, have visible companions already in the \textit{HST} and/or \textit{JWST} images alone, and the one classified as Non-Disk is, in most cases, the less luminous and less massive for those with available mass measurement.
The higher non-disk fraction among the CRISTAL kinematic sample compared to samples of typical MS SFGs at lower redshift may not be surprising in view of the increase in merger rates with redshift from both observational work \citep{Duncan2019,Romano2021,Duan2025,Puskas2025,Shibuya2025} and numerical simulations \citep{RodriguezGomez2015,Pillepich2019,OLeary2021}.  
It will be interesting in future work to investigate in more detail the environment of the galaxies to quantify, for instance, interaction strengths from neighbouring systems.  More complete redshift identification around the CRISTAL galaxies will be necessary for that purpose.

\section{Summary and Outlook}\label{sec:summary}
We presented the kinematics study of a sample of MS SFGs at redshift $4$\,$<$\,$z$\,$<$\,$6$\ from the ALMA-CRISTAL program. 
The angular resolution of the sensitive [\ion{C}{II}] ALMA observations enables us to characterise the kinematics of galaxies at a scale of $\sim$\,$1$\,kpc.
We classified the galaxies primarily based on their kinematic features traced by [\ion{C}{II}] using a variety of methods that have been applied in lower-redshift studies, supplemented by morphological information from \textit{HST} and \textit{JWST} imaging (\S~\ref{sec:kins_class}).
We find that $50\pm9\%$ of the galaxies in our sample are disk-dominated (\S~\ref{sec:diskfrac}), 
with over half of them located in systems of multiple components, 
which differs from the more isolated environment of lower-redshift disks. 
We fitted the kinematics of the disks using fully forward-modelled 3D kinematic models 
(\S~\ref{sec:diskkins}). 
Our kinematics modelling reveals the following important properties of the CRISTAL main sequence disks:

\begin{enumerate}
\item They have a high contribution of turbulence relative to rotational support, with a median disk velocity dispersion of $\sim$\,$70\,$${\rm km\,s^{-1}}$ and $V_{\rm rot}(R_{\rm e})/\sigma_0$ of $\sim$\,$2$ (\S~\ref{sec:turbulence_z}).
\item Their high-velocity dispersions are consistent with the predicted evolution based on Toomre theory and the extrapolated evolutionary trends based on detailed studies of cosmic noon to lower-redshift galaxies (\S~\ref{sec:turbulence_z}).
\item Their tentative correlation between gas mass and velocity dispersion provides hints that the high dispersion is sustained by gravitational instability, but it is unclear whether this is a local or global phenomenon (\S~\ref{sec:trends}). 
\item They tend to have a low dark matter fraction, with median $f_{\rm DM}(<R_{\rm e})$\,$\approx$\,${18\%}$, although spanning a significant range from $\sim$\,$5\%$ to $\sim$\,$60\%$; the median value is in broad agreement with the extrapolated trend based on studies of lower redshifts, albeit with large scatter (\S~\ref{sec:dm}).
\end{enumerate}

The deep, high resolution ALMA observations of [\ion{C}{II}]158$\mu$m line emission 
from the CRISTAL program enabled a first systematic census of the kinematics of typical MS star-forming galaxies at $z$\,$\sim$\,$4$--$6$. 
The brightness of [\ion{C}{II}] along with its sensitivity to the multi-phase ISM makes it an ideal probe of the gas motions (among other properties) over a large extent of galaxies encompassing cold molecular gas, ionised, and photodissociation regions, facilitating measurements reaching the outskirts of galaxies.  
As observations of rest-optical line emission originating from the warm ionised gas phase become available from \textit{JWST} IFU follow-up of CRISTAL targets, direct comparisons will become possible.  
This will be important especially with regard to the issue of gas turbulence, which, as discussed in this work (\S~\ref{sec:dispersion_props}) and in the literature \citep[e.g.][]{Girard2021,Wisnioski2025,Ejdetjarn22,Kohandel2024},
would greatly benefit from measurements in multiple tracers {\em for the same objects}. 
Another outlook enabled by CRISTAL-ALMA and \textit{JWST}-IFU synergies is the connection between spatial variations in gas-phase metallicity and kinematics, which will be the subject of a forthcoming study.

\begin{acknowledgements}
{We thank the anonymous referee for the constructive feedback which improves the clarity of the work.}
L.L.L. is thankful for the stimulating discussions with  M. Bureau, {Q. Fei}, S. Pastras, W. Maciejewski and {E. Wisnioski} at the various stages of this work.
L.L.L. thanks A. Nestor for providing the data tables for Fig.~\ref{fig:fdm_vc}. 
N.M.F.S. and J.C. acknowledge financial support from the European Research Council (ERC) Advanced Grant under the European Union's (EU's) Horizon Europe research and innovation programme (grant agreement AdG GALPHYS, No. 101055023).
H.Ü. acknowledges funding by the EU (ERC APEX, 101164796).
Views and opinions expressed are, however, those of the author(s)
only and do not necessarily reflect those of the EU or
the ERC. 
Neither the EU nor the granting authority can be held responsible for them.
R.H-C. thanks the Max Planck Society for support under the Partner Group project `The Baryon Cycle in Galaxies' between the Max Planck for Extraterrestrial Physics and the Universidad de Concepción.
M.A. and R.H-C. also gratefully acknowledge financial support from ANID - MILENIO - NCN2024\_112.
M.A., R.J.A., R.H-C., M.S. and K. Telikova acknowledge support from ANID BASAL FB210003.
R.J.A. was supported by FONDECYT grant number 1231718.
R.B. acknowledges support from an STFC Ernest Rutherford Fellowship [grant number ST/T003596/1].
R.L.D. is supported by the Australian Research Council through the Discovery Early Career Researcher Award (DECRA) Fellowship DE240100136 funded by the Australian Government.
T.D-S. acknowledges the research project was supported by the Hellenic Foundation for Research and Innovation (HFRI) under the `2nd Call for HFRI Research Projects to support Faculty Members \& Researchers' (Project Number: 03382)
I.D.L. acknowledges funding from the ERC under the EU's Horizon 2020 research and innovation program DustOrigin (ERC-2019-StG-851622) and from the Flemish Fund for Scientific Research (FWO-Vlaanderen) through the research project G0A1523N.
{R.I. is supported by Grants-in-Aid for Japan Society for the Promotion of Science (JSPS) Fellows (KAKENHI Grant Number 23KJ1006).}
T.N. acknowledges the support of the Deutsche Forschungsgemeinschaft (DFG, German Research Foundation) under Germany's Excellence Strategy - EXC-2094 - 390783311 of the DFG Cluster of Excellence `ORIGINS'.
M.S. was financially supported by Becas-ANID scholarship \#21221511.
K. Tadaki acknowledges support from JSPS KAKENHI Grant No. 23K03466.
K. Telikova was supported by ALMA ANID grant number 31220026.
 % and by the ANID BASAL project FB210003.
V.V. acknowledges support from the ALMA-ANID Postdoctoral Fellowship under the award ASTRO21-0062.
This paper makes use of the following ALMA data: 
ADS/JAO.ALMA\#2021.1.00280.L, 2017.1.00428.L, 2012.1.00523.S, 2018.1.01359.S, 2019.1.01075.S.
ALMA is a partnership of ESO (representing its member states), NSF (USA) and NINS (Japan), together with NRC (Canada), NSC and ASIAA (Taiwan), and KASI (Republic of Korea), in cooperation with the Republic of Chile. The Joint ALMA Observatory is operated by ESO, AUI/NRAO and NAOJ.
This work is based in part on observations made with the NASA/ESA/CSA James Webb Space Telescope and NASA/ESA Hubble Space Telescope. 
The data were obtained from the Mikulski Archive for Space Telescopes (MAST) at the Space Telescope Science Institute, which is operated by the Association of Universities for Research in Astronomy, Inc., under NASA contract NAS 5-03127 for JWST and NAS 5–26555 for HST. 
The specific observations analysed can be accessed via
\href{https://doi.org/10.17909/2gpc-vd24}{10.17909/2gpc-vd24}. 
Support to MAST for these data is provided by the NASA Office of Space Science via grant NAG5–7584 and by other grants and contracts.
Some of the data products presented herein were retrieved from the Dawn JWST Archive (DJA). DJA is an initiative of the Cosmic Dawn Center (DAWN), which is funded by the Danish National Research Foundation under grant DNRF140.
This work made use of the following Python packages:
\texttt{Astropy} \citep{Astropy2022}, 
\texttt{corner} \citep{Foreman-Mackey2016}, 
\texttt{DysmalPy} \citep{Davies2004a,Davies2004b,Davies2011,Cresci2009,Wuyts2016,Lang2017,Price2021,Lee2025}, 
\texttt{emcee} \citep{ForemanMackey2013}, 
\texttt{Imfit} \citep{erwin2015}, 
\texttt{Matplotlib} \citep{Hunter2007}, 
\texttt{Numpy} \citep{harris2020}, 
\texttt{pymccorrelation} \citep{Curran2014,Privon2020},
\texttt{Trilogy} \citep{Coe2012},
and \texttt{Scipy} \citep{Virtanen2020}.
\end{acknowledgements}

\bibliographystyle{aau}
\bibliography{ref}

\begin{thebibliography}{243}
\expandafter\ifx\csname natexlab\endcsname\relax\def\natexlab#1{#1}\fi

\bibitem[{{Amvrosiadis} {et~al.}(2025){Amvrosiadis}, {Lange}, {Nightingale},
  {He}, {Frenk}, {Oman}, {Smail}, {Swinbank}, {Fragkoudi}, {Gadotti}, {Cole},
  {Borsato}, {Robertson}, {Massey}, {Cao}, \& {Li}}]{Amvrosiadis2025}
{Amvrosiadis}, A., {Lange}, S., {Nightingale}, J.~W., {et~al.} 2025,
  \href{http://dx.doi.org/10.1093/mnras/staf048}{\color{blue}\mnras},
  \href{https://ui.adsabs.harvard.edu/abs/2025MNRAS.537.1163A}{537, 1163}

\bibitem[{{Astropy Collaboration} {et~al.}(2022){Astropy Collaboration},
  {Price-Whelan}, {Lim}, {Earl}, {Starkman}, {Bradley}, {Shupe}, {Patil},
  {Corrales}, {Brasseur}, {N{\"o}the}, {Donath}, {Tollerud}, {Morris},
  {Ginsburg}, {Vaher}, {Weaver}, {Tocknell}, {Jamieson}, {van Kerkwijk},
  {Robitaille}, {Merry}, {Bachetti}, {G{\"u}nther}, {Aldcroft},
  {Alvarado-Montes}, {Archibald}, {B{\'o}di}, {Bapat}, {Barentsen},
  {Baz{\'a}n}, {Biswas}, {Boquien}, {Burke}, {Cara}, {Cara}, {Conroy},
  {Conseil}, {Craig}, {Cross}, {Cruz}, {D'Eugenio}, {Dencheva}, {Devillepoix},
  {Dietrich}, {Eigenbrot}, {Erben}, {Ferreira}, {Foreman-Mackey}, {Fox},
  {Freij}, {Garg}, {Geda}, {Glattly}, {Gondhalekar}, {Gordon}, {Grant},
  {Greenfield}, {Groener}, {Guest}, {Gurovich}, {Handberg}, {Hart},
  {Hatfield-Dodds}, {Homeier}, {Hosseinzadeh}, {Jenness}, {Jones}, {Joseph},
  {Kalmbach}, {Karamehmetoglu}, {Ka{\l}uszy{\'n}ski}, {Kelley}, {Kern},
  {Kerzendorf}, {Koch}, {Kulumani}, {Lee}, {Ly}, {Ma}, {MacBride}, {Maljaars},
  {Muna}, {Murphy}, {Norman}, {O'Steen}, {Oman}, {Pacifici}, {Pascual},
  {Pascual-Granado}, {Patil}, {Perren}, {Pickering}, {Rastogi}, {Roulston},
  {Ryan}, {Rykoff}, {Sabater}, {Sakurikar}, {Salgado}, {Sanghi}, {Saunders},
  {Savchenko}, {Schwardt}, {Seifert-Eckert}, {Shih}, {Jain}, {Shukla}, {Sick},
  {Simpson}, {Singanamalla}, {Singer}, {Singhal}, {Sinha}, {Sip{\H{o}}cz},
  {Spitler}, {Stansby}, {Streicher}, {{\v{S}}umak}, {Swinbank}, {Taranu},
  {Tewary}, {Tremblay}, {de Val-Borro}, {Van Kooten}, {Vasovi{\'c}}, {Verma},
  {de Miranda Cardoso}, {Williams}, {Wilson}, {Winkel}, {Wood-Vasey}, {Xue},
  {Yoachim}, {Zhang}, {Zonca}, \& {Astropy Project Contributors}}]{Astropy2022}
{Astropy Collaboration}, {Price-Whelan}, A.~M., {Lim}, P.~L., {et~al.} 2022,
  \href{http://dx.doi.org/10.3847/1538-4357/ac7c74}{\color{blue}\apj},
  \href{https://ui.adsabs.harvard.edu/abs/2022ApJ...935..167A}{935, 167}

\bibitem[{{Bari{\v{s}}i{\'c}} {et~al.}(2025){Bari{\v{s}}i{\'c}}, {Jones},
  {Mortensen}, {Nanayakkara}, {Chen}, {Sanders}, {Bullock}, {Bundy},
  {Faucher-Gigu{\`e}re}, {Glazebrook}, {Henry}, {Ju}, {Malkan}, {Morishita},
  {Obreschkow}, {Roy}, {Espejo Salcedo}, {Shapley}, {Treu}, {Wang}, \&
  {Westfall}}]{Barisic2025}
{Bari{\v{s}}i{\'c}}, I., {Jones}, T., {Mortensen}, K., {et~al.} 2025,
  \href{http://dx.doi.org/10.3847/1538-4357/ada617}{\color{blue}\apj},
  \href{https://ui.adsabs.harvard.edu/abs/2025ApJ...983..139B}{983, 139}

\bibitem[{{Beckers}(1982)}]{Beckers1982}
{Beckers}, J.~M. 1982,
  \href{http://dx.doi.org/10.1080/713820871}{\color{blue}Optica Acta},
  \href{https://ui.adsabs.harvard.edu/abs/1982AcOpt..29..361B}{29, 361}

\bibitem[{{Behrendt} {et~al.}(2015){Behrendt}, {Burkert}, \&
  {Schartmann}}]{Behrendt2015}
{Behrendt}, M., {Burkert}, A., \& {Schartmann}, M. 2015,
  \href{http://dx.doi.org/10.1093/mnras/stv027}{\color{blue}\mnras},
  \href{https://ui.adsabs.harvard.edu/abs/2015MNRAS.448.1007B}{448, 1007}

\bibitem[{{B{\'e}thermin} {et~al.}(2023){B{\'e}thermin}, {Accard}, {Guillaume},
  {Dessauges-Zavadsky}, {Ibar}, {Cassata}, {Devereaux}, {Faisst}, {Freundlich},
  {Jones}, {Kraljic}, {Algera}, {Amor{\'\i}n}, {Bardelli}, {Boquien}, {Buat},
  {Donghia}, {Dubois}, {Ferrara}, {Fudamoto}, {Ginolfi}, {Guillard},
  {Giavalisco}, {Gruppioni}, {Gururajan}, {Hathi}, {Hayward}, {Koekemoer},
  {Lemaux}, {Magdis}, {Molina}, {Narayanan}, {Mayer}, {Pozzi}, {Rizzo},
  {Romano}, {Tasca}, {Theul{\'e}}, {Vergani}, {Vallini}, {Zamorani}, {Zanella},
  \& {Zucca}}]{Bethermin2023}
{B{\'e}thermin}, M., {Accard}, C., {Guillaume}, C., {et~al.} 2023,
  \href{http://dx.doi.org/10.1051/0004-6361/202348115}{\color{blue}\aap},
  \href{https://ui.adsabs.harvard.edu/abs/2023A&A...680L...8B}{680, L8}

\bibitem[{{B{\'e}thermin} {et~al.}(2015){B{\'e}thermin}, {Daddi}, {Magdis},
  {Lagos}, {Sargent}, {Albrecht}, {Aussel}, {Bertoldi}, {Buat}, {Galametz},
  {Heinis}, {Ilbert}, {Karim}, {Koekemoer}, {Lacey}, {Le Floc'h}, {Navarrete},
  {Pannella}, {Schreiber}, {Smol{\v{c}}i{\'c}}, {Symeonidis}, \&
  {Viero}}]{Bethermin2015}
{B{\'e}thermin}, M., {Daddi}, E., {Magdis}, G., {et~al.} 2015,
  \href{http://dx.doi.org/10.1051/0004-6361/201425031}{\color{blue}\aap},
  \href{https://ui.adsabs.harvard.edu/abs/2015A&A...573A.113B}{573, A113}

\bibitem[{{B{\'e}thermin} {et~al.}(2020){B{\'e}thermin}, {Fudamoto}, {Ginolfi},
  {Loiacono}, {Khusanova}, {Capak}, {Cassata}, {Faisst}, {Le F{\`e}vre},
  {Schaerer}, {Silverman}, {Yan}, {Amorin}, {Bardelli}, {Boquien}, {Cimatti},
  {Davidzon}, {Dessauges-Zavadsky}, {Fujimoto}, {Gruppioni}, {Hathi}, {Ibar},
  {Jones}, {Koekemoer}, {Lagache}, {Lemaux}, {Moreau}, {Oesch}, {Pozzi},
  {Riechers}, {Talia}, {Toft}, {Vallini}, {Vergani}, {Zamorani}, \&
  {Zucca}}]{Bethermin2020}
{B{\'e}thermin}, M., {Fudamoto}, Y., {Ginolfi}, M., {et~al.} 2020,
  \href{http://dx.doi.org/10.1051/0004-6361/202037649}{\color{blue}\aap},
  \href{https://ui.adsabs.harvard.edu/abs/2020A&A...643A...2B}{643, A2}

\bibitem[{{B{\'e}thermin} {et~al.}(2017){B{\'e}thermin}, {Wu}, {Lagache},
  {Davidzon}, {Ponthieu}, {Cousin}, {Wang}, {Dor{\'e}}, {Daddi}, \&
  {Lapi}}]{Bethermin2017}
{B{\'e}thermin}, M., {Wu}, H.-Y., {Lagache}, G., {et~al.} 2017,
  \href{http://dx.doi.org/10.1051/0004-6361/201730866}{\color{blue}\aap},
  \href{https://ui.adsabs.harvard.edu/abs/2017A&A...607A..89B}{607, A89}

\bibitem[{{Binney} \& {Tremaine}(2008)}]{BinneyTremaine2008}
{Binney}, J. \& {Tremaine}, S. 2008, {Galactic Dynamics: Second Edition}
  ({Princeton University Press})

\bibitem[{{Birkin} {et~al.}(2024){Birkin}, {Puglisi}, {Swinbank}, {Smail},
  {An}, {Chapman}, {Chen}, {Conselice}, {Dudzevi{\v{c}}i{\={u}}t{\.{e}}},
  {Farrah}, {Gullberg}, {Matsuda}, {Schinnerer}, {Scott}, {Wardlow}, \& {van
  der Werf}}]{Birkin2024}
{Birkin}, J.~E., {Puglisi}, A., {Swinbank}, A.~M., {et~al.} 2024,
  \href{http://dx.doi.org/10.1093/mnras/stae1089}{\color{blue}\mnras},
  \href{https://ui.adsabs.harvard.edu/abs/2024MNRAS.531...61B}{531, 61}

\bibitem[{{Birkin} {et~al.}(2025){Birkin}, {Spilker}, {Herrera-Camus},
  {Davies}, {Lee}, {Aravena}, {Assef}, {Barcos-Mu{\~n}oz}, {Bolatto},
  {Diaz-Santos}, {Faisst}, {Ferrara}, {Fisher}, {Gonz{\'a}lez-L{\'o}pez},
  {Ikeda}, {Knudsen}, {Li}, {Li}, {de Looze}, {Lutz}, {Mitsuhashi}, {Posses},
  {Rela{\~n}o}, {Solimano}, {Tadaki}, \& {Villanueva}}]{Birkin2025}
{Birkin}, J.~E., {Spilker}, J.~S., {Herrera-Camus}, R., {et~al.} 2025,
  \href{http://dx.doi.org/10.3847/1538-4357/adced3}{\color{blue}\apj},
  \href{https://ui.adsabs.harvard.edu/abs/2025ApJ...985..243B}{985, 243}

\bibitem[{{Bland-Hawthorn} \& {Gerhard}(2016)}]{BH2016}
{Bland-Hawthorn}, J. \& {Gerhard}, O. 2016,
  \href{http://dx.doi.org/10.1146/annurev-astro-081915-023441}{\color{blue}\araa},
  \href{https://ui.adsabs.harvard.edu/abs/2016ARA&A..54..529B}{54, 529}

\bibitem[{{Blumenthal} {et~al.}(1986){Blumenthal}, {Faber}, {Flores}, \&
  {Primack}}]{Blumenthal1986}
{Blumenthal}, G.~R., {Faber}, S.~M., {Flores}, R., \& {Primack}, J.~R. 1986,
  \href{http://dx.doi.org/10.1086/163867}{\color{blue}\apj},
  \href{https://ui.adsabs.harvard.edu/abs/1986ApJ...301...27B}{301, 27}

\bibitem[{{Bouch{\'e}} {et~al.}(2022){Bouch{\'e}}, {Bera}, {Krajnovi{\'c}},
  {Emsellem}, {Mercier}, {Schaye}, {Epinat}, {Richard}, {Zoutendijk},
  {Abril-Melgarejo}, {Brinchmann}, {Bacon}, {Contini}, {Boogaard}, {Wisotzki},
  {Maseda}, \& {Steinmetz}}]{Bouche2022}
{Bouch{\'e}}, N.~F., {Bera}, S., {Krajnovi{\'c}}, D., {et~al.} 2022,
  \href{http://dx.doi.org/10.1051/0004-6361/202141762}{\color{blue}\aap},
  \href{https://ui.adsabs.harvard.edu/abs/2022A&A...658A..76B}{658, A76}

\bibitem[{{Bouwens} {et~al.}(2022){Bouwens}, {Smit}, {Schouws}, {Stefanon},
  {Bowler}, {Endsley}, {Gonzalez}, {Inami}, {Stark}, {Oesch}, {Hodge},
  {Aravena}, {da Cunha}, {Dayal}, {de Looze}, {Ferrara}, {Fudamoto},
  {Graziani}, {Li}, {Nanayakkara}, {Pallottini}, {Schneider}, {Sommovigo},
  {Topping}, {van der Werf}, {Algera}, {Barrufet}, {Hygate}, {Labb{\'e}},
  {Riechers}, \& {Witstok}}]{Bouwens2022}
{Bouwens}, R.~J., {Smit}, R., {Schouws}, S., {et~al.} 2022,
  \href{http://dx.doi.org/10.3847/1538-4357/ac5a4a}{\color{blue}\apj},
  \href{https://ui.adsabs.harvard.edu/abs/2022ApJ...931..160B}{931, 160}

\bibitem[{{Bovy} \& {Rix}(2013)}]{Bovy2013}
{Bovy}, J. \& {Rix}, H.-W. 2013,
  \href{http://dx.doi.org/10.1088/0004-637X/779/2/115}{\color{blue}\apj},
  \href{https://ui.adsabs.harvard.edu/abs/2013ApJ...779..115B}{779, 115}

\bibitem[{{Burkert} {et~al.}(2010){Burkert}, {Genzel}, {Bouch{\'e}}, {Cresci},
  {Khochfar}, {Sommer-Larsen}, {Sternberg}, {Naab}, {F{\"o}rster Schreiber},
  {Tacconi}, {Shapiro}, {Hicks}, {Lutz}, {Davies}, {Buschkamp}, \&
  {Genel}}]{Burkert2010}
{Burkert}, A., {Genzel}, R., {Bouch{\'e}}, N., {et~al.} 2010,
  \href{http://dx.doi.org/10.1088/0004-637X/725/2/2324}{\color{blue}\apj},
  \href{https://ui.adsabs.harvard.edu/abs/2010ApJ...725.2324B}{725, 2324}

\bibitem[{{CASA Team} {et~al.}(2022){CASA Team}, {Bean}, {Bhatnagar}, {Castro},
  {Donovan Meyer}, {Emonts}, {Garcia}, {Garwood}, {Golap}, {Gonzalez Villalba},
  {Harris}, {Hayashi}, {Hoskins}, {Hsieh}, {Jagannathan}, {Kawasaki},
  {Keimpema}, {Kettenis}, {Lopez}, {Marvil}, {Masters}, {McNichols},
  {Mehringer}, {Miel}, {Moellenbrock}, {Montesino}, {Nakazato}, {Ott}, {Petry},
  {Pokorny}, {Raba}, {Rau}, {Schiebel}, {Schweighart}, {Sekhar}, {Shimada},
  {Small}, {Steeb}, {Sugimoto}, {Suoranta}, {Tsutsumi}, {van Bemmel},
  {Verkouter}, {Wells}, {Xiong}, {Szomoru}, {Griffith}, {Glendenning}, \&
  {Kern}}]{CASA2022}
{CASA Team}, {Bean}, B., {Bhatnagar}, S., {et~al.} 2022,
  \href{http://dx.doi.org/10.1088/1538-3873/ac9642}{\color{blue}\pasp},
  \href{https://ui.adsabs.harvard.edu/abs/2022PASP..134k4501C}{134, 114501}

\bibitem[{{Cathey} {et~al.}(2024){Cathey}, {Gonzalez}, {Lower}, {Phadke},
  {Spilker}, {Aravena}, {Bayliss}, {Birkin}, {Birrer}, {Chapman}, {Dahle},
  {Hayward}, {Hezaveh}, {Hill}, {Hutchison}, {Kim}, {Mahler}, {Marrone},
  {Narayanan}, {Navarre}, {Reuter}, {Rigby}, {Sharon}, {Solimano},
  {Sulzenauer}, {Vieira}, \& {Vizgan}}]{Cathey2024}
{Cathey}, J., {Gonzalez}, A.~H., {Lower}, S., {et~al.} 2024,
  \href{http://dx.doi.org/10.3847/1538-4357/ad33c9}{\color{blue}\apj},
  \href{https://ui.adsabs.harvard.edu/abs/2024ApJ...967...11C}{967, 11}

\bibitem[{{Christy} {et~al.}(1981){Christy}, {Wellnitz}, \&
  {Currie}}]{Christy1981}
{Christy}, J.~W., {Wellnitz}, D.~D., \& {Currie}, D.~G. 1981, Lowell
  Observatory Bulletin,
  \href{https://ui.adsabs.harvard.edu/abs/1983LowOB...9...28C}{9, 28}

\bibitem[{{Clark} {et~al.}(2019){Clark}, {Glover}, {Ragan}, \&
  {Duarte-Cabral}}]{Clark2019}
{Clark}, P.~C., {Glover}, S. C.~O., {Ragan}, S.~E., \& {Duarte-Cabral}, A.
  2019, \href{http://dx.doi.org/10.1093/mnras/stz1119}{\color{blue}\mnras},
  \href{https://ui.adsabs.harvard.edu/abs/2019MNRAS.486.4622C}{486, 4622}

\bibitem[{{Coe} {et~al.}(2012){Coe}, {Umetsu}, {Zitrin}, {Donahue},
  {Medezinski}, {Postman}, {Carrasco}, {Anguita}, {Geller}, {Rines},
  {Diaferio}, {Kurtz}, {Bradley}, {Koekemoer}, {Zheng}, {Nonino}, {Molino},
  {Mahdavi}, {Lemze}, {Infante}, {Ogaz}, {Melchior}, {Host}, {Ford}, {Grillo},
  {Rosati}, {Jim{\'e}nez-Teja}, {Moustakas}, {Broadhurst}, {Ascaso}, {Lahav},
  {Bartelmann}, {Ben{\'\i}tez}, {Bouwens}, {Graur}, {Graves}, {Jha}, {Jouvel},
  {Kelson}, {Moustakas}, {Maoz}, {Meneghetti}, {Merten}, {Riess}, {Rodney}, \&
  {Seitz}}]{Coe2012}
{Coe}, D., {Umetsu}, K., {Zitrin}, A., {et~al.} 2012,
  \href{http://dx.doi.org/10.1088/0004-637X/757/1/22}{\color{blue}\apj},
  \href{https://ui.adsabs.harvard.edu/abs/2012ApJ...757...22C}{757, 22}

\bibitem[{{Cole} {et~al.}(2025){Cole}, {Papovich}, {Finkelstein}, {Bagley},
  {Dickinson}, {Iyer}, {Yung}, {Ciesla}, {Amor{\'\i}n}, {Arrabal Haro},
  {Bhatawdekar}, {Calabr{\`o}}, {Cleri}, {de la Vega}, {Dekel}, {Endsley},
  {Gawiser}, {Giavalisco}, {Hathi}, {Hirschmann}, {Holwerda}, {Kartaltepe},
  {Koekemoer}, {Lucas}, {Mascia}, {Mobasher}, {P{\'e}rez-Gonz{\'a}lez},
  {Rodighiero}, {Ronayne}, {Tacchella}, {Weiner}, \& {Wilkins}}]{Cole2025}
{Cole}, J.~W., {Papovich}, C., {Finkelstein}, S.~L., {et~al.} 2025,
  \href{http://dx.doi.org/10.3847/1538-4357/ad9a6a}{\color{blue}\apj},
  \href{https://ui.adsabs.harvard.edu/abs/2025ApJ...979..193C}{979, 193}

\bibitem[{{Condon} {et~al.}(1998){Condon}, {Cotton}, {Greisen}, {Yin},
  {Perley}, {Taylor}, \& {Broderick}}]{Condon1998}
{Condon}, J.~J., {Cotton}, W.~D., {Greisen}, E.~W., {et~al.} 1998,
  \href{http://dx.doi.org/10.1086/300337}{\color{blue}\aj},
  \href{https://ui.adsabs.harvard.edu/abs/1998AJ....115.1693C}{115, 1693}

\bibitem[{{Cormier} {et~al.}(2019){Cormier}, {Abel}, {Hony}, {Lebouteiller},
  {Madden}, {Polles}, {Galliano}, {De Looze}, {Galametz}, \&
  {Lambert-Huyghe}}]{Cormier2019}
{Cormier}, D., {Abel}, N.~P., {Hony}, S., {et~al.} 2019,
  \href{http://dx.doi.org/10.1051/0004-6361/201834457}{\color{blue}\aap},
  \href{https://ui.adsabs.harvard.edu/abs/2019A&A...626A..23C}{626, A23}

\bibitem[{{Costantin} {et~al.}(2025){Costantin}, {Gillman}, {Boogaard},
  {P{\'e}rez-Gonz{\'a}lez}, {Iani}, {Rinaldi}, {Melinder}, {Crespo G{\'o}mez},
  {Colina}, {Greve}, {{\"O}stlin}, {Wright}, {Alonso-Herrero},
  {{\'A}lvarez-M{\'a}rquez}, {Annunziatella}, {Bik.}, {Caputi}, {Dicken},
  {Eckart}, {Hjorth}, {Ilbert}, {Jermann}, {Labiano}, {Langeroodi},
  {Pei{\ss}ker}, {Pye}, {Tikkanen}, {van der Werf}, {Walter}, {Ward},
  {G{\"u}del}, \& {Henning}}]{Costantin2025}
{Costantin}, L., {Gillman}, S., {Boogaard}, L.~A., {et~al.} 2025,
  \href{https://doi.org/10.1051/0004-6361/202451330}{\href{http://dx.doi.org/10.1051/0004-6361/202451330}{\color{blue}\aap},
  in press}

\bibitem[{{Courteau}(1997)}]{Courteau1997}
{Courteau}, S. 1997, \href{http://dx.doi.org/10.1086/118656}{\color{blue}\aj},
  \href{https://ui.adsabs.harvard.edu/abs/1997AJ....114.2402C}{114, 2402}

\bibitem[{{Courteau} \& {Dutton}(2015)}]{Courteau2015}
{Courteau}, S. \& {Dutton}, A.~A. 2015,
  \href{http://dx.doi.org/10.1088/2041-8205/801/2/L20}{\color{blue}\apjl},
  \href{https://ui.adsabs.harvard.edu/abs/2015ApJ...801L..20C}{801, L20}

\bibitem[{{Cresci} {et~al.}(2009){Cresci}, {Hicks}, {Genzel}, {F{\"o}rster
  Schreiber}, {Davies}, {Bouch{\'e}}, {Buschkamp}, {Genel}, {Shapiro},
  {Tacconi}, {Sommer-Larsen}, {Burkert}, {Eisenhauer}, {Gerhard}, {Lutz},
  {Naab}, {Sternberg}, {Cimatti}, {Daddi}, {Erb}, {Kurk}, {Lilly}, {Renzini},
  {Shapley}, {Steidel}, \& {Caputi}}]{Cresci2009}
{Cresci}, G., {Hicks}, E.~K.~S., {Genzel}, R., {et~al.} 2009,
  \href{http://dx.doi.org/10.1088/0004-637X/697/1/115}{\color{blue}\apj},
  \href{https://ui.adsabs.harvard.edu/abs/2009ApJ...697..115C}{697, 115}

\bibitem[{{Curran}(2014)}]{Curran2014}
{Curran}, P.~A. 2014,
  \href{https://ui.adsabs.harvard.edu/abs/2014arXiv1411.3816C}{\href{http://dx.doi.org/10.48550/arXiv.1411.3816}{\color{blue}arXiv
  e-prints}, arXiv:1411.3816}

\bibitem[{{Dalcanton} \& {Stilp}(2010)}]{DalcantonStilp2010}
{Dalcanton}, J.~J. \& {Stilp}, A.~M. 2010,
  \href{http://dx.doi.org/10.1088/0004-637X/721/1/547}{\color{blue}\apj},
  \href{https://ui.adsabs.harvard.edu/abs/2010ApJ...721..547D}{721, 547}

\bibitem[{{Danhaive} {et~al.}(2025){Danhaive}, {Tacchella}, {\"Ubler}, {de
  Graaff}, {Egami}, {Johnson}, {Sun}, {Arribas}, {Bunker}, {Carniani}, {Jones},
  {Maiolino}, {McClymont}, {Parlanti}, {Simmonds}, {Villanueva}, {Baker},
  {Jaffe}, {Eisenstein}, {Hainline}, {Helton}, {Ji}, {Lin},
  {Pusk\textbackslash'as}, {Rieke}, {Rinaldi}, {Robertson}, {Scholz},
  {Williams}, \& {Willmer}}]{Danhaive2025}
{Danhaive}, A.~L., {Tacchella}, S., {\"Ubler}, H., {et~al.} 2025,
  \href{https://ui.adsabs.harvard.edu/abs/2025arXiv250321863D}{\href{http://dx.doi.org/10.48550/arXiv.2503.21863}{\color{blue}arXiv
  e-prints}, arXiv:2503.21863}

\bibitem[{{Davies} {et~al.}(2011){Davies}, {F{\"o}rster Schreiber}, {Cresci},
  {Genzel}, {Bouch{\'e}}, {Burkert}, {Buschkamp}, {Genel}, {Hicks}, {Kurk},
  {Lutz}, {Newman}, {Shapiro}, {Sternberg}, {Tacconi}, \& {Wuyts}}]{Davies2011}
{Davies}, R., {F{\"o}rster Schreiber}, N.~M., {Cresci}, G., {et~al.} 2011,
  \href{http://dx.doi.org/10.1088/0004-637X/741/2/69}{\color{blue}\apj},
  \href{https://ui.adsabs.harvard.edu/abs/2011ApJ...741...69D}{741, 69}

\bibitem[{{Davies} {et~al.}(2004{\natexlab{a}}){Davies}, {Tacconi}, \&
  {Genzel}}]{Davies2004b}
{Davies}, R.~I., {Tacconi}, L.~J., \& {Genzel}, R. 2004{\natexlab{a}},
  \href{http://dx.doi.org/10.1086/423315}{\color{blue}\apj},
  \href{https://ui.adsabs.harvard.edu/abs/2004ApJ...613..781D}{613, 781}

\bibitem[{{Davies} {et~al.}(2004{\natexlab{b}}){Davies}, {Tacconi}, \&
  {Genzel}}]{Davies2004a}
{Davies}, R.~I., {Tacconi}, L.~J., \& {Genzel}, R. 2004{\natexlab{b}},
  \href{http://dx.doi.org/10.1086/380995}{\color{blue}\apj},
  \href{https://ui.adsabs.harvard.edu/abs/2004ApJ...602..148D}{602, 148}

\bibitem[{{Davies} {et~al.}(2025){Davies}, {Fisher}, {Herrera-Camus}, {Davies},
  {Fisher}, {Herrera-Camus}, \& {et al.}}]{Davies2025}
{Davies}, R.~L., {Fisher}, D.~B., {Herrera-Camus}, R., {et~al.} 2025, Nature
  Astronomy, submitted

\bibitem[{{Davis} {et~al.}(2013){Davis}, {Alatalo}, {Bureau}, {Cappellari},
  {Scott}, {Young}, {Blitz}, {Crocker}, {Bayet}, {Bois}, {Bournaud}, {Davies},
  {de Zeeuw}, {Duc}, {Emsellem}, {Khochfar}, {Krajnovi{\'c}}, {Kuntschner},
  {Lablanche}, {McDermid}, {Morganti}, {Naab}, {Oosterloo}, {Sarzi}, {Serra},
  \& {Weijmans}}]{Davis2013}
{Davis}, T.~A., {Alatalo}, K., {Bureau}, M., {et~al.} 2013,
  \href{http://dx.doi.org/10.1093/mnras/sts353}{\color{blue}\mnras},
  \href{https://ui.adsabs.harvard.edu/abs/2013MNRAS.429..534D}{429, 534}

\bibitem[{{Davis} {et~al.}(2017){Davis}, {Bureau}, {Onishi}, {Cappellari},
  {Iguchi}, \& {Sarzi}}]{Davis2017}
{Davis}, T.~A., {Bureau}, M., {Onishi}, K., {et~al.} 2017,
  \href{http://dx.doi.org/10.1093/mnras/stw3217}{\color{blue}\mnras},
  \href{https://ui.adsabs.harvard.edu/abs/2017MNRAS.468.4675D}{468, 4675}

\bibitem[{{de Blok} {et~al.}(2024){de Blok}, {Healy}, {Maccagni}, {Pisano},
  {Bosma}, {English}, {Jarrett}, {Marasco}, {Meurer}, {Veronese}, {Bigiel},
  {Chemin}, {Fraternali}, {Holwerda}, {Kamphuis}, {Kl{\"o}ckner}, {Kleiner},
  {Leroy}, {Mogotsi}, {Oman}, {Schinnerer}, {Verdes-Montenegro}, {Westmeier},
  {Wong}, {Zabel}, {Amram}, {Carignan}, {Combes}, {Brinks}, {Dettmar},
  {Gibson}, {Jozsa}, {Koribalski}, {McGaugh}, {Oosterloo}, {Spekkens},
  {Schr{\"o}der}, {Adams}, {Athanassoula}, {Bershady}, {Beswick}, {Blyth},
  {Elson}, {Frank}, {Heald}, {Henning}, {Kurapati}, {Loubser}, {Lucero},
  {Meyer}, {Namumba}, {Oh}, {Sardone}, {Sheth}, {Smith}, {Sorgho}, {Walter},
  {Williams}, {Woudt}, \& {Zijlstra}}]{deBlok2024}
{de Blok}, W.~J.~G., {Healy}, J., {Maccagni}, F.~M., {et~al.} 2024,
  \href{http://dx.doi.org/10.1051/0004-6361/202348297}{\color{blue}\aap},
  \href{https://ui.adsabs.harvard.edu/abs/2024A&A...688A.109D}{688, A109}

\bibitem[{{de Graaff} {et~al.}(2024){de Graaff}, {Rix}, {Carniani}, {Suess},
  {Charlot}, {Curtis-Lake}, {Arribas}, {Baker}, {Boyett}, {Bunker}, {Cameron},
  {Chevallard}, {Curti}, {Eisenstein}, {Franx}, {Hainline}, {Hausen}, {Ji},
  {Johnson}, {Jones}, {Maiolino}, {Maseda}, {Nelson}, {Parlanti}, {Rawle},
  {Robertson}, {Tacchella}, {{\"U}bler}, {Williams}, {Willmer}, \&
  {Willott}}]{deGraaff2024}
{de Graaff}, A., {Rix}, H.-W., {Carniani}, S., {et~al.} 2024,
  \href{http://dx.doi.org/10.1051/0004-6361/202347755}{\color{blue}\aap},
  \href{https://ui.adsabs.harvard.edu/abs/2024A&A...684A..87D}{684, A87}

\bibitem[{{Dekel} {et~al.}(2009){Dekel}, {Birnboim}, {Engel}, {Freundlich},
  {Goerdt}, {Mumcuoglu}, {Neistein}, {Pichon}, {Teyssier}, \&
  {Zinger}}]{Dekel2009}
{Dekel}, A., {Birnboim}, Y., {Engel}, G., {et~al.} 2009,
  \href{http://dx.doi.org/10.1038/nature07648}{\color{blue}\nat},
  \href{https://ui.adsabs.harvard.edu/abs/2009Natur.457..451D}{457, 451}

\bibitem[{{Dekel} \& {Burkert}(2014)}]{DekelBurkert2014}
{Dekel}, A. \& {Burkert}, A. 2014,
  \href{http://dx.doi.org/10.1093/mnras/stt2331}{\color{blue}\mnras},
  \href{https://ui.adsabs.harvard.edu/abs/2014MNRAS.438.1870D}{438, 1870}

\bibitem[{{Dekel} {et~al.}(2020){Dekel}, {Ginzburg}, {Jiang}, {Freundlich},
  {Lapiner}, {Ceverino}, \& {Primack}}]{Dekel2020}
{Dekel}, A., {Ginzburg}, O., {Jiang}, F., {et~al.} 2020,
  \href{http://dx.doi.org/10.1093/mnras/staa470}{\color{blue}\mnras},
  \href{https://ui.adsabs.harvard.edu/abs/2020MNRAS.493.4126D}{493, 4126}

\bibitem[{{Dessauges-Zavadsky} {et~al.}(2020){Dessauges-Zavadsky}, {Ginolfi},
  {Pozzi}, {B{\'e}thermin}, {Le F{\`e}vre}, {Fujimoto}, {Silverman}, {Jones},
  {Vallini}, {Schaerer}, {Faisst}, {Khusanova}, {Fudamoto}, {Cassata},
  {Loiacono}, {Capak}, {Yan}, {Amorin}, {Bardelli}, {Boquien}, {Cimatti},
  {Gruppioni}, {Hathi}, {Ibar}, {Koekemoer}, {Lemaux}, {Narayanan}, {Oesch},
  {Rodighiero}, {Romano}, {Talia}, {Toft}, {Vergani}, {Zamorani}, \&
  {Zucca}}]{Dessauges-Zavadsky2020}
{Dessauges-Zavadsky}, M., {Ginolfi}, M., {Pozzi}, F., {et~al.} 2020,
  \href{http://dx.doi.org/10.1051/0004-6361/202038231}{\color{blue}\aap},
  \href{https://ui.adsabs.harvard.edu/abs/2020A&A...643A...5D}{643, A5}

\bibitem[{{Devereaux} {et~al.}(2024){Devereaux}, {Cassata}, {Ibar}, {Accard},
  {Guillaume}, {B{\'e}thermin}, {Dessauges-Zavadsky}, {Faisst}, {Jones},
  {Zanella}, {Bardelli}, {Boquien}, {D'Onghia}, {Giavalisco}, {Ginolfi},
  {Gobat}, {Hayward}, {Koekemoer}, {Lemaux}, {Magdis}, {Mendez-Hernandez},
  {Molina}, {Pozzi}, {Romano}, {Tasca}, {Vergani}, {Zamorani}, \&
  {Zucca}}]{Devereaux2024}
{Devereaux}, T., {Cassata}, P., {Ibar}, E., {et~al.} 2024,
  \href{http://dx.doi.org/10.1051/0004-6361/202348511}{\color{blue}\aap},
  \href{https://ui.adsabs.harvard.edu/abs/2024A&A...686A.156D}{686, A156}

\bibitem[{{Di Teodoro} \& {Fraternali}(2015)}]{DiTeodoro2015}
{Di Teodoro}, E.~M. \& {Fraternali}, F. 2015,
  \href{http://dx.doi.org/10.1093/mnras/stv1213}{\color{blue}\mnras},
  \href{https://ui.adsabs.harvard.edu/abs/2015MNRAS.451.3021D}{451, 3021}

\bibitem[{{Di Teodoro} {et~al.}(2016){Di Teodoro}, {Fraternali}, \&
  {Miller}}]{DiTeodoro2016}
{Di Teodoro}, E.~M., {Fraternali}, F., \& {Miller}, S.~H. 2016,
  \href{http://dx.doi.org/10.1051/0004-6361/201628315}{\color{blue}\aap},
  \href{https://ui.adsabs.harvard.edu/abs/2016A&A...594A..77D}{594, A77}

\bibitem[{{Duan} {et~al.}(2025){Duan}, {Conselice}, {Li}, {Austin}, {Harvey},
  {Adams}, {Duncan}, {Trussler}, {Ferreira}, {Westcott}, {Harris}, {Windhorst},
  {Holwerda}, {Broadhurst}, {Coe}, {Cohen}, {Du}, {Driver}, {Frye}, {Grogin},
  {Hathi}, {Jansen}, {Koekemoer}, {Marshall}, {Nonino}, {Ortiz}, {Pirzkal},
  {Robotham}, {Ryan}, {Summers}, {D'Silva}, {Willmer}, \& {Yan}}]{Duan2025}
{Duan}, Q., {Conselice}, C.~J., {Li}, Q., {et~al.} 2025,
  \href{https://ui.adsabs.harvard.edu/abs/2025MNRAS.tmp..606D}{\href{http://dx.doi.org/10.1093/mnras/staf638}{\color{blue}\mnras},
  staf638}

\bibitem[{{Duncan} {et~al.}(2019){Duncan}, {Conselice}, {Mundy}, {Bell},
  {Donley}, {Galametz}, {Guo}, {Grogin}, {Hathi}, {Kartaltepe}, {Kocevski},
  {Koekemoer}, {P{\'e}rez-Gonz{\'a}lez}, {Mantha}, {Snyder}, \&
  {Stefanon}}]{Duncan2019}
{Duncan}, K., {Conselice}, C.~J., {Mundy}, C., {et~al.} 2019,
  \href{http://dx.doi.org/10.3847/1538-4357/ab148a}{\color{blue}\apj},
  \href{https://ui.adsabs.harvard.edu/abs/2019ApJ...876..110D}{876, 110}

\bibitem[{{Dutton} \& {Macci{\`o}}(2014)}]{Dutton2014}
{Dutton}, A.~A. \& {Macci{\`o}}, A.~V. 2014,
  \href{http://dx.doi.org/10.1093/mnras/stu742}{\color{blue}\mnras},
  \href{https://ui.adsabs.harvard.edu/abs/2014MNRAS.441.3359D}{441, 3359}

\bibitem[{{Eisenhauer} {et~al.}(2003){Eisenhauer}, {Tecza}, {Thatte}, {Genzel},
  {Abuter}, {Iserlohe}, {Schreiber}, {Huber}, {Roehrle}, {Horrobin},
  {Schegerer}, {Baker}, {Bender}, {Davies}, {Lehnert}, {Lutz}, {Nesvadba},
  {Ott}, {Seitz}, {Schoedel}, {Tacconi}, {Bonnet}, {Castillo}, {Conzelmann},
  {Donaldson}, {Finger}, {Gillet}, {Hubin}, {Kissler-Patig}, {Lizon}, {Monnet},
  \& {Stroebele}}]{Eisenhauer2003}
{Eisenhauer}, F., {Tecza}, M., {Thatte}, N., {et~al.} 2003, The Messenger,
  \href{https://ui.adsabs.harvard.edu/abs/2003Msngr.113...17E}{113, 17}

\bibitem[{{Ejdetj{\"a}rn} {et~al.}(2022){Ejdetj{\"a}rn}, {Agertz},
  {{\"O}stlin}, {Renaud}, \& {Romeo}}]{Ejdetjarn22}
{Ejdetj{\"a}rn}, T., {Agertz}, O., {{\"O}stlin}, G., {Renaud}, F., \& {Romeo},
  A.~B. 2022,
  \href{http://dx.doi.org/10.1093/mnras/stac1414}{\color{blue}\mnras},
  \href{https://ui.adsabs.harvard.edu/abs/2022MNRAS.514..480E}{514, 480}

\bibitem[{{El-Zant} {et~al.}(2001){El-Zant}, {Shlosman}, \&
  {Hoffman}}]{ElZant2001}
{El-Zant}, A., {Shlosman}, I., \& {Hoffman}, Y. 2001,
  \href{http://dx.doi.org/10.1086/322516}{\color{blue}\apj},
  \href{https://ui.adsabs.harvard.edu/abs/2001ApJ...560..636E}{560, 636}

\bibitem[{{Elmegreen} {et~al.}(2022){Elmegreen}, {Martinez}, \&
  {Hunter}}]{Elmegreen2022}
{Elmegreen}, B.~G., {Martinez}, Z., \& {Hunter}, D.~A. 2022,
  \href{http://dx.doi.org/10.3847/1538-4357/ac559c}{\color{blue}\apj},
  \href{https://ui.adsabs.harvard.edu/abs/2022ApJ...928..143E}{928, 143}

\bibitem[{{Epinat} {et~al.}(2008){Epinat}, {Amram}, \& {Marcelin}}]{Epinat2008}
{Epinat}, B., {Amram}, P., \& {Marcelin}, M. 2008,
  \href{http://dx.doi.org/10.1111/j.1365-2966.2008.13796.x}{\color{blue}\mnras},
  \href{https://ui.adsabs.harvard.edu/abs/2008MNRAS.390..466E}{390, 466}

\bibitem[{{Epinat} {et~al.}(2009){Epinat}, {Contini}, {Le F{\`e}vre},
  {Vergani}, {Garilli}, {Amram}, {Queyrel}, {Tasca}, \& {Tresse}}]{Epinat2009}
{Epinat}, B., {Contini}, T., {Le F{\`e}vre}, O., {et~al.} 2009,
  \href{http://dx.doi.org/10.1051/0004-6361/200911995}{\color{blue}\aap},
  \href{https://ui.adsabs.harvard.edu/abs/2009A&A...504..789E}{504, 789}

\bibitem[{{Erwin}(2015)}]{erwin2015}
{Erwin}, P. 2015,
  \href{http://dx.doi.org/10.1088/0004-637X/799/2/226}{\color{blue}\apj},
  \href{https://ui.adsabs.harvard.edu/abs/2015ApJ...799..226E}{799, 226}

\bibitem[{{Faisst} {et~al.}(2020{\natexlab{a}}){Faisst}, {Fudamoto}, {Oesch},
  {Scoville}, {Riechers}, {Pavesi}, \& {Capak}}]{Faisst2020b}
{Faisst}, A.~L., {Fudamoto}, Y., {Oesch}, P.~A., {et~al.} 2020{\natexlab{a}},
  \href{http://dx.doi.org/10.1093/mnras/staa2545}{\color{blue}\mnras},
  \href{https://ui.adsabs.harvard.edu/abs/2020MNRAS.498.4192F}{498, 4192}

\bibitem[{{Faisst} {et~al.}(2020{\natexlab{b}}){Faisst}, {Schaerer}, {Lemaux},
  {Oesch}, {Fudamoto}, {Cassata}, {B{\'e}thermin}, {Capak}, {Le F{\`e}vre},
  {Silverman}, {Yan}, {Ginolfi}, {Koekemoer}, {Morselli}, {Amor{\'\i}n},
  {Bardelli}, {Boquien}, {Brammer}, {Cimatti}, {Dessauges-Zavadsky},
  {Fujimoto}, {Gruppioni}, {Hathi}, {Hemmati}, {Ibar}, {Jones}, {Khusanova},
  {Loiacono}, {Pozzi}, {Talia}, {Tasca}, {Riechers}, {Rodighiero}, {Romano},
  {Scoville}, {Toft}, {Vallini}, {Vergani}, {Zamorani}, \&
  {Zucca}}]{Faisst2020a}
{Faisst}, A.~L., {Schaerer}, D., {Lemaux}, B.~C., {et~al.} 2020{\natexlab{b}},
  \href{http://dx.doi.org/10.3847/1538-4365/ab7ccd}{\color{blue}\apjs},
  \href{https://ui.adsabs.harvard.edu/abs/2020ApJS..247...61F}{247, 61}

\bibitem[{{Fei} {et~al.}(2025){Fei}, {Silverman}, {Fujimoto}, {Wang}, {Ho},
  {Bischetti}, {Carniani}, {Ginolfi}, {Jones}, {Maiolino}, {Rujopakarn},
  {F{\"o}rster Schreiber}, {Espejo Salcedo}, \& {Lee}}]{Fei2025}
{Fei}, Q., {Silverman}, J.~D., {Fujimoto}, S., {et~al.} 2025,
  \href{http://dx.doi.org/10.3847/1538-4357/ada145}{\color{blue}\apj},
  \href{https://ui.adsabs.harvard.edu/abs/2025ApJ...980...84F}{980, 84}

\bibitem[{{Ferreira} {et~al.}(2023){Ferreira}, {Conselice}, {Sazonova},
  {Ferrari}, {Caruana}, {Tohill}, {Lucatelli}, {Adams}, {Irodotou}, {Marshall},
  {Roper}, {Lovell}, {Verma}, {Austin}, {Trussler}, \&
  {Wilkins}}]{Ferreira2023}
{Ferreira}, L., {Conselice}, C.~J., {Sazonova}, E., {et~al.} 2023,
  \href{http://dx.doi.org/10.3847/1538-4357/acec76}{\color{blue}\apj},
  \href{https://ui.adsabs.harvard.edu/abs/2023ApJ...955...94F}{955, 94}

\bibitem[{{Foreman-Mackey}(2016)}]{Foreman-Mackey2016}
{Foreman-Mackey}, D. 2016,
  \href{http://dx.doi.org/10.21105/joss.00024}{\color{blue}The Journal of Open
  Source Software},
  \href{https://ui.adsabs.harvard.edu/abs/2016JOSS....1...24F}{1, 24}

\bibitem[{{Foreman-Mackey} {et~al.}(2013){Foreman-Mackey}, {Hogg}, {Lang}, \&
  {Goodman}}]{ForemanMackey2013}
{Foreman-Mackey}, D., {Hogg}, D.~W., {Lang}, D., \& {Goodman}, J. 2013,
  \href{http://dx.doi.org/10.1086/670067}{\color{blue}\pasp},
  \href{https://ui.adsabs.harvard.edu/abs/2013PASP..125..306F}{125, 306}

\bibitem[{{F{\"o}rster Schreiber} {et~al.}(2009){F{\"o}rster Schreiber},
  {Genzel}, {Bouch{\'e}}, {Cresci}, {Davies}, {Buschkamp}, {Shapiro},
  {Tacconi}, {Hicks}, {Genel}, {Shapley}, {Erb}, {Steidel}, {Lutz},
  {Eisenhauer}, {Gillessen}, {Sternberg}, {Renzini}, {Cimatti}, {Daddi},
  {Kurk}, {Lilly}, {Kong}, {Lehnert}, {Nesvadba}, {Verma}, {McCracken},
  {Arimoto}, {Mignoli}, \& {Onodera}}]{nmfs2009}
{F{\"o}rster Schreiber}, N.~M., {Genzel}, R., {Bouch{\'e}}, N., {et~al.} 2009,
  \href{http://dx.doi.org/10.1088/0004-637X/706/2/1364}{\color{blue}\apj},
  \href{https://ui.adsabs.harvard.edu/abs/2009ApJ...706.1364F}{706, 1364}

\bibitem[{{F{\"o}rster Schreiber} {et~al.}(2018){F{\"o}rster Schreiber},
  {Renzini}, {Mancini}, {Genzel}, {Bouch{\'e}}, {Cresci}, {Hicks}, {Lilly},
  {Peng}, {Burkert}, {Carollo}, {Cimatti}, {Daddi}, {Davies}, {Genel}, {Kurk},
  {Lang}, {Lutz}, {Mainieri}, {McCracken}, {Mignoli}, {Naab}, {Oesch},
  {Pozzetti}, {Scodeggio}, {Shapiro Griffin}, {Shapley}, {Sternberg},
  {Tacchella}, {Tacconi}, {Wuyts}, \& {Zamorani}}]{nmfs2018}
{F{\"o}rster Schreiber}, N.~M., {Renzini}, A., {Mancini}, C., {et~al.} 2018,
  \href{http://dx.doi.org/10.3847/1538-4365/aadd49}{\color{blue}\apjs},
  \href{https://ui.adsabs.harvard.edu/abs/2018ApJS..238...21F}{238, 21}

\bibitem[{{F{\"o}rster Schreiber} \& {Wuyts}(2020)}]{fs2020}
{F{\"o}rster Schreiber}, N.~M. \& {Wuyts}, S. 2020,
  \href{http://dx.doi.org/10.1146/annurev-astro-032620-021910}{\color{blue}\araa},
  \href{https://ui.adsabs.harvard.edu/abs/2020ARA&A..58..661F}{58, 661}

\bibitem[{{Fraternali} {et~al.}(2021){Fraternali}, {Karim}, {Magnelli},
  {G{\'o}mez-Guijarro}, {Jim{\'e}nez-Andrade}, \& {Posses}}]{Fraternali2021}
{Fraternali}, F., {Karim}, A., {Magnelli}, B., {et~al.} 2021,
  \href{http://dx.doi.org/10.1051/0004-6361/202039807}{\color{blue}\aap},
  \href{https://ui.adsabs.harvard.edu/abs/2021A&A...647A.194F}{647, A194}

\bibitem[{{Freundlich} {et~al.}(2020){Freundlich}, {Dekel}, {Jiang}, {Ishai},
  {Cornuault}, {Lapiner}, {Dutton}, \& {Macci{\`o}}}]{Freundlich2020}
{Freundlich}, J., {Dekel}, A., {Jiang}, F., {et~al.} 2020,
  \href{http://dx.doi.org/10.1093/mnras/stz3306}{\color{blue}\mnras},
  \href{https://ui.adsabs.harvard.edu/abs/2020MNRAS.491.4523F}{491, 4523}

\bibitem[{{Fujimoto} {et~al.}(2021){Fujimoto}, {Oguri}, {Brammer}, {Yoshimura},
  {Laporte}, {Gonz{\'a}lez-L{\'o}pez}, {Caminha}, {Kohno}, {Zitrin}, {Richard},
  {Ouchi}, {Bauer}, {Smail}, {Hatsukade}, {Ono}, {Kokorev}, {Umehata},
  {Schaerer}, {Knudsen}, {Sun}, {Magdis}, {Valentino}, {Ao}, {Toft},
  {Dessauges-Zavadsky}, {Shimasaku}, {Caputi}, {Kusakabe}, {Morokuma-Matsui},
  {Shotaro}, {Egami}, {Lee}, {Rawle}, \& {Espada}}]{Fujimoto2021}
{Fujimoto}, S., {Oguri}, M., {Brammer}, G., {et~al.} 2021,
  \href{http://dx.doi.org/10.3847/1538-4357/abd7ec}{\color{blue}\apj},
  \href{https://ui.adsabs.harvard.edu/abs/2021ApJ...911...99F}{911, 99}

\bibitem[{{Fujimoto} {et~al.}(2024){Fujimoto}, {Ouchi}, {Kohno}, {Valentino},
  {Gim{\'e}nez-Arteaga}, {Brammer}, {Furtak}, {Kohandel}, {Oguri},
  {Pallottini}, {Richard}, {Zitrin}, {Bauer}, {Boylan-Kolchin},
  {Dessauges-Zavadsky}, {Egami}, {Finkelstein}, {Ma}, {Smail}, {Watson},
  {Hutchison}, {Rigby}, {Welch}, {Ao}, {Bradley}, {Caminha}, {Caputi},
  {Espada}, {Endsley}, {Fudamoto}, {Gonz{\'a}lez-L{\'o}pez}, {Hatsukade},
  {Koekemoer}, {Kokorev}, {Laporte}, {Lee}, {Magdis}, {Ono}, {Rizzo},
  {Shibuya}, {Shimasaku}, {Sun}, {Toft}, {Umehata}, {Wang}, \&
  {Yajima}}]{Fujimoto2024}
{Fujimoto}, S., {Ouchi}, M., {Kohno}, K., {et~al.} 2024,
  \href{https://ui.adsabs.harvard.edu/abs/2024arXiv240218543F}{\href{http://dx.doi.org/10.48550/arXiv.2402.18543}{\color{blue}arXiv
  e-prints}, arXiv:2402.18543}

\bibitem[{{Gatto} {et~al.}(2015){Gatto}, {Walch}, {Low}, {Naab}, {Girichidis},
  {Glover}, {W{\"u}nsch}, {Klessen}, {Clark}, {Baczynski}, {Peters},
  {Ostriker}, {Ib{\'a}{\~n}ez-Mej{\'\i}a}, \& {Haid}}]{Gatto2015}
{Gatto}, A., {Walch}, S., {Low}, M. M.~M., {et~al.} 2015,
  \href{http://dx.doi.org/10.1093/mnras/stv324}{\color{blue}\mnras},
  \href{https://ui.adsabs.harvard.edu/abs/2015MNRAS.449.1057G}{449, 1057}

\bibitem[{{Genzel} {et~al.}(2008){Genzel}, {Burkert}, {Bouch{\'e}}, {Cresci},
  {F{\"o}rster Schreiber}, {Shapley}, {Shapiro}, {Tacconi}, {Buschkamp},
  {Cimatti}, {Daddi}, {Davies}, {Eisenhauer}, {Erb}, {Genel}, {Gerhard},
  {Hicks}, {Lutz}, {Naab}, {Ott}, {Rabien}, {Renzini}, {Steidel}, {Sternberg},
  \& {Lilly}}]{Genzel2008}
{Genzel}, R., {Burkert}, A., {Bouch{\'e}}, N., {et~al.} 2008,
  \href{http://dx.doi.org/10.1086/591840}{\color{blue}\apj},
  \href{https://ui.adsabs.harvard.edu/abs/2008ApJ...687...59G}{687, 59}

\bibitem[{{Genzel} {et~al.}(2014){Genzel}, {F{\"o}rster Schreiber}, {Lang},
  {Tacchella}, {Tacconi}, {Wuyts}, {Bandara}, {Burkert}, {Buschkamp},
  {Carollo}, {Cresci}, {Davies}, {Eisenhauer}, {Hicks}, {Kurk}, {Lilly},
  {Lutz}, {Mancini}, {Naab}, {Newman}, {Peng}, {Renzini}, {Shapiro Griffin},
  {Sternberg}, {Vergani}, {Wisnioski}, {Wuyts}, \& {Zamorani}}]{Genzel2014a}
{Genzel}, R., {F{\"o}rster Schreiber}, N.~M., {Lang}, P., {et~al.} 2014,
  \href{http://dx.doi.org/10.1088/0004-637X/785/1/75}{\color{blue}\apj},
  \href{https://ui.adsabs.harvard.edu/abs/2014ApJ...785...75G}{785, 75}

\bibitem[{{Genzel} {et~al.}(2017){Genzel}, {F{\"o}rster Schreiber},
  {{\"U}bler}, {Lang}, {Naab}, {Bender}, {Tacconi}, {Wisnioski}, {Wuyts},
  {Alexander}, {Beifiori}, {Belli}, {Brammer}, {Burkert}, {Carollo}, {Chan},
  {Davies}, {Fossati}, {Galametz}, {Genel}, {Gerhard}, {Lutz}, {Mendel},
  {Momcheva}, {Nelson}, {Renzini}, {Saglia}, {Sternberg}, {Tacchella},
  {Tadaki}, \& {Wilman}}]{Genzel2017}
{Genzel}, R., {F{\"o}rster Schreiber}, N.~M., {{\"U}bler}, H., {et~al.} 2017,
  \href{http://dx.doi.org/10.1038/nature21685}{\color{blue}\nat},
  \href{https://ui.adsabs.harvard.edu/abs/2017Natur.543..397G}{543, 397}

\bibitem[{{Genzel} {et~al.}(2023){Genzel}, {Jolly}, {Liu}, {Price}, {Lee},
  {F{\"o}rster Schreiber}, {Tacconi}, {Herrera-Camus}, {Barfety}, {Burkert},
  {Cao}, {Davies}, {Dekel}, {Lee}, {Lutz}, {Naab}, {Neri}, {Nestor Shachar},
  {Pastras}, {Pulsoni}, {Renzini}, {Schuster}, {Shimizu}, {Stanley},
  {Sternberg}, \& {{\"U}bler}}]{Genzel2023}
{Genzel}, R., {Jolly}, J.~B., {Liu}, D., {et~al.} 2023,
  \href{http://dx.doi.org/10.3847/1538-4357/acef1a}{\color{blue}\apj},
  \href{https://ui.adsabs.harvard.edu/abs/2023ApJ...957...48G}{957, 48}

\bibitem[{{Genzel} {et~al.}(2011){Genzel}, {Newman}, {Jones}, {F{\"o}rster
  Schreiber}, {Shapiro}, {Genel}, {Lilly}, {Renzini}, {Tacconi}, {Bouch{\'e}},
  {Burkert}, {Cresci}, {Buschkamp}, {Carollo}, {Ceverino}, {Davies}, {Dekel},
  {Eisenhauer}, {Hicks}, {Kurk}, {Lutz}, {Mancini}, {Naab}, {Peng},
  {Sternberg}, {Vergani}, \& {Zamorani}}]{Genzel2011}
{Genzel}, R., {Newman}, S., {Jones}, T., {et~al.} 2011,
  \href{http://dx.doi.org/10.1088/0004-637X/733/2/101}{\color{blue}\apj},
  \href{https://ui.adsabs.harvard.edu/abs/2011ApJ...733..101G}{733, 101}

\bibitem[{{Genzel} {et~al.}(2020){Genzel}, {Price}, {{\"U}bler}, {F{\"o}rster
  Schreiber}, {Shimizu}, {Tacconi}, {Bender}, {Burkert}, {Contursi}, {Coogan},
  {Davies}, {Davies}, {Dekel}, {Herrera-Camus}, {Lee}, {Lutz}, {Naab}, {Neri},
  {Nestor}, {Renzini}, {Saglia}, {Schuster}, {Sternberg}, {Wisnioski}, \&
  {Wuyts}}]{Genzel2020}
{Genzel}, R., {Price}, S.~H., {{\"U}bler}, H., {et~al.} 2020,
  \href{http://dx.doi.org/10.3847/1538-4357/abb0ea}{\color{blue}\apj},
  \href{https://ui.adsabs.harvard.edu/abs/2020ApJ...902...98G}{902, 98}

\bibitem[{{Genzel} {et~al.}(2013){Genzel}, {Tacconi}, {Kurk}, {Wuyts},
  {Combes}, {Freundlich}, {Bolatto}, {Cooper}, {Neri}, {Nordon}, {Bournaud},
  {Burkert}, {Comerford}, {Cox}, {Davis}, {F{\"o}rster Schreiber},
  {Garc{\'\i}a-Burillo}, {Gracia-Carpio}, {Lutz}, {Naab}, {Newman},
  {Saintonge}, {Shapiro Griffin}, {Shapley}, {Sternberg}, \&
  {Weiner}}]{Genzel2013}
{Genzel}, R., {Tacconi}, L.~J., {Kurk}, J., {et~al.} 2013,
  \href{http://dx.doi.org/10.1088/0004-637X/773/1/68}{\color{blue}\apj},
  \href{https://ui.adsabs.harvard.edu/abs/2013ApJ...773...68G}{773, 68}

\bibitem[{{Genzel} {et~al.}(2015){Genzel}, {Tacconi}, {Lutz}, {Saintonge},
  {Berta}, {Magnelli}, {Combes}, {Garc{\'\i}a-Burillo}, {Neri}, {Bolatto},
  {Contini}, {Lilly}, {Boissier}, {Boone}, {Bouch{\'e}}, {Bournaud}, {Burkert},
  {Carollo}, {Colina}, {Cooper}, {Cox}, {Feruglio}, {F{\"o}rster Schreiber},
  {Freundlich}, {Gracia-Carpio}, {Juneau}, {Kovac}, {Lippa}, {Naab}, {Salome},
  {Renzini}, {Sternberg}, {Walter}, {Weiner}, {Weiss}, \& {Wuyts}}]{Genzel2015}
{Genzel}, R., {Tacconi}, L.~J., {Lutz}, D., {et~al.} 2015,
  \href{http://dx.doi.org/10.1088/0004-637X/800/1/20}{\color{blue}\apj},
  \href{https://ui.adsabs.harvard.edu/abs/2015ApJ...800...20G}{800, 20}

\bibitem[{{Ginzburg} {et~al.}(2022){Ginzburg}, {Dekel}, {Mandelker}, \&
  {Krumholz}}]{Ginzburg2022}
{Ginzburg}, O., {Dekel}, A., {Mandelker}, N., \& {Krumholz}, M.~R. 2022,
  \href{http://dx.doi.org/10.1093/mnras/stac1324}{\color{blue}\mnras},
  \href{https://ui.adsabs.harvard.edu/abs/2022MNRAS.513.6177G}{513, 6177}

\bibitem[{{Girard} {et~al.}(2019){Girard}, {Dessauges-Zavadsky}, {Combes},
  {Chisholm}, {Patr{\'\i}cio}, {Richard}, \& {Schaerer}}]{Girard2019}
{Girard}, M., {Dessauges-Zavadsky}, M., {Combes}, F., {et~al.} 2019,
  \href{http://dx.doi.org/10.1051/0004-6361/201935896}{\color{blue}\aap},
  \href{https://ui.adsabs.harvard.edu/abs/2019A&A...631A..91G}{631, A91}

\bibitem[{{Girard} {et~al.}(2018){Girard}, {Dessauges-Zavadsky}, {Schaerer},
  {Cirasuolo}, {Turner}, {Cava}, {Rodr{\'\i}guez-Mu{\~n}oz}, {Richard}, \&
  {P{\'e}rez-Gonz{\'a}lez}}]{Girard2018}
{Girard}, M., {Dessauges-Zavadsky}, M., {Schaerer}, D., {et~al.} 2018,
  \href{http://dx.doi.org/10.1051/0004-6361/201731988}{\color{blue}\aap},
  \href{https://ui.adsabs.harvard.edu/abs/2018A&A...613A..72G}{613, A72}

\bibitem[{{Girard} {et~al.}(2021){Girard}, {Fisher}, {Bolatto}, {Abraham},
  {Bassett}, {Glazebrook}, {Herrera-Camus}, {Jim{\'e}nez}, {Lenki{\'c}}, \&
  {Obreschkow}}]{Girard2021}
{Girard}, M., {Fisher}, D.~B., {Bolatto}, A.~D., {et~al.} 2021,
  \href{http://dx.doi.org/10.3847/1538-4357/abd5b9}{\color{blue}\apj},
  \href{https://ui.adsabs.harvard.edu/abs/2021ApJ...909...12G}{909, 12}

\bibitem[{{Glazebrook}(2013)}]{Glazebrook2013}
{Glazebrook}, K. 2013,
  \href{http://dx.doi.org/10.1017/pasa.2013.34}{\color{blue}\pasa},
  \href{https://ui.adsabs.harvard.edu/abs/2013PASA...30...56G}{30, e056}

\bibitem[{{Gnerucci} {et~al.}(2010){Gnerucci}, {Marconi}, {Capetti}, {Axon}, \&
  {Robinson}}]{Gnerucci2010}
{Gnerucci}, A., {Marconi}, A., {Capetti}, A., {Axon}, D.~J., \& {Robinson}, A.
  2010, \href{http://dx.doi.org/10.1051/0004-6361/200912530}{\color{blue}\aap},
  \href{https://ui.adsabs.harvard.edu/abs/2010A&A...511A..19G}{511, A19}

\bibitem[{{Gnerucci} {et~al.}(2011){Gnerucci}, {Marconi}, {Cresci}, {Maiolino},
  {Mannucci}, {Calura}, {Cimatti}, {Cocchia}, {Grazian}, {Matteucci}, {Nagao},
  {Pozzetti}, \& {Troncoso}}]{Gnerucci2011}
{Gnerucci}, A., {Marconi}, A., {Cresci}, G., {et~al.} 2011,
  \href{http://dx.doi.org/10.1051/0004-6361/201015465}{\color{blue}\aap},
  \href{https://ui.adsabs.harvard.edu/abs/2011A&A...528A..88G}{528, A88}

\bibitem[{Harris {et~al.}(2020)Harris, Millman, van~der Walt, Gommers,
  Virtanen, Cournapeau, Wieser, Taylor, Berg, Smith, Kern, Picus, Hoyer, van
  Kerkwijk, Brett, Haldane, del R{\'{i}}o, Wiebe, Peterson,
  G{\'{e}}rard-Marchant, Sheppard, Reddy, Weckesser, Abbasi, Gohlke, \&
  Oliphant}]{harris2020}
Harris, C.~R., Millman, K.~J., van~der Walt, S.~J., {et~al.} 2020,
  \href{http://dx.doi.org/10.1038/s41586-020-2649-2}{\color{blue}Nature}, 585,
  357

\bibitem[{{Hayward} \& {Hopkins}(2017)}]{HaywardHopkins2017}
{Hayward}, C.~C. \& {Hopkins}, P.~F. 2017,
  \href{http://dx.doi.org/10.1093/mnras/stw2888}{\color{blue}\mnras},
  \href{https://ui.adsabs.harvard.edu/abs/2017MNRAS.465.1682H}{465, 1682}

\bibitem[{{Hayward} {et~al.}(2021){Hayward}, {Sparre}, {Chapman}, {Hernquist},
  {Nelson}, {Pakmor}, {Pillepich}, {Springel}, {Torrey}, {Vogelsberger}, \&
  {Weinberger}}]{Hayward2021}
{Hayward}, C.~C., {Sparre}, M., {Chapman}, S.~C., {et~al.} 2021,
  \href{http://dx.doi.org/10.1093/mnras/stab246}{\color{blue}\mnras},
  \href{https://ui.adsabs.harvard.edu/abs/2021MNRAS.502.2922H}{502, 2922}

\bibitem[{{Herrera-Camus} {et~al.}(2022){Herrera-Camus}, {F{\"o}rster
  Schreiber}, {Price}, {{\"U}bler}, {Bolatto}, {Davies}, {Fisher}, {Genzel},
  {Lutz}, {Naab}, {Nestor}, {Shimizu}, {Sternberg}, {Tacconi}, \&
  {Tadaki}}]{Herrera-Camus2022}
{Herrera-Camus}, R., {F{\"o}rster Schreiber}, N.~M., {Price}, S.~H., {et~al.}
  2022, \href{http://dx.doi.org/10.1051/0004-6361/202142562}{\color{blue}\aap},
  \href{https://ui.adsabs.harvard.edu/abs/2022A&A...665L...8H}{665, L8}

\bibitem[{{Herrera-Camus} {et~al.}(2025){Herrera-Camus},
  {Gonz{\'a}lez-L{\'o}pez}, {F{\"o}rster Schreiber}, {Aravena}, {de Looze},
  {Spilker}, {Tadaki}, {Barcos-Mu{\~n}oz}, {Assef}, {Birkin}, {Bolatto},
  {Bouwens}, {Bovino}, {Bowler}, {Calistro Rivera}, {da Cunha}, {Davies},
  {Davies}, {D{\'\i}az-Santos}, {Ferrara}, {Fisher}, {Genzel}, {Hodge},
  {Ikeda}, {Killi}, {Lee}, {Li}, {Li}, {Liu}, {Lutz}, {Mitsuhashi},
  {Narayanan}, {Naab}, {Palla}, {Price}, {Posses}, {Rela{\~n}o}, {Smit},
  {Solimano}, {Sternberg}, {Tacconi}, {Telikova}, {{\"U}bler}, {van der
  Giessen}, {Veilleux}, {Villanueva}, \& {Baeza-Garay}}]{HerreraCamus2025}
{Herrera-Camus}, R., {Gonz{\'a}lez-L{\'o}pez}, J., {F{\"o}rster Schreiber}, N.,
  {et~al.} 2025,
  \href{http://dx.doi.org/10.1051/0004-6361/202553896}{\color{blue}\aap},
  \href{https://doi.org/10.1051/0004-6361/202553896}{699, A80}

\bibitem[{{Hirtenstein} {et~al.}(2019){Hirtenstein}, {Jones}, {Wang}, {Wetzel},
  {El-Badry}, {Hoag}, {Treu}, {Brada{\v{c}}}, \& {Morishita}}]{Hirtenstein2019}
{Hirtenstein}, J., {Jones}, T., {Wang}, X., {et~al.} 2019,
  \href{http://dx.doi.org/10.3847/1538-4357/ab113e}{\color{blue}\apj},
  \href{https://ui.adsabs.harvard.edu/abs/2019ApJ...880...54H}{880, 54}

\bibitem[{{Hodge} {et~al.}(2012){Hodge}, {Carilli}, {Walter}, {de Blok},
  {Riechers}, {Daddi}, \& {Lentati}}]{Hodge2012}
{Hodge}, J.~A., {Carilli}, C.~L., {Walter}, F., {et~al.} 2012,
  \href{http://dx.doi.org/10.1088/0004-637X/760/1/11}{\color{blue}\apj},
  \href{https://ui.adsabs.harvard.edu/abs/2012ApJ...760...11H}{760, 11}

\bibitem[{{Hogan} {et~al.}(2021){Hogan}, {Rigopoulou}, {Magdis},
  {Pereira-Santaella}, {Garc{\'\i}a-Bernete}, {Thatte}, {Grisdale}, \&
  {Huang}}]{Hogan2021}
{Hogan}, L., {Rigopoulou}, D., {Magdis}, G.~E., {et~al.} 2021,
  \href{http://dx.doi.org/10.1093/mnras/stab527}{\color{blue}\mnras},
  \href{https://ui.adsabs.harvard.edu/abs/2021MNRAS.503.5329H}{503, 5329}

\bibitem[{{Huang} {et~al.}(2023){Huang}, {Kawabe}, {Kohno}, {Saito},
  {Mizukoshi}, {Iono}, {Michiyama}, {Tamura}, {Hayward}, \&
  {Umehata}}]{Huang2023}
{Huang}, S., {Kawabe}, R., {Kohno}, K., {et~al.} 2023,
  \href{http://dx.doi.org/10.3847/2041-8213/acff63}{\color{blue}\apjl},
  \href{https://ui.adsabs.harvard.edu/abs/2023ApJ...958L..26H}{958, L26}

\bibitem[{{Huertas-Company} {et~al.}(2024){Huertas-Company}, {Iyer},
  {Angeloudi}, {Bagley}, {Finkelstein}, {Kartaltepe}, {McGrath}, {Sarmiento},
  {Vega-Ferrero}, {Arrabal Haro}, {Behroozi}, {Buitrago}, {Cheng}, {Costantin},
  {Dekel}, {Dickinson}, {Elbaz}, {Grogin}, {Hathi}, {Holwerda}, {Koekemoer},
  {Lucas}, {Papovich}, {P{\'e}rez-Gonz{\'a}lez}, {Pirzkal}, {Seill{\'e}}, {de
  la Vega}, {Wuyts}, {Yang}, \& {Yung}}]{Huertas-Company2024}
{Huertas-Company}, M., {Iyer}, K.~G., {Angeloudi}, E., {et~al.} 2024,
  \href{http://dx.doi.org/10.1051/0004-6361/202346800}{\color{blue}\aap},
  \href{https://ui.adsabs.harvard.edu/abs/2024A&A...685A..48H}{685, A48}

\bibitem[{{Huertas-Company} {et~al.}(2025){Huertas-Company}, {Shuntov}, {Dong},
  {Walmsley}, {Ilbert}, {McCracken}, {Akins}, {Allen}, {Casey}, {Costantin},
  {Daddi}, {Dekel}, {Franco}, {Garland}, {G{\'e}ron}, {Gozaliasl},
  {Hirschmann}, {Kartaltepe}, {Koekemoer}, {Lintott}, {Liu}, {Lucas},
  {Masters}, {Pacucci}, {Paquereau}, {P'erez-Gonz'alez}, {Rhodes}, {Robertson},
  {Simmons}, {Smethurst}, {Toft}, \& {Yang}}]{HuertasCompany2025}
{Huertas-Company}, M., {Shuntov}, M., {Dong}, Y., {et~al.} 2025,
  \href{https://ui.adsabs.harvard.edu/abs/2025arXiv250203532H}{\href{http://dx.doi.org/10.48550/arXiv.2502.03532}{\color{blue}arXiv
  e-prints}, arXiv:2502.03532}

\bibitem[{{Hunter}(2007)}]{Hunter2007}
{Hunter}, J.~D. 2007,
  \href{http://dx.doi.org/10.1109/MCSE.2007.55}{\color{blue}Computing in
  Science and Engineering},
  \href{https://ui.adsabs.harvard.edu/abs/2007CSE.....9...90H}{9, 90}

\bibitem[{{Ikeda} {et~al.}(2025){Ikeda}, {Tadaki}, {Mitsuhashi}, {Aravena}, {De
  Looze}, {F{\"o}rster Schreiber}, {Gonz{\'a}lez-L{\'o}pez}, {Herrera-Camus},
  {Spilker}, {Barcos-Mu{\~n}oz}, {Bowler}, {Calistro Rivera}, {da Cunha},
  {Davies}, {D{\'\i}az-Santos}, {Ferrara}, {Killi}, {Lee}, {Li}, {Lutz},
  {Posses}, {Smit}, {Solimano}, {Telikova}, {{\"U}bler}, {Veilleux}, \&
  {Villanueva}}]{Ikeda2025}
{Ikeda}, R., {Tadaki}, K.-i., {Mitsuhashi}, I., {et~al.} 2025,
  \href{http://dx.doi.org/10.1051/0004-6361/202451811}{\color{blue}\aap},
  \href{https://ui.adsabs.harvard.edu/abs/2025A&A...693A.237I}{693, A237}

\bibitem[{{Jacobs} {et~al.}(2023){Jacobs}, {Glazebrook}, {Calabr{\`o}}, {Treu},
  {Nannayakkara}, {Jones}, {Merlin}, {Abraham}, {Stevens}, {Vulcani}, {Yang},
  {Bonchi}, {Boyett}, {Brada{\v{c}}}, {Castellano}, {Fontana}, {Marchesini},
  {Malkan}, {Mason}, {Morishita}, {Paris}, {Santini}, {Trenti}, \&
  {Wang}}]{Jacobs2023}
{Jacobs}, C., {Glazebrook}, K., {Calabr{\`o}}, A., {et~al.} 2023,
  \href{http://dx.doi.org/10.3847/2041-8213/accd6d}{\color{blue}\apjl},
  \href{https://ui.adsabs.harvard.edu/abs/2023ApJ...948L..13J}{948, L13}

\bibitem[{{Jim{\'e}nez} {et~al.}(2023){Jim{\'e}nez}, {Lagos}, {Ludlow}, \&
  {Wisnioski}}]{Jimenez2023}
{Jim{\'e}nez}, E., {Lagos}, C. d.~P., {Ludlow}, A.~D., \& {Wisnioski}, E. 2023,
  \href{http://dx.doi.org/10.1093/mnras/stad2119}{\color{blue}\mnras},
  \href{https://ui.adsabs.harvard.edu/abs/2023MNRAS.524.4346J}{524, 4346}

\bibitem[{{Johnson} {et~al.}(2018){Johnson}, {Harrison}, {Swinbank}, {Tiley},
  {Stott}, {Bower}, {Smail}, {Bunker}, {Sobral}, {Turner}, {Best}, {Bureau},
  {Cirasuolo}, {Jarvis}, {Magdis}, {Sharples}, {Bland-Hawthorn}, {Catinella},
  {Cortese}, {Croom}, {Federrath}, {Glazebrook}, {Sweet}, {Bryant}, {Goodwin},
  {Konstantopoulos}, {Lawrence}, {Medling}, {Owers}, \&
  {Richards}}]{Johnson2018}
{Johnson}, H.~L., {Harrison}, C.~M., {Swinbank}, A.~M., {et~al.} 2018,
  \href{http://dx.doi.org/10.1093/mnras/stx3016}{\color{blue}\mnras},
  \href{https://ui.adsabs.harvard.edu/abs/2018MNRAS.474.5076J}{474, 5076}

\bibitem[{{Jones} {et~al.}(2020){Jones}, {B{\'e}thermin}, {Fudamoto},
  {Ginolfi}, {Capak}, {Cassata}, {Faisst}, {Le F{\`e}vre}, {Schaerer},
  {Silverman}, {Yan}, {Bardelli}, {Boquien}, {Cimatti}, {Dessauges-Zavadsky},
  {Giavalisco}, {Gruppioni}, {Ibar}, {Khusanova}, {Koekemoer}, {Lemaux},
  {Loiacono}, {Maiolino}, {Oesch}, {Pozzi}, {Riechers}, {Rodighiero}, {Talia},
  {Vallini}, {Vergani}, {Zamorani}, \& {Zucca}}]{Jones2020}
{Jones}, G.~C., {B{\'e}thermin}, M., {Fudamoto}, Y., {et~al.} 2020,
  \href{http://dx.doi.org/10.1093/mnrasl/slz154}{\color{blue}\mnras},
  \href{https://ui.adsabs.harvard.edu/abs/2020MNRAS.491L..18J}{491, L18}

\bibitem[{{Jones} {et~al.}(2021){Jones}, {Vergani}, {Romano}, {Ginolfi},
  {Fudamoto}, {B{\'e}thermin}, {Fujimoto}, {Lemaux}, {Morselli}, {Capak},
  {Cassata}, {Faisst}, {Le F{\`e}vre}, {Schaerer}, {Silverman}, {Yan},
  {Boquien}, {Cimatti}, {Dessauges-Zavadsky}, {Ibar}, {Maiolino}, {Rizzo},
  {Talia}, \& {Zamorani}}]{Jones2021}
{Jones}, G.~C., {Vergani}, D., {Romano}, M., {et~al.} 2021,
  \href{http://dx.doi.org/10.1093/mnras/stab2226}{\color{blue}\mnras},
  \href{https://ui.adsabs.harvard.edu/abs/2021MNRAS.507.3540J}{507, 3540}

\bibitem[{{Jones} {et~al.}(2010){Jones}, {Swinbank}, {Ellis}, {Richard}, \&
  {Stark}}]{Jones2010}
{Jones}, T.~A., {Swinbank}, A.~M., {Ellis}, R.~S., {Richard}, J., \& {Stark},
  D.~P. 2010,
  \href{http://dx.doi.org/10.1111/j.1365-2966.2010.16378.x}{\color{blue}\mnras},
  \href{https://ui.adsabs.harvard.edu/abs/2010MNRAS.404.1247J}{404, 1247}

\bibitem[{{Jorsater} \& {van Moorsel}(1995)}]{JvM1995}
{Jorsater}, S. \& {van Moorsel}, G.~A. 1995,
  \href{http://dx.doi.org/10.1086/117668}{\color{blue}\aj},
  \href{https://ui.adsabs.harvard.edu/abs/1995AJ....110.2037J}{110, 2037}

\bibitem[{Kaasinen {et~al.}(2019)Kaasinen, Scoville, Walter, Cunha, Popping,
  Pavesi, Darvish, Casey, Riechers, \& Glover}]{Kaasinen2019}
Kaasinen, M., Scoville, N., Walter, F., {et~al.} 2019,
  \href{http://dx.doi.org/10.3847/1538-4357/ab253b}{\color{blue}\apj}, 880, 15

\bibitem[{{Kaasinen} {et~al.}(2020){Kaasinen}, {Walter}, {Novak}, {Neeleman},
  {Smail}, {Boogaard}, {Cunha}, {Weiss}, {Liu}, {Decarli}, {Popping},
  {Diaz-Santos}, {Cort{\'e}s}, {Aravena}, {Werf}, {Riechers}, {Inami}, {Hodge},
  {Rix}, \& {Cox}}]{Kaasinen2020}
{Kaasinen}, M., {Walter}, F., {Novak}, M., {et~al.} 2020,
  \href{http://dx.doi.org/10.3847/1538-4357/aba438}{\color{blue}\apj},
  \href{https://ui.adsabs.harvard.edu/abs/2020ApJ...899...37K}{899, 37}

\bibitem[{{Kartaltepe} {et~al.}(2023){Kartaltepe}, {Rose}, {Vanderhoof},
  {McGrath}, {Costantin}, {Cox}, {Yung}, {Kocevski}, {Wuyts}, {Ferguson},
  {Bagley}, {Finkelstein}, {Amor{\'\i}n}, {Andrews}, {Arrabal Haro},
  {Backhaus}, {Behroozi}, {Bisigello}, {Calabr{\`o}}, {Casey}, {Coogan},
  {Cooper}, {Croton}, {de la Vega}, {Dickinson}, {Fontana}, {Franco},
  {Grazian}, {Grogin}, {Hathi}, {Holwerda}, {Huertas-Company}, {Iyer}, {Jogee},
  {Jung}, {Kewley}, {Kirkpatrick}, {Koekemoer}, {Liu}, {Lotz}, {Lucas},
  {Newman}, {Pacifici}, {Pandya}, {Papovich}, {Pentericci},
  {P{\'e}rez-Gonz{\'a}lez}, {Petersen}, {Pirzkal}, {Rafelski}, {Ravindranath},
  {Simons}, {Snyder}, {Somerville}, {Stanway}, {Straughn}, {Tacchella},
  {Trump}, {Vega-Ferrero}, {Wilkins}, {Yang}, \& {Zavala}}]{Kartaltepe2023}
{Kartaltepe}, J.~S., {Rose}, C., {Vanderhoof}, B.~N., {et~al.} 2023,
  \href{http://dx.doi.org/10.3847/2041-8213/acad01}{\color{blue}\apjl},
  \href{https://ui.adsabs.harvard.edu/abs/2023ApJ...946L..15K}{946, L15}

\bibitem[{{Kassin} {et~al.}(2012){Kassin}, {Weiner}, {Faber}, {Gardner},
  {Willmer}, {Coil}, {Cooper}, {Devriendt}, {Dutton}, {Guhathakurta}, {Koo},
  {Metevier}, {Noeske}, \& {Primack}}]{Kassin2012}
{Kassin}, S.~A., {Weiner}, B.~J., {Faber}, S.~M., {et~al.} 2012,
  \href{http://dx.doi.org/10.1088/0004-637X/758/2/106}{\color{blue}\apj},
  \href{https://ui.adsabs.harvard.edu/abs/2012ApJ...758..106K}{758, 106}

\bibitem[{{Kim} \& {Ostriker}(2018)}]{Kim2018}
{Kim}, C.-G. \& {Ostriker}, E.~C. 2018,
  \href{http://dx.doi.org/10.3847/1538-4357/aaa5ff}{\color{blue}\apj},
  \href{https://ui.adsabs.harvard.edu/abs/2018ApJ...853..173K}{853, 173}

\bibitem[{{Kohandel} {et~al.}(2024){Kohandel}, {Pallottini}, {Ferrara},
  {Zanella}, {Rizzo}, \& {Carniani}}]{Kohandel2024}
{Kohandel}, M., {Pallottini}, A., {Ferrara}, A., {et~al.} 2024,
  \href{http://dx.doi.org/10.1051/0004-6361/202348209}{\color{blue}\aap},
  \href{https://ui.adsabs.harvard.edu/abs/2024A&A...685A..72K}{685, A72}

\bibitem[{{Koprowski} {et~al.}(2024){Koprowski}, {Wijesekera}, {Dunlop},
  {McLeod}, {Micha{\l}owski}, {Lisiecki}, \& {McLure}}]{Koprowski2024}
{Koprowski}, M.~P., {Wijesekera}, J.~V., {Dunlop}, J.~S., {et~al.} 2024,
  \href{http://dx.doi.org/10.1051/0004-6361/202449948}{\color{blue}\aap},
  \href{https://ui.adsabs.harvard.edu/abs/2024A&A...691A.164K}{691, A164}

\bibitem[{{Krajnovi{\'c}} {et~al.}(2006){Krajnovi{\'c}}, {Cappellari}, {de
  Zeeuw}, \& {Copin}}]{Krajnovic2006}
{Krajnovi{\'c}}, D., {Cappellari}, M., {de Zeeuw}, P.~T., \& {Copin}, Y. 2006,
  \href{http://dx.doi.org/10.1111/j.1365-2966.2005.09902.x}{\color{blue}\mnras},
  \href{https://ui.adsabs.harvard.edu/abs/2006MNRAS.366..787K}{366, 787}

\bibitem[{{Kretschmer} {et~al.}(2021){Kretschmer}, {Dekel}, {Freundlich},
  {Lapiner}, {Ceverino}, \& {Primack}}]{Kretschmer2021}
{Kretschmer}, M., {Dekel}, A., {Freundlich}, J., {et~al.} 2021,
  \href{http://dx.doi.org/10.1093/mnras/stab833}{\color{blue}\mnras},
  \href{https://ui.adsabs.harvard.edu/abs/2021MNRAS.503.5238K}{503, 5238}

\bibitem[{{Krumholz} {et~al.}(2018){Krumholz}, {Burkhart}, {Forbes}, \&
  {Crocker}}]{Krumholz2018}
{Krumholz}, M.~R., {Burkhart}, B., {Forbes}, J.~C., \& {Crocker}, R.~M. 2018,
  \href{http://dx.doi.org/10.1093/mnras/sty852}{\color{blue}\mnras},
  \href{https://ui.adsabs.harvard.edu/abs/2018MNRAS.477.2716K}{477, 2716}

\bibitem[{{Lambert} {et~al.}(2023){Lambert}, {Posses}, {Aravena},
  {G{\'o}nzalez-L{\'o}pez}, {Assef}, {D{\'\i}az-Santos}, {Brisbin}, {Decarli},
  {Herrera-Camus}, {Mej{\'\i}a}, \& {Ricci}}]{Lambert2023}
{Lambert}, T.~S., {Posses}, A., {Aravena}, M., {et~al.} 2023,
  \href{http://dx.doi.org/10.1093/mnras/stac3016}{\color{blue}\mnras},
  \href{https://ui.adsabs.harvard.edu/abs/2023MNRAS.518.3183L}{518, 3183}

\bibitem[{{Lang} {et~al.}(2017){Lang}, {F{\"o}rster Schreiber}, {Genzel},
  {Wuyts}, {Wisnioski}, {Beifiori}, {Belli}, {Bender}, {Brammer}, {Burkert},
  {Chan}, {Davies}, {Fossati}, {Galametz}, {Kulkarni}, {Lutz}, {Mendel},
  {Momcheva}, {Naab}, {Nelson}, {Saglia}, {Seitz}, {Tacchella}, {Tacconi},
  {Tadaki}, {{\"U}bler}, {van Dokkum}, \& {Wilman}}]{Lang2017}
{Lang}, P., {F{\"o}rster Schreiber}, N.~M., {Genzel}, R., {et~al.} 2017,
  \href{http://dx.doi.org/10.3847/1538-4357/aa6d82}{\color{blue}\apj},
  \href{https://ui.adsabs.harvard.edu/abs/2017ApJ...840...92L}{840, 92}

\bibitem[{{Law} {et~al.}(2009){Law}, {Steidel}, {Erb}, {Larkin}, {Pettini},
  {Shapley}, \& {Wright}}]{Law2009}
{Law}, D.~R., {Steidel}, C.~C., {Erb}, D.~K., {et~al.} 2009,
  \href{http://dx.doi.org/10.1088/0004-637X/697/2/2057}{\color{blue}\apj},
  \href{https://ui.adsabs.harvard.edu/abs/2009ApJ...697.2057L}{697, 2057}

\bibitem[{{Le F{\`e}vre} {et~al.}(2020){Le F{\`e}vre}, {B{\'e}thermin},
  {Faisst}, {Jones}, {Capak}, {Cassata}, {Silverman}, {Schaerer}, {Yan},
  {Amorin}, {Bardelli}, {Boquien}, {Cimatti}, {Dessauges-Zavadsky},
  {Giavalisco}, {Hathi}, {Fudamoto}, {Fujimoto}, {Ginolfi}, {Gruppioni},
  {Hemmati}, {Ibar}, {Koekemoer}, {Khusanova}, {Lagache}, {Lemaux}, {Loiacono},
  {Maiolino}, {Mancini}, {Narayanan}, {Morselli}, {M{\'e}ndez-Hern{\`a}ndez},
  {Oesch}, {Pozzi}, {Romano}, {Riechers}, {Scoville}, {Talia}, {Tasca},
  {Thomas}, {Toft}, {Vallini}, {Vergani}, {Walter}, {Zamorani}, \&
  {Zucca}}]{LeFevre2020}
{Le F{\`e}vre}, O., {B{\'e}thermin}, M., {Faisst}, A., {et~al.} 2020,
  \href{http://dx.doi.org/10.1051/0004-6361/201936965}{\color{blue}\aap},
  \href{https://ui.adsabs.harvard.edu/abs/2020A&A...643A...1L}{643, A1}

\bibitem[{{Lee} {et~al.}(2024){Lee}, {Park}, {Hwang}, \& {Kwon}}]{JHLee2024}
{Lee}, J.~H., {Park}, C., {Hwang}, H.~S., \& {Kwon}, M. 2024,
  \href{http://dx.doi.org/10.3847/1538-4357/ad3448}{\color{blue}\apj},
  \href{https://ui.adsabs.harvard.edu/abs/2024ApJ...966..113L}{966, 113}

\bibitem[{{Lee} {et~al.}(2025){Lee}, {F{\"o}rster Schreiber}, {Price}, {Liu},
  {Genzel}, {Davies}, {Tacconi}, {Shimizu}, {Nestor Shachar}, {Espejo Salcedo},
  {Pastras}, {Wuyts}, {Lutz}, {Renzini}, {{\"U}bler}, {Herrera-Camus}, \&
  {Sternberg}}]{Lee2025}
{Lee}, L.~L., {F{\"o}rster Schreiber}, N.~M., {Price}, S.~H., {et~al.} 2025,
  \href{http://dx.doi.org/10.3847/1538-4357/ad90b5}{\color{blue}\apj},
  \href{https://ui.adsabs.harvard.edu/abs/2025ApJ...978...14L}{978, 14}

\bibitem[{{Leethochawalit} {et~al.}(2016){Leethochawalit}, {Jones}, {Ellis},
  {Stark}, {Richard}, {Zitrin}, \& {Auger}}]{Leethochawalit2016}
{Leethochawalit}, N., {Jones}, T.~A., {Ellis}, R.~S., {et~al.} 2016,
  \href{http://dx.doi.org/10.3847/0004-637X/820/2/84}{\color{blue}\apj},
  \href{https://ui.adsabs.harvard.edu/abs/2016ApJ...820...84L}{820, 84}

\bibitem[{{Lelli} {et~al.}(2018){Lelli}, {De Breuck}, {Falkendal},
  {Fraternali}, {Man}, {Nesvadba}, \& {Lehnert}}]{Lelli2018}
{Lelli}, F., {De Breuck}, C., {Falkendal}, T., {et~al.} 2018,
  \href{http://dx.doi.org/10.1093/mnras/sty1795}{\color{blue}\mnras},
  \href{https://ui.adsabs.harvard.edu/abs/2018MNRAS.479.5440L}{479, 5440}

\bibitem[{{Lelli} {et~al.}(2021){Lelli}, {Di Teodoro}, {Fraternali}, {Man},
  {Zhang}, {De Breuck}, {Davis}, \& {Maiolino}}]{Lelli2021}
{Lelli}, F., {Di Teodoro}, E.~M., {Fraternali}, F., {et~al.} 2021,
  \href{http://dx.doi.org/10.1126/science.abc1893}{\color{blue}Science},
  \href{https://ui.adsabs.harvard.edu/abs/2021Sci...371..713L}{371, 713}

\bibitem[{{Lemaux} {et~al.}(2018){Lemaux}, {Le F{\`e}vre}, {Cucciati},
  {Ribeiro}, {Tasca}, {Zamorani}, {Ilbert}, {Thomas}, {Bardelli}, {Cassata},
  {Hathi}, {Pforr}, {Smol{\v{c}}i{\'c}}, {Delvecchio}, {Novak}, {Berta},
  {McCracken}, {Koekemoer}, {Amor{\'\i}n}, {Garilli}, {Maccagni}, {Schaerer},
  \& {Zucca}}]{Lemaux2018}
{Lemaux}, B.~C., {Le F{\`e}vre}, O., {Cucciati}, O., {et~al.} 2018,
  \href{http://dx.doi.org/10.1051/0004-6361/201730870}{\color{blue}\aap},
  \href{https://ui.adsabs.harvard.edu/abs/2018A&A...615A..77L}{615, A77}

\bibitem[{{Leroy} {et~al.}(2011){Leroy}, {Bolatto}, {Gordon}, {Sandstrom},
  {Gratier}, {Rosolowsky}, {Engelbracht}, {Mizuno}, {Corbelli}, {Fukui}, \&
  {Kawamura}}]{Leroy2011}
{Leroy}, A.~K., {Bolatto}, A., {Gordon}, K., {et~al.} 2011,
  \href{http://dx.doi.org/10.1088/0004-637X/737/1/12}{\color{blue}\apj},
  \href{https://ui.adsabs.harvard.edu/abs/2011ApJ...737...12L}{737, 12}

\bibitem[{{Li} {et~al.}(2024){Li}, {Da Cunha}, {Gonz{\'a}lez-L{\'o}pez},
  {Aravena}, {De Looze}, {F{\"o}rster Schreiber}, {Herrera-Camus}, {Spilker},
  {Tadaki}, {Barcos-Munoz}, {Battisti}, {Birkin}, {Bowler}, {Davies},
  {D{\'\i}az-Santos}, {Ferrara}, {Fisher}, {Hodge}, {Ikeda}, {Killi}, {Lee},
  {Liu}, {Lutz}, {Mitsuhashi}, {Naab}, {Posses}, {Rela{\~n}o}, {Solimano},
  {{\"U}bler}, {van der Giessen}, \& {Villanueva}}]{Li2024}
{Li}, J., {Da Cunha}, E., {Gonz{\'a}lez-L{\'o}pez}, J., {et~al.} 2024,
  \href{http://dx.doi.org/10.3847/1538-4357/ad7fee}{\color{blue}\apj},
  \href{https://ui.adsabs.harvard.edu/abs/2024ApJ...976...70L}{976, 70}

\bibitem[{{Lines} {et~al.}(2025){Lines}, {Bowler}, {Adams}, {Fisher},
  {Varadaraj}, {Nakazato}, {Aravena}, {Assef}, {Birkin}, {Ceverino}, {da
  Cunha}, {Cullen}, {De Looze}, {Donnan}, {Dunlop}, {Ferrara}, {Grogin},
  {Herrera-Camus}, {Ikeda}, {Koekemoer}, {Killi}, {Li}, {McLeod}, {McLure},
  {Mitsuhashi}, {P{\'e}rez-Gonz{\'a}lez}, {Relano}, {Solimano}, {Spilker},
  {Villanueva}, \& {Yoshida}}]{Lines2025}
{Lines}, N.~E.~P., {Bowler}, R.~A.~A., {Adams}, N.~J., {et~al.} 2025,
  \href{https://ui.adsabs.harvard.edu/abs/2025MNRAS.tmp..598L}{\href{http://dx.doi.org/10.1093/mnras/staf627}{\color{blue}\mnras},
  staf627}

\bibitem[{{Liu} {et~al.}(2023){Liu}, {F{\"o}rster Schreiber}, {Genzel}, {Lutz},
  {Price}, {Lee}, {Baker}, {Burkert}, {Coogan}, {Davies}, {Davies},
  {Herrera-Camus}, {Kodama}, {Lee}, {Nestor}, {Pulsoni}, {Renzini}, {Sharon},
  {Shimizu}, {Tacconi}, {Tadaki}, \& {{\"U}bler}}]{Liu2023}
{Liu}, D., {F{\"o}rster Schreiber}, N.~M., {Genzel}, R., {et~al.} 2023,
  \href{http://dx.doi.org/10.3847/1538-4357/aca46b}{\color{blue}\apj},
  \href{https://ui.adsabs.harvard.edu/abs/2023ApJ...942...98L}{942, 98}

\bibitem[{{Liu} {et~al.}(2024{\natexlab{a}}){Liu}, {F{\"o}rster Schreiber},
  {Harrington}, {Lee}, {Kamieneski}, {Davies}, {Lutz}, {Renzini}, {Wuyts},
  {Tacconi}, {Genzel}, {Burkert}, {Herrera-Camus}, {Alcalde Pampliega},
  {Vishwas}, {Kaasinen}, {Wang}, {Jim{\'e}nez-Andrade}, {Lowenthal}, {Foo},
  {Frye}, {Shangguan}, {Cao}, {Agapito}, {Berbel}, {Barfety}, {Baruffolo},
  {Berman}, {Black}, {Bonaglia}, {Briguglio}, {Carbonaro}, {Chapman}, {Chen},
  {Cikota}, {Concas}, {Cooper}, {Cresci}, {Dallilar}, {Deysenroth}, {Di
  Antonio}, {Di Cianno}, {Di Rico}, {Doelman}, {Dolci}, {Eisenhauer}, {Espejo},
  {Esposito}, {Fantinel}, {Ferruzzi}, {Feuchtgruber}, {Gao}, {Garcia Diaz},
  {Gillessen}, {Grani}, {Hartl}, {Henry}, {Huber}, {Jolly}, {Keller},
  {Kenworthy}, {Kravchenko}, {Lee}, {Lightfoot}, {Lunney}, {Macintosh},
  {Mannucci}, {Ott}, {Pascale}, {Pastras}, {Pearson}, {Puglisi}, {Pulsoni},
  {Rabien}, {Rau}, {Riccardi}, {Salasnich}, {Shimizu}, {Snik}, {Sturm},
  {Taylor}, {Valentini}, {Waring}, {Wiezorrek}, {Xompero}, \& {Yun}}]{Liu2024}
{Liu}, D., {F{\"o}rster Schreiber}, N.~M., {Harrington}, K.~C., {et~al.}
  2024{\natexlab{a}},
  \href{https://ui.adsabs.harvard.edu/abs/2024NatAs.tmp..129L}{\href{http://dx.doi.org/10.1038/s41550-024-02296-7}{\color{blue}Nature
  Astronomy}}

\bibitem[{{Liu} {et~al.}(2025){Liu}, {Kodama}, {Morishita}, {Lee}, {Sun},
  {Kubo}, {Cai}, {Wu}, \& {Li}}]{Liu2025}
{Liu}, Z., {Kodama}, T., {Morishita}, T., {et~al.} 2025,
  \href{http://dx.doi.org/10.3847/1538-4357/ada937}{\color{blue}\apj},
  \href{https://ui.adsabs.harvard.edu/abs/2025ApJ...980...69L}{980, 69}

\bibitem[{{Liu} {et~al.}(2024{\natexlab{b}}){Liu}, {Silverman}, {Daddi},
  {Puglisi}, {Renzini}, {Kalita}, {Kartaltepe}, {Kashino}, {Rodighiero},
  {Rujopakarn}, {Suzuki}, {Tanaka}, {Valentino}, {Andika}, {Casey}, {Faisst},
  {Franco}, {Gozaliasl}, {Gillman}, {Hayward}, {Koekemoer}, {Kokorev},
  {Lambrides}, {Lee}, {Magdis}, {Harish}, {McCracken}, {Rhodes}, {Shuntov}, \&
  {Ding}}]{ZXLiu2024}
{Liu}, Z., {Silverman}, J.~D., {Daddi}, E., {et~al.} 2024{\natexlab{b}},
  \href{http://dx.doi.org/10.3847/1538-4357/ad4096}{\color{blue}\apj},
  \href{https://ui.adsabs.harvard.edu/abs/2024ApJ...968...15L}{968, 15}

\bibitem[{{Livermore} {et~al.}(2015){Livermore}, {Jones}, {Richard}, {Bower},
  {Swinbank}, {Yuan}, {Edge}, {Ellis}, {Kewley}, {Smail}, {Coppin}, \&
  {Ebeling}}]{Livermore2015}
{Livermore}, R.~C., {Jones}, T.~A., {Richard}, J., {et~al.} 2015,
  \href{http://dx.doi.org/10.1093/mnras/stv686}{\color{blue}\mnras},
  \href{https://ui.adsabs.harvard.edu/abs/2015MNRAS.450.1812L}{450, 1812}

\bibitem[{{Lovell} {et~al.}(2018){Lovell}, {Pillepich}, {Genel}, {Nelson},
  {Springel}, {Pakmor}, {Marinacci}, {Weinberger}, {Torrey}, {Vogelsberger},
  {Alabi}, \& {Hernquist}}]{Lovell2018}
{Lovell}, M.~R., {Pillepich}, A., {Genel}, S., {et~al.} 2018,
  \href{http://dx.doi.org/10.1093/mnras/sty2339}{\color{blue}\mnras},
  \href{https://ui.adsabs.harvard.edu/abs/2018MNRAS.481.1950L}{481, 1950}

\bibitem[{{Madau} \& {Dickinson}(2014)}]{MD2014}
{Madau}, P. \& {Dickinson}, M. 2014,
  \href{http://dx.doi.org/10.1146/annurev-astro-081811-125615}{\color{blue}\araa},
  \href{https://ui.adsabs.harvard.edu/abs/2014ARA&A..52..415M}{52, 415}

\bibitem[{{Martinsson} {et~al.}(2013{\natexlab{a}}){Martinsson}, {Verheijen},
  {Westfall}, {Bershady}, {Andersen}, \& {Swaters}}]{Martinsson2013a}
{Martinsson}, T. P.~K., {Verheijen}, M. A.~W., {Westfall}, K.~B., {et~al.}
  2013{\natexlab{a}},
  \href{http://dx.doi.org/10.1051/0004-6361/201321390}{\color{blue}\aap},
  \href{https://ui.adsabs.harvard.edu/abs/2013A&A...557A.131M}{557, A131}

\bibitem[{{Martinsson} {et~al.}(2013{\natexlab{b}}){Martinsson}, {Verheijen},
  {Westfall}, {Bershady}, {Schechtman-Rook}, {Andersen}, \&
  {Swaters}}]{Martinsson2013b}
{Martinsson}, T. P.~K., {Verheijen}, M. A.~W., {Westfall}, K.~B., {et~al.}
  2013{\natexlab{b}},
  \href{http://dx.doi.org/10.1051/0004-6361/201220515}{\color{blue}\aap},
  \href{https://ui.adsabs.harvard.edu/abs/2013A&A...557A.130M}{557, A130}

\bibitem[{{Mason} {et~al.}(2017){Mason}, {Treu}, {Fontana}, {Jones},
  {Morishita}, {Amorin}, {Brada{\v{c}}}, {Quinn Finney}, {Grillo}, {Henry},
  {Hoag}, {Huang}, {Schmidt}, {Trenti}, \& {Vulcani}}]{Mason2017}
{Mason}, C.~A., {Treu}, T., {Fontana}, A., {et~al.} 2017,
  \href{http://dx.doi.org/10.3847/1538-4357/aa60c4}{\color{blue}\apj},
  \href{https://ui.adsabs.harvard.edu/abs/2017ApJ...838...14M}{838, 14}

\bibitem[{{Miller} {et~al.}(2024){Miller}, {Suess}, {Setton}, {Price}, {Labbe},
  {Bezanson}, {Brammer}, {Cutler}, {Furtak}, {Leja}, {Pan}, {Wang}, {Weaver},
  {Whitaker}, {Dayal}, {de Graaff}, {Feldmann}, {Greene}, {Fujimoto}, {Maseda},
  {Nanayakkara}, {Nelson}, {van Dokkum}, \& {Zitrin}}]{Miller2024}
{Miller}, T.~B., {Suess}, K.~A., {Setton}, D.~J., {et~al.} 2024,
  \href{https://ui.adsabs.harvard.edu/abs/2024arXiv241206957M}{\href{http://dx.doi.org/10.48550/arXiv.2412.06957}{\color{blue}arXiv
  e-prints}, arXiv:2412.06957}

\bibitem[{{Mitsuhashi} {et~al.}(2024){Mitsuhashi}, {Tadaki}, {Ikeda},
  {Herrera-Camus}, {Aravena}, {De Looze}, {F{\"o}rster Schreiber},
  {Gonz{\'a}lez-L{\'o}pez}, {Spilker}, {Assef}, {Bouwens}, {Barcos-Munoz},
  {Birkin}, {Bowler}, {Rivera}, {Davies}, {Da Cunha}, {D{\'\i}az-Santos},
  {Ferrara}, {Fisher}, {Lee}, {Li}, {Lutz}, {Rela{\~n}o}, {Naab}, {Palla},
  {Posses}, {Solimano}, {Tacconi}, {{\"U}bler}, {van der Giessen}, \&
  {Veilleux}}]{Mitsuhashi2024b}
{Mitsuhashi}, I., {Tadaki}, K.-i., {Ikeda}, R., {et~al.} 2024,
  \href{http://dx.doi.org/10.1051/0004-6361/202348782}{\color{blue}\aap},
  \href{https://ui.adsabs.harvard.edu/abs/2024A&A...690A.197M}{690, A197}

\bibitem[{{Mizener} {et~al.}(2024){Mizener}, {Pope}, {McKinney}, {Kamieneski},
  {Whitaker}, {Battisti}, \& {Murphy}}]{Mizener2024}
{Mizener}, A., {Pope}, A., {McKinney}, J., {et~al.} 2024,
  \href{http://dx.doi.org/10.3847/1538-4357/ad4965}{\color{blue}\apj},
  \href{https://ui.adsabs.harvard.edu/abs/2024ApJ...970...30M}{970, 30}

\bibitem[{{Mogotsi} {et~al.}(2016){Mogotsi}, {de Blok}, {Cald{\'u}-Primo},
  {Walter}, {Ianjamasimanana}, \& {Leroy}}]{Mogotsi2016}
{Mogotsi}, K.~M., {de Blok}, W.~J.~G., {Cald{\'u}-Primo}, A., {et~al.} 2016,
  \href{http://dx.doi.org/10.3847/0004-6256/151/1/15}{\color{blue}\aj},
  \href{https://ui.adsabs.harvard.edu/abs/2016AJ....151...15M}{151, 15}

\bibitem[{{Moster} {et~al.}(2018){Moster}, {Naab}, \& {White}}]{Moster2018}
{Moster}, B.~P., {Naab}, T., \& {White}, S. D.~M. 2018,
  \href{http://dx.doi.org/10.1093/mnras/sty655}{\color{blue}\mnras},
  \href{https://ui.adsabs.harvard.edu/abs/2018MNRAS.477.1822M}{477, 1822}

\bibitem[{{Nakajima} {et~al.}(2023){Nakajima}, {Ouchi}, {Isobe}, {Harikane},
  {Zhang}, {Ono}, {Umeda}, \& {Oguri}}]{Nakajima2023}
{Nakajima}, K., {Ouchi}, M., {Isobe}, Y., {et~al.} 2023,
  \href{http://dx.doi.org/10.3847/1538-4365/acd556}{\color{blue}\apjs},
  \href{https://ui.adsabs.harvard.edu/abs/2023ApJS..269...33N}{269, 33}

\bibitem[{{Navarro} {et~al.}(1996){Navarro}, {Frenk}, \& {White}}]{Navarro1996}
{Navarro}, J.~F., {Frenk}, C.~S., \& {White}, S. D.~M. 1996,
  \href{http://dx.doi.org/10.1086/177173}{\color{blue}\apj},
  \href{https://ui.adsabs.harvard.edu/abs/1996ApJ...462..563N}{462, 563}

\bibitem[{{Neeleman} {et~al.}(2021){Neeleman}, {Novak}, {Venemans}, {Walter},
  {Decarli}, {Kaasinen}, {Schindler}, {Ba{\~n}ados}, {Carilli}, {Drake}, {Fan},
  \& {Rix}}]{Neeleman2021}
{Neeleman}, M., {Novak}, M., {Venemans}, B.~P., {et~al.} 2021,
  \href{http://dx.doi.org/10.3847/1538-4357/abe70f}{\color{blue}\apj},
  \href{https://ui.adsabs.harvard.edu/abs/2021ApJ...911..141N}{911, 141}

\bibitem[{{Neeleman} {et~al.}(2020){Neeleman}, {Prochaska}, {Kanekar}, \&
  {Rafelski}}]{Neeleman2020}
{Neeleman}, M., {Prochaska}, J.~X., {Kanekar}, N., \& {Rafelski}, M. 2020,
  \href{http://dx.doi.org/10.1038/s41586-020-2276-y}{\color{blue}\nat},
  \href{https://ui.adsabs.harvard.edu/abs/2020Natur.581..269N}{581, 269}

\bibitem[{{Neeleman} {et~al.}(2023){Neeleman}, {Walter}, {Decarli}, {Drake},
  {Eilers}, {Meyer}, \& {Venemans}}]{Neeleman2023}
{Neeleman}, M., {Walter}, F., {Decarli}, R., {et~al.} 2023,
  \href{http://dx.doi.org/10.3847/1538-4357/ad05d2}{\color{blue}\apj},
  \href{https://ui.adsabs.harvard.edu/abs/2023ApJ...958..132N}{958, 132}

\bibitem[{{Nestor Shachar} {et~al.}(2023){Nestor Shachar}, {Price},
  {F{\"o}rster Schreiber}, {Genzel}, {Shimizu}, {Tacconi}, {{\"U}bler},
  {Burkert}, {Davies}, {Dekel}, {Herrera-Camus}, {Lee}, {Liu}, {Lutz}, {Naab},
  {Neri}, {Renzini}, {Saglia}, {Schuster}, {Sternberg}, {Wisnioski}, \&
  {Wuyts}}]{Nestor2023}
{Nestor Shachar}, A., {Price}, S.~H., {F{\"o}rster Schreiber}, N.~M., {et~al.}
  2023, \href{http://dx.doi.org/10.3847/1538-4357/aca9cf}{\color{blue}\apj},
  \href{https://ui.adsabs.harvard.edu/abs/2023ApJ...944...78N}{944, 78}

\bibitem[{{Noordermeer}(2008)}]{Noordermeer2008}
{Noordermeer}, E. 2008,
  \href{http://dx.doi.org/10.1111/j.1365-2966.2008.12837.x}{\color{blue}\mnras},
  \href{https://ui.adsabs.harvard.edu/abs/2008MNRAS.385.1359N}{385, 1359}

\bibitem[{{O'Leary} {et~al.}(2021){O'Leary}, {Moster}, {Naab}, \&
  {Somerville}}]{OLeary2021}
{O'Leary}, J.~A., {Moster}, B.~P., {Naab}, T., \& {Somerville}, R.~S. 2021,
  \href{http://dx.doi.org/10.1093/mnras/staa3746}{\color{blue}\mnras},
  \href{https://ui.adsabs.harvard.edu/abs/2021MNRAS.501.3215O}{501, 3215}

\bibitem[{{Orr} {et~al.}(2020){Orr}, {Hayward}, {Medling}, {Gurvich},
  {Hopkins}, {Murray}, {Pineda}, {Faucher-Gigu{\`e}re}, {Kere{\v{s}}},
  {Wetzel}, \& {Su}}]{Orr2020}
{Orr}, M.~E., {Hayward}, C.~C., {Medling}, A.~M., {et~al.} 2020,
  \href{http://dx.doi.org/10.1093/mnras/staa1619}{\color{blue}\mnras},
  \href{https://ui.adsabs.harvard.edu/abs/2020MNRAS.496.1620O}{496, 1620}

\bibitem[{{Pandya} {et~al.}(2024){Pandya}, {Zhang}, {Huertas-Company}, {Iyer},
  {McGrath}, {Barro}, {Finkelstein}, {K{\"u}mmel}, {Hartley}, {Ferguson},
  {Kartaltepe}, {Primack}, {Dekel}, {Faber}, {Koo}, {Bryan}, {Somerville},
  {Amor{\'\i}n}, {Arrabal Haro}, {Bagley}, {Bell}, {Bertin}, {Costantin},
  {Dav{\'e}}, {Dickinson}, {Feldmann}, {Fontana}, {Gavazzi}, {Giavalisco},
  {Grazian}, {Grogin}, {Guo}, {Hahn}, {Holwerda}, {Kewley}, {Kirkpatrick},
  {Kocevski}, {Koekemoer}, {Lotz}, {Lucas}, {Papovich}, {Pentericci},
  {P{\'e}rez-Gonz{\'a}lez}, {Pirzkal}, {Ravindranath}, {Rose}, {Schefer},
  {Simons}, {Straughn}, {Tacchella}, {Trump}, {de la Vega}, {Wilkins}, {Wuyts},
  {Yang}, \& {Yung}}]{Pandya2024}
{Pandya}, V., {Zhang}, H., {Huertas-Company}, M., {et~al.} 2024,
  \href{http://dx.doi.org/10.3847/1538-4357/ad1a13}{\color{blue}\apj},
  \href{https://ui.adsabs.harvard.edu/abs/2024ApJ...963...54P}{963, 54}

\bibitem[{{Parlanti} {et~al.}(2023){Parlanti}, {Carniani}, {Pallottini},
  {Cignoni}, {Cresci}, {Kohandel}, {Mannucci}, \& {Marconi}}]{Parlanti2023}
{Parlanti}, E., {Carniani}, S., {Pallottini}, A., {et~al.} 2023,
  \href{http://dx.doi.org/10.1051/0004-6361/202245603}{\color{blue}\aap},
  \href{https://ui.adsabs.harvard.edu/abs/2023A&A...673A.153P}{673, A153}

\bibitem[{{Parlanti} {et~al.}(2024){Parlanti}, {Carniani}, {{\"U}bler},
  {Venturi}, {Circosta}, {D'Eugenio}, {Arribas}, {Bunker}, {Charlot},
  {L{\"u}tzgendorf}, {Maiolino}, {Perna}, {Rodr{\'\i}guez Del Pino}, {Willott},
  {B{\"o}ker}, {Cameron}, {Chevallard}, {Cresci}, {Jones}, {Kumari},
  {Lamperti}, \& {Scholtz}}]{Parlanti2024a}
{Parlanti}, E., {Carniani}, S., {{\"U}bler}, H., {et~al.} 2024,
  \href{http://dx.doi.org/10.1051/0004-6361/202347914}{\color{blue}\aap},
  \href{https://ui.adsabs.harvard.edu/abs/2024A&A...684A..24P}{684, A24}

\bibitem[{{Parlanti} {et~al.}(2025){Parlanti}, {Carniani}, {Venturi},
  {Herrera-Camus}, {Arribas}, {Bunker}, {Charlot}, {D'Eugenio}, {Maiolino},
  {Perna}, {{\"U}bler}, {B{\"o}ker}, {Cresci}, {Curti}, {Jones}, {Lamperti},
  {P{\'e}rez-Gonz{\'a}lez}, {Del Pino}, \& {Zamora}}]{Parlanti2025}
{Parlanti}, E., {Carniani}, S., {Venturi}, G., {et~al.} 2025,
  \href{http://dx.doi.org/10.1051/0004-6361/202451692}{\color{blue}\aap},
  \href{https://ui.adsabs.harvard.edu/abs/2025A&A...695A...6P}{695, A6}

\bibitem[{{Patr{\'\i}cio} {et~al.}(2018){Patr{\'\i}cio}, {Richard}, {Carton},
  {Contini}, {Epinat}, {Brinchmann}, {Schmidt}, {Krajnovi{\'c}}, {Bouch{\'e}},
  {Weilbacher}, {Pell{\'o}}, {Caruana}, {Maseda}, {Finley}, {Bauer},
  {Martinez}, {Mahler}, {Lagattuta}, {Cl{\'e}ment}, {Soucail}, \&
  {Wisotzki}}]{Patricio2018}
{Patr{\'\i}cio}, V., {Richard}, J., {Carton}, D., {et~al.} 2018,
  \href{http://dx.doi.org/10.1093/mnras/sty555}{\color{blue}\mnras},
  \href{https://ui.adsabs.harvard.edu/abs/2018MNRAS.477...18P}{477, 18}

\bibitem[{{Pavesi} {et~al.}(2019){Pavesi}, {Riechers}, {Faisst}, {Stacey}, \&
  {Capak}}]{Pavesi2019}
{Pavesi}, R., {Riechers}, D.~A., {Faisst}, A.~L., {Stacey}, G.~J., \& {Capak},
  P.~L. 2019,
  \href{http://dx.doi.org/10.3847/1538-4357/ab3a46}{\color{blue}\apj},
  \href{https://ui.adsabs.harvard.edu/abs/2019ApJ...882..168P}{882, 168}

\bibitem[{{Perna} {et~al.}(2025){Perna}, {Arribas}, {Ji}, {Marconcini},
  {Lamperti}, {Bertola}, {Circosta}, {D'Eugenio}, {{\"U}bler}, {B{\"o}ker},
  {Maiolino}, {Bunker}, {Carniani}, {Charlot}, {Willott}, {Cresci}, {Marconi},
  {Parlanti}, {Rodr{\'\i}guez Del Pino}, {Scholtz}, \& {Venturi}}]{Perna2025}
{Perna}, M., {Arribas}, S., {Ji}, X., {et~al.} 2025,
  \href{http://dx.doi.org/10.1051/0004-6361/202453090}{\color{blue}\aap},
  \href{https://ui.adsabs.harvard.edu/abs/2025A&A...694A.170P}{694, A170}

\bibitem[{{Peschken} {et~al.}(2020){Peschken}, {{\L}okas}, \&
  {Athanassoula}}]{Peschken2020}
{Peschken}, N., {{\L}okas}, E.~L., \& {Athanassoula}, E. 2020,
  \href{http://dx.doi.org/10.1093/mnras/staa299}{\color{blue}\mnras},
  \href{https://ui.adsabs.harvard.edu/abs/2020MNRAS.493.1375P}{493, 1375}

\bibitem[{{Pillepich} {et~al.}(2019){Pillepich}, {Nelson}, {Springel},
  {Pakmor}, {Torrey}, {Weinberger}, {Vogelsberger}, {Marinacci}, {Genel}, {van
  der Wel}, \& {Hernquist}}]{Pillepich2019}
{Pillepich}, A., {Nelson}, D., {Springel}, V., {et~al.} 2019,
  \href{http://dx.doi.org/10.1093/mnras/stz2338}{\color{blue}\mnras},
  \href{https://ui.adsabs.harvard.edu/abs/2019MNRAS.490.3196P}{490, 3196}

\bibitem[{{Planck Collaboration} {et~al.}(2011){Planck Collaboration},
  {Abergel}, {Ade}, {Aghanim}, {Arnaud}, {Ashdown}, {Aumont}, {Baccigalupi},
  {Balbi}, {Banday}, {Barreiro}, {Bartlett}, {Battaner}, {Benabed},
  {Beno{\^\i}t}, {Bernard}, {Bersanelli}, {Bhatia}, {Bock}, {Bonaldi}, {Bond},
  {Borrill}, {Bouchet}, {Boulanger}, {Bucher}, {Burigana}, {Cabella},
  {Cardoso}, {Catalano}, {Cay{\'o}n}, {Challinor}, {Chamballu}, {Chiang},
  {Chiang}, {Christensen}, {Colombi}, {Couchot}, {Coulais}, {Crill}, {Cuttaia},
  {Dame}, {Danese}, {Davies}, {Davis}, {de Bernardis}, {de Gasperis}, {de
  Rosa}, {de Zotti}, {Delabrouille}, {Delouis}, {D{\'e}sert}, {Dickinson},
  {Donzelli}, {Dor{\'e}}, {D{\"o}rl}, {Douspis}, {Dupac}, {Efstathiou},
  {En{\ss}lin}, {Finelli}, {Forni}, {Frailis}, {Franceschi}, {Galeotta},
  {Ganga}, {Giard}, {Giardino}, {Giraud-H{\'e}raud}, {Gonz{\'a}lez-Nuevo},
  {G{\'o}rski}, {Gratton}, {Gregorio}, {Grenier}, {Gruppuso}, {Hansen},
  {Harrison}, {Henrot-Versill{\'e}}, {Herranz}, {Hildebrandt}, {Hivon},
  {Hobson}, {Holmes}, {Hovest}, {Hoyland}, {Huffenberger}, {Jaffe}, {Jaffe},
  {Jones}, {Juvela}, {Keih{\"a}nen}, {Keskitalo}, {Kisner}, {Kneissl}, {Knox},
  {Kurki-Suonio}, {Lagache}, {L{\"a}hteenm{\"a}ki}, {Lamarre}, {Lasenby},
  {Laureijs}, {Lawrence}, {Leach}, {Leonardi}, {Leroy}, {Lilje},
  {Linden-V{\o}rnle}, {L{\'o}pez-Caniego}, {Lubin}, {Mac{\'\i}as-P{\'e}rez},
  {MacTavish}, {Maffei}, {Mandolesi}, {Mann}, {Maris}, {Marshall},
  {Mart{\'\i}nez-Gonz{\'a}lez}, {Masi}, {Matarrese}, {Matthai}, {Mazzotta},
  {McGehee}, {Meinhold}, {Melchiorri}, {Mendes}, {Mennella},
  {Miville-Desch{\^e}nes}, {Moneti}, {Montier}, {Morgante}, {Mortlock},
  {Munshi}, {Murphy}, {Naselsky}, {Natoli}, {Netterfield},
  {N{\o}rgaard-Nielsen}, {Noviello}, {Novikov}, {Novikov}, {Osborne}, {Pajot},
  {Paladini}, {Pasian}, {Patanchon}, {Perdereau}, {Perotto}, {Perrotta},
  {Piacentini}, {Piat}, {Plaszczynski}, {Pointecouteau}, {Polenta}, {Ponthieu},
  {Poutanen}, {Pr{\'e}zeau}, {Prunet}, {Puget}, {Rachen}, {Reach}, {Rebolo},
  {Reich}, {Renault}, {Ricciardi}, {Riller}, {Ristorcelli}, {Rocha}, {Rosset},
  {Rubi{\~n}o-Mart{\'\i}n}, {Rusholme}, {Sandri}, {Santos}, {Savini}, {Scott},
  {Seiffert}, {Shellard}, {Smoot}, {Starck}, {Stivoli}, {Stolyarov}, {Stompor},
  {Sudiwala}, {Sygnet}, {Tauber}, {Terenzi}, {Toffolatti}, {Tomasi}, {Torre},
  {Tristram}, {Tuovinen}, {Umana}, {Valenziano}, {Varis}, {Vielva}, {Villa},
  {Vittorio}, {Wade}, {Wandelt}, {Wilkinson}, {Ysard}, {Yvon}, {Zacchei}, \&
  {Zonca}}]{Planck2011}
{Planck Collaboration}, {Abergel}, A., {Ade}, P.~A.~R., {et~al.} 2011,
  \href{http://dx.doi.org/10.1051/0004-6361/201116455}{\color{blue}\aap},
  \href{https://ui.adsabs.harvard.edu/abs/2011A&A...536A..21P}{536, A21}

\bibitem[{{Pope} {et~al.}(2023){Pope}, {McKinney}, {Kamieneski}, {Battisti},
  {Aretxaga}, {Brammer}, {Diego}, {Hughes}, {Keller}, {Marchesini}, {Mizener},
  {Monta{\~n}a}, {Murphy}, {Whitaker}, {Wilson}, \& {Yun}}]{Pope2023}
{Pope}, A., {McKinney}, J., {Kamieneski}, P., {et~al.} 2023,
  \href{http://dx.doi.org/10.3847/2041-8213/acdf5a}{\color{blue}\apjl},
  \href{https://ui.adsabs.harvard.edu/abs/2023ApJ...951L..46P}{951, L46}

\bibitem[{{Posses} {et~al.}(2025){Posses}, {Aravena}, {Gonz{\'a}lez-L{\'o}pez},
  {F{\"o}rster Schreiber}, {Liu}, {Lee}, {Solimano}, {D{\'\i}az-Santos},
  {Assef}, {Barcos-Mu{\~n}oz}, {Bovino}, {Bowler}, {Calistro Rivera}, {da
  Cunha}, {Davies}, {Killi}, {De Looze}, {Ferrara}, {Fisher}, {Herrera-Camus},
  {Ikeda}, {Lambert}, {Li}, {Lutz}, {Mitsuhashi}, {Palla}, {Rela{\~n}o},
  {Spilker}, {Naab}, {Tadaki}, {Telikova}, {{\"U}bler}, {van der Giessen}, \&
  {Villanueva}}]{Posses2025}
{Posses}, A., {Aravena}, M., {Gonz{\'a}lez-L{\'o}pez}, J., {et~al.} 2025,
  \href{https://doi.org/10.1051/0004-6361/202449843}{\href{http://dx.doi.org/10.1051/0004-6361/202449843}{\color{blue}\aap},
  in press}

\bibitem[{{Posses} {et~al.}(2023){Posses}, {Aravena}, {Gonz{\'a}lez-L{\'o}pez},
  {Assef}, {Lambert}, {Jones}, {Bouwens}, {Brisbin}, {D{\'\i}az-Santos},
  {Herrera-Camus}, {Ricci}, \& {Smit}}]{Posses2023}
{Posses}, A.~C., {Aravena}, M., {Gonz{\'a}lez-L{\'o}pez}, J., {et~al.} 2023,
  \href{http://dx.doi.org/10.1051/0004-6361/202243399}{\color{blue}\aap},
  \href{https://ui.adsabs.harvard.edu/abs/2023A&A...669A..46P}{669, A46}

\bibitem[{{Price} {et~al.}(2020){Price}, {Kriek}, {Barro}, {Shapley}, {Reddy},
  {Freeman}, {Coil}, {Shivaei}, {Azadi}, {de Groot}, {Siana}, {Mobasher},
  {Sanders}, {Leung}, {Fetherolf}, {Zick}, {{\"U}bler}, \& {F{\"o}rster
  Schreiber}}]{Price2020}
{Price}, S.~H., {Kriek}, M., {Barro}, G., {et~al.} 2020,
  \href{http://dx.doi.org/10.3847/1538-4357/ab7990}{\color{blue}\apj},
  \href{https://ui.adsabs.harvard.edu/abs/2020ApJ...894...91P}{894, 91}

\bibitem[{{Price} {et~al.}(2021){Price}, {Shimizu}, {Genzel}, {{\"U}bler},
  {F{\"o}rster Schreiber}, {Tacconi}, {Davies}, {Coogan}, {Lutz}, {Wuyts},
  {Wisnioski}, {Nestor}, {Sternberg}, {Burkert}, {Bender}, {Contursi},
  {Davies}, {Herrera-Camus}, {Lee}, {Naab}, {Neri}, {Renzini}, {Saglia},
  {Schruba}, \& {Schuster}}]{Price2021}
{Price}, S.~H., {Shimizu}, T.~T., {Genzel}, R., {et~al.} 2021,
  \href{http://dx.doi.org/10.3847/1538-4357/ac22ad}{\color{blue}\apj},
  \href{https://ui.adsabs.harvard.edu/abs/2021ApJ...922..143P}{922, 143}

\bibitem[{{Price} {et~al.}(2022){Price}, {{\"U}bler}, {F{\"o}rster Schreiber},
  {de Zeeuw}, {Burkert}, {Genzel}, {Tacconi}, {Davies}, \& {Price}}]{Price2022}
{Price}, S.~H., {{\"U}bler}, H., {F{\"o}rster Schreiber}, N.~M., {et~al.} 2022,
  \href{http://dx.doi.org/10.1051/0004-6361/202244143}{\color{blue}\aap},
  \href{https://ui.adsabs.harvard.edu/abs/2022A&A...665A.159P}{665, A159}

\bibitem[{{Privon} {et~al.}(2020){Privon}, {Ricci}, {Aalto}, {Viti}, {Armus},
  {D{\'\i}az-Santos}, {Gonz{\'a}lez-Alfonso}, {Iwasawa}, {Jeff}, {Treister},
  {Bauer}, {Evans}, {Garg}, {Herrero-Illana}, {Mazzarella}, {Larson}, {Blecha},
  {Barcos-Mu{\~n}oz}, {Charmandaris}, {Stierwalt}, \&
  {P{\'e}rez-Torres}}]{Privon2020}
{Privon}, G.~C., {Ricci}, C., {Aalto}, S., {et~al.} 2020,
  \href{http://dx.doi.org/10.3847/1538-4357/ab8015}{\color{blue}\apj},
  \href{https://ui.adsabs.harvard.edu/abs/2020ApJ...893..149P}{893, 149}

\bibitem[{{Puglisi} {et~al.}(2023){Puglisi}, {Dudzevi{\v{c}}i{\={u}}t{\.{e}}},
  {Swinbank}, {Gillman}, {Tiley}, {Bower}, {Cirasuolo}, {Cortese},
  {Glazebrook}, {Harrison}, {Ibar}, {Molina}, {Obreschkow}, {Oman}, {Schaller},
  {Shankar}, \& {Sharples}}]{Puglisi2023}
{Puglisi}, A., {Dudzevi{\v{c}}i{\={u}}t{\.{e}}}, U., {Swinbank}, M., {et~al.}
  2023, \href{http://dx.doi.org/10.1093/mnras/stad1966}{\color{blue}\mnras},
  \href{https://ui.adsabs.harvard.edu/abs/2023MNRAS.524.2814P}{524, 2814}

\bibitem[{{Pusk{\'a}s} {et~al.}(2025){Pusk{\'a}s}, {Tacchella}, {Simmonds},
  {Hainline}, {D'Eugenio}, {Alberts}, {Arribas}, {Baker}, {Bunker}, {Carniani},
  {Charlot}, {Duan}, {Eisenstein}, {Ji}, {Johnson}, {Jones}, {Maiolino},
  {McClymont}, {Rieke}, {Rinaldi}, {Robertson}, {{\"U}bler}, {Williams},
  {Willmer}, {Willott}, \& {Witstok}}]{Puskas2025}
{Pusk{\'a}s}, D., {Tacchella}, S., {Simmonds}, C., {et~al.} 2025,
  \href{https://ui.adsabs.harvard.edu/abs/2025arXiv250201721P}{\href{http://dx.doi.org/10.48550/arXiv.2502.01721}{\color{blue}arXiv
  e-prints}, arXiv:2502.01721}

\bibitem[{{Rathjen} {et~al.}(2023){Rathjen}, {Naab}, {Walch}, {Seifried},
  {Girichidis}, \& {W{\"u}nsch}}]{Rathjen2023}
{Rathjen}, T.-E., {Naab}, T., {Walch}, S., {et~al.} 2023,
  \href{http://dx.doi.org/10.1093/mnras/stad1104}{\color{blue}\mnras},
  \href{https://ui.adsabs.harvard.edu/abs/2023MNRAS.522.1843R}{522, 1843}

\bibitem[{{Rhoades} {et~al.}(2025){Rhoades}, {Jones}, {Keerthi Vasan G.},
  {Chen}, {Leethochawalit}, {Ellis}, {Shajib}, {Glazebrook}, {Mortensen}, \&
  {Sanders}}]{Rhoades2025}
{Rhoades}, S., {Jones}, T., {Keerthi Vasan G.}, C., {et~al.} 2025,
  \href{https://ui.adsabs.harvard.edu/abs/2025arXiv250322039R}{\href{http://dx.doi.org/10.48550/arXiv.2503.22039}{\color{blue}arXiv
  e-prints}, arXiv:2503.22039}

\bibitem[{{Riechers} {et~al.}(2014){Riechers}, {Carilli}, {Capak}, {Scoville},
  {Smol{\v{c}}i{\'c}}, {Schinnerer}, {Yun}, {Cox}, {Bertoldi}, {Karim}, \&
  {Yan}}]{Riechers2014}
{Riechers}, D.~A., {Carilli}, C.~L., {Capak}, P.~L., {et~al.} 2014,
  \href{http://dx.doi.org/10.1088/0004-637X/796/2/84}{\color{blue}\apj},
  \href{https://ui.adsabs.harvard.edu/abs/2014ApJ...796...84R}{796, 84}

\bibitem[{{Rizzo} {et~al.}(2023){Rizzo}, {Roman-Oliveira}, {Fraternali},
  {Frickmann}, {Valentino}, {Brammer}, {Zanella}, {Kokorev}, {Popping},
  {Whitaker}, {Kohandel}, {Magdis}, {Di Mascolo}, {Ikeda}, {Jin}, \&
  {Toft}}]{Rizzo2023}
{Rizzo}, F., {Roman-Oliveira}, F., {Fraternali}, F., {et~al.} 2023,
  \href{http://dx.doi.org/10.1051/0004-6361/202346444}{\color{blue}\aap},
  \href{https://ui.adsabs.harvard.edu/abs/2023A&A...679A.129R}{679, A129}

\bibitem[{{Rizzo} {et~al.}(2021){Rizzo}, {Vegetti}, {Fraternali}, {Stacey}, \&
  {Powell}}]{Rizzo2021}
{Rizzo}, F., {Vegetti}, S., {Fraternali}, F., {Stacey}, H.~R., \& {Powell}, D.
  2021, \href{http://dx.doi.org/10.1093/mnras/stab2295}{\color{blue}\mnras},
  \href{https://ui.adsabs.harvard.edu/abs/2021MNRAS.507.3952R}{507, 3952}

\bibitem[{{Rizzo} {et~al.}(2020){Rizzo}, {Vegetti}, {Powell}, {Fraternali},
  {McKean}, {Stacey}, \& {White}}]{Rizzo2020}
{Rizzo}, F., {Vegetti}, S., {Powell}, D., {et~al.} 2020,
  \href{http://dx.doi.org/10.1038/s41586-020-2572-6}{\color{blue}\nat},
  \href{https://ui.adsabs.harvard.edu/abs/2020Natur.584..201R}{584, 201}

\bibitem[{{Robertson} {et~al.}(2006){Robertson}, {Bullock}, {Cox}, {Di Matteo},
  {Hernquist}, {Springel}, \& {Yoshida}}]{Robertson2006}
{Robertson}, B., {Bullock}, J.~S., {Cox}, T.~J., {et~al.} 2006,
  \href{http://dx.doi.org/10.1086/504412}{\color{blue}\apj},
  \href{https://ui.adsabs.harvard.edu/abs/2006ApJ...645..986R}{645, 986}

\bibitem[{{Rodriguez-Gomez} {et~al.}(2015){Rodriguez-Gomez}, {Genel},
  {Vogelsberger}, {Sijacki}, {Pillepich}, {Sales}, {Torrey}, {Snyder},
  {Nelson}, {Springel}, {Ma}, \& {Hernquist}}]{RodriguezGomez2015}
{Rodriguez-Gomez}, V., {Genel}, S., {Vogelsberger}, M., {et~al.} 2015,
  \href{http://dx.doi.org/10.1093/mnras/stv264}{\color{blue}\mnras},
  \href{https://ui.adsabs.harvard.edu/abs/2015MNRAS.449...49R}{449, 49}

\bibitem[{{Roman-Oliveira} {et~al.}(2023){Roman-Oliveira}, {Fraternali}, \&
  {Rizzo}}]{RomanOliveira2023}
{Roman-Oliveira}, F., {Fraternali}, F., \& {Rizzo}, F. 2023,
  \href{http://dx.doi.org/10.1093/mnras/stad530}{\color{blue}\mnras},
  \href{https://ui.adsabs.harvard.edu/abs/2023MNRAS.521.1045R}{521, 1045}

\bibitem[{{Romano} {et~al.}(2021){Romano}, {Cassata}, {Morselli}, {Jones},
  {Ginolfi}, {Zanella}, {B{\'e}thermin}, {Capak}, {Faisst}, {Le F{\`e}vre},
  {Schaerer}, {Silverman}, {Yan}, {Bardelli}, {Boquien}, {Cimatti},
  {Dessauges-Zavadsky}, {Enia}, {Fujimoto}, {Gruppioni}, {Hathi}, {Ibar},
  {Koekemoer}, {Lemaux}, {Rodighiero}, {Vergani}, {Zamorani}, \&
  {Zucca}}]{Romano2021}
{Romano}, M., {Cassata}, P., {Morselli}, L., {et~al.} 2021,
  \href{http://dx.doi.org/10.1051/0004-6361/202141306}{\color{blue}\aap},
  \href{https://ui.adsabs.harvard.edu/abs/2021A&A...653A.111R}{653, A111}

\bibitem[{{Rowland} {et~al.}(2024){Rowland}, {Hodge}, {Bouwens}, {Pi{\~n}a},
  {Hygate}, {Algera}, {Aravena}, {Bowler}, {da Cunha}, {Dayal}, {Ferrara},
  {Herard-Demanche}, {Inami}, {van Leeuwen}, {de Looze}, {Oesch}, {Pallottini},
  {Phillips}, {Rybak}, {Schouws}, {Smit}, {Sommovigo}, {Stefanon}, \& {van der
  Werf}}]{Rowland2024}
{Rowland}, L.~E., {Hodge}, J., {Bouwens}, R., {et~al.} 2024,
  \href{http://dx.doi.org/10.1093/mnras/stae2217}{\color{blue}\mnras},
  \href{https://ui.adsabs.harvard.edu/abs/2024MNRAS.535.2068R}{535, 2068}

\bibitem[{{Scoville} {et~al.}(2017){Scoville}, {Lee}, {Vanden Bout},
  {Diaz-Santos}, {Sanders}, {Darvish}, {Bongiorno}, {Casey}, {Murchikova},
  {Koda}, {Capak}, {Vlahakis}, {Ilbert}, {Sheth}, {Morokuma-Matsui}, {Ivison},
  {Aussel}, {Laigle}, {McCracken}, {Armus}, {Pope}, {Toft}, \&
  {Masters}}]{Scoville2017}
{Scoville}, N., {Lee}, N., {Vanden Bout}, P., {et~al.} 2017,
  \href{http://dx.doi.org/10.3847/1538-4357/aa61a0}{\color{blue}\apj},
  \href{https://ui.adsabs.harvard.edu/abs/2017ApJ...837..150S}{837, 150}

\bibitem[{{Scoville} {et~al.}(2016){Scoville}, {Sheth}, {Aussel}, {Vanden
  Bout}, {Capak}, {Bongiorno}, {Casey}, {Murchikova}, {Koda},
  {{\'A}lvarez-M{\'a}rquez}, {Lee}, {Laigle}, {McCracken}, {Ilbert}, {Pope},
  {Sanders}, {Chu}, {Toft}, {Ivison}, \& {Manohar}}]{Scoville2016}
{Scoville}, N., {Sheth}, K., {Aussel}, H., {et~al.} 2016,
  \href{http://dx.doi.org/10.3847/0004-637X/820/2/83}{\color{blue}\apj},
  \href{https://ui.adsabs.harvard.edu/abs/2016ApJ...820...83S}{820, 83}

\bibitem[{{Sedov}(1946)}]{Sedov1946}
{Sedov}, L.~I. 1946, Journal of Applied Mathematics and Mechanics,
  \href{https://ui.adsabs.harvard.edu/abs/1946JApMM..10..241S}{10, 241}

\bibitem[{{Sedov}(1959)}]{Sedov1959}
{Sedov}, L.~I. 1959, {Similarity and Dimensional Methods in Mechanics}

\bibitem[{{S\'ersic}(1968)}]{Sersic1968}
{S\'ersic}, J.~L. 1968, {Atlas de Galaxias Australes} ({Observatorio
  Astronomico, Universidad Nacional de Cordoba})

\bibitem[{{Shao} {et~al.}(2022){Shao}, {Wang}, {Weiss}, {Wagg}, {Carilli},
  {Strauss}, {Walter}, {Cox}, {Fan}, {Menten}, {Narayanan}, {Riechers},
  {Bertoldi}, {Omont}, \& {Jiang}}]{Shao2022}
{Shao}, Y., {Wang}, R., {Weiss}, A., {et~al.} 2022,
  \href{http://dx.doi.org/10.1051/0004-6361/202244610}{\color{blue}\aap},
  \href{https://ui.adsabs.harvard.edu/abs/2022A&A...668A.121S}{668, A121}

\bibitem[{{Shapiro} {et~al.}(2008){Shapiro}, {Genzel}, {F{\"o}rster Schreiber},
  {Tacconi}, {Bouch{\'e}}, {Cresci}, {Davies}, {Eisenhauer}, {Johansson},
  {Krajnovi{\'c}}, {Lutz}, {Naab}, {Arimoto}, {Arribas}, {Cimatti}, {Colina},
  {Daddi}, {Daigle}, {Erb}, {Hernandez}, {Kong}, {Mignoli}, {Onodera},
  {Renzini}, {Shapley}, \& {Steidel}}]{Shapiro2008}
{Shapiro}, K.~L., {Genzel}, R., {F{\"o}rster Schreiber}, N.~M., {et~al.} 2008,
  \href{http://dx.doi.org/10.1086/587133}{\color{blue}\apj},
  \href{https://ui.adsabs.harvard.edu/abs/2008ApJ...682..231S}{682, 231}

\bibitem[{{Sharda} {et~al.}(2019){Sharda}, {da Cunha}, {Federrath},
  {Wisnioski}, {Di Teodoro}, {Tadaki}, {Yun}, {Aretxaga}, \&
  {Kawabe}}]{Sharda2019}
{Sharda}, P., {da Cunha}, E., {Federrath}, C., {et~al.} 2019,
  \href{http://dx.doi.org/10.1093/mnras/stz1543}{\color{blue}\mnras},
  \href{https://ui.adsabs.harvard.edu/abs/2019MNRAS.487.4305S}{487, 4305}

\bibitem[{{Sharma} {et~al.}(2021){Sharma}, {Salucci}, {Harrison}, {van de Ven},
  \& {Lapi}}]{Sharma2021}
{Sharma}, G., {Salucci}, P., {Harrison}, C.~M., {van de Ven}, G., \& {Lapi}, A.
  2021, \href{http://dx.doi.org/10.1093/mnras/stab249}{\color{blue}\mnras},
  \href{https://ui.adsabs.harvard.edu/abs/2021MNRAS.503.1753S}{503, 1753}

\bibitem[{{Sharples} {et~al.}(2013){Sharples}, {Bender}, {Agudo Berbel},
  {Bezawada}, {Castillo}, {Cirasuolo}, {Davidson}, {Davies}, {Dubbeldam},
  {Fairley}, {Finger}, {F{\"o}rster Schreiber}, {Gonte}, {Hess}, {Jung},
  {Lewis}, {Lizon}, {Muschielok}, {Pasquini}, {Pirard}, {Popovic}, {Ramsay},
  {Rees}, {Richter}, {Riquelme}, {Rodrigues}, {Saviane}, {Schlichter},
  {Schmidtobreick}, {Segovia}, {Smette}, {Szeifert}, {van Kesteren}, {Wegner},
  \& {Wiezorrek}}]{Sharples2013}
{Sharples}, R., {Bender}, R., {Agudo Berbel}, A., {et~al.} 2013, The Messenger,
  \href{https://ui.adsabs.harvard.edu/abs/2013Msngr.151...21S}{151, 21}

\bibitem[{{Shetty} \& {Ostriker}(2012)}]{Shetty2012}
{Shetty}, R. \& {Ostriker}, E.~C. 2012,
  \href{http://dx.doi.org/10.1088/0004-637X/754/1/2}{\color{blue}\apj},
  \href{https://ui.adsabs.harvard.edu/abs/2012ApJ...754....2S}{754, 2}

\bibitem[{{Shibuya} {et~al.}(2025){Shibuya}, {Ito}, {Asai}, {Kirihara},
  {Fujimoto}, {Toba}, {Miura}, {Umayahara}, {Iwadate}, {Ali}, \&
  {Kodama}}]{Shibuya2025}
{Shibuya}, T., {Ito}, Y., {Asai}, K., {et~al.} 2025,
  \href{http://dx.doi.org/10.1093/pasj/psae096}{\color{blue}\pasj},
  \href{https://ui.adsabs.harvard.edu/abs/2025PASJ...77...21S}{77, 21}

\bibitem[{{Smit} {et~al.}(2018){Smit}, {Bouwens}, {Carniani}, {Oesch},
  {Labb{\'e}}, {Illingworth}, {van der Werf}, {Bradley}, {Gonzalez}, {Hodge},
  {Holwerda}, {Maiolino}, \& {Zheng}}]{Smit2018}
{Smit}, R., {Bouwens}, R.~J., {Carniani}, S., {et~al.} 2018,
  \href{http://dx.doi.org/10.1038/nature24631}{\color{blue}\nat},
  \href{https://ui.adsabs.harvard.edu/abs/2018Natur.553..178S}{553, 178}

\bibitem[{{Solimano} {et~al.}(2025){Solimano}, {Gonz{\'a}lez-L{\'o}pez},
  {Aravena}, {Alcalde Pampliega}, {Assef}, {B{\'e}thermin}, {Boquien},
  {Bovino}, {Casey}, {Cassata}, {da Cunha}, {Davies}, {De Looze}, {Ding},
  {D{\'\i}az-Santos}, {Faisst}, {Ferrara}, {Fisher}, {F{\"o}rster-Schreiber},
  {Fujimoto}, {Ginolfi}, {Gruppioni}, {Guaita}, {Hathi}, {Herrera-Camus},
  {Ibar}, {Inami}, {Jones}, {Koekemoer}, {Lee}, {Li}, {Liu}, {Liu}, {Molina},
  {Ogle}, {Posses}, {Pozzi}, {Rela{\~n}o}, {Riechers}, {Romano}, {Spilker},
  {Sulzenauer}, {Telikova}, {Vallini}, {Vasan}, {Veilleux}, {Vergani},
  {Villanueva}, {Wang}, {Yan}, \& {Zamorani}}]{Solimano2025}
{Solimano}, M., {Gonz{\'a}lez-L{\'o}pez}, J., {Aravena}, M., {et~al.} 2025,
  \href{http://dx.doi.org/10.1051/0004-6361/202451551}{\color{blue}\aap},
  \href{https://ui.adsabs.harvard.edu/abs/2025A&A...693A..70S}{693, A70}

\bibitem[{{Sommovigo} {et~al.}(2022){Sommovigo}, {Ferrara}, {Carniani},
  {Pallottini}, {Dayal}, {Pizzati}, {Ginolfi}, {Markov}, \&
  {Faisst}}]{Sommovigo2022}
{Sommovigo}, L., {Ferrara}, A., {Carniani}, S., {et~al.} 2022,
  \href{http://dx.doi.org/10.1093/mnras/stac2997}{\color{blue}\mnras},
  \href{https://ui.adsabs.harvard.edu/abs/2022MNRAS.517.5930S}{517, 5930}

\bibitem[{{Sotillo-Ramos} {et~al.}(2022){Sotillo-Ramos}, {Pillepich},
  {Donnari}, {Nelson}, {Eisert}, {Rodriguez-Gomez}, {Joshi}, {Vogelsberger}, \&
  {Hernquist}}]{SotilloRamos2022}
{Sotillo-Ramos}, D., {Pillepich}, A., {Donnari}, M., {et~al.} 2022,
  \href{http://dx.doi.org/10.1093/mnras/stac2586}{\color{blue}\mnras},
  \href{https://ui.adsabs.harvard.edu/abs/2022MNRAS.516.5404S}{516, 5404}

\bibitem[{{Speagle} {et~al.}(2014){Speagle}, {Steinhardt}, {Capak}, \&
  {Silverman}}]{Speagle2014}
{Speagle}, J.~S., {Steinhardt}, C.~L., {Capak}, P.~L., \& {Silverman}, J.~D.
  2014, \href{http://dx.doi.org/10.1088/0067-0049/214/2/15}{\color{blue}\apjs},
  \href{https://ui.adsabs.harvard.edu/abs/2014ApJS..214...15S}{214, 15}

\bibitem[{{Springel} \& {Hernquist}(2005)}]{SpringelHernquist2005}
{Springel}, V. \& {Hernquist}, L. 2005,
  \href{http://dx.doi.org/10.1086/429486}{\color{blue}\apjl},
  \href{https://ui.adsabs.harvard.edu/abs/2005ApJ...622L...9S}{622, L9}

\bibitem[{{Stach} {et~al.}(2018){Stach}, {Smail}, {Swinbank}, {Simpson},
  {Geach}, {An}, {Almaini}, {Arumugam}, {Blain}, {Chapman}, {Chen},
  {Conselice}, {Cooke}, {Coppin}, {Dunlop}, {Farrah}, {Gullberg}, {Hartley},
  {Ivison}, {Maltby}, {Micha{\l}owski}, {Scott}, {Simpson}, {Thomson},
  {Wardlow}, \& {van der Werf}}]{Stach2018}
{Stach}, S.~M., {Smail}, I., {Swinbank}, A.~M., {et~al.} 2018,
  \href{http://dx.doi.org/10.3847/1538-4357/aac5e5}{\color{blue}\apj},
  \href{https://ui.adsabs.harvard.edu/abs/2018ApJ...860..161S}{860, 161}

\bibitem[{{Stark} {et~al.}(2008){Stark}, {Swinbank}, {Ellis}, {Dye}, {Smail},
  \& {Richard}}]{Stark2008}
{Stark}, D.~P., {Swinbank}, A.~M., {Ellis}, R.~S., {et~al.} 2008,
  \href{http://dx.doi.org/10.1038/nature07294}{\color{blue}\nat},
  \href{https://ui.adsabs.harvard.edu/abs/2008Natur.455..775S}{455, 775}

\bibitem[{{Stefanon} {et~al.}(2017){Stefanon}, {Yan}, {Mobasher}, {Barro},
  {Donley}, {Fontana}, {Hemmati}, {Koekemoer}, {Lee}, {Lee}, {Nayyeri}, {Peth},
  {Pforr}, {Salvato}, {Wiklind}, {Wuyts}, {Ashby}, {Castellano}, {Conselice},
  {Cooper}, {Cooray}, {Dolch}, {Ferguson}, {Galametz}, {Giavalisco}, {Guo},
  {Willner}, {Dickinson}, {Faber}, {Fazio}, {Gardner}, {Gawiser}, {Grazian},
  {Grogin}, {Kocevski}, {Koo}, {Lee}, {Lucas}, {McGrath}, {Nandra}, {Newman},
  \& {van der Wel}}]{Stefanon2017}
{Stefanon}, M., {Yan}, H., {Mobasher}, B., {et~al.} 2017,
  \href{http://dx.doi.org/10.3847/1538-4365/aa66cb}{\color{blue}\apjs},
  \href{https://ui.adsabs.harvard.edu/abs/2017ApJS..229...32S}{229, 32}

\bibitem[{{Swinbank} {et~al.}(2011){Swinbank}, {Papadopoulos}, {Cox}, {Krips},
  {Ivison}, {Smail}, {Thomson}, {Neri}, {Richard}, \& {Ebeling}}]{Swinbank2011}
{Swinbank}, A.~M., {Papadopoulos}, P.~P., {Cox}, P., {et~al.} 2011,
  \href{http://dx.doi.org/10.1088/0004-637X/742/1/11}{\color{blue}\apj},
  \href{https://ui.adsabs.harvard.edu/abs/2011ApJ...742...11S}{742, 11}

\bibitem[{{Swinbank} {et~al.}(2012){Swinbank}, {Smail}, {Sobral}, {Theuns},
  {Best}, \& {Geach}}]{Swinbank2012}
{Swinbank}, A.~M., {Smail}, I., {Sobral}, D., {et~al.} 2012,
  \href{http://dx.doi.org/10.1088/0004-637X/760/2/130}{\color{blue}\apj},
  \href{https://ui.adsabs.harvard.edu/abs/2012ApJ...760..130S}{760, 130}

\bibitem[{{Tacconi} {et~al.}(2010){Tacconi}, {Genzel}, {Neri}, {Cox}, {Cooper},
  {Shapiro}, {Bolatto}, {Bouch{\'e}}, {Bournaud}, {Burkert}, {Combes},
  {Comerford}, {Davis}, {F{\"o}rster Schreiber}, {Garcia-Burillo},
  {Gracia-Carpio}, {Lutz}, {Naab}, {Omont}, {Shapley}, {Sternberg}, \&
  {Weiner}}]{Tacconi2010}
{Tacconi}, L.~J., {Genzel}, R., {Neri}, R., {et~al.} 2010,
  \href{http://dx.doi.org/10.1038/nature08773}{\color{blue}\nat},
  \href{https://ui.adsabs.harvard.edu/abs/2010Natur.463..781T}{463, 781}

\bibitem[{{Tacconi} {et~al.}(2018){Tacconi}, {Genzel}, {Saintonge}, {Combes},
  {Garc{\'\i}a-Burillo}, {Neri}, {Bolatto}, {Contini}, {F{\"o}rster Schreiber},
  {Lilly}, {Lutz}, {Wuyts}, {Accurso}, {Boissier}, {Boone}, {Bouch{\'e}},
  {Bournaud}, {Burkert}, {Carollo}, {Cooper}, {Cox}, {Feruglio}, {Freundlich},
  {Herrera-Camus}, {Juneau}, {Lippa}, {Naab}, {Renzini}, {Salome}, {Sternberg},
  {Tadaki}, {{\"U}bler}, {Walter}, {Weiner}, \& {Weiss}}]{Tacconi2018}
{Tacconi}, L.~J., {Genzel}, R., {Saintonge}, A., {et~al.} 2018,
  \href{http://dx.doi.org/10.3847/1538-4357/aaa4b4}{\color{blue}\apj},
  \href{https://ui.adsabs.harvard.edu/abs/2018ApJ...853..179T}{853, 179}

\bibitem[{Tacconi {et~al.}(2020)Tacconi, Genzel, \& Sternberg}]{Tacconi2020}
Tacconi, L.~J., Genzel, R., \& Sternberg, A. 2020,
  \href{http://dx.doi.org/10.1146/annurev-astro-082812-141034}{\color{blue}Annual
  Review of Astronomy and Astrophysics}, 58, 157

\bibitem[{{Tadaki} {et~al.}(2018){Tadaki}, {Iono}, {Yun}, {Aretxaga},
  {Hatsukade}, {Hughes}, {Ikarashi}, {Izumi}, {Kawabe}, {Kohno}, {Lee},
  {Matsuda}, {Nakanishi}, {Saito}, {Tamura}, {Ueda}, {Umehata}, {Wilson},
  {Michiyama}, {Ando}, \& {Kamieneski}}]{Tadaki2018}
{Tadaki}, K., {Iono}, D., {Yun}, M.~S., {et~al.} 2018,
  \href{http://dx.doi.org/10.1038/s41586-018-0443-1}{\color{blue}\nat},
  \href{https://ui.adsabs.harvard.edu/abs/2018Natur.560..613T}{560, 613}

\bibitem[{{Tanaka} {et~al.}(2024){Tanaka}, {Silverman}, {Nakazato}, {Onoue},
  {Shimasaku}, {Fudamoto}, {Fujimoto}, {Ding}, {Faisst}, {Valentino}, {Jin},
  {Hayward}, {Kokorev}, {Ceverino}, {Kalita}, {Casey}, {Liu}, {Kaminsky},
  {Fei}, {Andika}, {Lambrides}, {Akins}, {Kartaltepe}, {Koekemoer},
  {McCracken}, {Rhodes}, {Robertson}, {Franco}, {Liu}, {Chartab}, {Gillman},
  {Gozaliasl}, {Hirschmann}, {Huertas-Company}, {Massey}, {Roy}, {Sattari},
  {Shuntov}, {Sterling}, {Toft}, {Trakhtenbrot}, {Yoshida}, \&
  {Zavala}}]{Tanaka2024}
{Tanaka}, T.~S., {Silverman}, J.~D., {Nakazato}, Y., {et~al.} 2024,
  \href{http://dx.doi.org/10.1093/pasj/psae091}{\color{blue}\pasj},
  \href{https://ui.adsabs.harvard.edu/abs/2024PASJ...76.1323T}{76, 1323}

\bibitem[{{Taylor}(1950)}]{Taylor1950}
{Taylor}, G. 1950,
  \href{http://dx.doi.org/10.1098/rspa.1950.0049}{\color{blue}Proceedings of
  the Royal Society of London Series A},
  \href{https://ui.adsabs.harvard.edu/abs/1950RSPSA.201..159T}{201, 159}

\bibitem[{{Telikova} {et~al.}(2025){Telikova}, {Gonz{\'a}lez-L{\'o}pez},
  {Aravena}, {Posses}, {Villanueva}, {Baeza-Garay}, {Jones}, {Solimano}, {Lee},
  {Assef}, {De Looze}, {Diaz Santos}, {Ferrara}, {Ikeda}, {Herrera-Camus},
  {{\"U}bler}, {Lamperti}, {Mitsuhashi}, {Relano}, {Perna}, \&
  {Tadaki}}]{Telikova2025}
{Telikova}, K., {Gonz{\'a}lez-L{\'o}pez}, J., {Aravena}, M., {et~al.} 2025,
  \href{http://dx.doi.org/10.1051/0004-6361/202452990}{\color{blue}\aap},
  \href{https://ui.adsabs.harvard.edu/abs/2024arXiv241109033T}{669, A5}

\bibitem[{{Tiley} {et~al.}(2019){Tiley}, {Swinbank}, {Harrison}, {Smail},
  {Turner}, {Schaller}, {Stott}, {Sobral}, {Theuns}, {Sharples}, {Gillman},
  {Bower}, {Bunker}, {Best}, {Richard}, {Bacon}, {Bureau}, {Cirasuolo}, \&
  {Magdis}}]{Tiley2019b}
{Tiley}, A.~L., {Swinbank}, A.~M., {Harrison}, C.~M., {et~al.} 2019,
  \href{http://dx.doi.org/10.1093/mnras/stz428}{\color{blue}\mnras},
  \href{https://ui.adsabs.harvard.edu/abs/2019MNRAS.485..934T}{485, 934}

\bibitem[{{Tohill} {et~al.}(2024){Tohill}, {Bamford}, {Conselice}, {Ferreira},
  {Harvey}, {Adams}, \& {Austin}}]{Tohill2024}
{Tohill}, C., {Bamford}, S.~P., {Conselice}, C.~J., {et~al.} 2024,
  \href{http://dx.doi.org/10.3847/1538-4357/ad17b8}{\color{blue}\apj},
  \href{https://ui.adsabs.harvard.edu/abs/2024ApJ...962..164T}{962, 164}

\bibitem[{{Toomre}(1964)}]{Toomre64}
{Toomre}, A. 1964, \href{http://dx.doi.org/10.1086/147861}{\color{blue}\apj},
  \href{https://ui.adsabs.harvard.edu/abs/1964ApJ...139.1217T}{139, 1217}

\bibitem[{{Tsukui} \& {Iguchi}(2021)}]{Tsukui2021}
{Tsukui}, T. \& {Iguchi}, S. 2021,
  \href{http://dx.doi.org/10.1126/science.abe9680}{\color{blue}Science},
  \href{https://ui.adsabs.harvard.edu/abs/2021Sci...372.1201T}{372, 1201}

\bibitem[{{Turner} {et~al.}(2017){Turner}, {Cirasuolo}, {Harrison}, {McLure},
  {Dunlop}, {Swinbank}, {Johnson}, {Sobral}, {Matthee}, \&
  {Sharples}}]{Turner2017}
{Turner}, O.~J., {Cirasuolo}, M., {Harrison}, C.~M., {et~al.} 2017,
  \href{http://dx.doi.org/10.1093/mnras/stx1366}{\color{blue}\mnras},
  \href{https://ui.adsabs.harvard.edu/abs/2017MNRAS.471.1280T}{471, 1280}

\bibitem[{{{\"U}bler} {et~al.}(2024){{\"U}bler}, {D'Eugenio}, {Perna},
  {Arribas}, {Jones}, {Bunker}, {Carniani}, {Charlot}, {Maiolino},
  {Rodr{\'\i}guez del Pino}, {Willott}, {B{\"o}ker}, {Cresci}, {Kumari},
  {Lamperti}, {Parlanti}, {Scholtz}, \& {Venturi}}]{Uebler2024b}
{{\"U}bler}, H., {D'Eugenio}, F., {Perna}, M., {et~al.} 2024,
  \href{http://dx.doi.org/10.1093/mnras/stae1993}{\color{blue}\mnras},
  \href{https://ui.adsabs.harvard.edu/abs/2024MNRAS.533.4287U}{533, 4287}

\bibitem[{{{\"U}bler} {et~al.}(2021){{\"U}bler}, {Genel}, {Sternberg},
  {Genzel}, {Price}, {F{\"o}rster Schreiber}, {Shimizu}, {Pillepich}, {Nelson},
  {Burkert}, {Davies}, {Hernquist}, {Lang}, {Lutz}, {Pakmor}, \&
  {Tacconi}}]{Uebler2021}
{{\"U}bler}, H., {Genel}, S., {Sternberg}, A., {et~al.} 2021,
  \href{http://dx.doi.org/10.1093/mnras/staa3464}{\color{blue}\mnras},
  \href{https://ui.adsabs.harvard.edu/abs/2021MNRAS.500.4597U}{500, 4597}

\bibitem[{{{\"U}bler} {et~al.}(2018){{\"U}bler}, {Genzel}, {Tacconi},
  {F{\"o}rster Schreiber}, {Neri}, {Contursi}, {Belli}, {Nelson}, {Lang},
  {Shimizu}, {Davies}, {Herrera-Camus}, {Lutz}, {Plewa}, {Price}, {Schuster},
  {Sternberg}, {Tadaki}, {Wisnioski}, \& {Wuyts}}]{Uebler2018}
{{\"U}bler}, H., {Genzel}, R., {Tacconi}, L.~J., {et~al.} 2018,
  \href{http://dx.doi.org/10.3847/2041-8213/aaacfa}{\color{blue}\apjl},
  \href{https://ui.adsabs.harvard.edu/abs/2018ApJ...854L..24U}{854, L24}

\bibitem[{{{\"U}bler} {et~al.}(2019){{\"U}bler}, {Genzel}, {Wisnioski},
  {F{\"o}rster Schreiber}, {Shimizu}, {Price}, {Tacconi}, {Belli}, {Wilman},
  {Fossati}, {Mendel}, {Davies}, {Beifiori}, {Bender}, {Brammer}, {Burkert},
  {Chan}, {Davies}, {Fabricius}, {Galametz}, {Herrera-Camus}, {Lang}, {Lutz},
  {Momcheva}, {Naab}, {Nelson}, {Saglia}, {Tadaki}, {van Dokkum}, \&
  {Wuyts}}]{Uebler2019}
{{\"U}bler}, H., {Genzel}, R., {Wisnioski}, E., {et~al.} 2019,
  \href{http://dx.doi.org/10.3847/1538-4357/ab27cc}{\color{blue}\apj},
  \href{https://ui.adsabs.harvard.edu/abs/2019ApJ...880...48U}{880, 48}

\bibitem[{{{\"U}bler} {et~al.}(2014){{\"U}bler}, {Naab}, {Oser}, {Aumer},
  {Sales}, \& {White}}]{Ubler2014}
{{\"U}bler}, H., {Naab}, T., {Oser}, L., {et~al.} 2014,
  \href{http://dx.doi.org/10.1093/mnras/stu1275}{\color{blue}\mnras},
  \href{https://ui.adsabs.harvard.edu/abs/2014MNRAS.443.2092U}{443, 2092}

\bibitem[{{Umehata} {et~al.}(2025){Umehata}, {Steidel}, {Smail}, {Swinbank},
  {Monson}, {Rosario}, {Lehmer}, {Nakanishi}, {Kubo}, {Iono}, {Alexander},
  {Kohno}, {Tamura}, {Ivison}, {Saito}, {Mitsuhashi}, {Huang}, \&
  {Matsuda}}]{Umehata2025}
{Umehata}, H., {Steidel}, C.~C., {Smail}, I., {et~al.} 2025,
  \href{http://dx.doi.org/10.1093/pasj/psaf010}{\color{blue}\pasj},
  \href{https://ui.adsabs.harvard.edu/abs/2025PASJ...77..432U}{77, 432}

\bibitem[{{Vallini} {et~al.}(2015){Vallini}, {Gallerani}, {Ferrara},
  {Pallottini}, \& {Yue}}]{Vallini2015}
{Vallini}, L., {Gallerani}, S., {Ferrara}, A., {Pallottini}, A., \& {Yue}, B.
  2015, \href{http://dx.doi.org/10.1088/0004-637X/813/1/36}{\color{blue}\apj},
  \href{https://ui.adsabs.harvard.edu/abs/2015ApJ...813...36V}{813, 36}

\bibitem[{{van Albada} {et~al.}(1985){van Albada}, {Bahcall}, {Begeman}, \&
  {Sancisi}}]{vanAlbada1985}
{van Albada}, T.~S., {Bahcall}, J.~N., {Begeman}, K., \& {Sancisi}, R. 1985,
  \href{http://dx.doi.org/10.1086/163375}{\color{blue}\apj},
  \href{https://ui.adsabs.harvard.edu/abs/1985ApJ...295..305V}{295, 305}

\bibitem[{{van der Kruit} \& {Allen}(1978)}]{vanderKruit1978}
{van der Kruit}, P.~C. \& {Allen}, R.~J. 1978,
  \href{http://dx.doi.org/10.1146/annurev.aa.16.090178.000535}{\color{blue}\araa},
  \href{https://ui.adsabs.harvard.edu/abs/1978ARA&A..16..103V}{16, 103}

\bibitem[{{van der Wel} {et~al.}(2014){van der Wel}, {Franx}, {van Dokkum},
  {Skelton}, {Momcheva}, {Whitaker}, {Brammer}, {Bell}, {Rix}, {Wuyts},
  {Ferguson}, {Holden}, {Barro}, {Koekemoer}, {Chang}, {McGrath},
  {H{\"a}ussler}, {Dekel}, {Behroozi}, {Fumagalli}, {Leja}, {Lundgren},
  {Maseda}, {Nelson}, {Wake}, {Patel}, {Labb{\'e}}, {Faber}, {Grogin}, \&
  {Kocevski}}]{vanderWel2014}
{van der Wel}, A., {Franx}, M., {van Dokkum}, P.~G., {et~al.} 2014,
  \href{http://dx.doi.org/10.1088/0004-637X/788/1/28}{\color{blue}\apj},
  \href{https://ui.adsabs.harvard.edu/abs/2014ApJ...788...28V}{788, 28}

\bibitem[{{Varadaraj} {et~al.}(2024){Varadaraj}, {Bowler}, {Jarvis}, {Adams},
  {Choustikov}, {Koekemoer}, {Carnall}, {McLeod}, {Dunlop}, {Donnan}, \&
  {Grogin}}]{Varadaraj2024}
{Varadaraj}, R.~G., {Bowler}, R.~A.~A., {Jarvis}, M.~J., {et~al.} 2024,
  \href{http://dx.doi.org/10.1093/mnras/stae2022}{\color{blue}\mnras},
  \href{https://ui.adsabs.harvard.edu/abs/2024MNRAS.533.3724V}{533, 3724}

\bibitem[{{Venkateshwaran} {et~al.}(2024){Venkateshwaran}, {Weiss},
  {Sulzenauer}, {Menten}, {Aravena}, {Chapman}, {Gonzalez}, {Gururajan},
  {Hayward}, {Hill}, {Reuter}, {Spilker}, \& {Vieira}}]{Venkateshwaran2024}
{Venkateshwaran}, A., {Weiss}, A., {Sulzenauer}, N., {et~al.} 2024,
  \href{http://dx.doi.org/10.3847/1538-4357/ad7bb4}{\color{blue}\apj},
  \href{https://ui.adsabs.harvard.edu/abs/2024ApJ...977..161V}{977, 161}

\bibitem[{{Villanueva} {et~al.}(2024){Villanueva}, {Herrera-Camus},
  {Gonz{\'a}lez-L{\'o}pez}, {Aravena}, {Assef}, {Baeza-Garay},
  {Barcos-Mu{\~n}oz}, {Bovino}, {Bowler}, {da Cunha}, {De Looze},
  {Diaz-Santos}, {Ferrara}, {F{\"o}rster Schreiber}, {Algera}, {Ikeda},
  {Killi}, {Mitsuhashi}, {Naab}, {Relano}, {Spilker}, {Solimano}, {Palla},
  {Price}, {Posses}, {Tadaki}, {Telikova}, \& {{\"U}bler}}]{Villanueva2024}
{Villanueva}, V., {Herrera-Camus}, R., {Gonz{\'a}lez-L{\'o}pez}, J., {et~al.}
  2024, \href{http://dx.doi.org/10.1051/0004-6361/202451490}{\color{blue}\aap},
  \href{https://ui.adsabs.harvard.edu/abs/2024A&A...691A.133V}{691, A133}

\bibitem[{{Virtanen} {et~al.}(2020){Virtanen}, {Gommers}, {Oliphant},
  {Haberland}, {Reddy}, {Cournapeau}, {Burovski}, {Peterson}, {Weckesser},
  {Bright}, {van der Walt}, {Brett}, {Wilson}, {Millman}, {Mayorov}, {Nelson},
  {Jones}, {Kern}, {Larson}, {Carey}, {Polat}, {Feng}, {Moore}, {VanderPlas},
  {Laxalde}, {Perktold}, {Cimrman}, {Henriksen}, {Quintero}, {Harris},
  {Archibald}, {Ribeiro}, {Pedregosa}, {van Mulbregt}, \& {SciPy 1. 0
  Contributors}}]{Virtanen2020}
{Virtanen}, P., {Gommers}, R., {Oliphant}, T.~E., {et~al.} 2020,
  \href{http://dx.doi.org/10.1038/s41592-019-0686-2}{\color{blue}Nature
  Methods}, \href{https://ui.adsabs.harvard.edu/abs/2020NatMe..17..261V}{17,
  261}

\bibitem[{{Wang} {et~al.}(2025){Wang}, {Cantalupo}, {Pensabene}, {Galbiati},
  {Travascio}, \& et~al.}]{WWang2025}
{Wang}, W., {Cantalupo}, S., {Pensabene}, A., {et~al.} 2025,
  \href{http://dx.doi.org/10.1038/s41550-025-02500-2}{\color{blue}Nature
  Astronomy}

\bibitem[{{Wisnioski} {et~al.}(2019){Wisnioski}, {F{\"o}rster Schreiber},
  {Fossati}, {Mendel}, {Wilman}, {Genzel}, {Bender}, {Wuyts}, {Davies},
  {{\"U}bler}, {Bandara}, {Beifiori}, {Belli}, {Brammer}, {Chan}, {Davies},
  {Fabricius}, {Galametz}, {Lang}, {Lutz}, {Nelson}, {Momcheva}, {Price},
  {Rosario}, {Saglia}, {Seitz}, {Shimizu}, {Tacconi}, {Tadaki}, {van Dokkum},
  \& {Wuyts}}]{Wisnioski2019}
{Wisnioski}, E., {F{\"o}rster Schreiber}, N.~M., {Fossati}, M., {et~al.} 2019,
  \href{http://dx.doi.org/10.3847/1538-4357/ab4db8}{\color{blue}\apj},
  \href{https://ui.adsabs.harvard.edu/abs/2019ApJ...886..124W}{886, 124}

\bibitem[{{Wisnioski} {et~al.}(2015){Wisnioski}, {F{\"o}rster Schreiber},
  {Wuyts}, {Wuyts}, {Bandara}, {Wilman}, {Genzel}, {Bender}, {Davies},
  {Fossati}, {Lang}, {Mendel}, {Beifiori}, {Brammer}, {Chan}, {Fabricius},
  {Fudamoto}, {Kulkarni}, {Kurk}, {Lutz}, {Nelson}, {Momcheva}, {Rosario},
  {Saglia}, {Seitz}, {Tacconi}, \& {van Dokkum}}]{Wisnioski2015}
{Wisnioski}, E., {F{\"o}rster Schreiber}, N.~M., {Wuyts}, S., {et~al.} 2015,
  \href{http://dx.doi.org/10.1088/0004-637X/799/2/209}{\color{blue}\apj},
  \href{https://ui.adsabs.harvard.edu/abs/2015ApJ...799..209W}{799, 209}

\bibitem[{{Wisnioski} {et~al.}(2025){Wisnioski}, {Mendel}, {Leaman}, {Tsukui},
  {{\"U}bler}, \& {F{\"o}rster Schreiber}}]{Wisnioski2025}
{Wisnioski}, E., {Mendel}, J.~T., {Leaman}, R., {et~al.} 2025,
  \href{https://ui.adsabs.harvard.edu/abs/2025arXiv250524129W}{\href{http://dx.doi.org/10.48550/arXiv.2505.24129}{\color{blue}arXiv
  e-prints}, arXiv:2505.24129}

\bibitem[{{Wolfire} {et~al.}(2003){Wolfire}, {McKee}, {Hollenbach}, \&
  {Tielens}}]{Wolfire2003}
{Wolfire}, M.~G., {McKee}, C.~F., {Hollenbach}, D., \& {Tielens}, A.~G.~G.~M.
  2003, \href{http://dx.doi.org/10.1086/368016}{\color{blue}\apj},
  \href{https://ui.adsabs.harvard.edu/abs/2003ApJ...587..278W}{587, 278}

\bibitem[{{Wolfire} {et~al.}(2022){Wolfire}, {Vallini}, \&
  {Chevance}}]{Wolfire2022}
{Wolfire}, M.~G., {Vallini}, L., \& {Chevance}, M. 2022,
  \href{http://dx.doi.org/10.1146/annurev-astro-052920-010254}{\color{blue}\araa},
  \href{https://ui.adsabs.harvard.edu/abs/2022ARA&A..60..247W}{60, 247}

\bibitem[{{Wuyts} {et~al.}(2016){Wuyts}, {F{\"o}rster Schreiber}, {Wisnioski},
  {Genzel}, {Burkert}, {Bandara}, {Beifiori}, {Belli}, {Bender}, {Brammer},
  {Chan}, {Davies}, {Fossati}, {Galametz}, {Kulkarni}, {Lang}, {Lutz},
  {Mendel}, {Momcheva}, {Naab}, {Nelson}, {Saglia}, {Seitz}, {Tacconi},
  {Tadaki}, {{\"U}bler}, {van Dokkum}, {Wilman}, \& {Wuyts}}]{Wuyts2016}
{Wuyts}, S., {F{\"o}rster Schreiber}, N.~M., {Wisnioski}, E., {et~al.} 2016,
  \href{http://dx.doi.org/10.3847/0004-637X/831/2/149}{\color{blue}\apj},
  \href{https://ui.adsabs.harvard.edu/abs/2016ApJ...831..149W}{831, 149}

\bibitem[{{Xu} \& {Yu}(2024)}]{XuD2024}
{Xu}, D. \& {Yu}, S.-Y. 2024,
  \href{http://dx.doi.org/10.1051/0004-6361/202449252}{\color{blue}\aap},
  \href{https://ui.adsabs.harvard.edu/abs/2024A&A...682L..17X}{682, L17}

\bibitem[{{Zanella} {et~al.}(2018){Zanella}, {Daddi}, {Magdis}, {Diaz Santos},
  {Cormier}, {Liu}, {Cibinel}, {Gobat}, {Dickinson}, {Sargent}, {Popping},
  {Madden}, {Bethermin}, {Hughes}, {Valentino}, {Rujopakarn}, {Pannella},
  {Bournaud}, {Walter}, {Wang}, {Elbaz}, \& {Coogan}}]{Zanella2018}
{Zanella}, A., {Daddi}, E., {Magdis}, G., {et~al.} 2018,
  \href{http://dx.doi.org/10.1093/mnras/sty2394}{\color{blue}\mnras},
  \href{https://ui.adsabs.harvard.edu/abs/2018MNRAS.481.1976Z}{481, 1976}

\bibitem[{{Zolotov} {et~al.}(2015){Zolotov}, {Dekel}, {Mandelker}, {Tweed},
  {Inoue}, {DeGraf}, {Ceverino}, {Primack}, {Barro}, \& {Faber}}]{Zolotov2015}
{Zolotov}, A., {Dekel}, A., {Mandelker}, N., {et~al.} 2015,
  \href{http://dx.doi.org/10.1093/mnras/stv740}{\color{blue}\mnras},
  \href{https://ui.adsabs.harvard.edu/abs/2015MNRAS.450.2327Z}{450, 2327}

\end{thebibliography}

\begin{appendix}

\section{Global properties of individual galaxies}\label{app:globalprops}

For each galaxy individually, we briefly comment below on its global properties and its overall kinematic classification, based on the methods described in \S~\ref{sec:kins_class}.
Detailed properties of CRISTAL-05 and 22 are presented in \citet{Posses2025} and \citet{Telikova2025}, respectively. 
Figs.~\ref{fig:jwst_hst_gallery_best_disk} to \ref{fig:jwst_hst_gallery_nondisk2}, from left to right, display the \textit{JWST}, \textit{HST} colour-composite images, the [\ion{C}{II}] line emission, velocity and velocity dispersion maps. 
The velocity locus measured with SA in \S~\ref{subsec:sa} are overlaid on either the \textit{JWST} or \textit{HST} colour images in the rightmost column. The sizes and transparencies of these overlays are proportional to the positional uncertainty error. 
Figs.~\ref{fig:kins_map_bestdisk} to \ref{fig:kins_map_nondisk3} shows additionally the p-v diagrams along the kinematic major and minor axes and the integrated spectrum.
Where applicable, the quoted distance between companions is always projected.

\paragraph{{CRISTAL}-01a:} Non-Disk. 
Identified as `Multi-UV' in \citet{Ikeda2025} with \textit{HST} data, \textit{JWST} further supports that 
it is interacting with a massive neighbour SMG\,J1000+0234\ and a companion at its north-east. It is also consistent with the multiple peaks in the integrated spectrum.
The velocity field displays a gradient along the morphological minor axis observed in imaging data. 
This could suggest that this system is in its later phase of the merger, and the misaligned gradient may be a signature of gas inflow triggered by the merger.
The interaction of the members in the CRISTAL-01 system is studied in more detail in \citet{Solimano2025}.

\paragraph{{CRISTAL}-01b:} Non-Disk. 
A pair of interacting systems, 
situated approximately $\sim$\,$48$\,kpc south-east of CRISTAL-01a.
The binary nature is conspicuous in both \textit{HST} and \textit{JWST} images, 
with the two components separated by a distance of $\sim$\,$7$\,kpc. 
The [\ion{C}{II}] line map shows emission south of the eastern companion,
and there is an apparent spatial offset between the [\ion{C}{II}] emission 
and the stellar component of the western companion, possibly originating from the gas that is being stripped away in the interaction. 
Although the eastern companion exhibits a velocity gradient 
transitioning from north to south, 
its dispersion appears disrupted, 
with elevated peaks in the north-eastern region.

\begin{figure*}
    \centering
    \includegraphics[width=\textwidth]{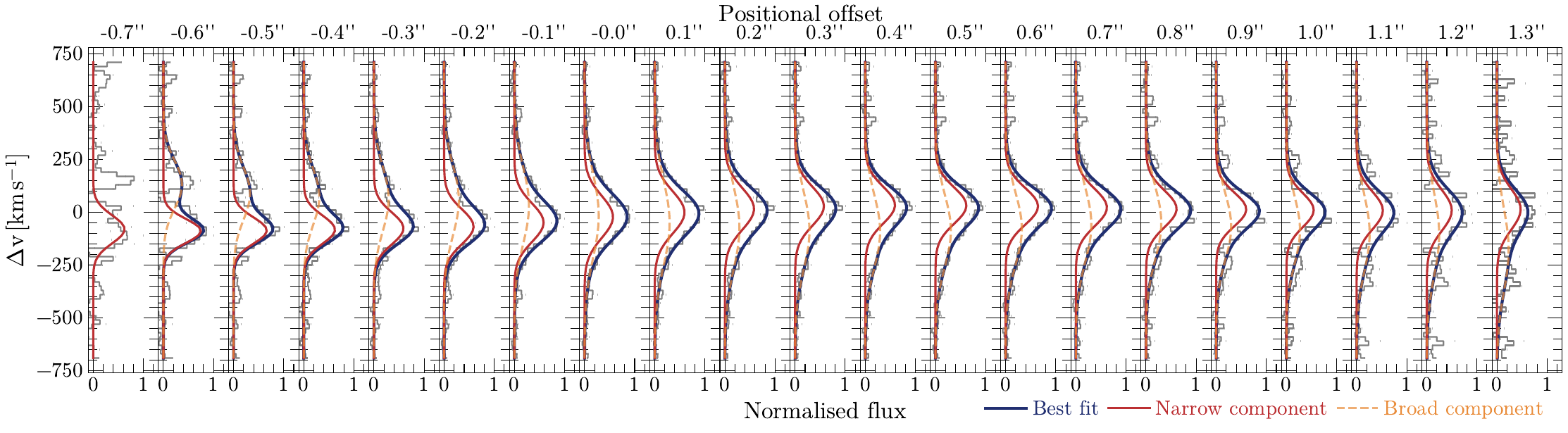}
    \caption{Two-component Gaussian fit  (blue thick line) of CRISTAL-02 emission profile along each column in the position-velocity (p-v) diagram (grey lines) extracted along the kinematic major axis. 
    The underlying p-v diagram from which the profiles are extracted is shown in Figure~\ref{fig:kins_map_disk} in the Appendix~\ref{app:globalprops}. 
    We assume the narrow component (red) to trace the bulk rotation motion of the gas, which will be used for subsequent kinematics modelling, 
    while the broad component (orange) is associated with the outflow and the interacting companion in the north (at offset $\gtrsim0\farcs8$). For other galaxies, we assume a single Gaussian profile using the same extraction method.
    }
    \label{fig:c2_pvfit}
\end{figure*}

\begin{figure*}{}
\centering
     \includegraphics[width=\textwidth]{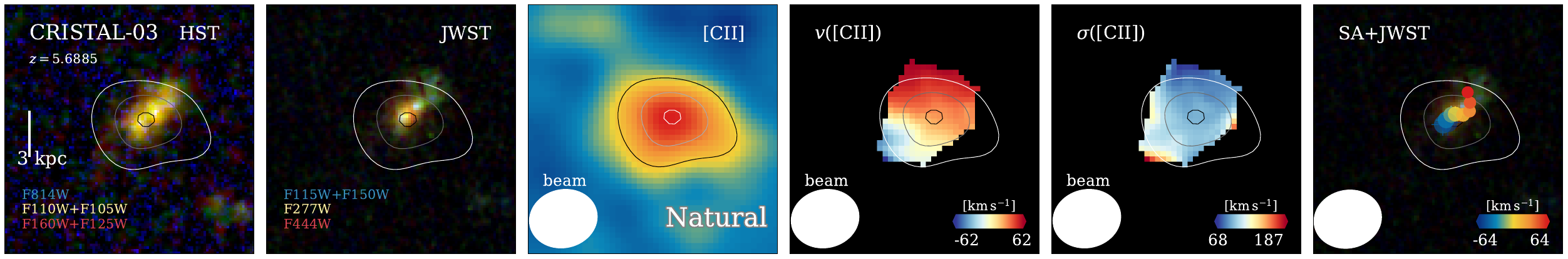}\\
    \includegraphics[width=\textwidth]{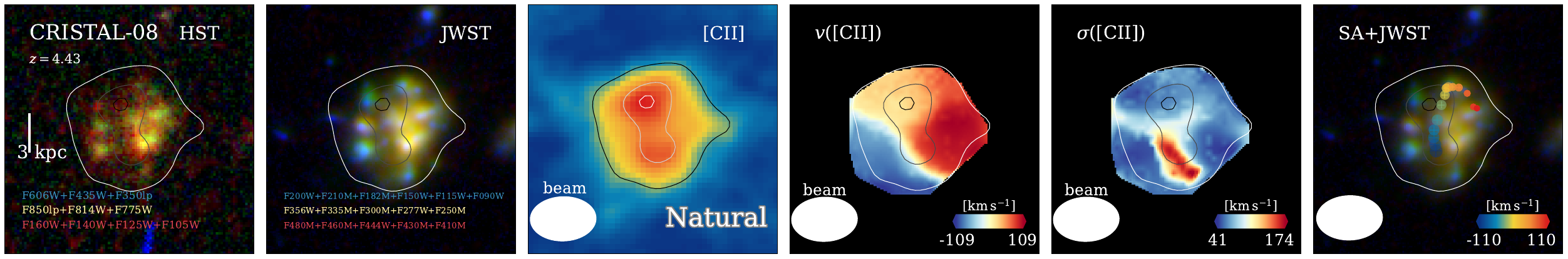}\\
      \includegraphics[width=\textwidth]{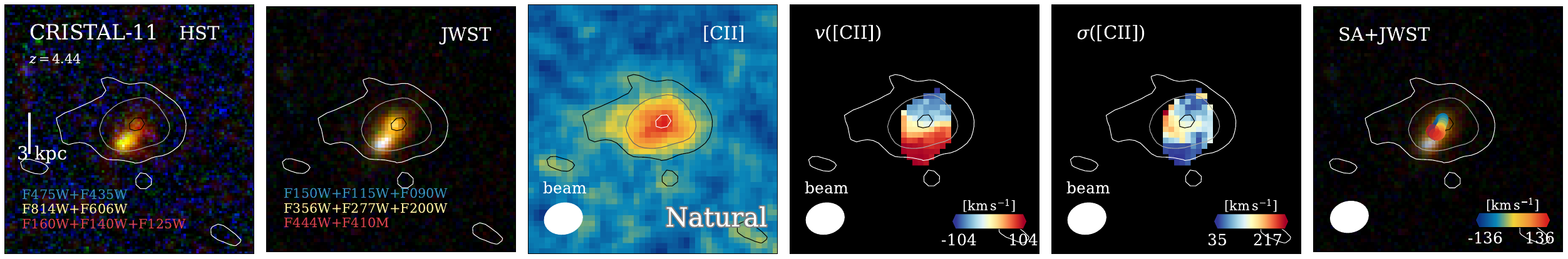}\\
      \includegraphics[width=\textwidth]{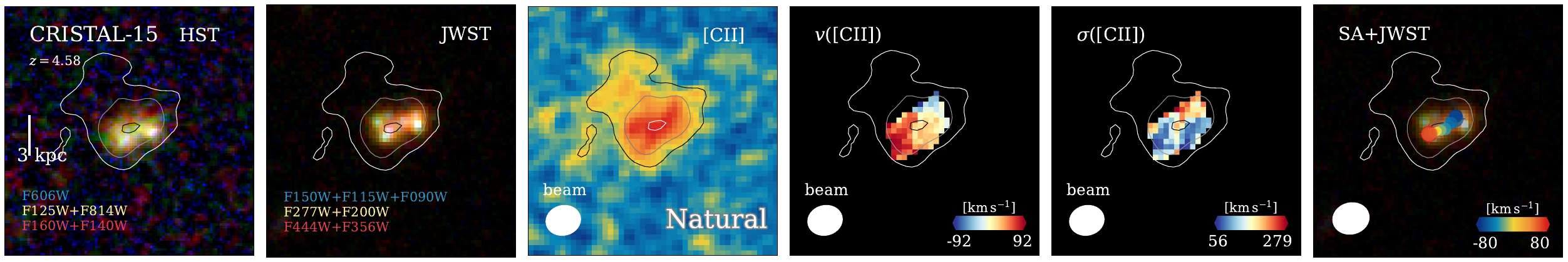}\\
      \includegraphics[width=\textwidth]{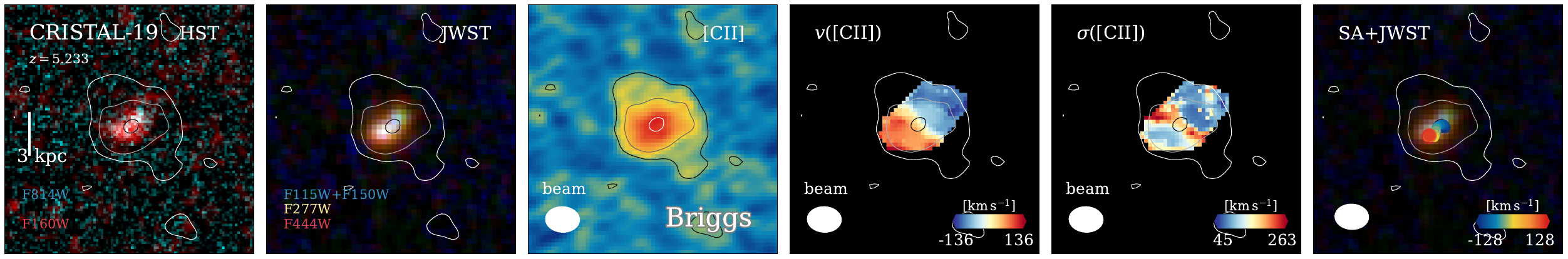}\\
      \includegraphics[width=\textwidth]{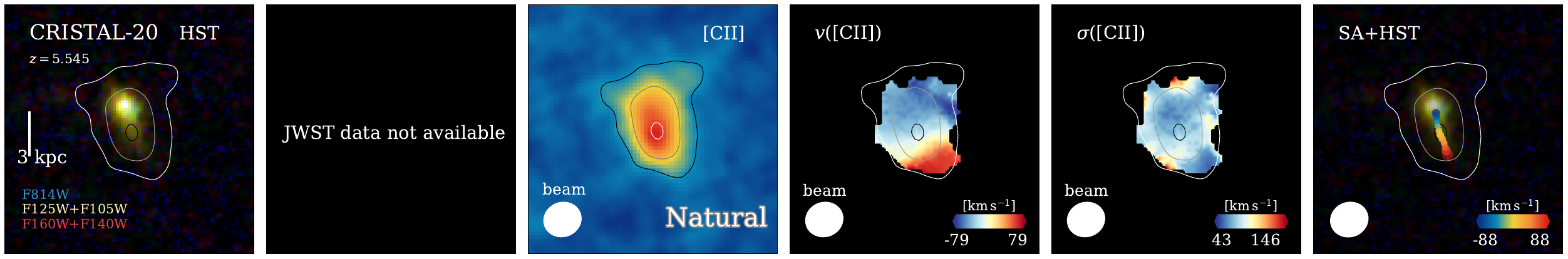}\\
      \includegraphics[width=\textwidth]{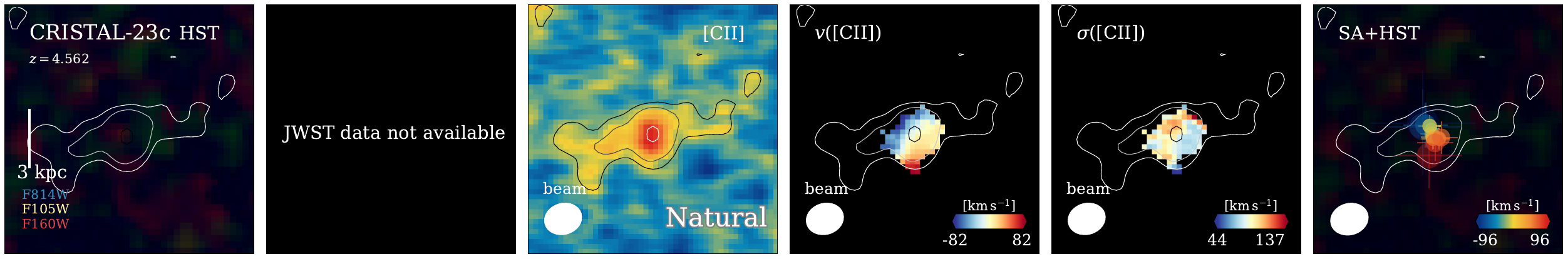}\\
      \caption{Gallery showing the multi-wavelength properties of the Best Disk. 
      The first column displays the colour-composite images from \textit{HST}, and if available, NIRCam/\textit{JWST} images are shown in the second column. The third to fifth columns are ALMA images of [\ion{C}{II}] emission, shown as line intensity, velocity, and velocity dispersion maps. The velocity and velocity dispersion maps are obtained by pixel-by-pixel single Gaussian fitting to the natural-weighted line cube, binned at $\Delta V = 20{\,\rm km\,s^{-1}}$ spectral resolution. 
      They are \textit{not} corrected for spectral broadening, beam-smearing and projection. The last column shows the spectro-astrometry (SA) measurement overlaid on \textit{HST} (or NIRCam/\textit{JWST} if available) colour images, 
      in which the sizes and transparencies of the points are proportional to the positional uncertainties derived using Equation~1 in \citet{Condon1998}. For all panels, the contours correspond to the intensities of the {[\ion{C}{II}]} line map, plotted from $1.5\sigma$ to $3\sigma.$
      }
      \label{fig:jwst_hst_gallery_best_disk}
\end{figure*}

\begin{figure*}{}
\centering
     \includegraphics[width=\textwidth]{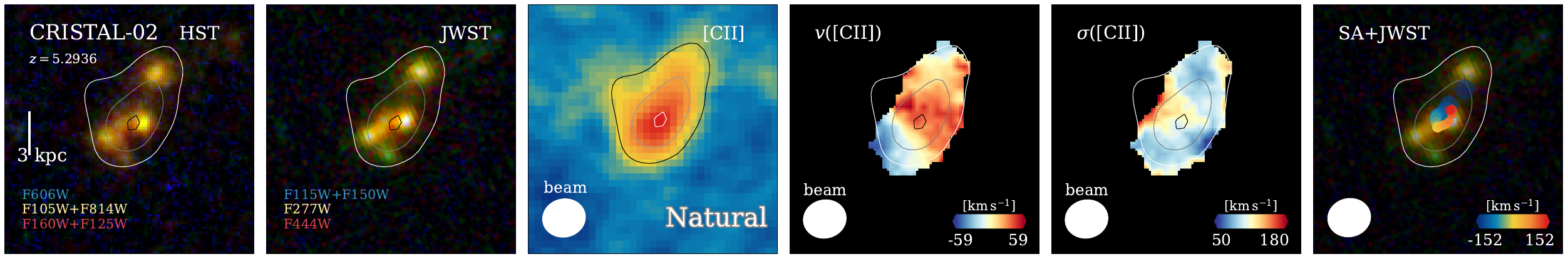}\\
    \includegraphics[width=\textwidth]{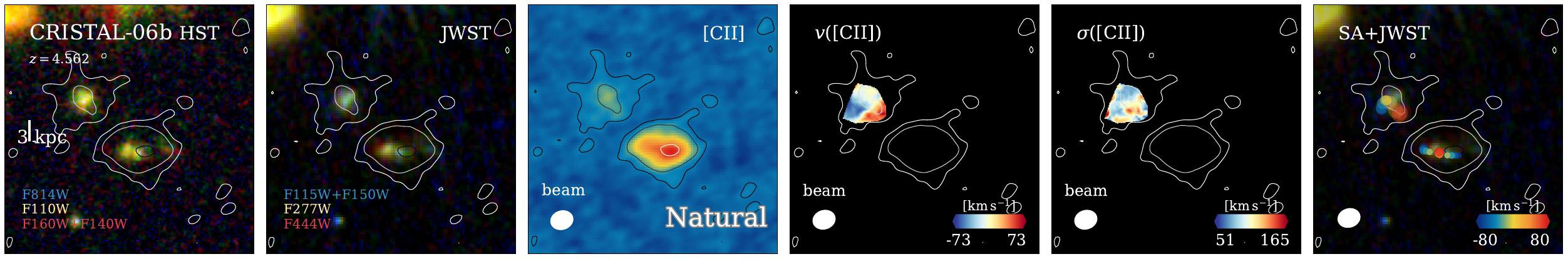}\\
      \includegraphics[width=\textwidth]{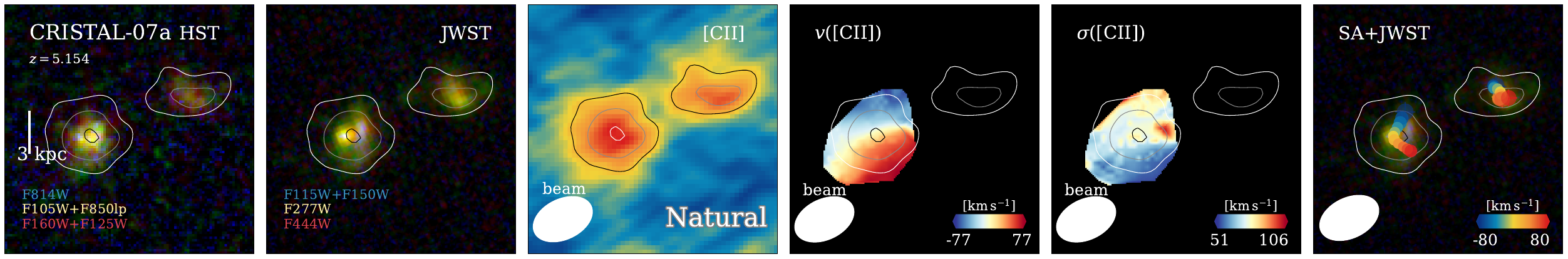}\\
      \includegraphics[width=\textwidth]{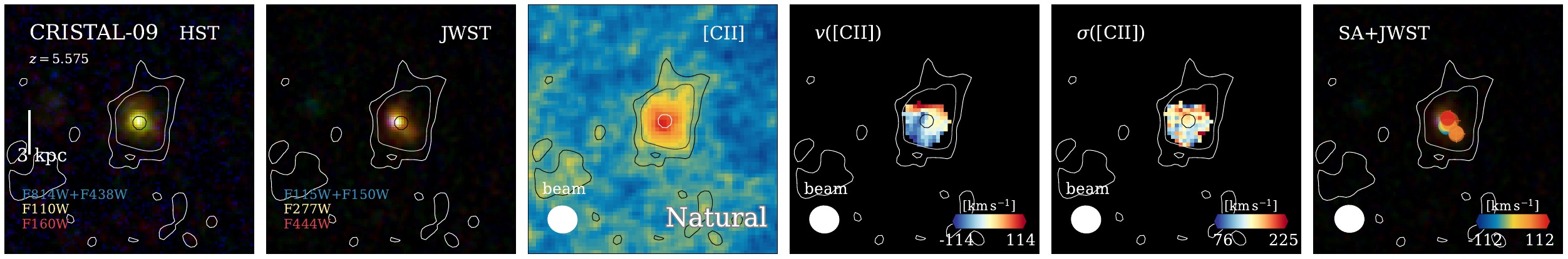}\\
      \includegraphics[width=\textwidth]{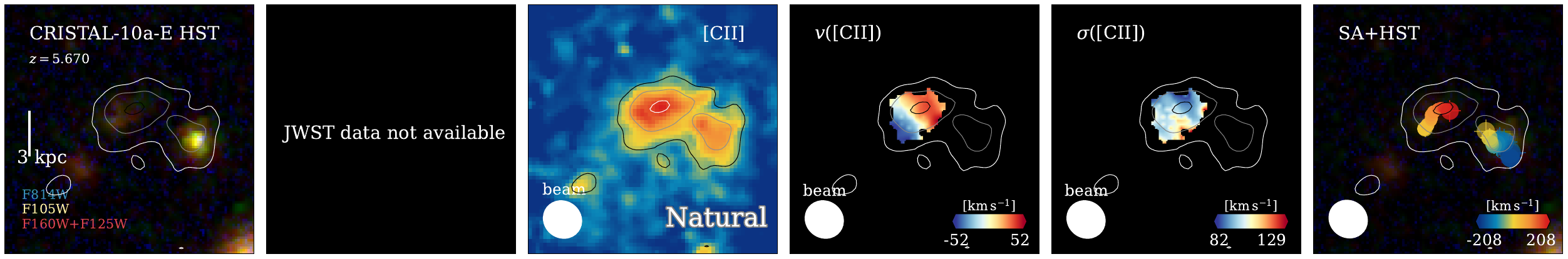}\\
      \includegraphics[width=\textwidth]{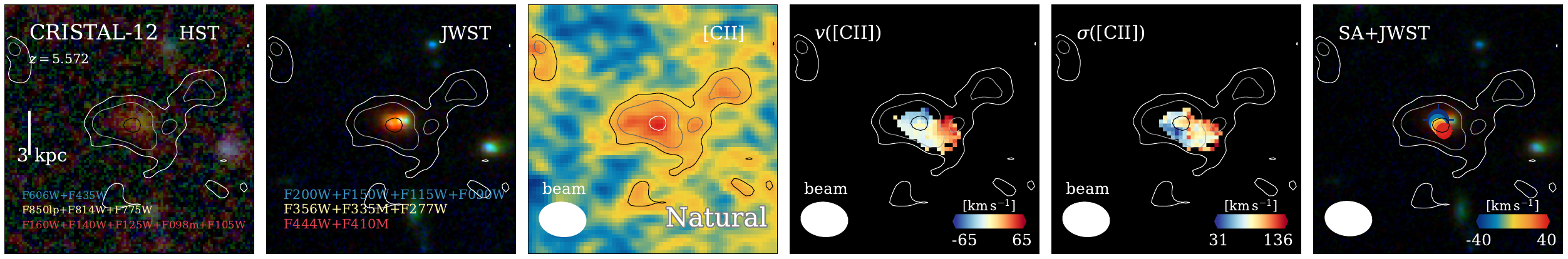}\\
      \includegraphics[width=\textwidth]{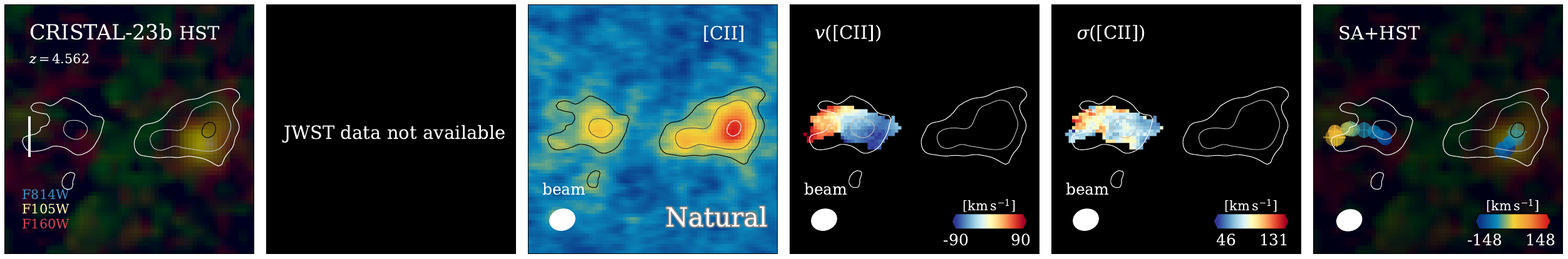}\\
      \caption{Same as Figure~\ref{fig:jwst_hst_gallery_best_disk}, but for the Disk.
       }
      \label{fig:jwst_hst_gallery_disk}
\end{figure*}

\paragraph{{CRISTAL}-02:} Disk. This system is also known as the LBG-1 in \citet{Riechers2014}.
The SMG AzTEC-3 ($z$\,$=$\,$5.298$) $\sim$\,$90\,$kpc away is also detected in [\ion{C}{II}] at the edge of FoV.
There is an overall isovelocity pattern consistent with disk rotation. 
The receding side is influenced by the interaction with the lower-mass satellite galaxy, 
resulting in a velocity reversal at the very north. 
The p-v diagram shows excess emission in high-velocity wings near the centre, 
which is associated with an outflow as evidenced in the integrated spectrum in Fig.~\ref{fig:kins_map_disk}. 
NIRSpec/\textit{JWST} data ($R$\,$\sim$\,$1000$) further confirms the biconical outflow signature 
\citep{Davies2025}. 
Considering the outflow and the interacting companion in the north, 
we therefore fit the emission in the p-v diagram (method detailed in \S~\ref{subsec:pv}) with a broad and a narrow component (Fig.~\ref{fig:c2_pvfit}). 
We attribute the narrower line to the bulk rotation motion, and the velocity and dispersion profiles derived from it are used in the analysis in \S~\ref{subsec:dysmalpy}.
By removing the broad component, we effectively minimise the contributions from the outflow and the satellite galaxy.

\paragraph{CRISTAL-03:} Best Disk; 
Velocity gradient is along the overall morphological PA of the \textit{HST} and \textit{JWST} images, 
but not with that of [\ion{C}{II}], especially at the outer isophotes due to the protrusion at the south-west.
Although this galaxy is the most poorly resolved in our sample, the NIRSpec/\textit{JWST} data with 5$\times$ better angular resolution reveals a consistent rotational pattern in H${\alpha}$\ (W. Ren et al., in prep.). 
This galaxy is also flagged as the most evident candidate for an AGN, 
which could potentially account for the protruded [\ion{C}{II}] emission towards the south-west, and the slight deviation from the uni-directional SA locus in Fig.~\ref{fig:jwst_hst_gallery_best_disk}.

\paragraph{CRISTAL-04a:} Non-Disk; 
Have a velocity gradient running from south to north and elevated dispersion along the zeroth velocity contour. However, the velocity isocontours deviate substantially from a spider diagram, and the p-v diagram along the major axis is primarily flat. 
The perturbation is most likely due to its recent minor-merger interaction with CRISTAL-04b (mass ratio $\sim$17:1, \citealt{HerreraCamus2025}).
CRISTAL-04a and CRISTAL-04b are identified as `Pair' in \citet{Ikeda2025}.

\paragraph{CRISTAL-04b:} Non-Disk; A blob of faint {[\ion{C}{II}]} emission $\sim$\,$10{\,\rm kpc}$ north of CRISTAL-04a that does not show a clear rotating signature. 
It could comprise two smaller substructures, as evidenced by the [\ion{C}{II}] line map and the JWST images.

\paragraph{CRISTAL-06a:} Non-Disk. Possibly a counter-rotating binary merger of two disks, 
where we observe a reversal in velocity gradient along the major axis, 
yet the individual components 
remain spatially unresolved at the current resolution. 
This system was also classified as a pair-merger in association with CRISTAL-06b in \citet{LeFevre2020} 
and is the only multiple-[\ion{C}{II}] system in \citet{Ikeda2025}.
Alternatively, it could be a single disk with the approaching side at the east perturbed by CRISTAL-06b, as hinted by a velocity dispersion that peaks where the gradient is steepest.
\citet{Ikeda2025} identified the multi-[\ion{C}{II}] nature within CRISTAL-06a, while CRISTAL-06a and CRISTAL-06b are identified as `Pair' in \citet{Ikeda2025}.

\paragraph{CRISTAL-06b:} Disk; it is located $\sim$\,$10{\,\rm kpc}$ north-east of CRISTAL-06a with its velocity and dispersion fields consistent with a rotating disk. The slightly twisted velocity isocontours is likely a result of the perturbation from CRISTAL-06a, which is 8$\times$ more massive \citep{HerreraCamus2025}.

\paragraph{CRISTAL-07a:} Disk. 
It exhibits a clear velocity gradient from north to south, 
but the northern part is perturbed by its interaction with CRISTAL-07b (mass ratio $\sim$2:1) 
in the early phase. 
In the kinematics modelling, 
we symmetrise the velocity dispersion profile from the receding and approaching sides (following \citealt{Posses2025}) to reduce the impact of the interaction on the intrinsic dispersion, as the receding side is much less perturbed. 

\paragraph{CRISTAL-07b:} Non-Disk, an interacting neighbour with CRISTAL-07a separated by $\sim$\,$8\,$kpc; there is an overall velocity gradient stretching from east to west, but the isovelocity contours are more disrupted compared to CRISTAL-07a due to its lower mass. 
CRISTAL-07a and CRISTAL-07b are identified as `Pair' in \citet{Ikeda2025}, while
CRISTAL-07 as a whole was classified as a pair-merger in \citet{LeFevre2020}. 

\paragraph{CRISTAL-07c:} Non-Disk; an interacting system of comparable masses locates at $140\,{\rm kpc}$ west of CRISTAL-07a and CRISTAL-07b.

\paragraph{CRISTAL-08:} Disk. 
It exhibits a remarkably smooth velocity gradient, 
although there is a change of isovelocity contours PA from south to west (visible also in the SA locus), 
likely due to non-circular motion along the minor axis 
(H. Herrera-Camus et al. 2025c, in prep.).
The velocity dispersion peaks at the south-west corner, 
away from the morphological centre and peak of the {[\ion{C}{II}]} line emission; 
The JWST colour image shows a very clumpy appearance (see also the F444W residual image in Fig.~\ref{fig:f444w_imfit}) similar to the galaxy in \citet{Tanaka2024} at $z$\,$=$\,$4.91$, 
with clumps of various colours distributed from 8 o'clock to 2 o'clock positions. 
The disk classification is consistent with the `rotating disk' classification in \citet{LeFevre2020}. 
\citet{HerreraCamus2025} will present the detailed properties of the clumps.

\paragraph{CRISTAL-09:} Disk. 
Observed as a compact object in the \textit{HST} image, \textit{JWST} reveals extended stellar light at the south-west at faint levels. 
The extended stellar component is more apparent in the residual left after subtracting a S\'ersic model (Fig.~\ref{fig:f444w_imfit}). [\ion{C}{II}] data provides tentative evidence of disk rotation 
but is still limited by the angular resolution and S/N. This system was classified as a pair-merger in \citet{LeFevre2020}.

\paragraph{CRISTAL-10a:} Non-Disk. 
No \textit{JWST} image is available for this system. Based on the \textit{HST} data, \citet{Ikeda2025} classified it as a Multi-UV system, with the two UV components separated by $\sim5.5\,$kpc. 
[\ion{C}{II}] line map shows a consistent separation of the two components.
Consider the two components as one system, it exhibits a monotonic velocity gradient from north-east to south-west across the two components,
with a prominent dispersion peak where the velocity gradient is steepest. 
However, the Gaussian fit errors at the dispersion peak are large due to low S/N.
SA locus and p-v diagram reveals an abrupt velocity jump between the east (CRISTAL-10a-E) and west components, with a clear gap in the velocity structure, 
coinciding with that in the \textit{HST} image. 
This suggests that the velocity gradient is plausibly driven by the orbital motion of the two components. Notably, this system shows a strong [\ion{C}{II}]/FIR deficit \citep{HerreraCamus2025}, 
with prominent dust continuum emission where [\ion{C}{II}] emission is weakest, particularly around the southern parts of both components. 
While this could indicate that the CRISTAL-10a and CRISTAL-10b are substructures of a single larger system, 
the [\ion{C}{II}] kinematics does not support the interpretation for now, so we classify the system as Non-Disk.

\paragraph{CRISTAL-10a-E:} Disk. 
The east component of CRISTAL-10a. Despite being in a potentially interacting system, it shows a very promising disk-rotating signature in its own velocity and dispersion maps. The SA locus and p-v diagram resemble a disk-like system but also exhibit features indicative of interaction with the western component.

\paragraph{CRISTAL-11:} Disk. 
We observe a velocity gradient from south-east to north-west, with a dispersion peak near the morphological centre. The kinematic PA is aligned with the morphological PA in both \textit{JWST} and \textit{HST} images with a difference of $<10^\circ$. However, the north-west region appears more dust-attenuated and has a distinct colour compared to the south-east, suggesting a possible merger of the two systems. Higher resolution data would be required to discern the true nature of this system. Nevertheless, based on the available data, we maintain our classification as a Disk. \citet{LeFevre2020} classified this system as `extended and dispersion-dominated'.

\paragraph{CRISTAL-12:} Disk. 
The least massive system among CRISTAL with a shallow velocity gradient. 
It is classified as `compact' in \citet{LeFevre2020} and, even with our higher resolution data, 
it remains poorly resolved, with only a tentative velocity gradient from west to east. 
There is a faint [\ion{C}{II}] emission blob $\sim4\,$kpc away in the north-west, but no visible counterpart in \textit{HST} and \textit{JWST} images. 
This emission is linked to the elevated dispersion at the north-west.
This feature is also seen in other disk systems, CRISTAL-02, 15, 19 (and 03, although poorly resolved),  which may suggest an outflow origin of the gas. 
However, except for CRISTAL-02, deeper data will be required to confirm this speculation.

\paragraph{CRISTAL-13:} Non-Disk. 
Already noted by \citet{Ikeda2025} about its multi-UV appearance in \textit{HST}, 
\textit{JWST} images further reveal the intriguing structure of the eastern and western components.
The compact western source is dominated by an older stellar population \citep{Lines2025} situated adjacent to a clumpy, blue eastern tail. 
The clump properties are addressed in \citet{HerreraCamus2025}.
While both the velocity and dispersion maps exhibit characteristic signatures of a rotating disk, 
the p-v diagrams along both the major and minor axes show features that deviate substantially from typical disk-like rotation. 
SA indicates receding motion along the blue tail, suggesting that this system is likely the result of a merger between the western and eastern sources.

\paragraph{CRISTAL-14:} Non-Disk. We observe an apparent velocity gradient from north-east to south-west. 
However, a centralised dispersion peak is absent.
Furthermore, the p-v diagrams do not display the characteristic patterns associated with disk-like rotation.
Both \textit{HST} and \textit{JWST} images reveal a western component with a distinct colour that is faint in [\ion{C}{II}]. 
The [\ion{C}{II}] emission predominantly traces the eastern component, which contributes the majority of the velocity gradient. 
Therefore, it is likely that this system is similar to 
CRISTAL-13, with the eastern source being in a close encounter with the western source.

\paragraph{CRISTAL-15:} Best Disk. 
We observe a velocity gradient from south-east to north-west, 
which agrees with the uni-directional locus traced by SA. 
An extended [\ion{C}{II}] structure is present at the north-east of the kinematic minor axis, which could indicate slow outflowing gas (see also CRISTAL-12), 
as there is no significant broad emission component in the integrated spectrum. 
The `root' of this extended structure is also co-spatial with the elevated dispersion in the dispersion map. 
\textit{JWST} data reveals three clumps of bluer colour embedded within a redder disk, which is also more evident in the F444W residual image in Fig.~\ref{fig:f444w_imfit}.
We notice that the peak-to-peak velocity difference of CRISTAL-15 is only $50\,$${\rm km\,s^{-1}}$, and our ALMA data cube is binned to $20\,$${\rm km\,s^{-1}}$, which barely samples the velocity gradient. The data cube binned to $10\,$${\rm km\,s^{-1}}$\ is too low S/N for robust extraction of the RC and dispersion profile. As a result, 
the derived velocity dispersion in \S~\ref{sec:diskkins} for CRISTAL-15 may still be contaminated by the contribution of the unresolved velocity gradient.

\paragraph{CRISTAL-16a,b:} Non-Disk; 
Similar to CRISTAL-01b and CRISTAL-06, this system is an interacting pair consisting of at least two companions, CRISTAL-16a and CRISTAL-16b, separated by a distance of $\sim5.3$\,kpc. 
The eastern companion, CRISTAL-16a, 
appears to be composed of two clumps separated by $\lesssim2\,$kpc, similar to CRISTAL-02, although it is unclear whether these clumps represent 
separate galaxies in a merger,
or multiple star formation clumps within a galaxy.
The [\ion{C}{II}] line map reveals two barely resolved emission peaks 
co-spatial with the two companions of CRISTAL-16a.
When considering CRISTAL-16a as a whole, the global velocity gradient is aligned with the morphological minor axis. 
Interestingly, the zeroth-velocity contour coincides with the dispersion peaks, 
suggesting a possible scenario in which two disk-like systems are approaching each other along their respective minor axes. 
The \textit{JWST}/F444W image (Fig.~\ref{fig:f444w_imfit}) reveals a conceivable connecting `bridge' between the north-east and south-west clumps within CRISTAL-16a, 
providing further evidence for an interaction. 
CRISTAL-16b is too faint in [\ion{C}{II}] for kinematics extraction.
Overall, CRISTAL-16a system bears similarities to CRISTAL-07a and 07b, 
but perhaps with a different orbital configuration. 
However, the individual components remain spatially unresolved at the current resolution.

\paragraph{{CRISTAL}-19:} Best Disk;
The velocity map, SA locus and p-v diagram all show a coherent velocity gradient with little deviation from symmetry. The dispersion peak coincides with the position of the steepest velocity gradient.
There is no evidence of a physically associated neighbour in the available data.
Despite having a high intrinsic dispersion ($\sigma_0=65$\,${\rm km\,s^{-1}}$), the system's $V_{\rm rot}/\sigma_0$\,$\approx$\,$2.2$ ratio indicates it is not dispersion-dominated, contrary to the classification in \citet{LeFevre2020}. 

\paragraph{CRISTAL-20:} Best Disk; 
It is one of the CRISTAL pilot galaxies first studied by \citet{Herrera-Camus2022}, which also presented its outflow properties. The velocity map shows the typical rotating pattern of a disk, while the dispersion map shows the expected outflow signature at the north-west. 
SA aligns uni-directionally with the observed gradient in the velocity map; p-v diagrams along the major axis show a clean S-shape, while along the minor axis, no velocity gradient is seen, indicative of a classical disk. 
\citet{Parlanti2025} presented the NIRSpec/\textit{JWST} data of the ionised gas outflow traced by [\ion{O}{III}]$\lambda5007$\,\AA\ and a merger scenario based on the ionised gas.

\paragraph{CRISTAL-21:} Non-Disk; 
The velocity gradient is from north to south, but the dispersion map shows multiple peaks. 
The SA locus also displays a chaotic distribution within the central region, which could indicate unresolved line-of-sight mergers. 
It is consistent with the very elongated shape of the p-v diagram, stretching from -500\,${\rm km\,s^{-1}}$\ to 500\,${\rm km\,s^{-1}}$\ within the central $\sim2\,$kpc region, indicative of an absence of a global velocity gradient. 
Although this could also be caused by a strong outflow at the centre, \citet{Lambert2023} found no evidence of outflows from the system, as indicated by the absence of a broad secondary component in the [\ion{C}{II}] spectrum. Instead, the observed disturbed gas and dust emission, combined with its kinematics structure, suggest that the system is in a late-stage merger. NIRCam/\textit{JWST} images reveal a blue `clump' located at the east side of the [\ion{C}{II}] light centre, which could be the merger companion.

\paragraph{CRISTAL-23a:} Non-Disk.
One of the companions (source C) in the `triple-merger' system was first studied in detail by \citet{Jones2021}. 
As noted by \citet{Jones2021}, it is in proximity to the 
massive protocluster PC1\,J1001+0220 \citep{Lemaux2018}. 
Since it lies along the major axis of the protocluster and is only 
$\sim$\,3.5\,Mpc from the north-east component of this protocluster, 
it could be associated with the system in a filamentary structure.
Because of a different slit angle chosen, we observe a velocity gradient from north to south of this source, in contrary to \citet{Jones2021}. 
The p-v diagram however exhibits substantial broadening along the velocity axis in the central region, which is inexplicable by a disk origin (e.g. an outflow).

\paragraph{CRISTAL-23b:} Disk; 
Source E in \citet{Jones2020}. 
The velocity gradient exhibits a clear East-West orientation, 
consistent with SA analysis. 
The velocity dispersion is elevated around the zero-velocity contour. 
The p-v diagram along the major axis displays a characteristic disk profile, 
whereas the minor axis p-v diagram shows a slight deviation, 
likely attributed to the ongoing interaction with CRISTAL-23a.

\paragraph{CRISTAL-23c:} Disk; Noted also by \citet{Devereaux2024}, it is the least luminous of the three sources in the CRISTAL-23 system and is likely to be rotating. It is separated from CRISTAL-23a and 23b in velocity by $\sim$\,$300$\,${\rm km\,s^{-1}}$\ and spatially by 18\,kpc. 
The velocity and dispersion map of this galaxy showed signs of disk rotation but is slightly perturbed by its interaction with CRISTAL-23a and 23b. 
It is the source W in \citet{Jones2020}.

\paragraph{CRISTAL-24:} Non-Disk; Multiple dispersion peaks in the dispersion map; SA does not show a large-scale velocity gradient, with a closely spaced locus suggesting a line-of-sight merger, which is further supported by the almost vertical position-velocity diagram along the major axis.
Morphological and kinematic studies by \citet{Devereaux2024} rejected the outflow scenarios. This galaxy has the largest integrated line width $\sigma_{\rm int}$\,$\approx$\,$300\,$${\rm km\,s^{-1}}$. Although CRISTAL-07 is 130\,kpc (projected) away, it is not associated with CRISTAL-24 with a redshift difference of $\Delta z\approx0.6$.

\paragraph{CRISTAL-25:} Non-Disk;
Although \citet{Devereaux2024} previously dismissed the merger scenario based on the \textit{HST} morphology and dust continuum emission, the disjoint velocity structure in the p-v diagram, the well-defined velocity gradients of each component (Fig.~\ref{fig:kins_map_nondisk3}), with the northern component displaying a clear dispersion peak, could be indicative of a merger.

\clearpage

\begin{figure*}{}
\centering
     \includegraphics[width=0.98\textwidth]{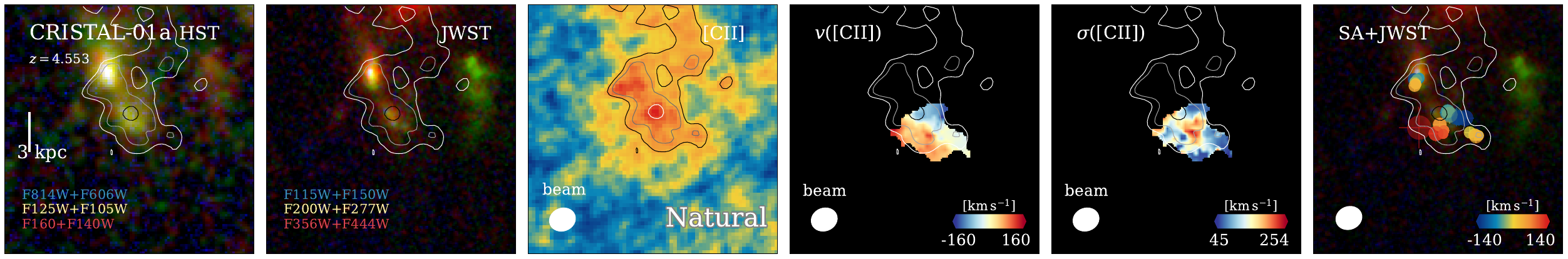}\\
    \includegraphics[width=0.98\textwidth]{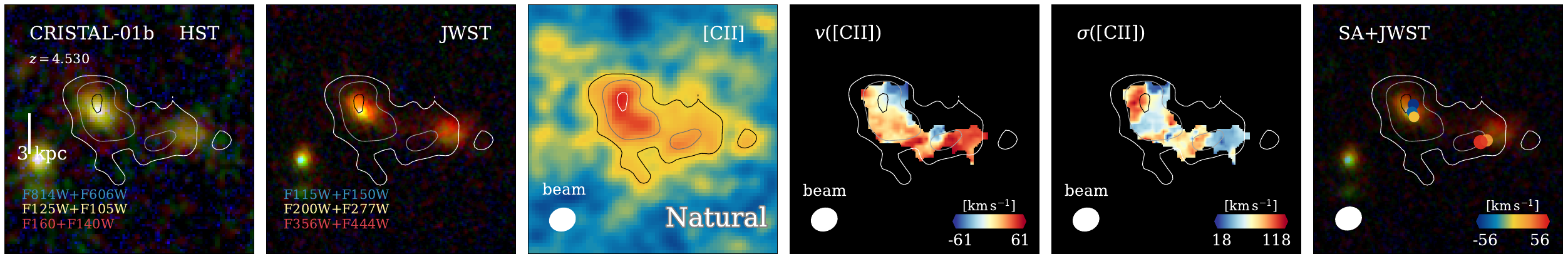}\\
      \includegraphics[width=0.98\textwidth]{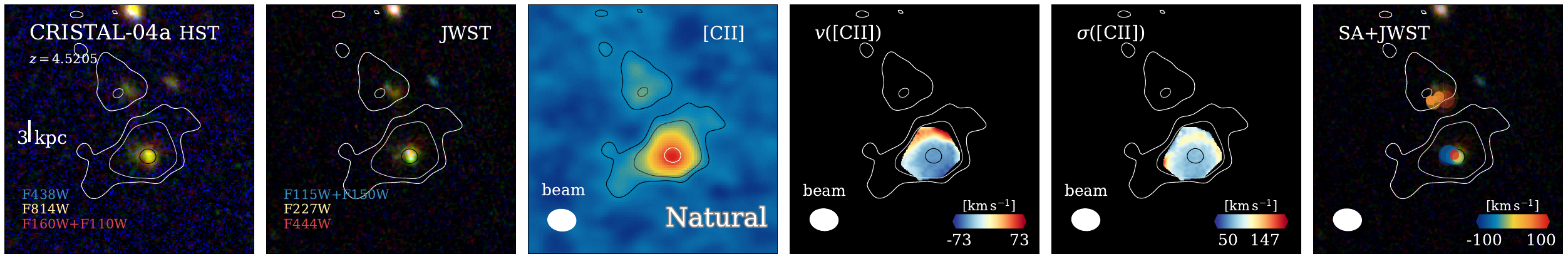}\\
      \includegraphics[width=0.98\textwidth]{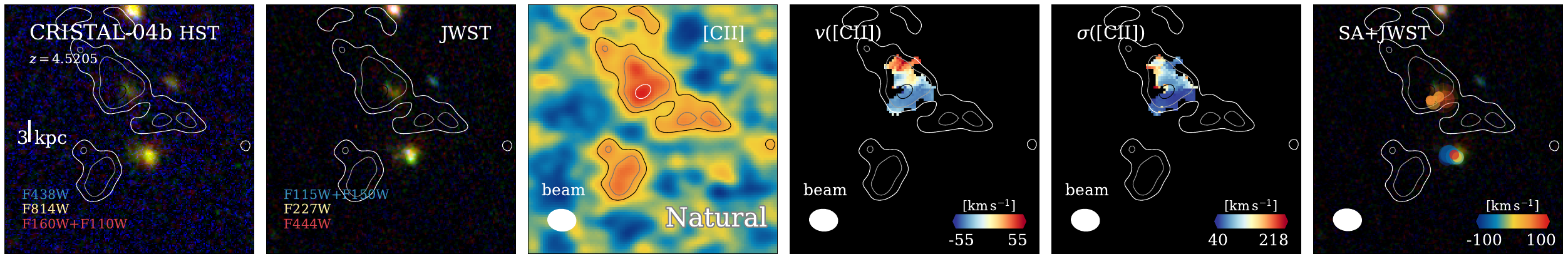}\\
      \includegraphics[width=0.98\textwidth]{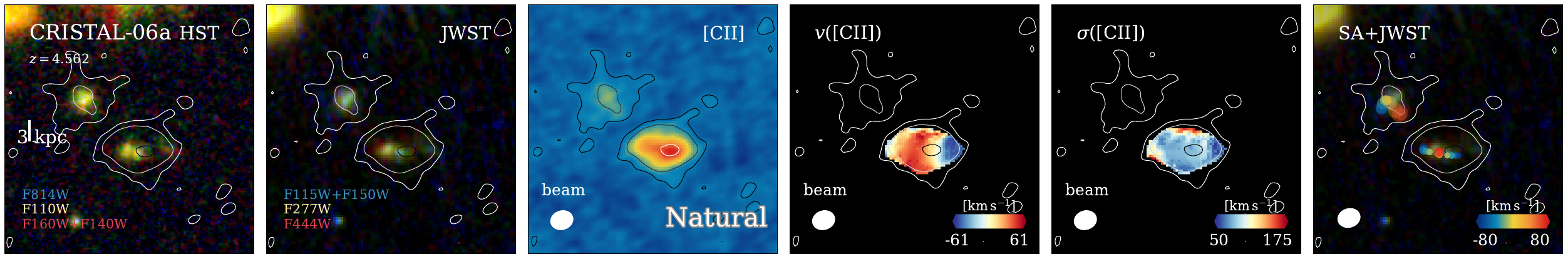}\\
      \includegraphics[width=0.98\textwidth]{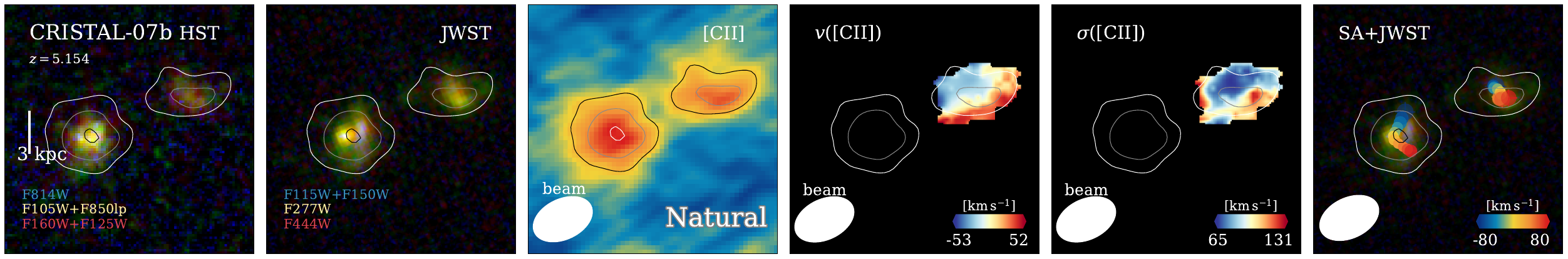}\\
      \includegraphics[width=0.98\textwidth]{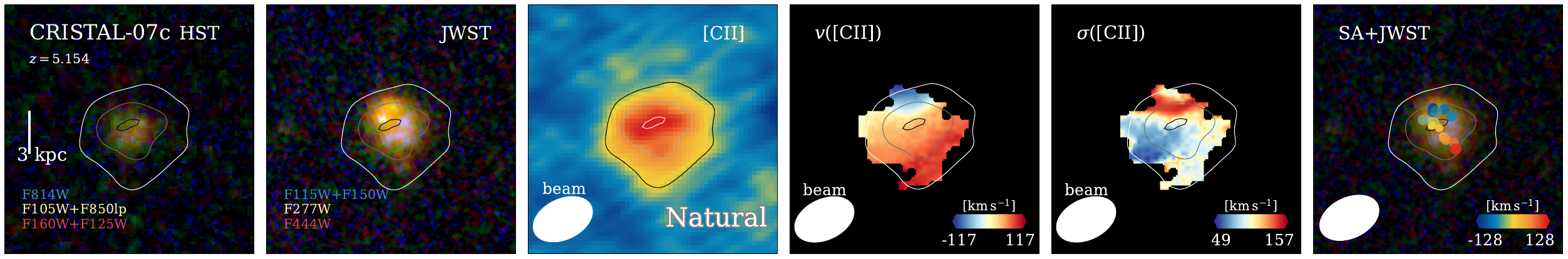}\\
      \includegraphics[width=0.98\textwidth]{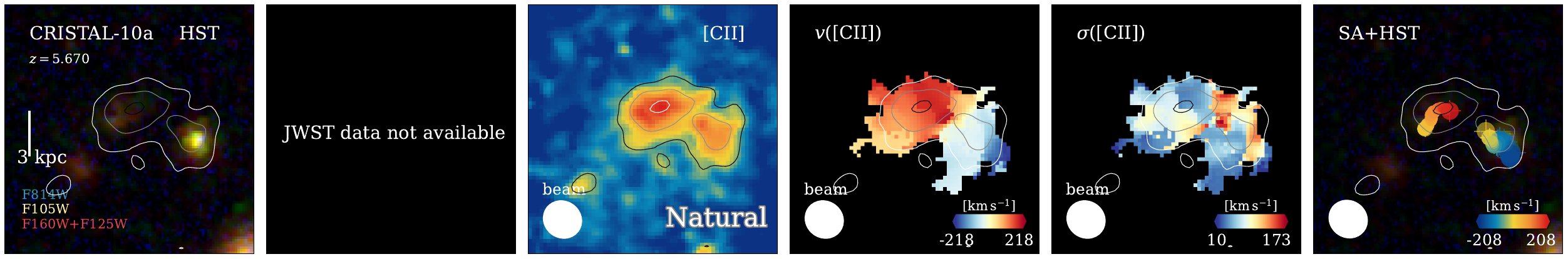}
      \caption{Same as Figure~\ref{fig:jwst_hst_gallery_best_disk}, but for the Non-Disk. 
      For CRISTAL-01a and 04b, we subtract the bright companions to improve visual contrast.
      }
      \label{fig:jwst_hst_gallery_nondisk}
\end{figure*}

\begin{figure*}[t]\ContinuedFloat
\centering
    \includegraphics[width=\textwidth]{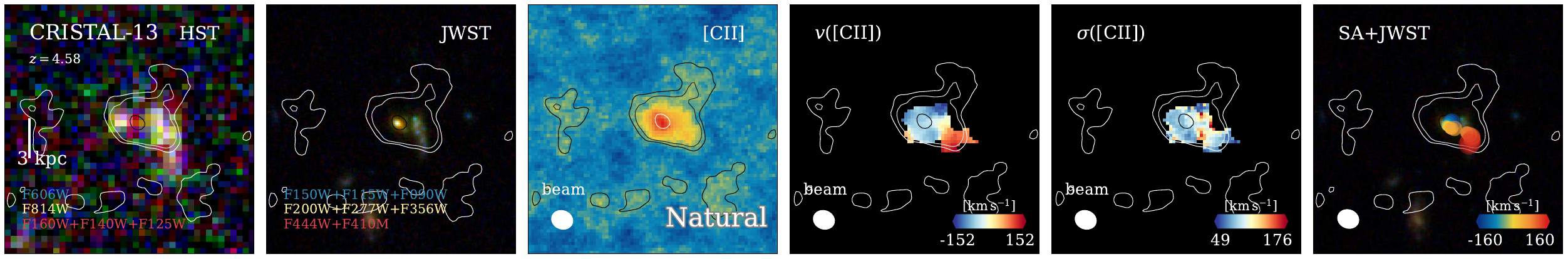}\\
    \includegraphics[width=\textwidth]{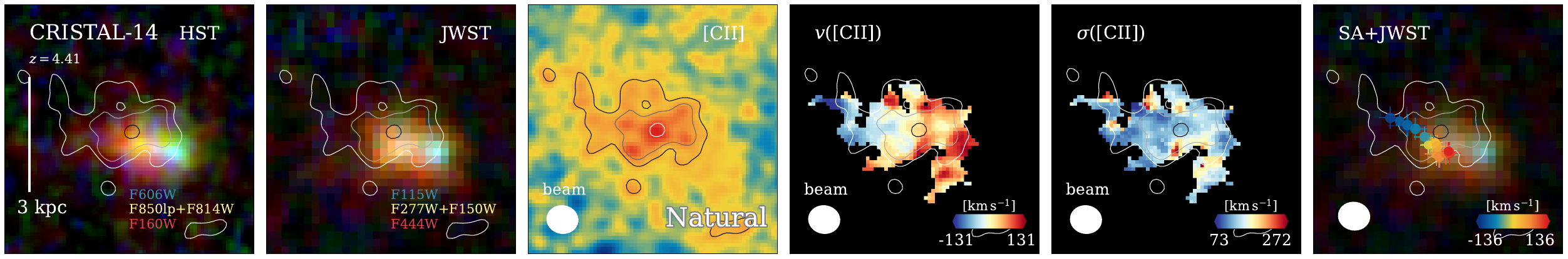}\\
    \includegraphics[width=\textwidth]{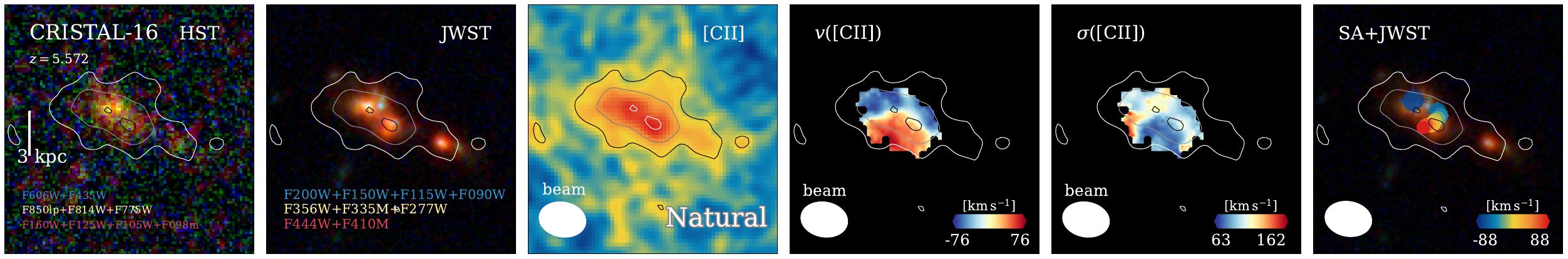}\\
    \includegraphics[width=\textwidth]{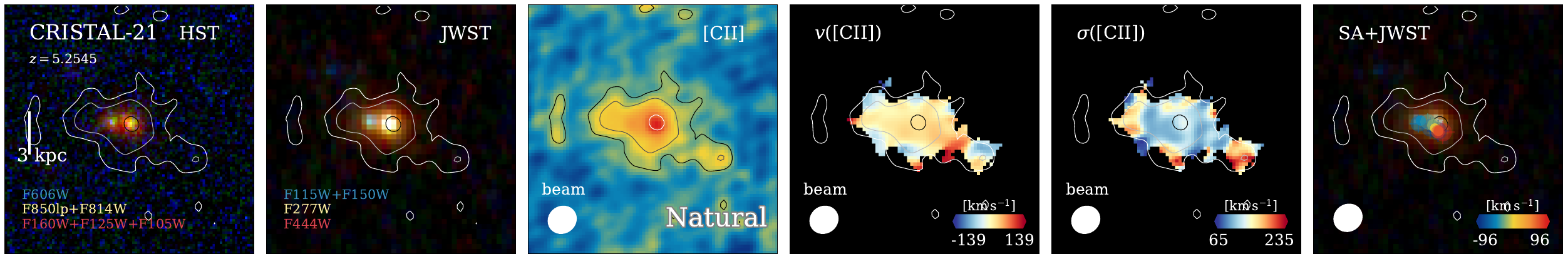}\\
    \includegraphics[width=\textwidth]{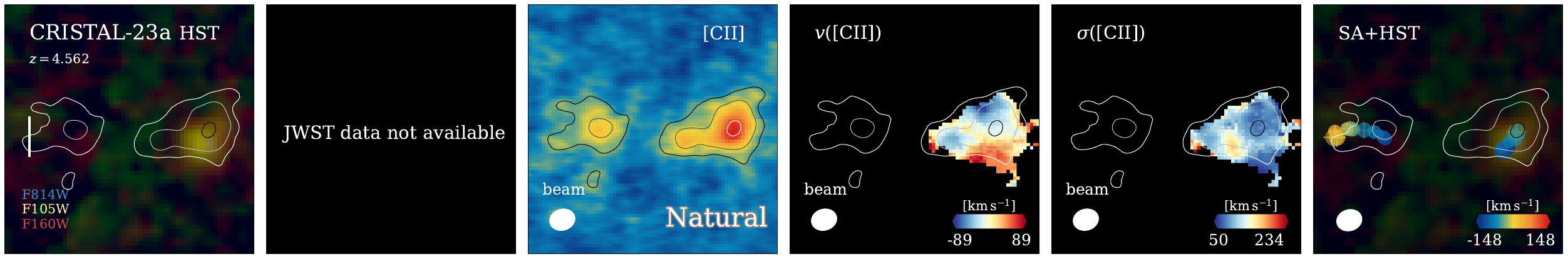}\\
    \includegraphics[width=\textwidth]{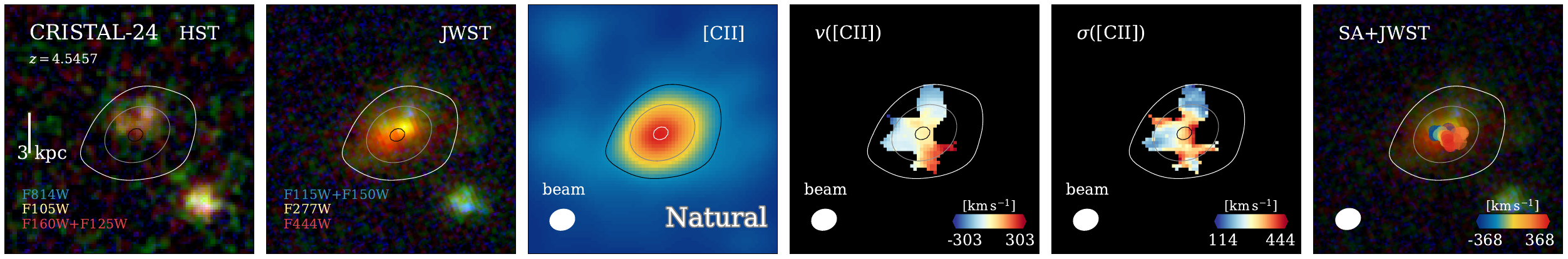}\\
     \includegraphics[width=\textwidth]{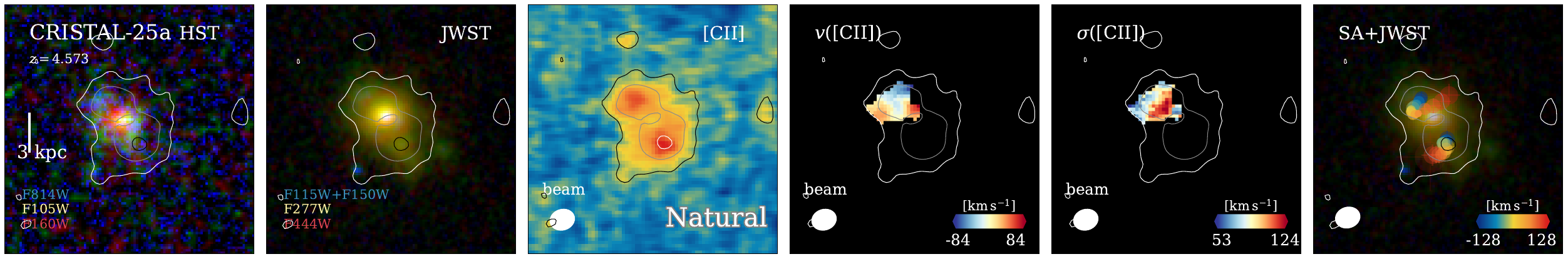}\\
     \includegraphics[width=\textwidth]{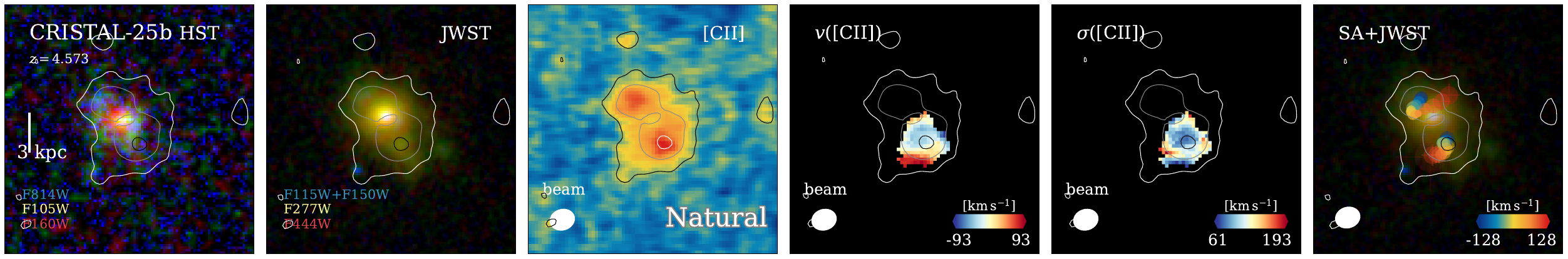}
      \caption{\textit{(Continued.)}}
      \label{fig:jwst_hst_gallery_nondisk2}
\end{figure*}

\begin{figure*}
\centering
    \includegraphics[width=1.05\textwidth]{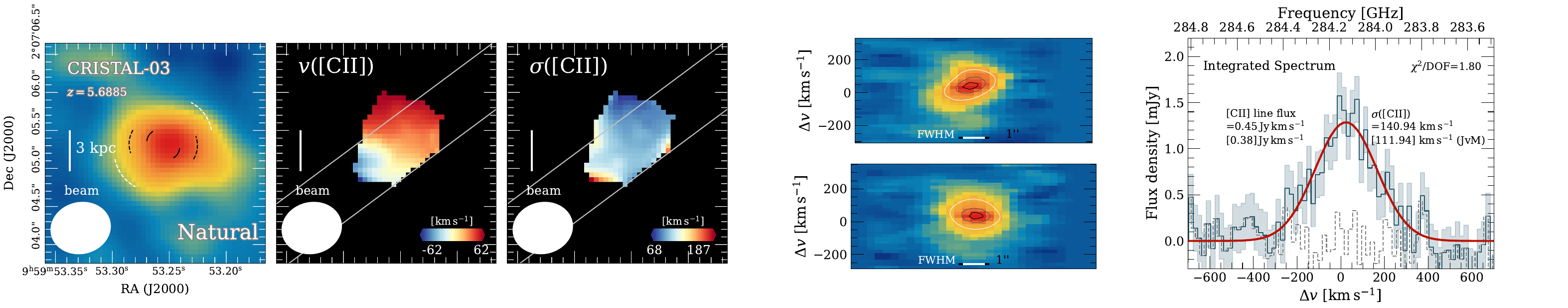}\\
    \includegraphics[width=1.05\textwidth]{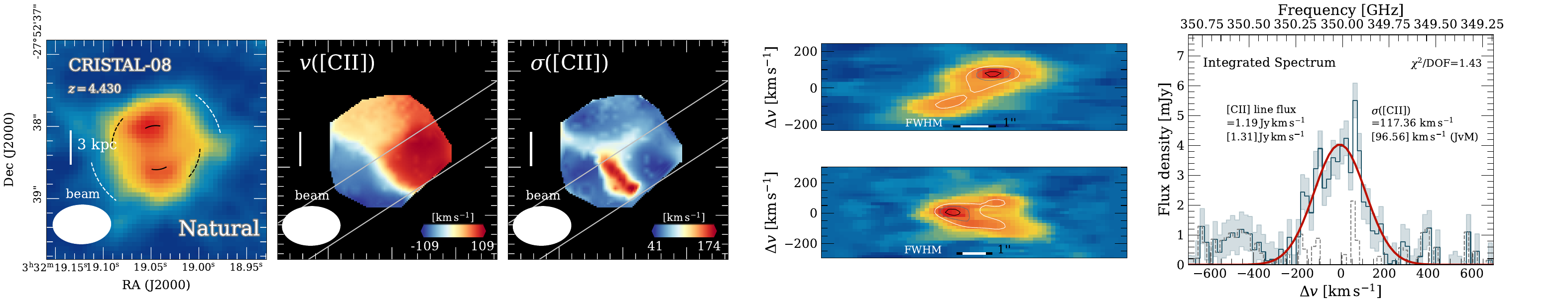}\\
    \includegraphics[width=1.05\textwidth]{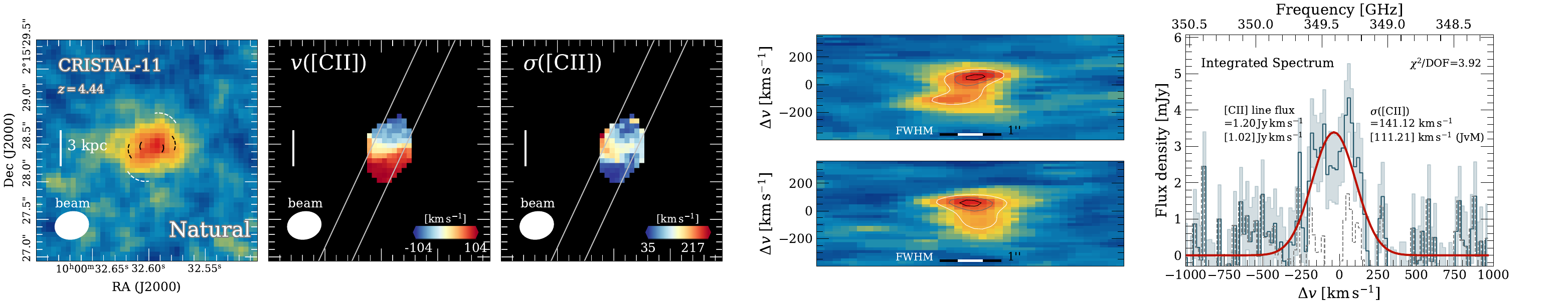}\\
    \includegraphics[width=1.05\textwidth]{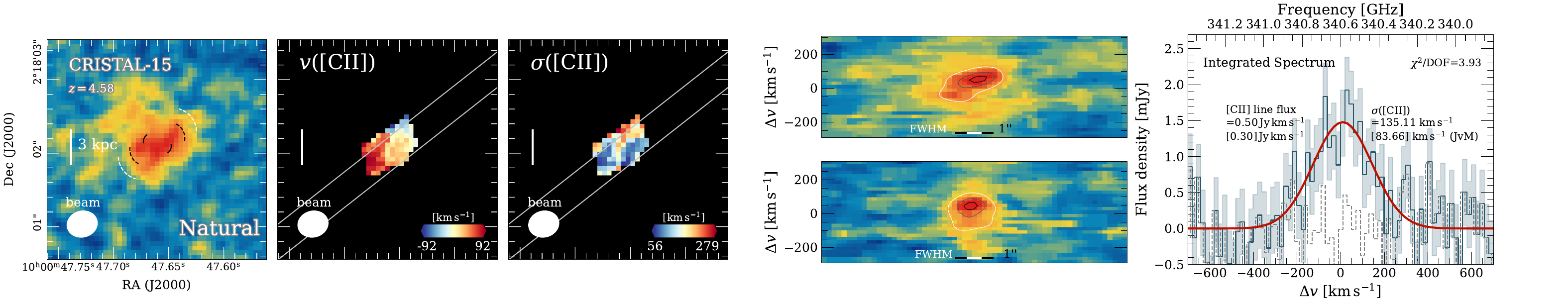}\\
     \includegraphics[width=1.05\textwidth]{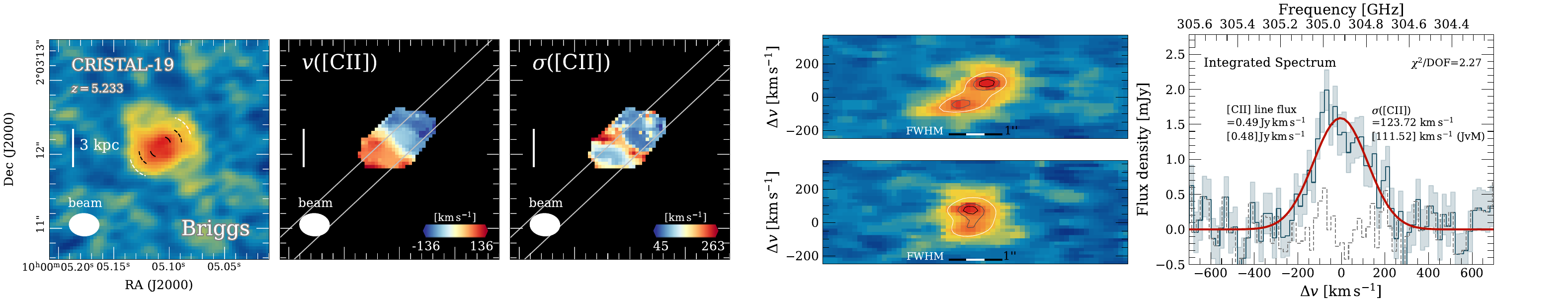}\\

      \caption{{[\ion{C}{II}]} line maps, velocity maps, dispersion maps, position-velocity (p-v) diagrams (with median filtering applied for visual purpose), integrated spectra and shifted spectra of the Disk.
      % Values of the integrated line widths are listed in Table.\,\ref{tab:intlineprops}. 
      In the {[\ion{C}{II}]} line maps, the dashed pairs of arcs delineating the emission indicate the apparent morphological position angle (${\rm PA_m}$, black) and the kinematic major axis (white). 
      The solid pair of arcs represent the intrinsic PA$_{\rm m}$ after correcting for the beam. In the p-v diagrams, the black and white horizontal bars correspond to $1\arcsec$ and the synthesised beam FWHM, respectively. 
      }
      \label{fig:kins_map_bestdisk}
\end{figure*}
\clearpage
\begin{figure*}[t]\ContinuedFloat
\centering
    \includegraphics[width=1.05\textwidth]{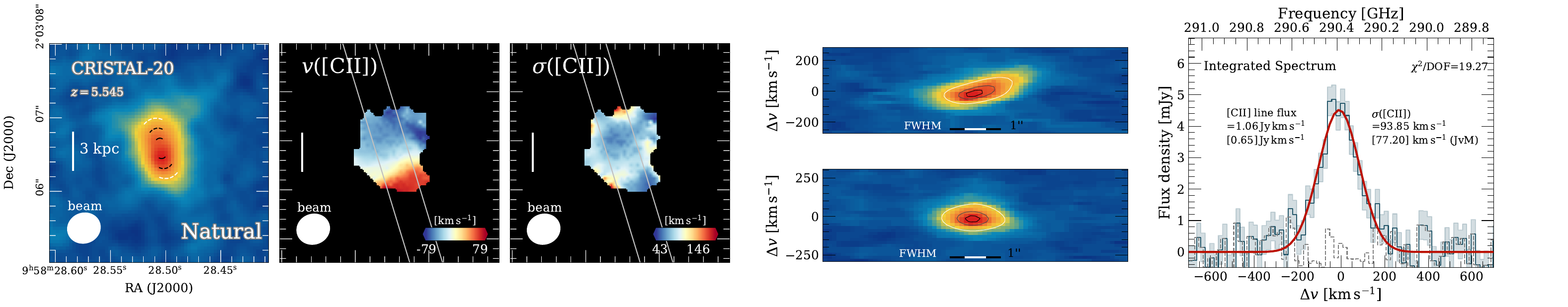}\\
    \includegraphics[width=1.05\textwidth]{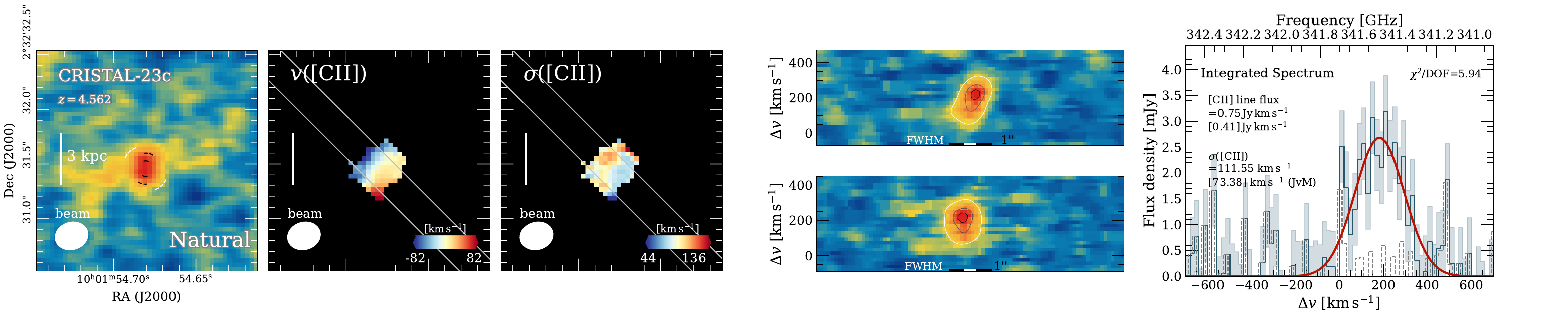}
      \caption{\textit{(Continued.)}
      }
      \label{fig:kins_map_bestdisk2}
\end{figure*}

\begin{figure*}{}
\centering
    \includegraphics[width=1.05\textwidth]{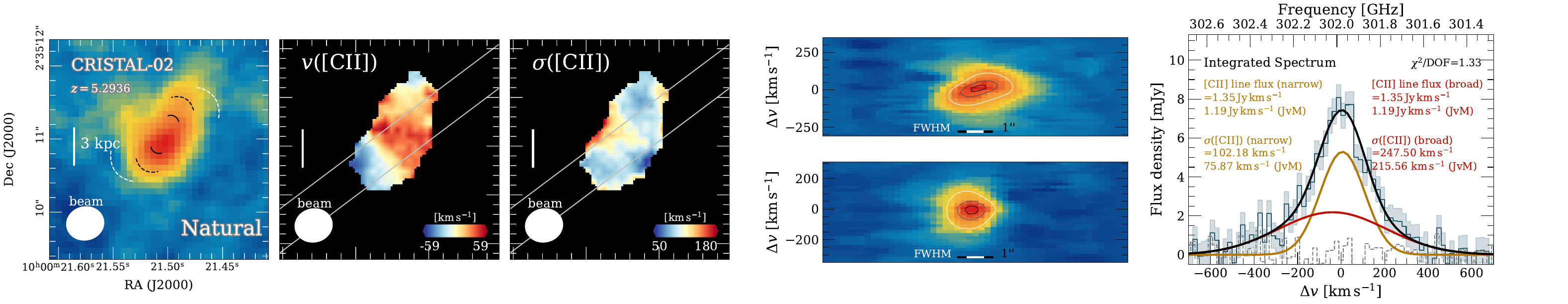}\\
    \includegraphics[width=1.05\textwidth]{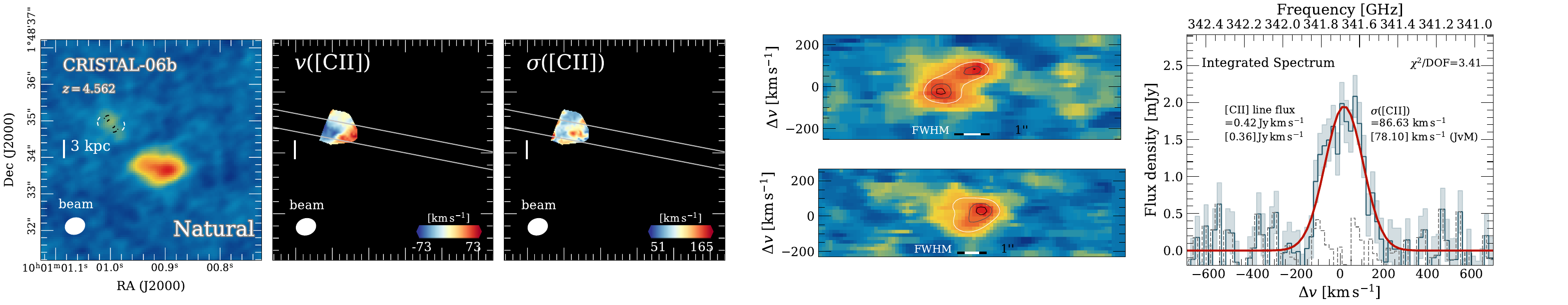}\\
    \includegraphics[width=1.05\textwidth]{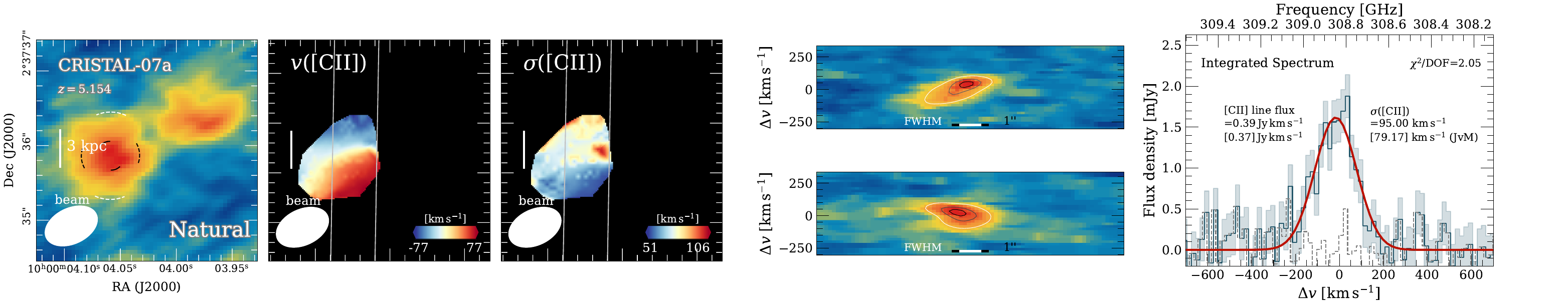}\\
    \includegraphics[width=1.05\textwidth]{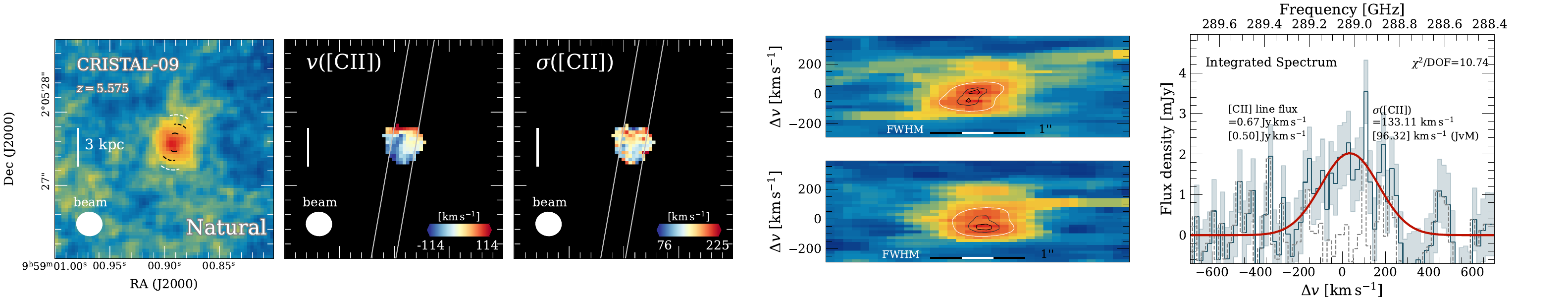}\\
    \includegraphics[width=1.05\textwidth]{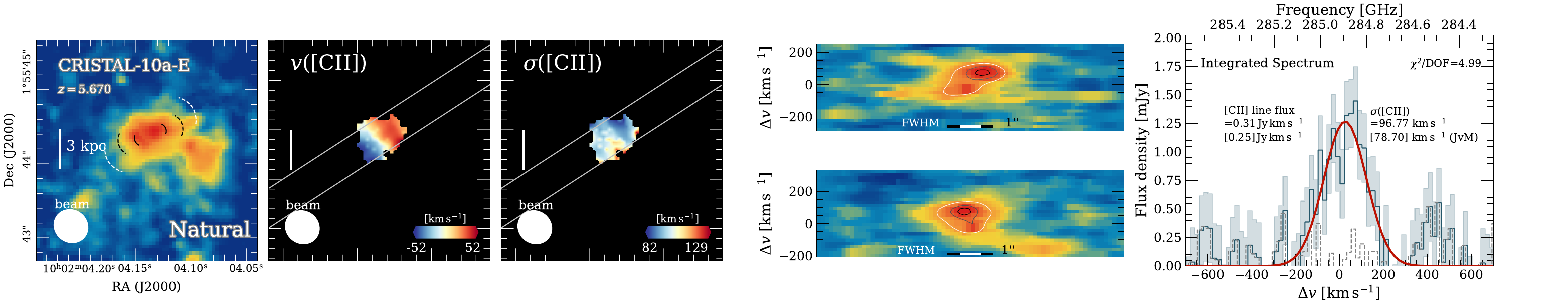}

      \caption{Same as Figure~\ref{fig:kins_map_bestdisk}, but for the Disk.
      }
      \label{fig:kins_map_disk}
\end{figure*}

\begin{figure*}[t]\ContinuedFloat
    \includegraphics[width=1.05\textwidth]{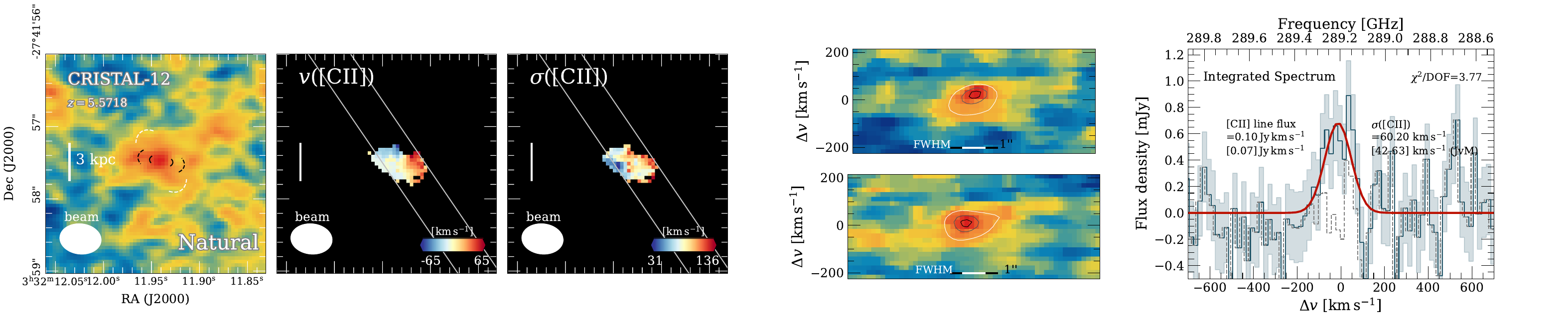}\\
    \includegraphics[width=1.05\textwidth]{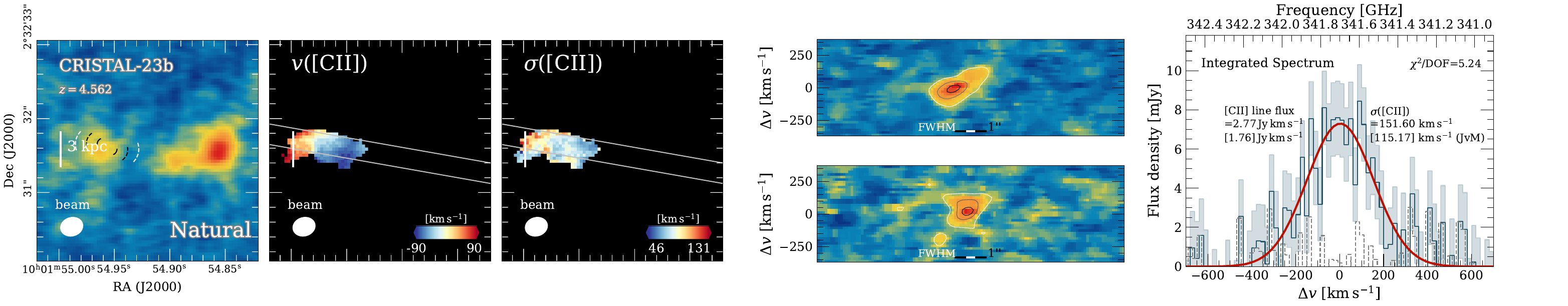}
      \caption{\textit{(Continued.)}
      }
      \label{fig:kins_map_disk2}
\end{figure*}

\begin{figure*}{}
\centering
    \includegraphics[width=1.05\textwidth]{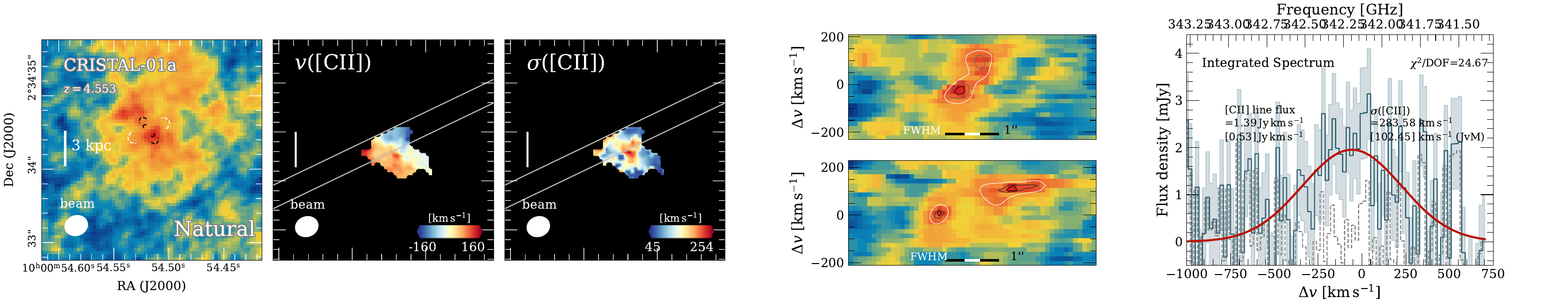}\\
    \includegraphics[width=1.05\textwidth]{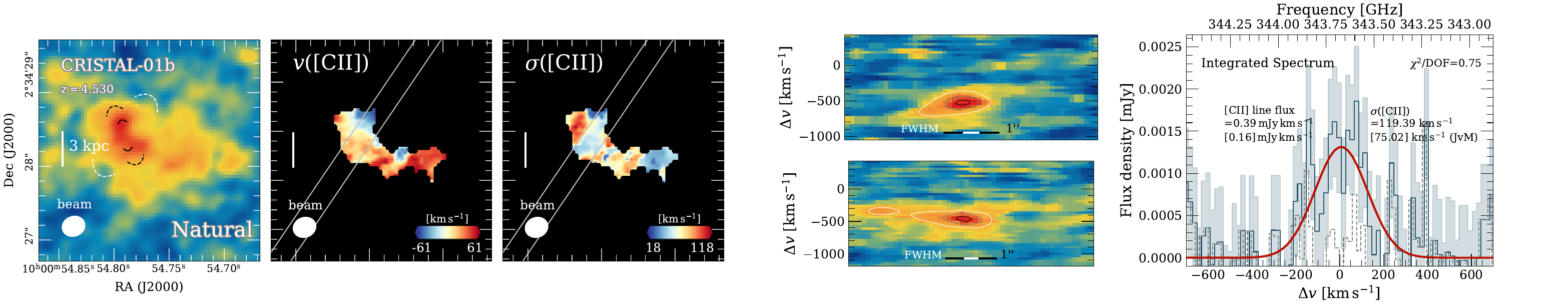}\\
    \includegraphics[width=1.05\textwidth]{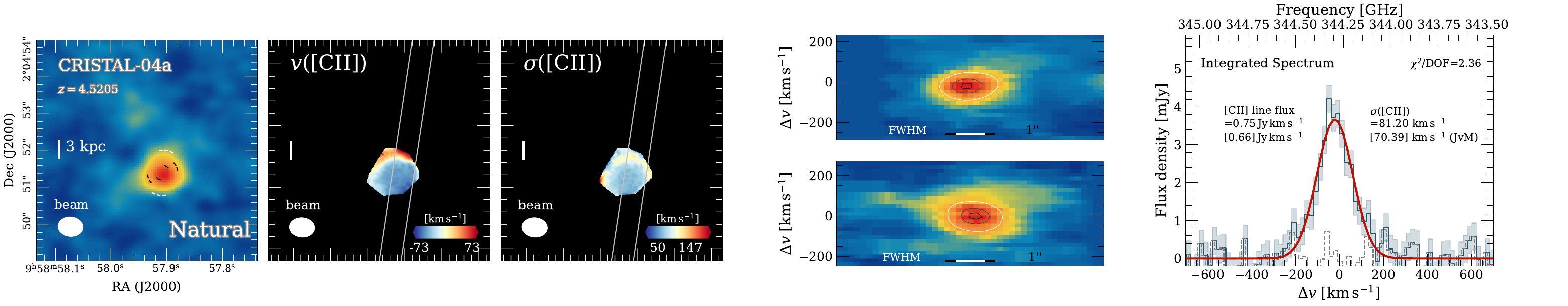}\\
    \includegraphics[width=1.05\textwidth]{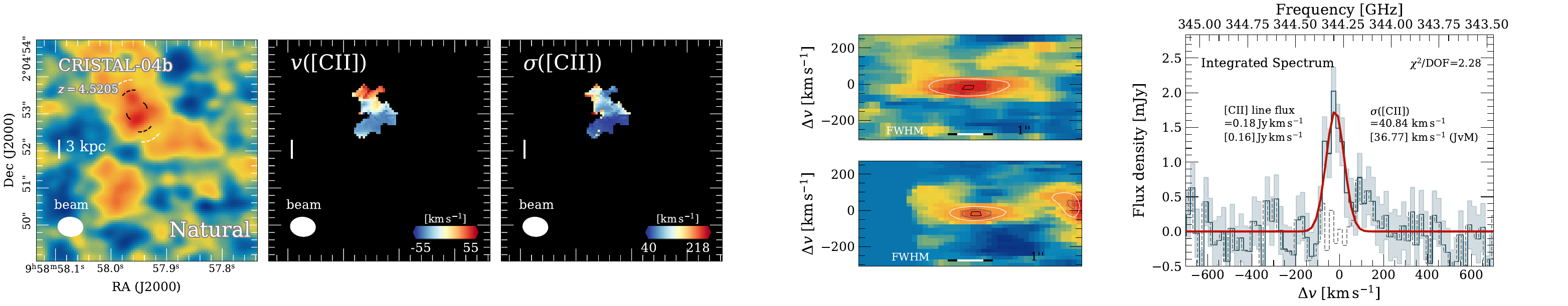}\\
    \includegraphics[width=1.05\textwidth]{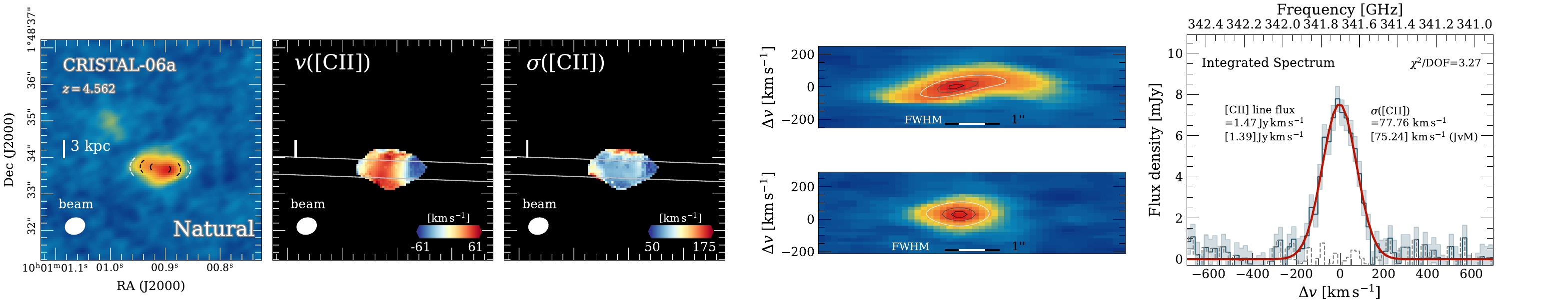}\\
    \includegraphics[width=1.05\textwidth]{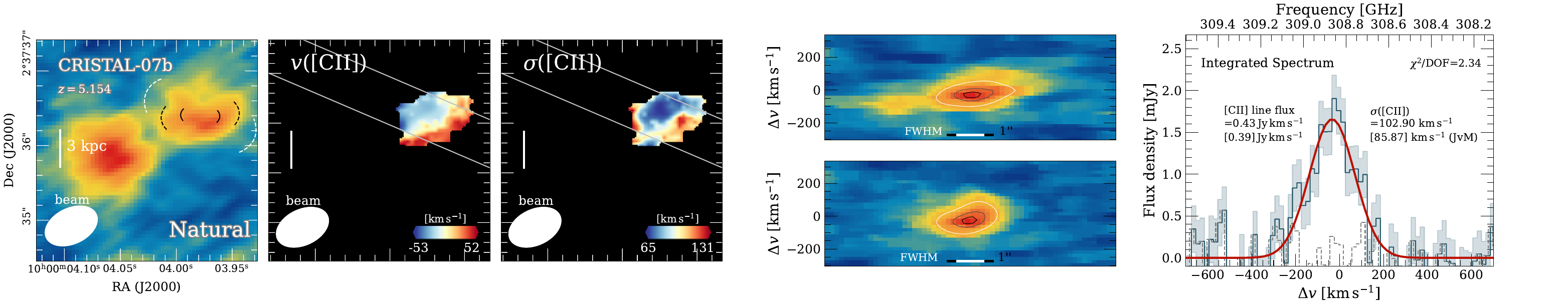}\\

      \caption{Same as Figure~\ref{fig:kins_map_bestdisk}, but for the Non-Disk. For CRISTAL-01a and CRISTAL-04b, the bright companions are subtracted from the line maps for better visual contrast.
      }
      \label{fig:kins_map_nondisk}
\end{figure*}

\begin{figure*}[t]\ContinuedFloat
\centering
    \includegraphics[width=1.05\textwidth]{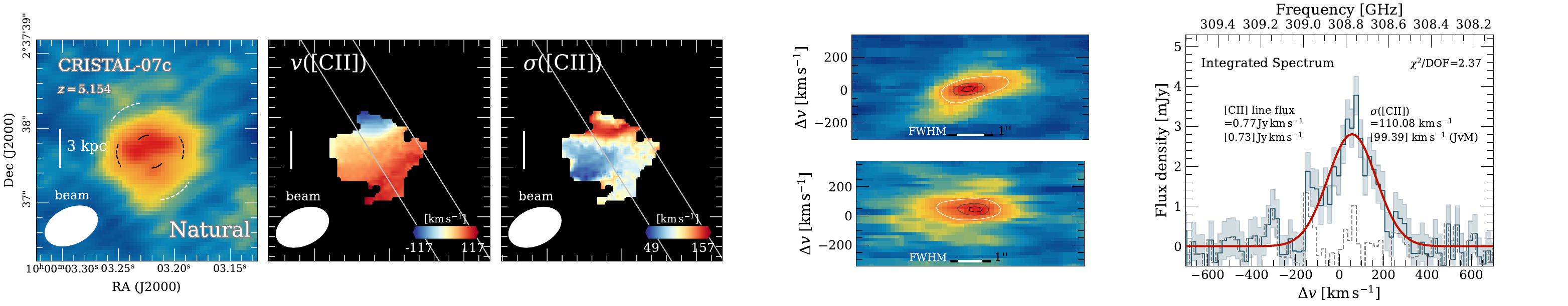}\\
    \includegraphics[width=1.05\textwidth]{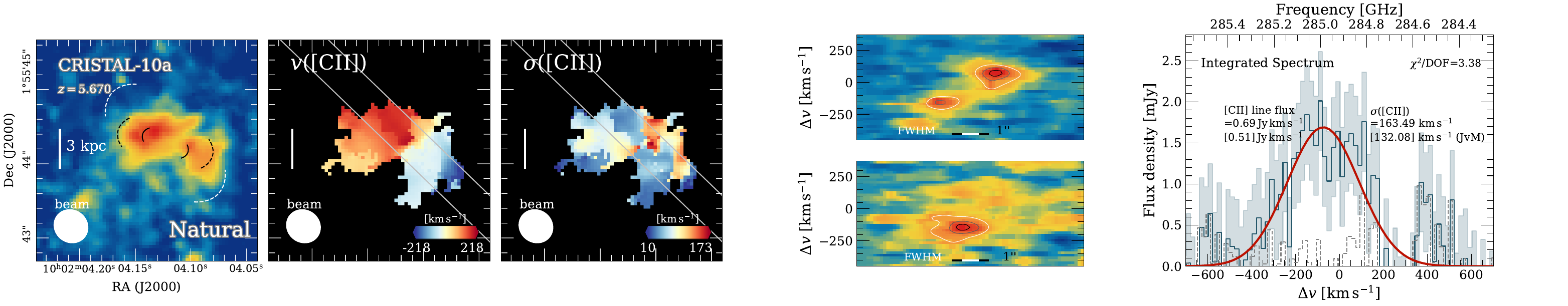}\\
    \includegraphics[width=1.05\textwidth]{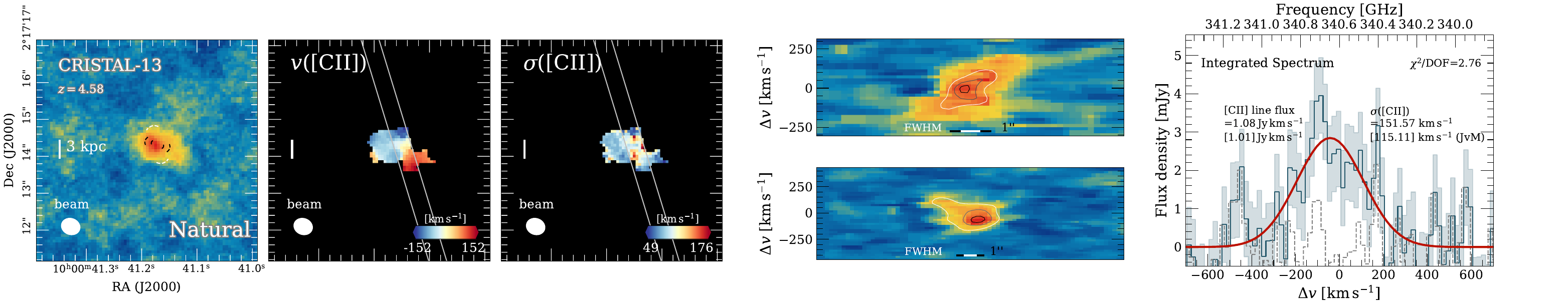}\\
    \includegraphics[width=1.05\textwidth]{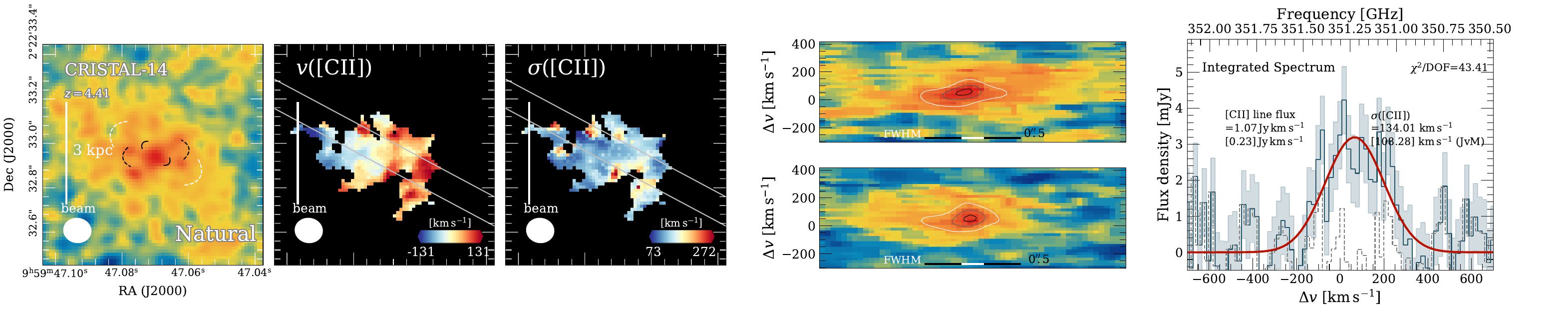}\\
    \includegraphics[width=1.05\textwidth]{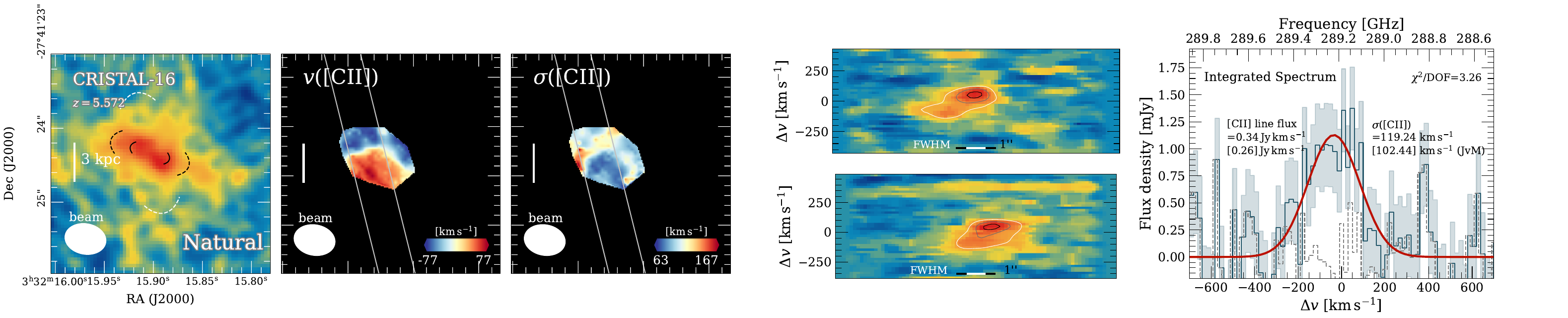}\\
    \includegraphics[width=1.05\textwidth]{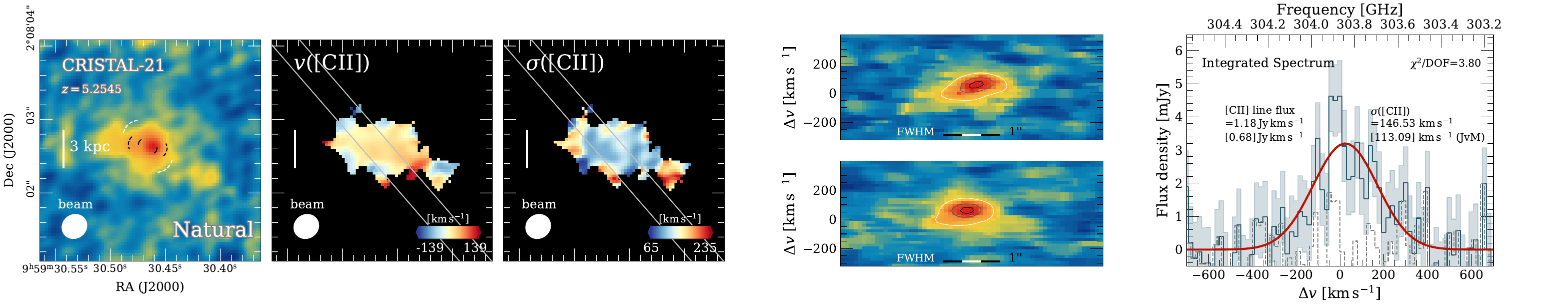}
      \caption{\textit{(Continued.)}
      }
      \label{fig:kins_map_nondisk2}
\end{figure*}

\begin{figure*}[t]\ContinuedFloat
\centering
    \includegraphics[width=1.05\textwidth]{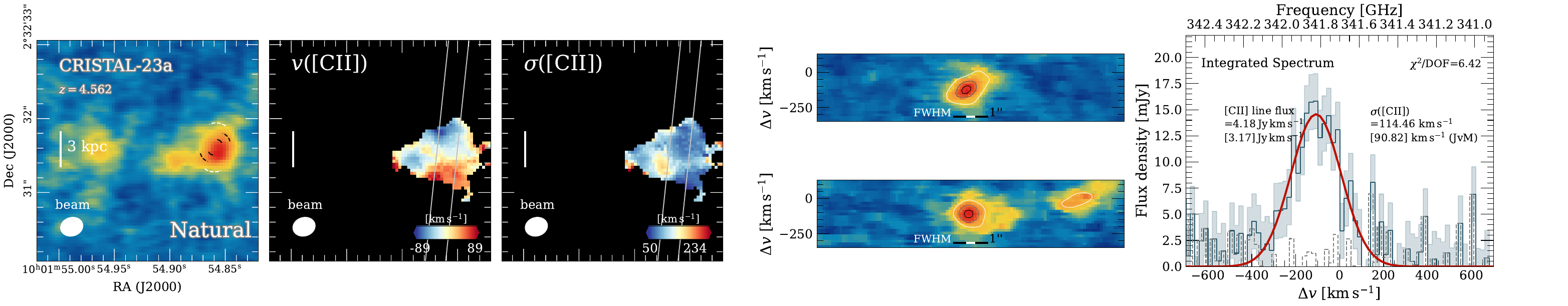}\\
    \includegraphics[width=1.05\textwidth]{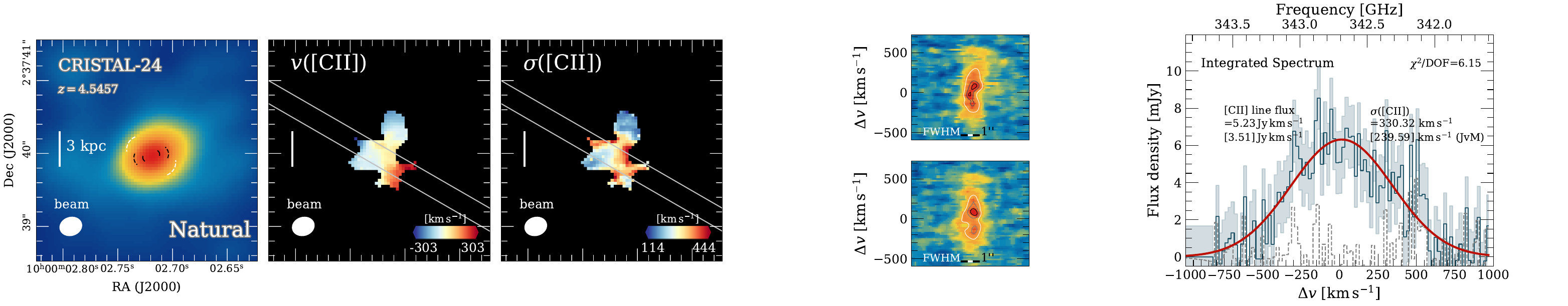}\\
    \includegraphics[width=1.05\textwidth]{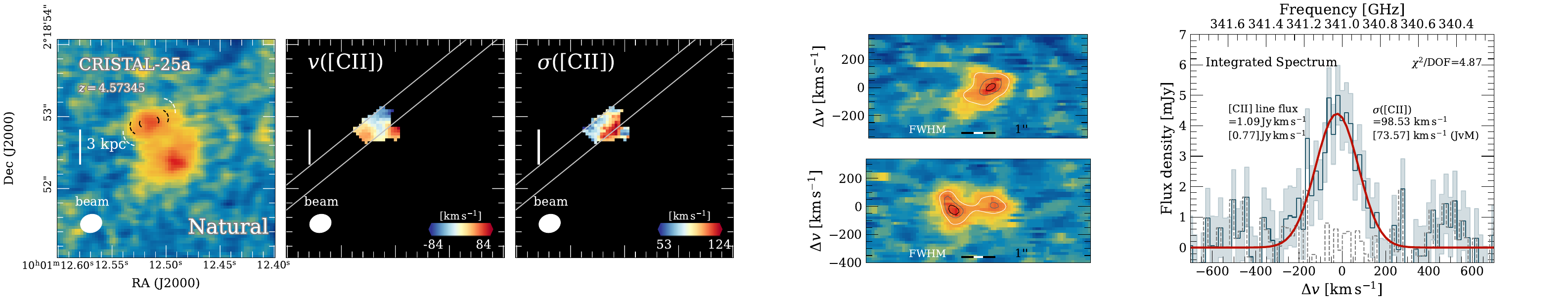}\\
    \includegraphics[width=1.05\textwidth]{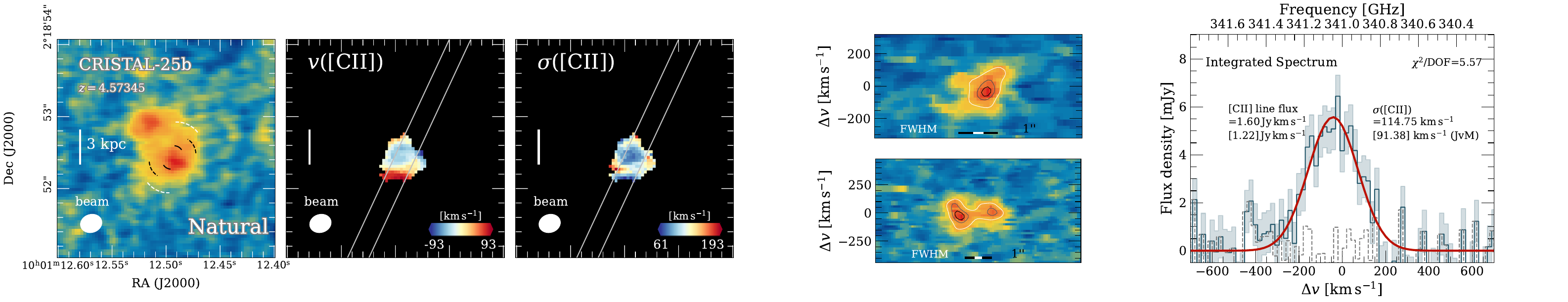}
      \caption{\textit{(Continued.)}
      }
      \label{fig:kins_map_nondisk3}
\end{figure*}

\clearpage
\section{Structural parameters fit of NIRCam/F444W images}\label{app:f444w_imfit}
We fit the 2D light distributions of the galaxies observed with NIRCam/\textit{JWST} F444W with \texttt{imfit} \citep{erwin2015}, accounting for the PSF.
The PSF model adopted is described in \citet{Li2024}.
We assume a S\'ersic profile with variable S\'ersic index $n$ and fixed $n=1$.
When let free, $n$ freely varying within $[0.5,6]$.
We also let the effective radius ($R_{\rm e}$), the ellipticity $e$\,$=$\,$1-b/a$, and the position angle (PA) of the major axis as free parameters.
Table~\ref{tab:imfit_table} lists the best-fit structural parameters of the galaxies for variable $n$.
Average S\'ersic index and effective radius are $\langle n \rangle$\,$=$\,$1.1\pm0.2$, 
$\langle R_{\rm e_{4.4\mu m}}\rangle$\,$=$\,$1.46\,\pm0.04$\,kpc, respectively.  
The effective radius is $\sim$\,$1.5\times$ that of the rest-frame UV emission 
$\langle R_{\rm e_{UV}} \rangle$\,$\approx$\,$0.98$\,kpc \citep[for the same sources considered,][]{Ikeda2025}. 
Our $\langle R_{\rm e_{4.4\mu m}}\rangle$ is comparable to that of the dust continuum emission $\langle R_{\rm e_{dust}} \rangle$\,$\approx$\,$1.78$\,kpc \citep{Mitsuhashi2024b}, 
but is smaller than the [\ion{C}{II}] sizes $\langle R_{\rm e_{[\ion{C}{II}]}}\rangle$\,$\approx$\,$1.9$\,kpc \citep{Ikeda2025}.
Fixing $n$\,$=$\,$1$ in our fit gives $\langle R_{\rm e_{4.4\mu m}}\rangle =1.77$\,kpc.
For consistency, we use this $n=1$ $R_{\rm e}$\ as the initial guess for the dynamical modelling.

In comparison, \citet{vanderWel2014}, which covers $0$\,$<$\,$z$\,$<$\,$3$ galaxies observed in the $H_{\rm 160}$ band, the expected value for SFGs extrapolated to $z$\,$=$\,$5$ and $\log{M_*/{\rm M_\odot}}=10$ is $2.4\,$kpc, our $R_{\rm e_{4.4\mu m}}$ is on average $40\%$ smaller, but is still within the $\sim0.2$\,dex scatter at $z\approx3$.

\begin{figure*}
    \centering
    \includegraphics[width=\textwidth]{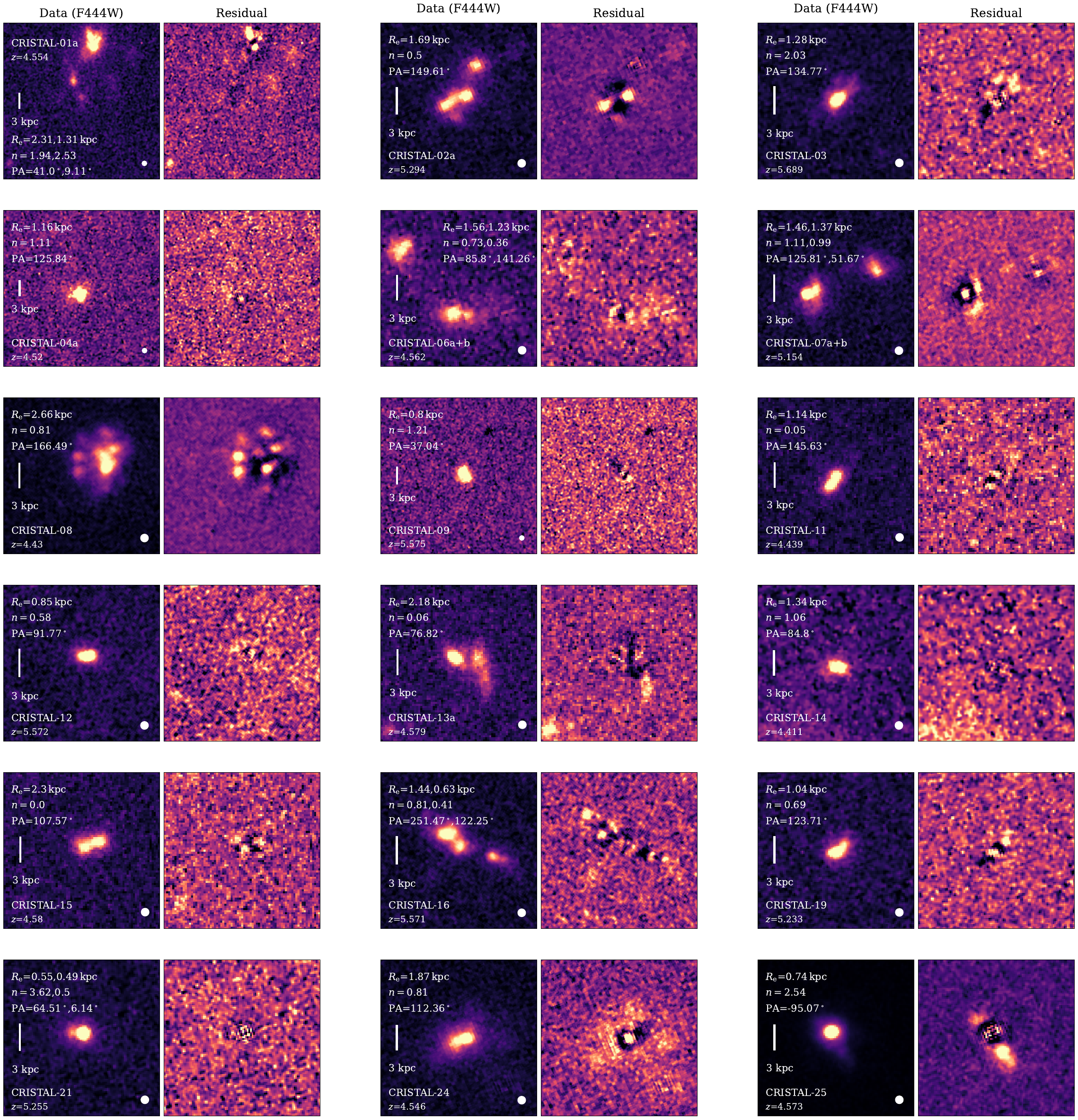}
    \caption{Examples of S\'ersic fits and residuals of CRISTAL galaxies observed in JWST/F444W (rest-frame $0.7$\,$\mu$m\ at $z\sim5$). North is up for all images. For galaxies with multiple components, we simultaneously fit all components, each characterised by a single S\'ersic profile. The resulting half-light radii ($R_{\rm e}$), S\'ersic indices ($n$), and position angles (PA, measured counter-clockwise from the north) are listed in Table~\ref{tab:imfit_table} and annotated in the figures in order of increasing $R_{\rm e}$. 
    Except in a few cases, such as CRISTAL-09, 12, and 14, the residuals exhibit clumpy structures and asymmetries in the light distributions due to multiple discrete clumps.
    }
    \label{fig:f444w_imfit}
\end{figure*}

\begin{table}
\caption{Rest-frame optical to near-IR (F444W) structural parameters fitted by \texttt{imfit}.}
\centering
\begin{tabular}{l c c c c c c c c }
\hline\hline  
  ID &
  $R_{\rm e}$~\tablefootmark{\scriptsize a}&
  $n$~\tablefootmark{\scriptsize b}&
  $e$~\tablefootmark{\scriptsize c}&
  PA~\tablefootmark{\scriptsize d}&
  $i$~\tablefootmark{\scriptsize e}\\
   &
  {\small (kpc)} &
   &
  &
  {\small ($^{\circ}$)} &
  {\small ($^{\circ}$)} \\ \hline
01a & 1.3 & 2.5 & 0.00 & \ldots\!\tablefootmark{$^\dagger$} & \ldots\!\tablefootmark{$^\dagger$} \\ 
    & 2.3& 1.9 & 0.51 & 9 & 64 \\ 
02 & 1.7 & 0.5 & 0.30 & 150  & 48 \\ 
03 & 1.3 & 2.0 & 0.57 & 135  & 69 \\ 
04a & 1.2 & 1.1 & 0.04 & \ldots\!\tablefootmark{$^\dagger$}  & \ldots\!\tablefootmark{$^\dagger$} \\ 
04b & 2.8 & 1.5 & 0.00 & \ldots\!\tablefootmark{$^\dagger$}  & \ldots\!\tablefootmark{$^\dagger$} \\ 
06a & 1.6 & 0.7 & 0.61 & 86  & 72 \\ 
06b & 1.2 & 0.4 & 0.25 & 141  & 43 \\ 
07a & 1.5 & 1.1 & 0.26 & 126  & 44 \\ 
07b & 1.4 & 1.0 & 0.27 & 52  & 45 \\ 
07c & 1.6 & 0.4 & 0.24 & 33  & 42 \\ 
08 & 2.7 & 0.8 & 0.29 & 166  & 46 \\ 
09 & 0.8 & 1.2 & 0.39 & 37  & 55 \\ 
11 & 1.1 & 0.1 & 0.71 & 146  & 82 \\ 
12 & 0.9 & 0.6 & 0.43 & 92  & 58 \\ 
13a & 2.2 & 0.1 & 0.64 & 77  & 75 \\ 
14 & 1.3 & 1.1 & 0.36 & 85  & 52 \\ 
15 & 2.3 & 0.0 & 0.62 & 108  & 73 \\ 
16a & 1.4 & 0.8 & 0.32 & 251 & 49 \\ 
   & 0.6& 0.4 & 0.22 & 122 & 40 \\ 
16b & 1.5 & 1.3 & 0.55 & 70 & 67 \\ 
19 & 1.0 & 0.7 & 0.54 & 124  & 66 \\ 
21 & 0.6 & 3.6 & 0.18 & 65 & 36 \\ 
   & 0.5& 0.5 & 0.10 & 6 & 26 \\ 
24 & 1.9 & 0.8 & 0.46 & 112  & 60 \\ 
25 & 0.7 & 2.5 & 0.20 & 85  & 39 \\ 
\hline
\end{tabular}
\tablefoot{
\tablefoottext{a}{Intrinsic half-light radius.}
\tablefoottext{b}{S\'ersic index.}
\tablefoottext{c}{Intrinsic ellipticity ($e = 1-b/a$).}
\tablefoottext{d}{Position angle (counter-clockwise from north).}
\tablefoottext{e}{Inclination inferred from Equation~\ref{eqn:i2e}.}
\tablefoottext{$^\dagger$}{The ellipticity is too low.}
Targets with two entries are fitted with a double-S\'ersic profile. CRISTAL-10, 20, 23 systems do not have NIRCam data available.
}
\label{tab:imfit_table}
\end{table}

\section{Molecular gas fractions $f_{\rm molgas}$}\label{app:dust2gas}

Previous studies of MS SFGs at $z$\,$\sim$\,$1$--$3$\ have shown that their gas kinematics are 
closely tied to their molecular gas content \citep[e.g.][]{Wisnioski2015,Uebler2019} inferred from scaling relations. 
We similarly explore the relationship between the kinematic properties of CRISTAL disks and their gas fractions 
in \S\S~\ref{sec:turbulence_z}--\ref{sec:trends}. 
As a first step, we need to estimate the gas content of our sample galaxies.

Measuring the gas content of CRISTAL galaxies using CO emission, which is a well-calibrated H$_2$ tracer at low-$z$, is challenging in our case due to the cosmic dimming effect and typically low metallicities of galaxies, exacerbated by the higher temperature of the cosmic microwave background for lower–$J$ transitions ($T_{\rm CMB}(z)$\,$=$\,$2.73(1 + z)\,$K). 
Furthermore, the lack of multi-band observations for all galaxies, 
which would be necessary to constrain the full far-infrared (FIR) SED shape \citep[e.g.][]{Bethermin2015,Kaasinen2019}, 
means that the only feasible estimator for most galaxies in CRISTALfootnote{
Data from Very Large Array (VLA) observations of CO(2-1) are available for CRISTAL-22 (HZ10) 
and CRISTAL-02 (LBG-1). 
} is to infer gas masses from the single-band dust continuum emission measured 
in the Rayleigh-Jeans (R-J) tail of the FIR SED following the method of \citet{Scoville2016,Scoville2017}.

With the rest frequencies probed by the Band-7 (343.5\,GHz) continuum data and if we adopt $\lambda_0$\,$=$\,$100$\,$\mu$m\ similar to \citet{Faisst2020b}, which is the wavelength where optical depth reaches unity, 
and single dust temperature of $T_{\rm dust}$\,$=$\,$50$\,K (see discussion below), then we can assume the R-J regime criterion is still satisfied and optically-thin approximation is valid. 
We measure the dust continuum fluxes on the image directly using the curve-of-growth method. They are in general good agreement with \citet{Mitsuhashi2024b}.

 We infer the gas mass $M_{\rm gas}$ from the observed flux density $S_{\nu_{\rm obs}}$ of the Band-7 dust continuum ($\lambda_{\rm obs}$\,$\approx$\,$850$--$1050$$\mu$m) 
 in the R-J regime following Eq.~(3) in \citet{Tacconi2020}:
\begin{equation}
 \begin{split}
     \Big(\frac{M_{\rm gas}}{1\times10^{10}M_{\rm \odot}}\Big) &= \Big(\frac{S_{\nu_{\rm obs}}D_{\rm L}^2}{{\rm mJy\, Gpc^2}}\Big)  \times (1+z)^{-(3+\beta)} \times\Big(\frac{\delta_{\rm gd}}{150}\Big) \\
                    & \times \Big(\frac{\nu_{\rm obs}}{352{\,\rm GHz}}\Big)^{-(2+\beta)}  
                    \times\Big(\frac{6.7\times 10^{19}}{\alpha_{\rm dust,0}}\Big) \\
                    &\times \frac{\Gamma_0}{\Gamma_{\rm R-J,\nu_{obs}}}
\label{eqn:dust2gas}
\end{split}
\end{equation}
We introduced the R-J departure coefficient $\Gamma_{\rm R-J}$ back to Eq.~(\ref{eqn:dust2gas}) to account for the deviation of dust temperature from $25\,$K, which was originally used in \citet{Tacconi2020}. $\Gamma_{\rm R-J}$ is defined as (Eq.~(6) in \citealt{Scoville2016})
\begin{equation}
\Gamma_{\rm R-J}(T_{\rm d},\nu_{\rm obs},z) = \frac{h\nu_{\rm obs}(1+z)/kT_{\rm d}}{e^{h\nu_{\rm obs}(1+z)/kT_{\rm d}}-1}.
\label{eqn:RJdeparture}
\end{equation}
The $\Gamma_{0}$ factor in Eq.~(\ref{eqn:dust2gas}) is taken at $z=0, T_{\rm d}=50{\,\rm K}$ (see below) and at the respective $\nu_{\rm obs}$ for different observations. 
On average, the coefficient $\langle\frac{\Gamma_0}{\Gamma_{\rm R-J,\nu_{obs}}}\rangle\sim2.4$.

We adopt a dust temperature\footnote{This temperature is luminosity-weighted, as the mass-weighted temperature is not available, see \citet{Scoville2016} and \citet{Scoville2017} for the discussion for the resulting difference on the dust mass between the two temperature.}  $T_{\rm d}$\,$=$\,$50 {\rm K}$, based on \citet{Villanueva2024} FIR SED modelling of CRISTAL-22, which benefits from the continuum data observed at multiple bands (but see also \citealt{Bethermin2020,Faisst2020b} and \citealt{Sommovigo2022}). 
We assume a dust emissivity index\footnote{$\beta$ is frequency-dependent but is not constrained by our data.} 
of $\beta$\,$=$\,$1.8$, the suggested value by \citet{Scoville2016} based on the findings of the \citet{Planck2011}. 
Such a choice is close to the value of $\beta$\,$=$\,$2$ measured by \citet{Villanueva2024}. 
We adopt the R-J luminosity-to-mass ratio
$\alpha_{\rm dust,0}=6.7\times10^{19}{\,\rm erg\,(s\,Hz\,{\rm M_{\odot}})^{-1}}$. 
Assuming $\alpha_{\rm dust,0}$\,$=$\,$8\times10^{19}{\,\rm erg\,(s\,Hz\,{\rm M_{\odot}})^{-1}}$ following \citet{Tacconi2020} 
\footnote{\citet{Tacconi2020} assumes lower $\alpha_{\rm CO}$ and $X_{\rm CO}$ than \citet{Scoville2016}'s.} 
would result in lower $M_{\rm gas}$\ by $\lesssim0.1$\,dex in median.

\begin{figure}
    \centering
    \includegraphics[width=0.5\textwidth]{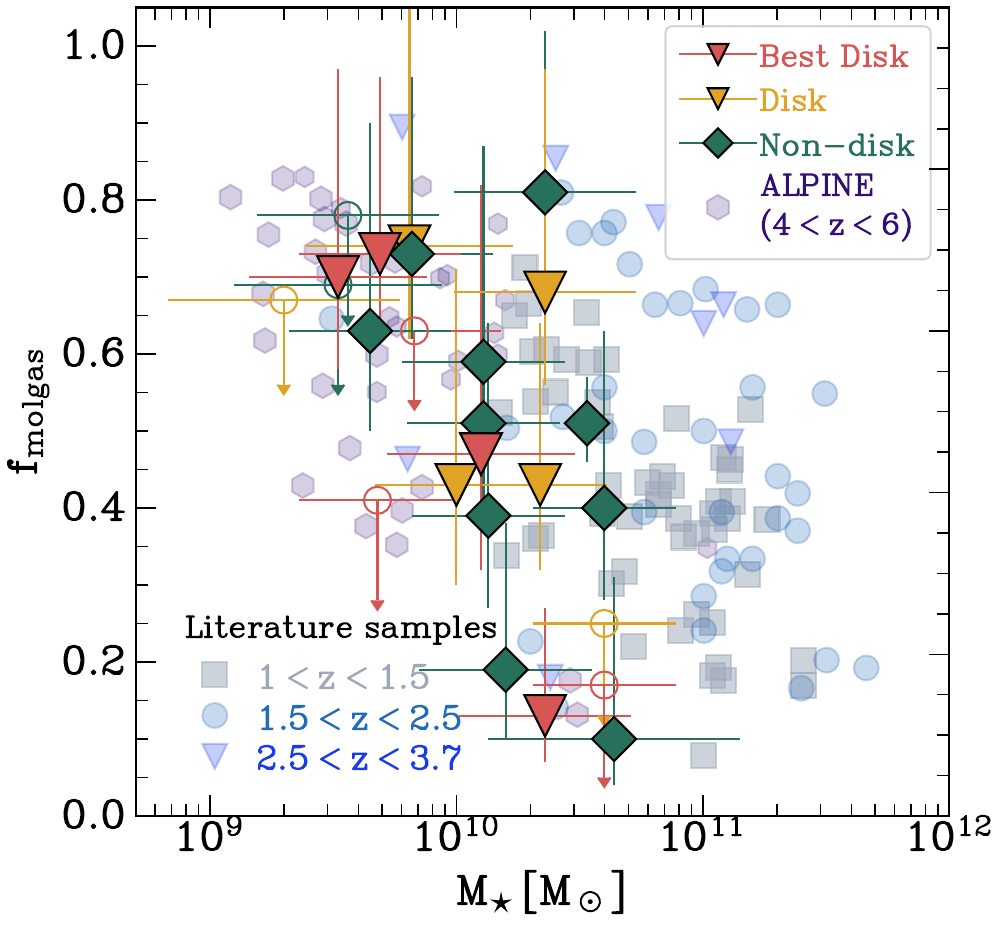}
    \caption{
    Molecular gas fraction $f_{\rm molgas}$\ (Equation~\ref{eqn:fgasdef}) as a function of stellar mass as inferred from the Band-7 dust continuum based on Equation~\ref{eqn:dust2gas}. The points are colour-coded by their classifications described in Section~\ref{sec:kins_class}.
    The empty circles represent upper limits. The median gas fraction of the CRISTAL sample is $\sim$\,$0.5$.
    Our measurements are in broad agreement with those reported by \citet{Dessauges-Zavadsky2020} for the ALPINE samples (purple hexagons), in which we both observe a similar slope of dependence of $f_{\rm molgas}$\ on $M_{\star}$, similar to that of the lower redshifts $1\lesssim$\,$z$\,$\lesssim6$ star-forming galaxies compiled by \citet{Dessauges-Zavadsky2020} and references therein (light grey to blue markers).
    }
    \label{fig:fgasdust_mstar}
\end{figure}
The gas-to-dust ratio, $\delta_{\rm gd}$\,$\coloneqq$\,$\frac{M_{\rm gas}}{M_{\rm dust}}$, anti-correlates with metallicity (Eq.~(10) in \citet{Genzel2015}, see also \citet{Leroy2011} and \citet{Tacconi2018}). 
We estimate metallicity using the 
mass-metallicity relation (MZR) 
inferred from NIRSpec/\textit{JWST} data of galaxies at $z$\,$=$\,$4$--$10$ by \citet{Nakajima2023},
which applies to galaxies with masses in the range $10^{7.5}$\,$<$\,$M_\star$\,$<$\,$10^{9.5}$. 
We extrapolate the MZR to cover the more massive galaxies in our sample.
Our sample's average metallicity 
$\langle12+\log({\rm O/H})\rangle$\,$=$\,$8.27\pm0.05$, 
and the corresponding gas-to-dust ratio coefficient is 
$\langle\delta_{\rm gd}/150\rangle = 2.20_{-0.24}^{+0.22}$. 
The uncertainties represent the scatter in the \citet{Nakajima2023}'s relation, which would translate to no more than $0.05$\,dex difference to the final $M_{\rm gas}$.

We note that individual gas-phase metallicities (except for CRISTAL-08, 21, and 23) can be determined from rest-frame optical strong line ratios using NIRSpec/\textit{JWST} data ($R$\,$=$\,$1000$), and the values are consistent with our adopted metallicities. 
A more detailed discussion of these results will be presented in A. Faisst (in prep.) and S. Fujimoto et al. (in prep.).
\citet{Parlanti2025} presented the metallicity measurement of CRISTAL-20 (HZ4) which is $12+\log({\rm O/H})$\,$\sim$\,$8.3$.

 Based on the $M_{\rm gas}$ derived above, Fig.~\ref{fig:fgasdust_mstar} and Table~\ref{tab:dysmalpy_results} show the gas fraction of the CRISTAL sample.
  % The data points are colour-coded by their classifications described in \S~\ref{sec:kins_class}.
 We define the gas fraction as
 \begin{equation}
 f_{\rm molgas} \coloneqq  \frac{M_{\rm gas}}{M_{\rm gas} + M_{\rm stars}}.
 \label{eqn:fgasdef}
 \end{equation}
 The median gas fraction $f_{\rm molgas}$\,$=$\,$0.51$. 
This is also consistent with 
the expected value from \citealt{Tacconi2020}'s relation at $z$\,$=$\,$5$ for $M_*$\,$=$\,$10^{10}M_{\rm \odot}$, which is $\sim$\,0.53.
 For CRISTAL-22, our measured $f_{\rm molgas}$\,$=0.71^{+0.28}_{-0.12}$ 
 is consistent with that inferred by \citet{Pavesi2019} from VLA CO(2-1) data.
For CRISTAL-01a, 08, 12, 14, 15, 16, 23b, 23c, the measured $f_{\rm molgas}$\ are upper-limits because the dust continuum fluxes are below the S/N threshold defined in \citet{Mitsuhashi2024b}.
There is insufficient co-spatial dust emission for CRISTAL-04b and 07b, so we have no $f_{\rm molgas}$\ measurement for those. The stellar mass $M_\star$ are based on SED fitting results presented in \citet{Li2024} for those sources with NIRCam/\textit{JWST} images; if not, then we adopt the values from \citet{Mitsuhashi2024b}.

Our $f_{\rm molgas}$\ measurements, regardless of their kinematics types, 
are in good agreement with \citet{Dessauges-Zavadsky2020} (purple hexagons in Fig.~\ref{fig:fgasdust_mstar}), which measured $f_{\rm molgas}$\ for the ALPINE sample based on 
(i) [\ion{C}{II}] emission following \citet{Zanella2018} 
(ii) rest-frame 850\,$\mu$m\ extrapolated from the FIR SED template of \citet{Bethermin2017} and 
(iii) dynamical mass assuming virialised spherical systems. 
We also recover a steep relation between $f_{\rm molgas}$\ and the stellar mass of the systems, similar to the lower-redshift studies following the same selection in \citet{Dessauges-Zavadsky2020} and references therein.

Inferring $M_{\rm gas}$ with the method described above is subject to several important caveats. 
Firstly, the assumption that the Band-7 continuum lies on the R-J tail is sensitive to the adopted single dust temperature, 
$T_{\rm d}$\,$=$\,$50\,$K. 
If $T_{\rm d}$ is lower, the R-J assumption breaks down, 
particularly for the highest-redshift sources in our sample. 
Using $T_{\rm d}$\,$=$\,$25\,$K would increase $M_{\rm gas}$ by $\sim$\,$0.5$\,dex.
Furthermore, we lack information on the mass-weighted $T_{\rm d}$.
The absence of longer wavelength data implies that we are neglecting a dominant colder component of the dust.
Additionally, 
the adopted values for $\alpha_{\rm dust,0}$ and $\delta_{\rm gd}$ 
are based on lower-redshift measurements and calibrations, 
although $M_{\rm gas}$ appears to be relatively insensitive to these values within the possible range.

An alternative method for estimating $M_{\rm gas}$ is by using the [\ion{C}{II}] luminosity.
If we apply the relation from \citet{Zanella2018} (their Eq.~(2)), 
we find that the resulting $M_{\rm gas}$ is systematically higher by $0.23$\,dex (in terms of the median), 
but still within the $0.3$\,dex scatter of the relation.
We also observe a significant correlation between the dust- and [\ion{C}{II}]-inferred $M_{\rm gas}$ values 
(with Kendall's $\tau=0.56^{+0.12}_{-0.14}$, $p\ll0.05$).

Overall, varying our assumptions and using different methods will very likely lead to higher $M_{\rm gas}$ and $f_{\rm molgas}$\ values, 
which, as a corollary, a lower Toomre $Q$ values, as inferred from Eq.~(\ref{eqn:toomreQ}) in \S~\ref{subsec:toomre}. 
This will not change the main findings that CRISTAL disks are gravitationally 
(un)stable and the correlations discussed in \S~\ref{sec:trends}. 
However, the statistical significance of the correlations would be weaker given the more uncertain $M_{\rm gas}$ and $f_{\rm molgas}$.

\section{Literature sample of $V_{\rm rot}$ and $\sigma$}\label{app:literature_refs}

Table~\ref{tab:literature_sample} lists the literature samples discussed in \S~\ref{sec:turbulence_z} and shown in Fig.~\ref{fig:s0z}.
\begin{table}
\small
\caption{Compilation of literature references for rotational velocity and velocity dispersion from
local to $z\lesssim8$ galaxies.}
\begin{tabular}{ll}
\hline \hline
Gas phase (Tracer(s)) & Reference  \\ \hline 
\multicolumn{2}{c}{$z\leqslant0.5$} \\ \hline
Cold/Atomic 
& \citealt{Mogotsi2016}  (THINGS) \tablefootmark{*}\\
(\ion{H}{I}, CO) & \citealt{Girard2021} (DYNAMO) \tablefootmark{**}\\
% \citealt{Leroy2009} (HERACLES)
% & \citealt{Bolatto2017} (EDGE-CALIFA) \\
% & \citealt{Levy2018} (EDGE-CALIFA)\\
% & \citealt{Girard2021} (DYNAMO) \tablefootmark{*} \\
Ionised
&\citealt{Epinat2008} (GHASP)\\
 (H${\alpha}$, [\ion{O}{II}]) & \citealt{Girard2021} (DYNAMO)\tablefootmark{**} \\
% & \citealt{Law2009} (OSIRIS)\\
% \citealt{Yang2008} (IMAGES) \\
% & \citealt{Mai2024} (MAGPI and KROSS)\\
\hline
\multicolumn{2}{c}{$0.5<z<4$} \\ 
\hline
Cold/Atomic & \citealt{Swinbank2011} \tablefootmark{$\dagger$} \\
 (CO, [\ion{C}{I}]) & \citealt{Lelli2018}\\
 &\citealt{Kaasinen2020} \\
 % &\citealt{Shao2022} \\
&\citealt{Huang2023} \\
&\citealt{Nestor2023} (RC100) \\
&\citealt{Rizzo2023} (ALPAKA) \\
% &\citealt{Liu2023} \tablefootmark{$\dagger$} \\
% &\citealt{Liu2024} \tablefootmark{$\dagger$} \\
% &\citealt{Mizener2024}\tablefootmark{$\dagger$}\\      
&\citealt{ZXLiu2024} \\
& \citealt{Liu2025} \\
Warm ([\ion{C}{II}])
&\citealt{Umehata2025} \\
Ionised 
&\citealt{Stark2008} \tablefootmark{$\dagger$} \\
(H${\alpha}$, [\ion{O}{III}], [\ion{O}{II}], \ion{C}{III}]) &\citealt{Epinat2009} (MASSIV)\\
&\citealt{Jones2010}\tablefootmark{$\dagger$} \\
&\citealt{Gnerucci2011} (AMAZE-LSD) \\
&\citealt{Livermore2015} \tablefootmark{$\dagger$} \\
&\citealt{DiTeodoro2016} \\
&\citealt{Leethochawalit2016} \tablefootmark{$\dagger$} \\
&\citealt{Mason2017} (KLASS) \tablefootmark{$\dagger$} \\
&\citealt{Turner2017} (KDS) \\
&\citealt{Girard2018} (KLENS) \tablefootmark{$\dagger$} \\
&\citealt{Patricio2018} \tablefootmark{$\dagger$}\\
&\citealt{Hirtenstein2019} \tablefootmark{$\dagger$} \\
&\citealt{Price2020} (MOSDEF) \\
&\citealt{Hogan2021} \\
% &\citealt{Liu2023} \tablefootmark{$\dagger$} \\
&\citealt{Nestor2023} (RC100) \\
&\citealt{Puglisi2023} (KURVS)  \\
&\citealt{Barisic2025} (MSA-3D)     \\
&\citealt{Birkin2024} (KAOSS) \\
% &\citealt{Liu2024} \tablefootmark{$\dagger$} \\
&\citealt{Rhoades2025} \tablefootmark{*,$\dagger$} \\
& \citealt{WWang2025}\\
\hline
\multicolumn{2}{c}{$z\geqslant4$} \\ \hline
Cold (CO)
&\citealt{Tadaki2018}  \\
 % & \\ %\citealt{Shao2022} 
Warm & \citealt{Rizzo2020} \tablefootmark{$\dagger$,$\dagger\dagger$} \\
([\ion{C}{II}]) & \citealt{Fujimoto2021} \tablefootmark{$\dagger$} \\
& \citealt{Jones2021}  \\
& \citealt{Neeleman2021}  \\
& \citealt{Rizzo2021}\tablefootmark{$\dagger$} \\
& \citealt{Tsukui2021} \\
% & \citealt{Shao2022}  \\
& \citealt{Neeleman2023}  \\
& \citealt{Parlanti2023}  \\
% & \citealt{Pope2023}\tablefootmark{$\dagger$}\\
& \citealt{Posses2023}  \\
& \citealt{RomanOliveira2023}  \\
& \citealt{Rowland2024}  \\
& \citealt{Venkateshwaran2024} \\
& \citealt{Amvrosiadis2025}\tablefootmark{$\dagger$} \\ 
& \citealt{Fei2025}\\
Ionised 
&\citealt{Parlanti2023}  \\
(H${\alpha}$, [\ion{O}{III}]) &\citealt{deGraaff2024} (JADES) \\
&\citealt{Danhaive2025} (`gold' only)\\
\hline
\end{tabular}
\tablefoot{
Sources listed in Table~\ref{tab:literature_sample_multiphase} 
are not duplicated here.
\tablefoottext{*}{Velocity dispersion only.} 
\tablefoottext{**}{Local analogues.} 
\tablefoottext{$\dagger$}{Lensed.} 
\tablefoottext{$\dagger\dagger$}{
The object is potentially 
a system undergoing a minor merger with mass ratios of $\sim$\,$1$\,$:$\,$6$ separated by $\sim$\,$4\,$kpc \citep{Cathey2024}. 
Still, the low dispersion of the system $\sim$\,$32\,$${\rm km\,s^{-1}}$\ suggests the interaction has not significantly perturbed the dynamics. 
With an updated stellar mass of $\log{(M_\star/{\rm M_\odot})}\approx10.39$ and a $\log[{\text{SFR}/({\rm M_\odot}\,{\rm yr}^{-1}})]\approx2.13$, 
the galaxy would be shifted closer to the main sequence, with a revised $\Delta\text{MS}=0.23$ at $z$\,$=$\,$4.2$.}
}
\label{tab:literature_sample}
\end{table}

\begin{table}
\small
\caption{Literature sample with multi-phase gas kinematics measurement.}
\begin{tabular}{llll}
\hline \hline
Object & $z$ &Tracer & Reference \\ 
\hline 
A521\tablefootmark{*,$\dagger$}  &1.0& CO(4-3) & \citealt{Girard2019}\\
     && [\ion{O}{II}] & \citealt{Girard2019}\\
ALESS073.1 & 4.8& [\ion{C}{II}]  & \citealt{Lelli2021}\\
             & &H${\alpha}$  & \citealt{Parlanti2024a}\\
BX610 & 2.2&CO(4-3) & \citealt{Genzel2023} \\ 
      & &H${\alpha}$ & \citealt{Nestor2023} \\ 
Cosmic Snake\tablefootmark{*,$\dagger$} &1.0&CO(4-3)& \citealt{Girard2019}\\
             &&[\ion{O}{II}]& \citealt{Girard2019}\\

EGS4-24985 &1.4 & CO(4-3) & \citealt{Uebler2018} \\ 
           && H${\alpha}$ & \citealt{Uebler2018} \\ 
GN20  & 4.1& CO(2-1)  & \citealt{Hodge2012}\\
       & &H${\alpha}$  & \citealt{Uebler2024b}\\  
J0235  & 6.1&[\ion{C}{II}] & \citealt{Parlanti2023}\\
        & &[\ion{O}{III}] & \citealt{Parlanti2023}\\
J0901\tablefootmark{$\dagger$} & 2.3&CO(4-3) & \citealt{Liu2023} \\ 
      & &H${\alpha}$ & \citealt{Liu2023} \\ 
J1211 & 6.0&[\ion{C}{II}] & \citealt{Parlanti2023}\\
      & &[\ion{O}{III}] & \citealt{Parlanti2023}\\
  J2310 & 6.0&CO(9-8) & \citealt{Shao2022}\\
  & &[\ion{C}{II}] & \citealt{Shao2022}\\
MACS0717\_Az9\tablefootmark{$\dagger$}  &4.3 & CO(4-3) & \citealt{Mizener2024}\\      
 & &[\ion{C}{II}] & \citealt{Pope2023}\\
PJ0116-24\tablefootmark{$\dagger$} & 1.4&CO(3-2) & \citealt{Liu2024} \\ 
          & &H${\alpha}$ & \citealt{Liu2024} \\
\hline
\end{tabular}
\tablefoot{
\tablefoottext{*}{Velocity dispersion only.} 
\tablefoottext{$\dagger$}{Lensed.} 
}
See also \citet{Genzel2013} and \citet{Fujimoto2024} for qualitative comparisons between kinematics in different gas phases.
\label{tab:literature_sample_multiphase}
\end{table}

\normalsize
\section{Model velocity profiles}\label{app:mdl_prof}
\begin{figure*}
    \centering
    \includegraphics[width=\textwidth]{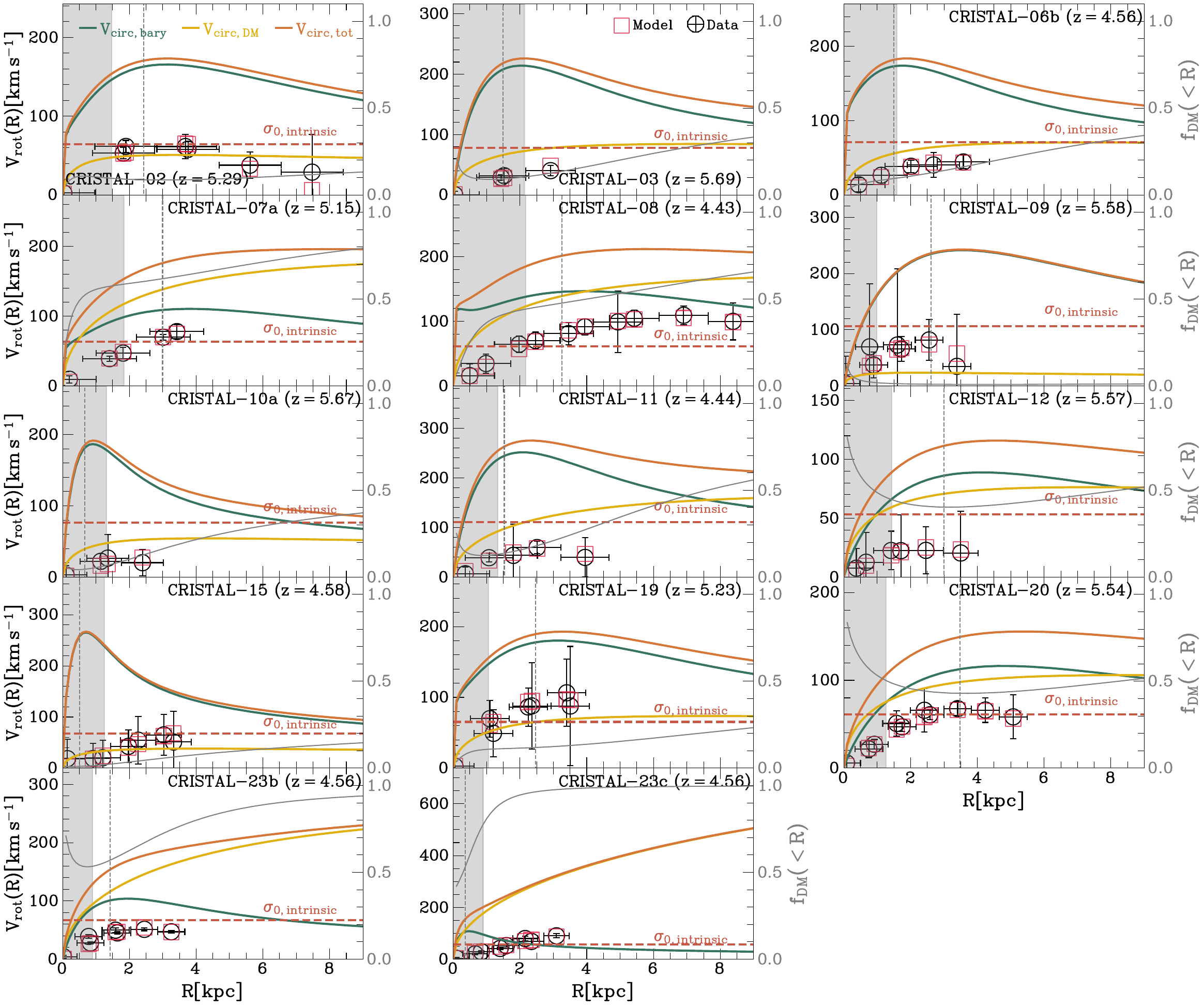}
    \caption{Intrinsic baryonic $V_{\rm circ,bary}$, dark matter $V_{\rm circ,DM}$, and total circular velocity $V_{\rm circ, tot}$ profiles of the CRISTAL disk samples with asymmetric drift correction.
    The grey circles mark the folded 1D observed velocity profiles, and the red squares show the observed profiles extracted from the \texttt{DysmalPy}-modelled cubes. The red dashed line indicates the intrinsic constant velocity dispersion $\sigma_{\rm 0,intrinsic}$.
    The dark matter fraction $f_{\rm DM}(<R)=V_\mathrm{circ,DM}^2(R)/V^2_\mathrm{circ,tot}(R)$ is shown in solid grey curves. The grey dashed vertical line marks the disk's effective radius. 
    }
    \label{fig:mass_profile}
\end{figure*}

Fig.~\ref{fig:mass_profile} shows the intrinsic $\sigma_0$, circular velocity profiles of the DM and the baryonic components. The DM fraction profile inferred from the circular velocities $f_{\rm DM}(<R)$\,$=$\,$V^2_{\rm circ, DM}(R)/V^2_{\rm circ,tot}(R)$ is shown as the secondary $y$-axis.

\section{Comparison to K18 model with total SFR}\label{app:krumholz18_SFR}
In {\S~\ref{subsec:KH18}}, we compare the $\sigma_0$--$\Sigma_{\rm SFR}$ with the \citet{Krumholz2018} analytic models. Here, we also compare the CRISTAL values with the models in $\sigma_0$--$\rm SFR$ space. Integrating Eq.~(\ref{eqn:sigma_g}) with radius, the total SFR is related to $\sigma_g$ by (Eq.~(60) in \citet{Krumholz2018}):
\begin{equation}
\begin{split}
\text{SFR} = &\sqrt{\frac{2}{1+\beta}}\frac{\phi_{a}f_{\rm SF}}{\pi GQ}f_{\rm g,Q}v_{\phi,{\rm out}}^2\sigma_g \\
&\cdot\text{max}\left[\sqrt{\frac{2(1+\beta)}{2f_{g,P}\phi_{\rm mp}}}\frac{8\epsilon_{\rm ff} f_{g,Q}}{Q},\frac{t_{\rm orb,out}}{t_{\rm sf, max}}\right],
\end{split}
\label{eqn:sigma_SFR}
\end{equation}
while for `feedback-only' (fixed $Q$) model (Eq.~(62) in \citealt{Krumholz2018}): 
\begin{equation}
\begin{split}
\text{SFR} = \frac{4\eta\sqrt{\phi_{\rm mp}\phi_{\rm nt}^3}\phi_Q\phi_a}{GQ^2\langle p_*/m* \rangle}\frac{f_{g,Q}^2}{f_{g,P}}v_{\phi,{\rm out}}^2\sigma_g^2.
\end{split}
\label{eqn:sigma_SFR_no_transport}
\end{equation}
The definitions of the symbols and values adopted are described in \S~\ref{sec:trends} and Table~\ref{tab:k18_fid_params}. Here the rotation velocity $v_\phi$ that is related to the radial gradient of the potential $\psi$, is set to be $v_\phi$\,$=$\,$\sqrt{r\frac{\partial \psi}{\partial r}}$\,$\in$\,$[100,350]\,$${\rm km\,s^{-1}}$, in increments of 50\,${\rm km\,s^{-1}}$.
 
\begin{table}
\caption{Fiducial values adopted for Equations~\ref{eqn:sigma_g}, \ref{eqn:sigma_sfr} and \ref{eqn:sigma_sfr_no_transport}.}
\centering
\begin{tabular}{ccc}
\hline \hline
Symbol & Value & Unit \\ \hline
  % colhead{Meaning} & 
$\eta$  &1.5 & \ldots \\%& Scaling factor for turbulent dissipation rate \\
$t_{\rm max}$  &2.0 & Gyr \\%& Maximum star formation time-scale \\
$Q_{\rm min}$   &1.0 & \ldots\\ %& Minimum possible disk stability parameter \\
$\epsilon_{\rm ff}$  &0.015 & \ldots \\%& Star formation efficiency per free-fall time \\
$\phi_{\rm mp}$  &1.4 & \ldots \\%& Ratio of total pressure to turbulent pressure at mid-plane \\
$\langle\frac{p_*}{m_*}\rangle$  &3000\tablefootmark{$\dagger$}  & ${\rm km\,s^{-1}}$\\%&  momentum per unit mass injected by feedback against the gas surface density
$\phi_Q$ & 2.0 & \ldots \\ \hline
\end{tabular}
\tablefoot{The definitions of each symbol can be found in Table~1 in \citet{Krumholz2018}.
\tablefoottext{$\dagger$}{{The momentum injection rate used is based on simulations \citep[e.g.][and references therein]{HaywardHopkins2017, Krumholz2018} and assumes that supernova remnants undergo an energy-conserving (Sedov-Taylor) phase \citep{Sedov1946,Sedov1959,Taylor1950}, where energy is converted into momentum. However, this value can vary by up to a factor of 4, which could sustain a higher velocity dispersion of the critical value $\sim$\,20\,${\rm km\,s^{-1}}$. To match the properties of CRISTAL disks using star-formation feedback alone, a factor of 10 larger than this adopted value would be required.
}}
} 
\label{tab:k18_fid_params}
\end{table}

\begin{figure}
    \centering
    \includegraphics[width=0.45\textwidth]{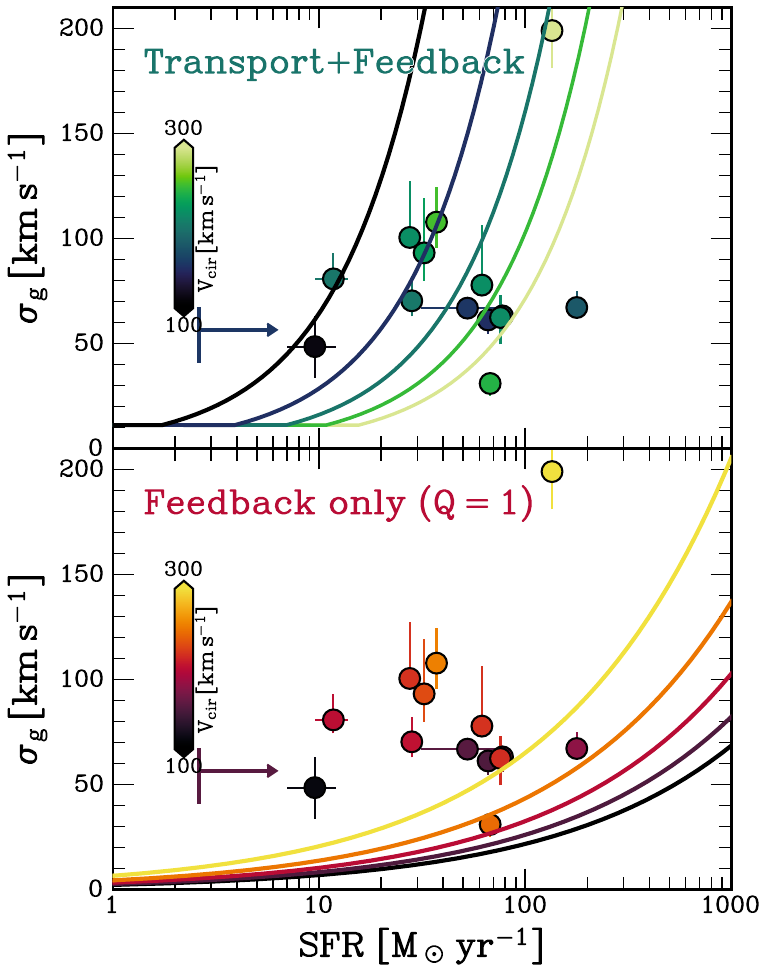}
    \caption{Same as Figure~\ref{fig:kh18_comp} but for $\sigma_0$ vs total star formation rate (SFR). The solid lines are analytical models from \citet{Krumholz2018} based on Equations~\ref{eqn:sigma_SFR} and \ref{eqn:sigma_SFR_no_transport}.
    }
    \label{fig:kh18_comp_SFR}
\end{figure}

\end{appendix}
\end{document}